\documentclass[aip,graphicx]{revtex4-1}
\usepackage{bm}
\usepackage{graphicx}
\usepackage{amsmath}
\usepackage{amsfonts}
\usepackage{color}
\usepackage{bigdelim,multirow}
\usepackage{subfigure}
\usepackage{appendix}
\usepackage{epstopdf}
\draft 

\begin{document}


\title{Dynamic mode decomposition using a Kalman filter for parameter estimation}


\author{Taku Nonomura}
\affiliation{Department of Aerospace Engineering, Graduate School of Engineering, Tohoku University, Sendai, Miyagi, 9808579, Japan}
\affiliation{Presto, JST, Sendai, Miyagi, 9808579, Japan}
\author{Hisaichi Shibata}
\affiliation{Institute of Space and Astronautical Science, Japan Aerospace Exploration Agency, Sagamihara, Kanagawa, 2525210, Japan}
\author{Ryoji Takaki}
\affiliation{Institute of Space and Astronautical Science, Japan Aerospace Exploration Agency, Sagamihara, Kanagawa, 2525210, Japan}

\date{\today}

\begin{abstract}
A novel dynamic mode decomposition (DMD) method based on a Kalman filter is proposed. This paper explains the fast algorithm of the proposed Kalman filter DMD (KFDMD) in combination with truncated proper orthogonal decomposition for many-degree-of-freedom problems. Numerical experiments reveal that KFDMD can estimate eigenmodes more precisely compared with standard DMD or total least-squares DMD (tlsDMD) methods for the severe noise condition if the nature of the observation noise is known, though tlsDMD works better than KFDMD in the low and medium noise level.  Moreover, KFDMD can track the eigenmodes precisely even when the system matrix varies with time similar to online DMD, and this extension is naturally conducted owing to the characteristics of the Kalman filter. In summary, the KFDMD is a promising tool with strong antinoise characteristics for analyzing sequential datasets.
\end{abstract}


\maketitle 

\section{Introduction}
Recently, fluid analysis has been conducted with high-resolution numerical simulations and experimental measurements. 
For such simulations and experiments provide large-scale data, it is necessary to understand and model essential phenomena from the data provided. Mode decomposition\cite{Taira2017} is a useful method for conducting such processes.

Proper orthogonal decomposition (POD) proposed by Lumrey has been applied to fluid analyses, especially turbulent analyses.\cite{Rowley2004,Berkooz1993}
Modes obtained by POD are known to be orthogonal to each other.
Furthermore, the original flow can be reconstructed with a limited number of modes.  
Proper orthogonal decomposition is equivalent to principal component analysis (PCA) and Karuhunen-Lo\'{e}ve expansion.
Note that fluid phenomena can be approximated and modeled by several methods using POD modes, e.g., the Galerkin projection method.

Proper orthogonal decomposition is optimum from the viewpoint of energy reconstruction with fewer modes, although the POD modes are not the solutions of the original fluid equations. Moreover, global linear stability analysis (GLSA)\cite{Theofilis2011, Shibata2015, Ohmichi2016} is a major method that can extract the eigenmodes of perturbations using governing equations (e.g., the Navier-Stokes equations) linearized around a nonlinear steady state. 
If GLSA is applied to the Navier-Stokes equations linearized on a steady state solution, the most unstable eigenmodes are extracted and are used to  judge whether the steady state solution is stable.
The modes obtained by GLSA satisfy the original linearized equation(s), although GLSA is more complicated than POD.
Note that these modes are generally not orthogonal.

Dynamic mode decomposition (DMD)\cite{Schmid2010} is developed as an intermediate method of POD and GLSA and is applied to numerous applications.\cite{Wang2017,Priebe2016,Ohmichi2017} In DMD, a sequential dataset of an unsteady flow solution, which is generally nonlinear, is given as input. 
This dataset is considered to be explained by a linear system ($\bm{x}_{k+1}=A\bm{x}_k$), and the eigenvalues and corresponding eigenvectors of $A$ are calculated for mode decomposition.
Although these modes are generally not orthogonal, they represent the single-frequency response with amplification or damping.
This means that physical phenomena of the DMD mode can be understood more simply than those of POD.
The original DMD\cite{Schmid2010} uses singular value decomposition (SVD) to compute a low-rank approximation of matrix $A$.
In exact DMD (EDMD), a Moore-Penrose pseudoinverse matrix is applied instead.\cite{Tu2014}
Various studies have recently been conducted in this field. With regard to DMD for noisy datasets, noise-corrected DMD (ncDMD),\cite{Dawson2016} forward/backward DMD (fbDMD),\cite{Dawson2016} and total least-squares DMD (tlsDMD),\cite{Hemati2016,Dawson2016,Hemati2017} have been proposed. A previous study showed that tlsDMD and fbDMD have the best performance for DMD mode estimation. 
The first method, ncDMD, estimates the eigenvalue and eigenmodes by including the noise covariance matrix in the computation. This method improves accuracy by compensating the expected error. 
The second method, fbDMD,\cite{Hemati2016} calculates the eigenvalue and eigenmodes using forward (or old-to-new) time-series data and inverse (backward or new-to-old) time-series data and blends them to compensate for the error. 

The third method, tlsDMD,\cite{Hemati2017} improves the estimation accuracy of the eigenvalues and eigenmodes using truncated SVD (POD) for the combined data of successive two snapshots that filter out the critical noise for estimation of dynamics. 
With regard to reconstruction of data using DMD modes, sparsity promoting DMD (spDMD)\cite{Jovanovic2014} chooses modes with which the original flows can be effectively reconstructed in the framework of DMD by introducing sparse modeling and compressing sensing ideas. 
Recently, online variants of DMD have been proposed: streaming DMD,\cite{Hemati2014} preconditioned DMD with online POD,\cite{Ohmichi2017a} on-the-fly DMD,\cite{Matsumoto2017} and on-line DMD.\cite{Zhang2017}  Among them, the online DMD method can estimate the time-varying systems with adopting forgetting factor.\cite{Zhang2017} As mentioned above, DMD is a more promising method for extracting the modes that can directly describe the system dynamics, as compared to POD, and further development is expected.

Following previous studies, DMD is reconsidered for parameter estimation of the matrix $A$ or system identification in the present study. In other words, if the matrix $A$ of DMD is considered to be a kind of filter, then the DMD problem is regarded as coefficient identification of the filter. Conventional approaches to solve this kind of problem are a recursive least-squares (RLS) method and a Kalman filter method.\cite{Kalman1960}
In the present study, we propose a novel method by which to use a Kalman filter to identify the matrix $A$. It should be noted that several studies have been conducted using a Kalman filter with the Koopman operator, or DMD. For instance, a Koopman Kalman filter has been applied as an observer for a nonlinear system.\cite{Surana2016,Surana2017} However, that study adopted the Kalman filter for state estimation whereas the present approach adopts the Kalman filter for only the parameter estimation of the $A$ matrix. 

The following advantages are expected when adopting a Kalman filter for the estimation of DMD modes as in the present study. 
\begin{itemize}
    \item More arbitrary treatment for denoising when the noise characteristics are known, and
    \item System identification of the transient system.
\end{itemize}
The two of the above advantages are demonstrated using the Kalman filter in the present paper. 
With regard to the first advantage, the system is considered to be estimated more precisely by the Kalman filter than by standard DMD methods if the observation noise covariance is known in advance. Data for space science, astronomy, and meteorology are contaminated by severe time-dependent noise, the characteristics of which are known, and system identification based on such observations appears to be useful. This type of problem, in which the time-varying noise level is known in advance, was effectively solved by the group of astrophysics using the information of noise.\cite{Bailey2012} Moreover, the proposed method will help in conducting DMD for extremely severe measurements at low-signal-noise ratios, such as the measurement of compressible turbulence. In addition, with regard to the second advantage, the Kalman filter can be adopted inherently for a time-variant system. Matrix $A$ is expected to be naturally identified, even if the matrix is time dependent, as shown in a previous study.\cite{Zhang2017}  

In Section II of the present paper, we introduce the Kalman filter for DMD and the fast algorithm in combination with POD to improve the poor computational efficiency of the straightforward implementation. In Sections III and IV, test problems are solved and two of the above-described advantages are demonstrated, comparing existing algorithms. Finally, Section V concludes this paper. 


%

\section{Kalman Filter}
\label{sec:KF}
\subsection{Proposed Algorithm}
\label{sec:PA}
In the present study, a discretized system in the temporal direction is considered, as is usual in standard DMD methods.
The subscript $k$ represents a $k$th quantity in discretized time, $k\Delta t$, where $\Delta t$ is the time interval of snapshots. 
A linear temporal evolution system is considered: 
\begin{eqnarray}
\bm{x}_{k+1} = A \bm{x}_k,
\end{eqnarray}
in a vector form or
\begin{eqnarray}
x_{i,k+1} = a_{ij} {x}_{j,k},
\end{eqnarray}
in a tensor form with the Einstein summation convention. 
Here, $A=(a_{ij}) \in \mathbb{R}^{n\times n}$
is a system matrix, $\bm{x}={x_i}$ are fluid variables, and $n$ is the dimension of the fluid variables. 
Only the snapshots of the system, i.e., a dataset assumed to be generated by $A$, can be observed. 
The snapshot data $\bm{x}$ can also be expressed as follows:
\begin{eqnarray}
	Z_{1:m} &=& \left[ \bm{x}_{1} \quad \bm{x}_{2} \quad  \cdots \quad  \bm{x}_{m-1} \quad  \bm{x}_{m}\right] \nonumber\\ 
	        &=& \left[ \bm{x}_{1} \quad A^{1}\bm{x}_{1} \quad  \cdots \quad  A^{m-2} \bm{x}_{1} \quad  A^{m-1} \bm{x}_{1}\right],\nonumber\\
	        X&=&Z_{1:m-1},\nonumber\\
	        Y&=&Z_{2:m},\label{eq:tsdata}
\end{eqnarray}
where the elements of $X$ and $Y$ are expressed as $\bm{x}_k$ and $\bm{y}_k$ ($k=1, \cdots , m-1$). Here subscript $\{1:m\}$ represents the-first-row-to-the-$m$th-row component of the matrix.

Or, our system can be used for the snapshot pairs of 
\begin{eqnarray}
\bm{y}_{k} = A \bm{x}_k. 
\end{eqnarray}
in a vector form and 
\begin{eqnarray}
y_{i,k} = a_{ij} {x}_{j,k},
\end{eqnarray}
in a tensor form, whereas $k$ represents $k$th pair. In this case, data matrices are defined as follows:
\begin{eqnarray}
	X &=& \left[ \bm{x}_{1} \quad  \bm{x}_{2} \quad  \cdots \quad  \bm{x}_{m-1} \quad  \bm{x}_{m}\right] \nonumber\\ 
	Y &=& \left[ \bm{y}_{1} \quad  \bm{y}_{2} \quad  \cdots \quad  \bm{y}_{m-1} \quad  \bm{y}_{m}\right] \nonumber\\ 
    Z &=& [X\quad Y]\label{eq:spdata}
\end{eqnarray}
A proposed algorithm can be used for both expression of data in Eqs. \ref{eq:tsdata} and \ref{eq:spdata}.

Then, we consider a system identification problem. Each element of matrix $A$ is estimated in the present study, and the parameter vector $\bm{\theta}$ is introduced as follows:
\begin{eqnarray}
\bm{\theta} = \mathrm{vec}\left(A^\text{T}\right)=\left.\left(
\begin{array}{*{20}{c}}
{{a_{11}}}\\{{a_{12}}}\\ \vdots \\{{a_{1n}}}\\{{a_{21}}}\\{{a_{22}}}\\ \vdots \\{{a_{nn}}}
\end{array} \right) \right\}\text{$n^2$ dimensions},
\end{eqnarray}
where $\bm{\theta}$ is considered to be a constant or slowly and randomly varying parameter vector according to the system noise;  these assumptions are similar to those for standard DMD or online DMD, respectively. Here, superscript $T$ represents a transverse matrix. The time evolution of $\bm{\theta}$ can be given as follows:
\begin{eqnarray}
	\bm{\theta}_{k+1} &=& F \bm{\theta}_{k} + \bm{v}_k\\
	                  &=&\bm{\theta}_{k} + \bm{v}_k, 
	\label{eqn:theta_time_domain}
\end{eqnarray}
where $F=I$ is an identity  matrix for a constant or slowly and randomly varying parameter vector, and $\bm{v}$ is a system noise.

An observation equation for the next input $\bm{y}_k=\bm{x}_{k+1}=A\bm{x}_{k}$ is given as follows:
\begin{eqnarray}
	\bm{y}_k = H_k \left( \bm{x}_{k} \right) \bm{\theta}_k + \bm{w}_k,
\end{eqnarray}
whereas
\begin{eqnarray}
H_k &=& \overset{\text{$n^2$ dimensions}}{\overbrace{\left.\left( 　
\begin{array}{ccccc}
\bm{x}_{k}^{\text{T}} & \bm{0} & \cdots & \cdots & \bm{0} \\
\bm{0} & \bm{x}_{k}^{\text{T}} & \bm{0} & \cdots & \bm{0} \\
\bm{0} & \bm{0} & \ddots & \bm{0} & \bm{0} \\
\bm{0} & \cdots & \bm{0} &\bm{x}_{k}^{\text{T}} & \bm{0}  \\		
\bm{0} & \cdots & \bm{0} & \bm{0} & \bm{x}_{k}^{\text{T}}  \\			
\end{array}
\right)   \right\}}} \text{\small{$n$ dimensions}}\\
&=&
\left(
\begin{array}{cccccccccccccccccr}
x_1& x_2& \dots& x_n& 0  &    &      &    &       &    &    &      &    &    &    & \dots&   0 \\
0  &    & \dots&   0& x_1& x_2& \dots& x_n&      0&    &    &      &    &    &    & \dots&   0 \\
0  &    &      &    &    &    &      &    & \ddots&    &    &      &    &    &    &      &   0 \\
0  &    & \dots&    &    &    &      &    &      0& x_1& x_2& \dots& x_n&   0&    & \dots&   0 \\
0  &    & \dots&    &    &    &      &    &       &    &    &      &   0& x_1& x_2& \dots& x_n  \\
\end{array}
\right),
\end{eqnarray}
and $\bm{w}_k$ is an observation noise. 
Since $H_k$ varies with the time step, the system is a linear time-variant system, and the resulting algorithm of the Kalman filter is standard for a linear time-variant system and not a special implementation. 
Based on these equations, a standard linear Kalman filter can be used with $\bm{\theta}$.

Following the theory of a Kalman filter, a covariance matrix regarding a priori estimation $P_{k|k-1}$ can be obtained using the covariance matrix of one step earlier, i.e., $P_{k-1|k-1}$,
\begin{eqnarray}
	P_{k|k-1} = F_k P_{k-1|k-1} F_k^\text{T} + Q_{k},
\end{eqnarray}
where $Q$ is a covariance matrix regarding system noise, and a system matrix $F_k$ becomes an identity matrix from Eq.~(\ref{eqn:theta_time_domain}).
In a priori estimation, $\bm{\theta}$ does not change because of the relationship:
\begin{eqnarray}
\bm{\theta}_{k|k-1}=F_k\bm{\theta}_{k|k}=I\bm{\theta}_{k|k}=\bm{\theta}_{k|k}.\label{eq:predicttheta} 
\end{eqnarray}
The state variables are updated by the Kalman gain when observation takes place.
A noise covariance matrix after observation, $S_k$, is given by,
\begin{eqnarray}
	S_k = R_k + {H}_k P_{k|k-1} {H}^\text{T}_k
\end{eqnarray}
where $R_k$ is a covariance matrix of observation noise and is generally time dependent.
The Kalman gain is then directly computed by
\begin{eqnarray}
K_k = P_{k|k-1} {H}^\text{T}_k S_k^{-1}.
\end{eqnarray}
The amount of modification of state variables $\bm{\theta}$ can be computed as follows:
\begin{eqnarray}
	\label{eqn:update_theta}
	\delta \bm{\theta}_{k|k} &=& K_k \left( \bm{y}_k - H_k \bm{\theta}_k \right)\\
	&=& K_k \left( \bm{y}_k - A \bm{x}_k \right), \label{eq:deltatheta}
\end{eqnarray}
where $A$ is generated from $\bm{\theta}_k$. Then, the state variable is updated as follows:
\begin{eqnarray}
	\bm{\theta}_{k|k} = \bm{\theta}_{k|k-1} + \delta \bm{\theta}_{k|k}. \label{eq:updatetheta}
\end{eqnarray}
Note that during this process, snapshot $\bm{y}_k$ is newly observed.
The quantity $\bm{y}_k$ observed here is used as $\bm{x}_k$ in the next time step to construct the observation matrix in the case of time-series data.

The covariance matrix after the observation is also updated by,
\begin{eqnarray}
	\label{eqn:cov_innovation}
	P_{k|k} = (I - K_k {H}_k) P_{k|k-1}.
\end{eqnarray}

The disadvantage of this formulation is that inversion of the large matrix $S_k$ with a dimension of $n^2$ is required.
In the next section, a novel algorithm with extremely low computational cost is introduced.
\subsection{Fast Algorithm}
\label{sec:fast_algo}
The periodicity and sparsity of the matrices appearing in the previous algorithm are used, and the problem is further simplified. 

The following assumptions are introduced for simplicity:
\begin{enumerate}
	\item{The initial covariance matrix $P$ is assumed to be a block diagonal matrix, and all of the diagonal matrices are identical.}
	\item{The covariance matrices of observation and system noises, $R$ and $Q$, are assumed to be block diagonal matrices and all of the diagonal matrices are identical (where $R_k=r_k I$ in this case)．}
\end{enumerate}
The above conditions are expressed as follows:
\begin{eqnarray}
P&=&
\left(
\begin{array}{ccccc}
P_{(1,1)} & \bm{0} & \cdots & \cdots  & \bm{0}  \\
\bm{0}  & P_{(2,2)}& \bm{0} & \cdots  & \bm{0}  \\
\bm{0}  & \bm{0} & \ddots & \bm{0}  & \bm{0}  \\
\bm{0}  & \cdots & \bm{0} & P_{(n-1,n-1)} & \bm{0}  \\		
\bm{0}  & \cdots & \bm{0} & \bm{0}  & P_{(n,n)} \\
\end{array}
\right)
=
\left(\begin{array}{ccccc}
P_{(1,1)} & \bm{0} & \cdots & \cdots  & \bm{0}  \\
\bm{0}  & P_{(1,1)}& \bm{0} & \cdots  & \bm{0}  \\
\bm{0}  & \bm{0} & \ddots & \bm{0}  & \bm{0}  \\
\bm{0}  & \cdots & \bm{0} & P_{(1,1)} & \bm{0}  \\		
\bm{0}  & \cdots & \bm{0} & \bm{0}  & P_{(1,1)} \\
\end{array}
\right),\\
Q&=&
\left(
\begin{array}{ccccc}
Q_{(1,1)} & \bm{0} & \cdots & \cdots  & \bm{0}  \\
\bm{0}  & Q_{(2,2)}& \bm{0} & \cdots  & \bm{0}  \\
\bm{0}  & \bm{0} & \ddots & \bm{0}  & \bm{0}  \\
\bm{0}  & \cdots & \bm{0} & Q_{(n-1,n-1)} & \bm{0}  \\		
\bm{0}  & \cdots & \bm{0} & \bm{0}  & Q_{(n,n)} \\
\end{array}
\right)
=
\left(\begin{array}{ccccc}
Q_{(1,1)} & \bm{0} & \cdots & \cdots  & \bm{0}  \\
\bm{0}  & Q_{(1,1)}& \bm{0} & \cdots  & \bm{0}  \\
\bm{0}  & \bm{0} & \ddots & \bm{0}  & \bm{0}  \\
\bm{0}  & \cdots & \bm{0} & Q_{(1,1)} & \bm{0}  \\		
\bm{0}  & \cdots & \bm{0} & \bm{0}  & Q_{(1,1)} \\
\end{array}
\right),\\
R&=&
\left(
\begin{array}{ccccc}
r_{(1,1)} & \bm{0} & \cdots & \cdots  & \bm{0}  \\
\bm{0}  & r_{(2,2)}& \bm{0} & \cdots  & \bm{0}  \\
\bm{0}  & \bm{0} & \ddots & \bm{0}  & \bm{0}  \\
\bm{0}  & \cdots & \bm{0} & r_{(n-1,n-1)} & \bm{0}  \\		
\bm{0}  & \cdots & \bm{0} & \bm{0}  & r_{(n,n)} \\
\end{array}
\right)
=r_{(1,1)}I,
\end{eqnarray}
where subscript $(1,1)$ represents the first block element in an original matrix.
In this case, $P_{(1,1)}=P_{(2,2)}=\dots=P_{(n,n)}$ is satisfied in the $k$th timestep because the update procedure of $P_{(1,1)}$ is exactly the same as others, though the rigorous proof is omitted for brevity. With regard to the second assumption, we usually assume that $R$ and $Q$ are identity matrices multiplied by a scalar and the second assumption is satisfied for those matrices. For the observation noise covarience matrix $R$, it seems to be reasonable if the sensors are independent and have the quality almost equivalent each other. On the other hand, for the system noise covariance matrix, it seems to be slightly broken because changes in $\bm{\theta}$ variables in the system identification problem might have relationship each other. However, even if this assumption is slightly broken up, the fast algorithm with assumption above works well as shown later. Therefore, we believe that this assumption is reasonable for the system identification.

With regard to a priori estimation we get 
\begin{eqnarray}
P_{k|k(1,1)}=P_{k|k-1(1,1)}+Q_{(1,1)} \label{eq:predictP11}.
\end{eqnarray}

With regard to update, we obtain the follwing equations. 
Here, $S_k=s_{k(1,1)} I$ is obtained, and its value $s_{k,(1,1)}$ is,
\begin{eqnarray}
	s_{k(1,1)} = r_{k(1,1)} + \bm{x}_{k}^\text{T}  P_{k|k-1\left(1,1\right)} \bm{x}_{k}.\label{eq:updates}
\end{eqnarray}

The Kalman gain becomes a vector and is expressed as follows:
\begin{eqnarray}
	K_{k\left(1\right)}  = P_{k|k-1\left(1,1\right)} \bm{x}_{k} s^{-1}.
\end{eqnarray}
Note that the dimensionality of the Kalman gain derived here is $n\times 1$.
When the Kalman gain matrix is multiplied by Eq.~(\ref{eqn:update_theta}), 
the element of the Kalman gain matrix is copied in the column direction, as follows:
\begin{eqnarray}
	K_k = \left( 
	\begin{array}{c}
K_{k\left(1\right)}\\
K_{k\left(1\right)}\\
\vdots\\
K_{k\left(1\right)}\\
K_{k\left(1\right)}
\end{array}\right)\label{eq:updatekg}
\end{eqnarray}
and its dimension is expanded to $n\times n$ as a result.
Here, the subscript $(1)$ indicates the first block-matrix column in the original matrix.

The covariance matrix after observation can be updated as follows:
\begin{eqnarray}
	P_{k|k\left(1,1\right)} = \left( I - K_{k\left(1\right)} \bm{x}_k^\text{T} \right)P_{k|k-1\left(1,1\right)}.\label{eq:updateP11}
\end{eqnarray}

 This algorithm can be applicable to a time-varying system because the assumptions on the $Q$ and $R$ matrices above are on their spatial distribution and not on their temporal behaviour. 

\subsection{Combination with truncated POD}
\label{sec:POD}
Although the use of the algorithm described in the previous subsection helps us to compute $\bm{\theta}$ quickly, matrix $P$ and state variables $\bm{\theta}$ require memories of $n^2$ variables, and for some fluid problems, it is impossible to store all of the matrix variables. Therefore, in the present study, truncated POD (truncated SVD) is used as a preconditioner and the number of degrees of freedom are reduced for applying the Kalman filter to the dataset of the fluid system. In the present study, 1) the batch POD is first applied, and 2) the proposed Kalman filter is then applied to the amplitude of each POD mode. Finally, the mode shape of the fluid system is recovered by multiplying the spatial POD modes.　

More concretely, the procedure is explained here. We assume that a data matrix $Z$ which contains $m$ temporal dimensions for time series data or $m$-pairs snapshot data as discussed in the previous sections.
First, we conduct SVD as follows:
\begin{eqnarray}
Z=U_ZD_ZV_Z^\text{T},\label{eq:svds}
\end{eqnarray}
whereas $U_Z$ and $V_Z$ contain the spatial and temporal POD modes, respectively, as row vectors. 
The $r$-rank approximation is obtained by applying the truncated POD, as follows:
\begin{eqnarray}
\tilde Z= \tilde U_\text{T}^Z \tilde D^\text{T}_Z \tilde V^\text{T}_{Z},
\end{eqnarray}
where the singular values (square roots of the eigenvalues of the covariance matrix) of the $r$-dimensional matrix $\tilde D$ are the same as the $r$ largest singular values of $D$. In addition, $\tilde U_Z$, and $\tilde V_Z$ have the same $r$ row vectors as $U_Z$ and $V_Z$. After this procedure, reduced-dimension data matrices $\tilde X$ and $\tilde Y$ are obtained as follows:
\begin{eqnarray}
\tilde X&=&\left(\tilde{U}_{Z}^\text{T} X\right)\label{eq:xtilde}\\
\tilde Y&=&\left(\tilde{U}_{Z}^\text{T} Y\right)\label{eq:ytilde}
\end{eqnarray}
and $\tilde X$ and $\tilde Y$ are treated similarly to $X$ and $Y$ in the Kalman filter DMD procedures.

In addition, for a more flexible online procedure , we can use the following formulation when the left singular vector (spatial mode) $\tilde U$ is known in advance: 
\begin{eqnarray}
\tilde x_k=\tilde U^\text{T} x_k,
\end{eqnarray}
where the left singular vector $\tilde U$ is assumed to be fixed. After obtaining the right eigenvector of the reduced system using the Kalman filter DMD, it is necessary to recover the original dimension by multiplying matrix $\tilde U$. In this case, the purely online algorithm is obtained. However, if the POD mode is not known in advance and needs to be estimated, online POD or other methods are required. Because spatial POD modes change with time, the time histories of the coefficients are not reliable. In addition, the online POD algorithm sometimes exchanges active modes and nonactive modes. Therefore, a straightforward extension of KFDMD with online POD is not trivial; this issue is left for the future study.

In addition, when the POD mode is used, the covarience matrix $R_\text{POD}$ of noise on the POD mode should be considered. If the noise level is spatially uniform and independent each other, the noise $\bm{w}$ in original space of $\bm{x}$ have following characteristics:
\begin{eqnarray}
E(\bm{w}^\text{T} \bm{w})=\sigma^2 I,
\end{eqnarray}
where $E$ and $\sigma^2$ represent an expected value and a variance of the noise, respectively. If we consider $R_\text{POD}$, noise on the POD mode amplitude becomes $\tilde{U}^\text{T}_Z\bm{w}$ and the expected value of the covariance matrix of them are 
\begin{eqnarray}
E(\tilde{U}_Z^\text{T}\bm{w} \bm{w}^\text{T} \tilde{U}_Z)&=&\tilde{U}_Z^\text{T} E(\bm{w}\bm{w}^\text{T})\tilde{U}_Z=\sigma^2 \tilde{U}_Z^\text{T} I \tilde{U}_Z,\\
&=&\sigma^2 I,
\end{eqnarray}
where $U_Z$ is assumed to consist of constant singular vectors. This result is not affected by a number of $r$ activated in trPOD. 

In the present paper, only Eqs. \ref{eq:xtilde} and \ref{eq:ytilde} are used for the truncated POD.   This procedure is adopted for many-degree-of-freedom problems ($n>10,000$) or denoising purpose, and it is not used unless otherwise mentioned. 

\subsection{Implementation of algorithm}
\label{sec:Imple}
Here, fast algorithm of Kalman filter DMD is briefly summarized. After initialization, prediction (a priori estimation) and update steps are alternately conducted. 
\subsubsection*{Initialization}
\begin{enumerate}
    \item If the degrees of freedom are large, trPOD is applied to data.
    \item Set $\bm{\theta}=vec(I)$ and $P_{0|0,(1,1)}=\gamma I$. Here, $\gamma$ is large value. (In the present study, we set $\gamma=1000$).
\end{enumerate}

\subsubsection*{Prediction step}
\begin{enumerate}
    \item $\theta_{k|k-1}$ is assumed to be the same as $\theta_{k-1|k-1}$ using Eq. \ref{eq:predicttheta}.
    \item $P_{k|k-1(1,1)}$ is predicted by Eq. \ref{eq:predictP11}
\end{enumerate}

\subsubsection*{Update step}
\begin{enumerate}
    \item Kalman gain $K$ is computed by Eqs. \ref{eq:updates} to \ref{eq:updatekg} and $A$ matrix is formed using $\theta_{k|k-1}$. 
    \item $\theta_{k|k}$ is updated by Eqs. \ref{eq:deltatheta} and \ref{eq:updatetheta}.
    \item $P_{k|k(1,1)}$ is updated by Eq. \ref{eq:updateP11}.
\end{enumerate}

\section{Numerical Experiments for Problems with Time Varying Noise }
\label{sec:NETVN}
The fast Kalman filter algorithm described in Section \ref{sec:fast_algo} is adopted in the numerical experiments below. 

\subsection{System Identification with Quasi-steady Noise}
\label{sec:RESSWON}
First, the performance of the Kalman-filter-based DMD (KFDMD) is investigated for the standard problem and is compared with those of the standard DMD and tlsDMD. 
The methods and dataset are almost similar to those used in a previous study.\cite{Hemati2017}
The eigenvalues are assumed to be positioned at $\lambda_1 = \exp\left( \pm 2\pi i \Delta t \right)$, $\lambda_2 = \exp\left( \pm 5\pi i \Delta t \right)$, and $\lambda_3 = \exp\left[\left( -0.3 \pm 11\pi i\right)\Delta t\right]$, where $\Delta t = 0.01$. The corresponding continuous eigenvalues are $\omega_1 =  \pm 2\pi i  $, $\omega_2 =  \pm 5\pi i $, and $\omega_3 = \left( -0.3 \pm 11\pi i\right)$. The number of degrees of freedom of this system is $d=6$. The original data $\bm{f}$ are computed as
\begin{eqnarray}
\frac{\text{d}\bm{f}}{\text{d}t}&=&B\bm{f} \label{eq:fcont}\\
B&=&\left(\begin{array}{cccccc}
 |\text{Re}(\omega_1)| & |\text{Im}(\omega_1)| & 0 & 0  & 0  & 0\\
-|\text{Im}(\omega_1)| & |\text{Re}(\omega_1)|  & 0 & 0  & 0  & 0\\
0  & 0& |\text{Re}(\omega_2)| & |\text{Im}(\omega_2)| & 0 & 0 \\
0  & 0&-|\text{Im}(\omega_2)| & |\text{Re}(\omega_2)|  & 0 & 0  \\
0 & 0  & 0  & 0&  |\text{Re}(\omega_3)| & |\text{Im}(\omega_3)|  \\
0 & 0  & 0  & 0& -|\text{Im}(\omega_3)| & |\text{Re}(\omega_3)|  \\
\end{array}
\right).
\end{eqnarray}
with the initial conditions of $\bm{f}_0$, each component of which is filled with $\mathcal{N}\left(1,0.1 \right)$.
Then, we construct snapshot data using QR decomposition.
A random matrix $T$ with dimensions $n\times d$ is initially filled with random numbers of $\mathcal{N}\left(0,1^2\right)$,  and then decomposed into $T=Q_{QR}R_{QR}$.
Finally, the original data $\bm{f}$ with dimension $d$ are transformed into $\bm{x}$ with dimension $n=200$ via the matrix $Q_{QR}$.
Quasi-steady white noise with $\mathcal{N}\left(0,\sigma^2\right)$ is added to the sequential snapshot data after this transformation.

For the initial adjustable parameters of the Kalman filter, the diagonal elements of the initial covariance matrix P are set to $10^3$ and all of the elements of $Q$ are set to zero. The diagonal elements of $R$ are correctly set using prior knowledge of noise, as discussed later. The assumption of $Q=0$ corresponds to the assumption that an identified system is time constant. 

\begin{figure}
	\subfigure[$\sigma^2=0.0001$]{\includegraphics[width=5cm]{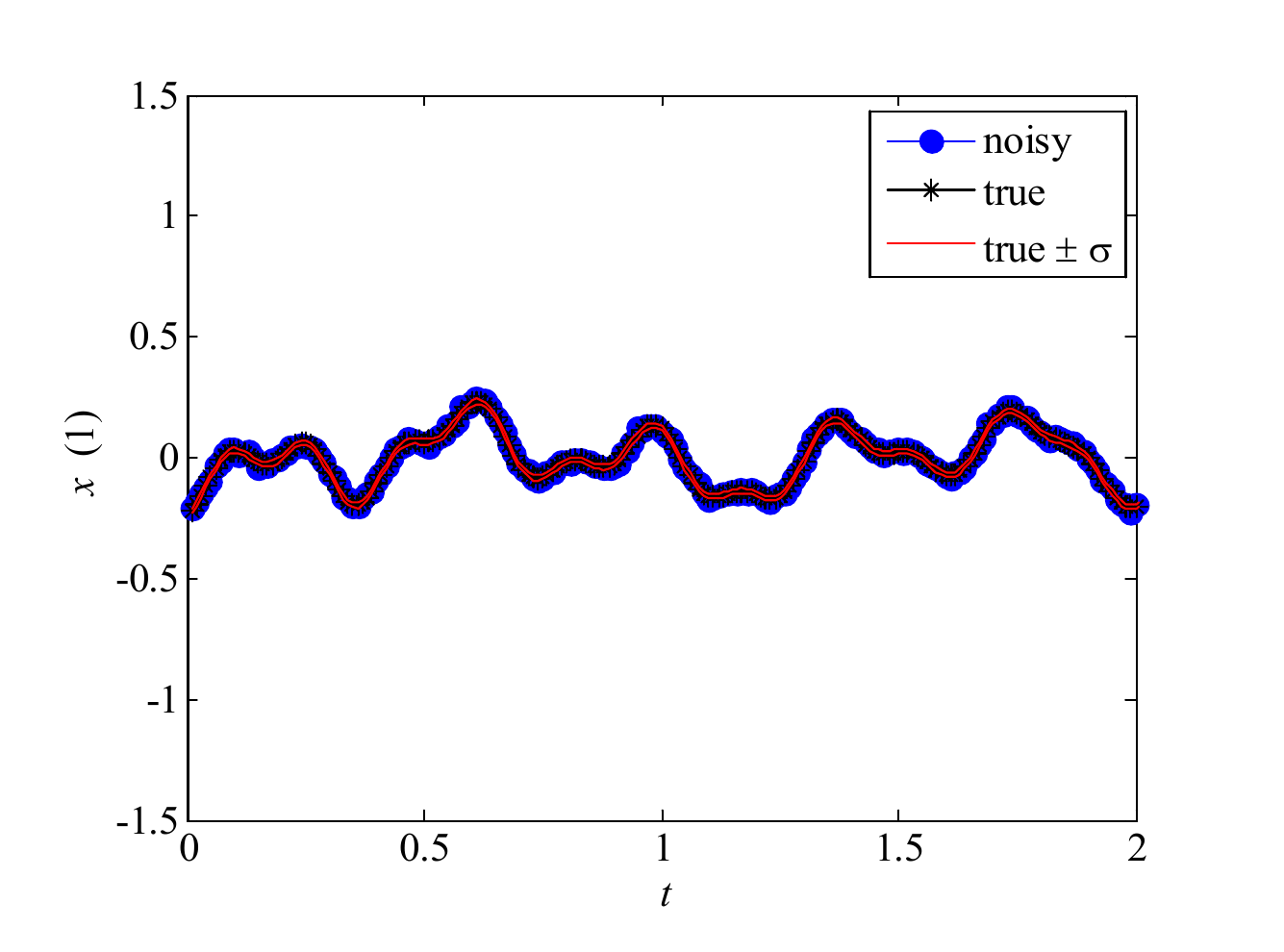}}
	\subfigure[$\sigma^2=0.001$ ]{\includegraphics[width=5cm]{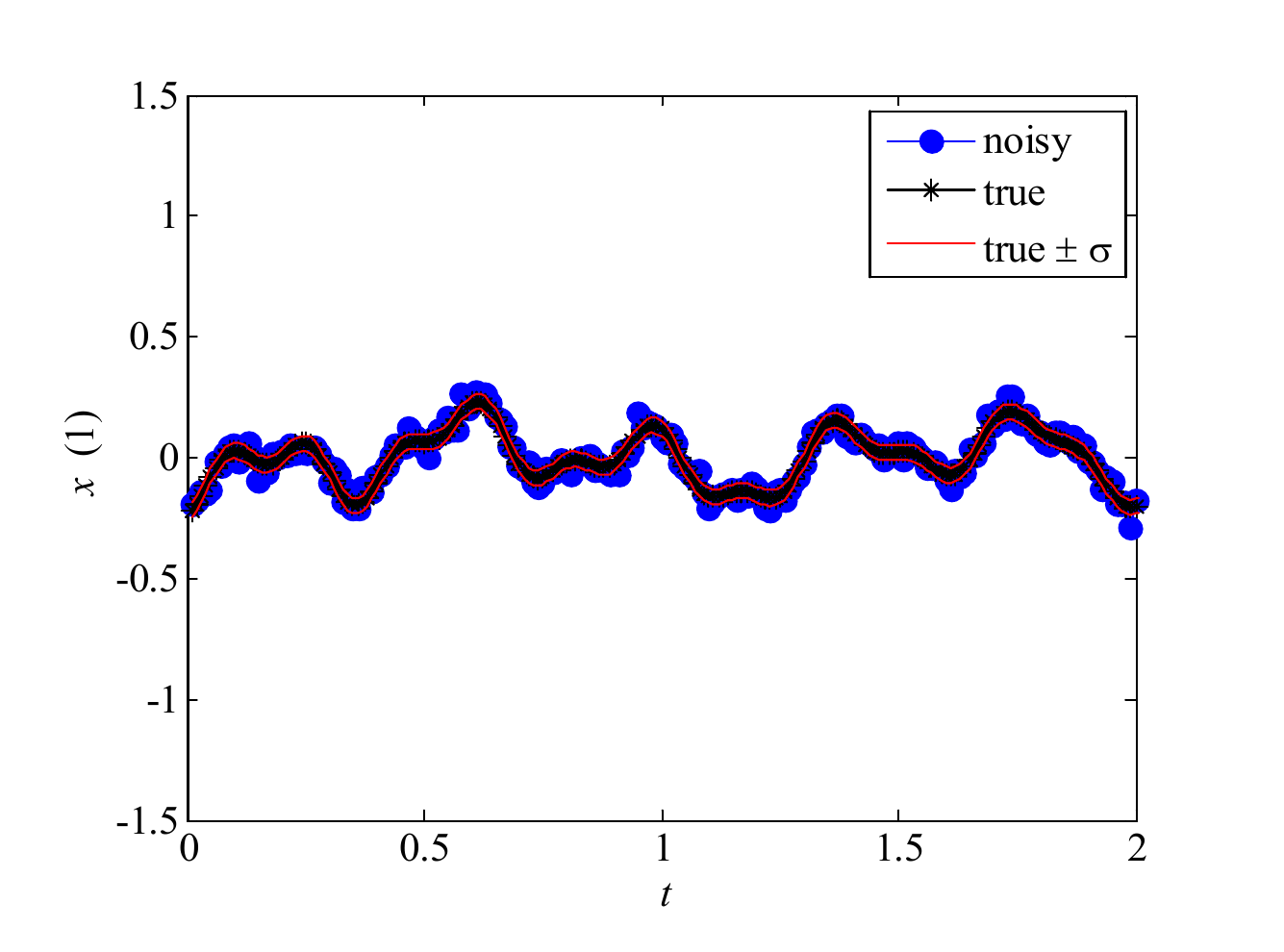}}\\
	\subfigure[$\sigma^2=0.01$  ]{\includegraphics[width=5cm]{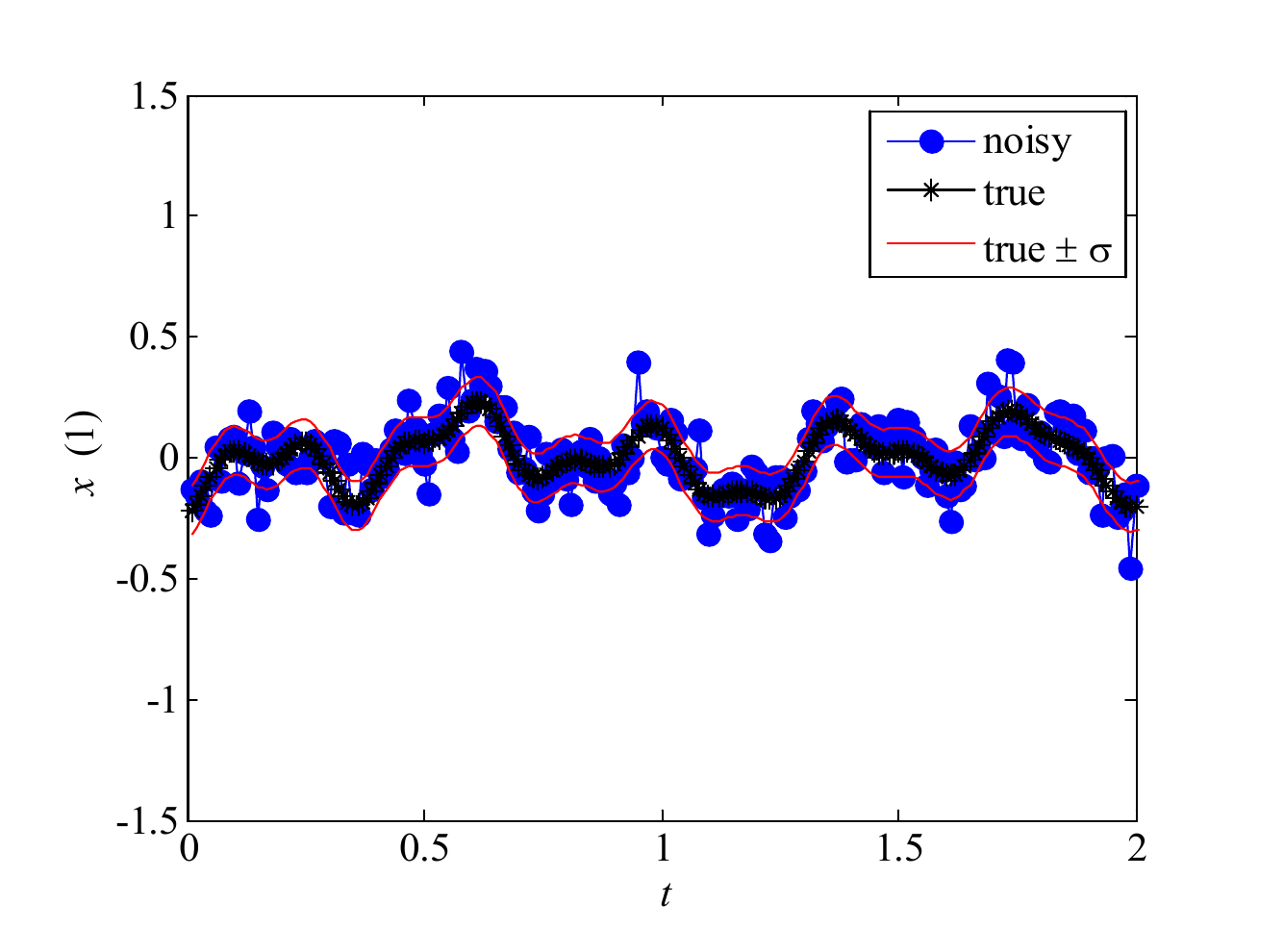}}
	\subfigure[$\sigma^2=0.1$   ]{\includegraphics[width=5cm]{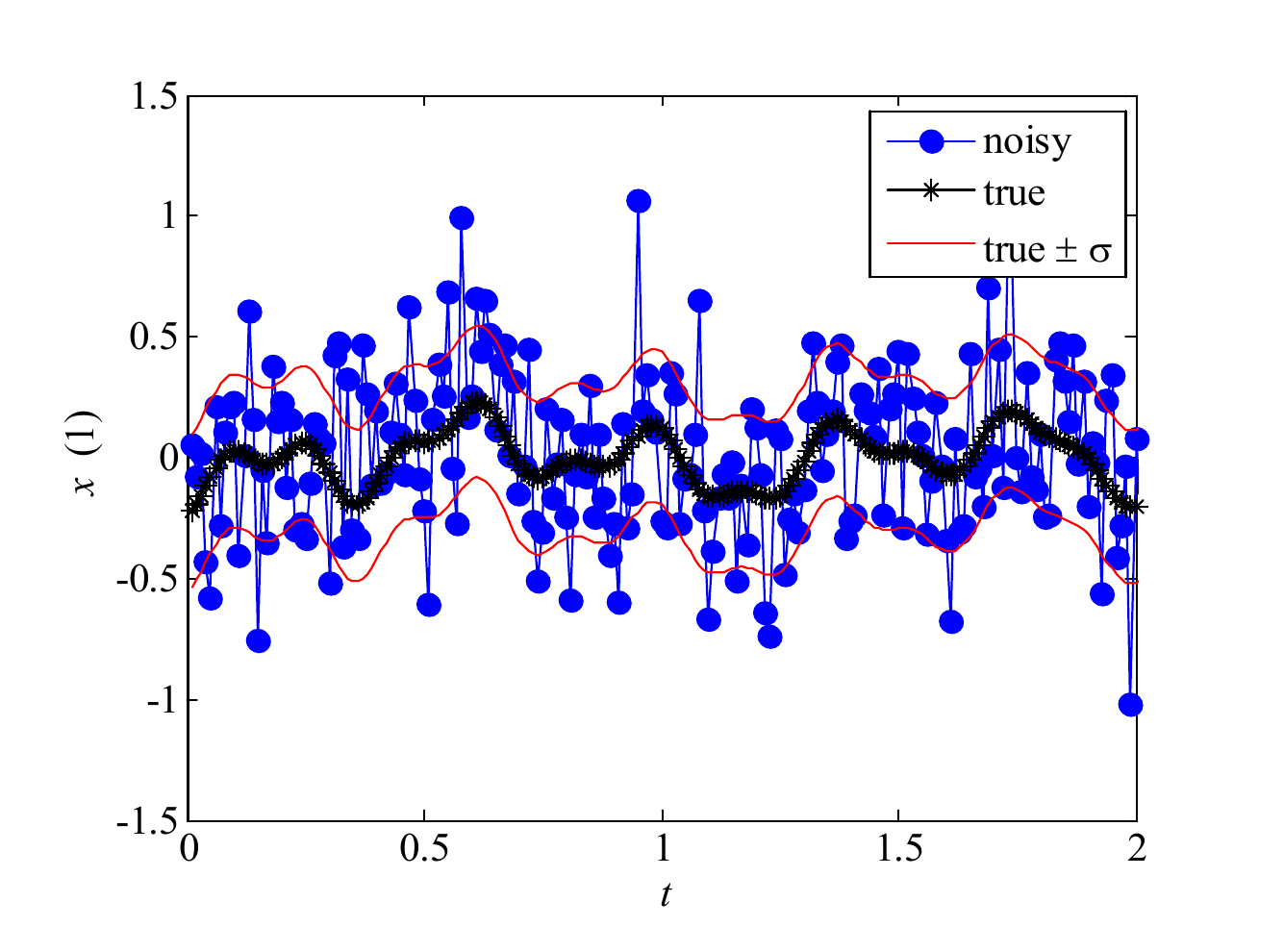}}
	\caption{The noisy and true time history of first node of data matrix for the test problem with quasi-steady noise. Here, only the first 200 steps of the entire data matrix is illustrated. }
	\label{fig:steadyhist}
\end{figure}

The noise strength is set to be $\sigma^2=0.0001, 0.001, 0.01$, and $0.1$, and the performance is evaluated for DMD, tlsDMD, and KFDMD. For KFDMD, observation error covariance is correctly given as $R=\sigma^2 I$. The other DMD variants for noisy data are compared in Appendix A. For tlsDMD and KFDMD, POD truncation in the algorithm is utilized and rank $r$ is set to be 6. The same computation is conducted 100 times using a different random number for each case. The noisy data of the first component of the data matrix $X$, corresponding to the true data, and standard deviation $\pm \sigma$ are shown in Fig. \ref{fig:steadyhist}.  Figure \ref{fig:steadyhist} shows the strength of the noise for various noise levels.  

\begin{figure}
	\subfigure[$\sigma^2=0.0001$]{\includegraphics[width=5cm]{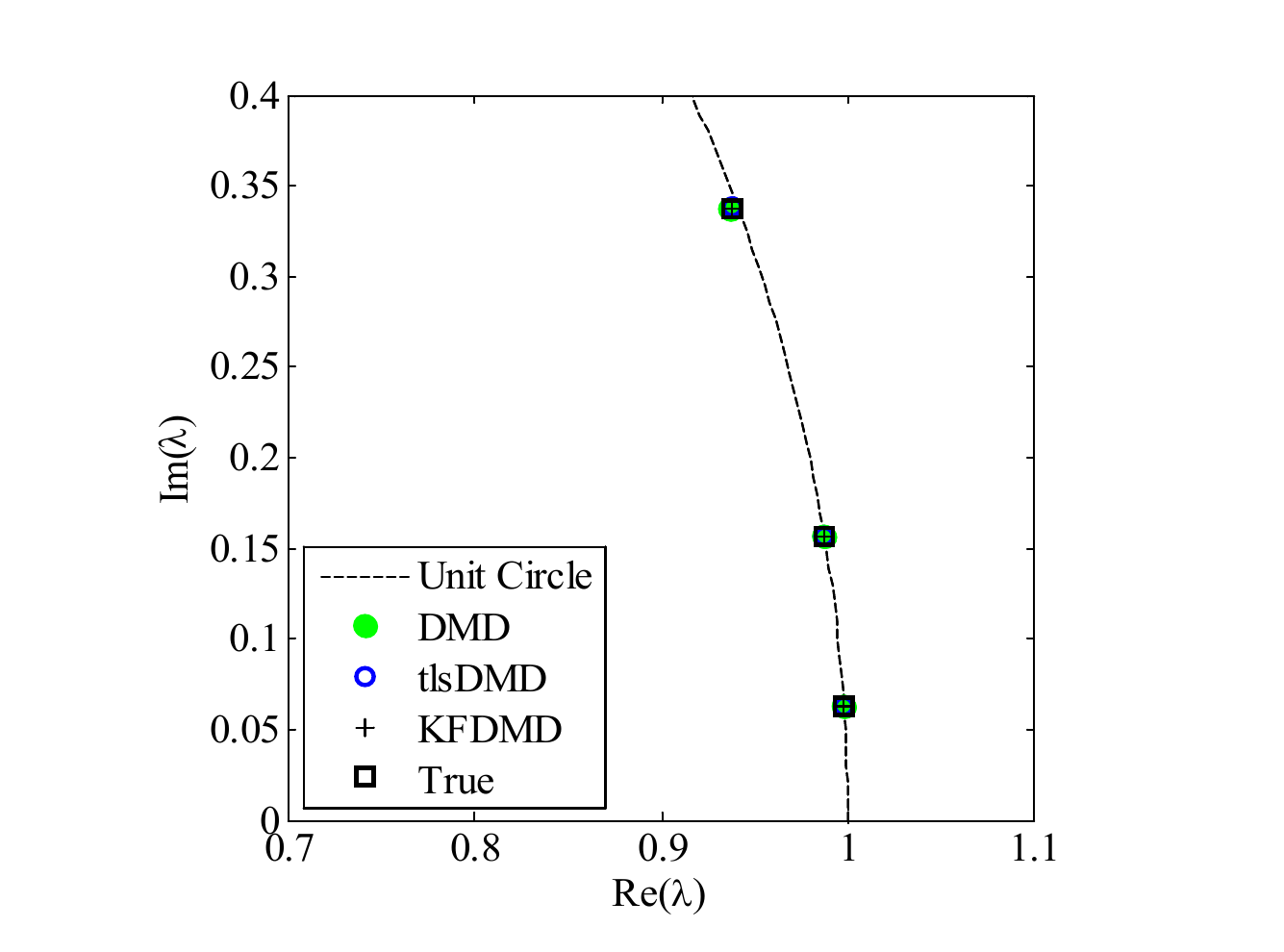}}
	\subfigure[$\sigma^2=0.001$ ]{\includegraphics[width=5cm]{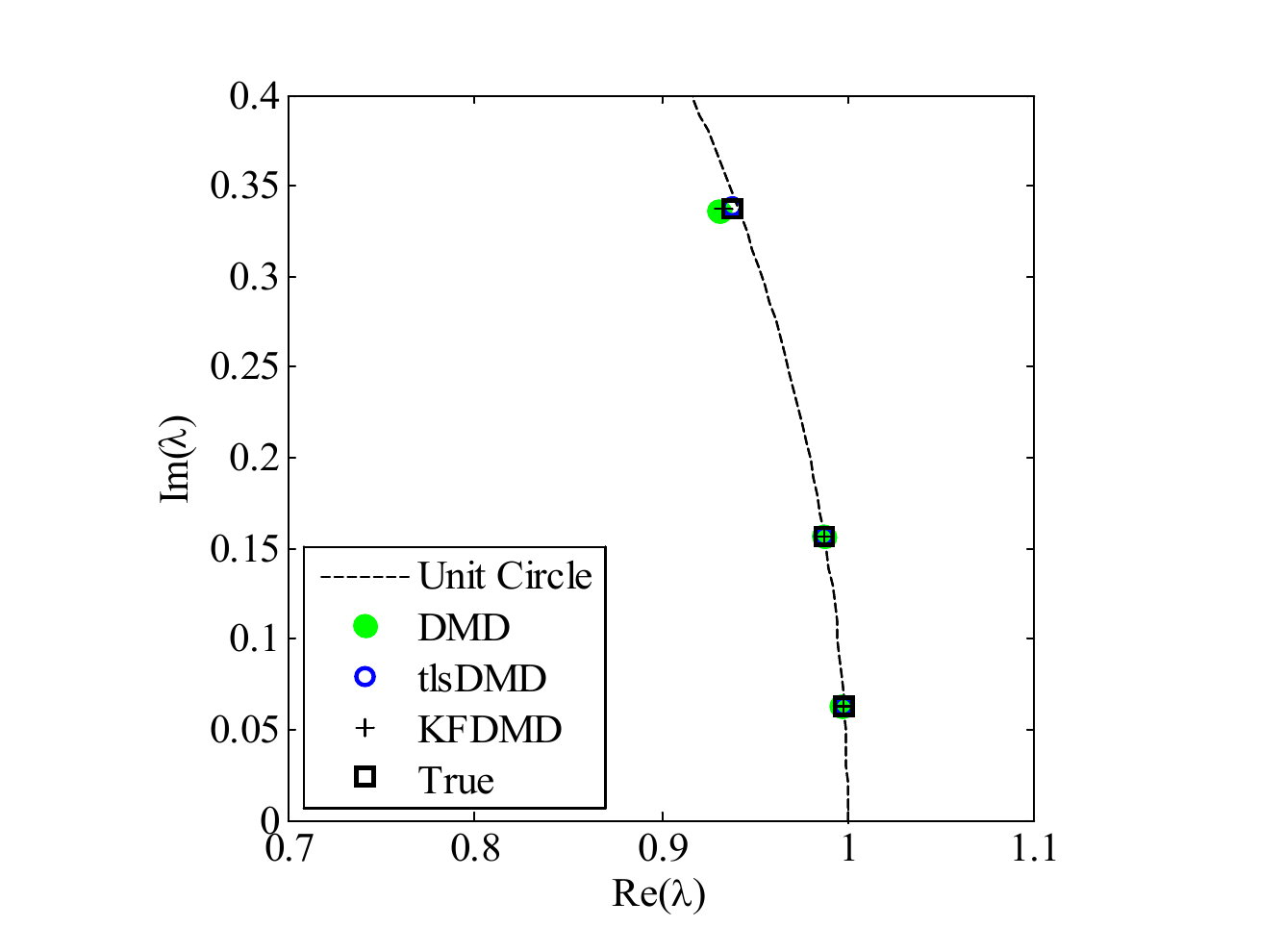}}\\
	\subfigure[$\sigma^2=0.01$  ]{\includegraphics[width=5cm]{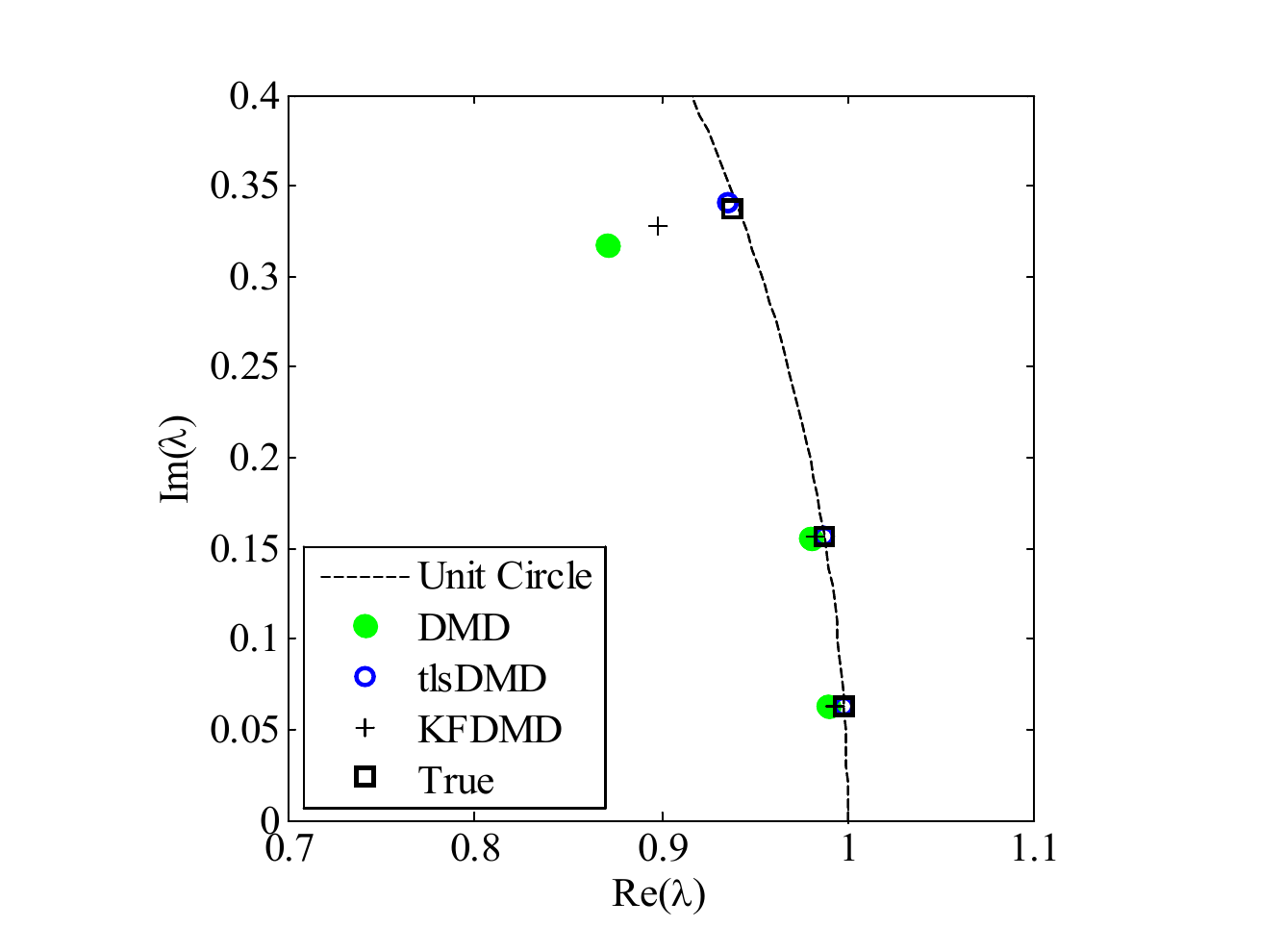}}
	\subfigure[$\sigma^2=0.1$   ]{\includegraphics[width=5cm]{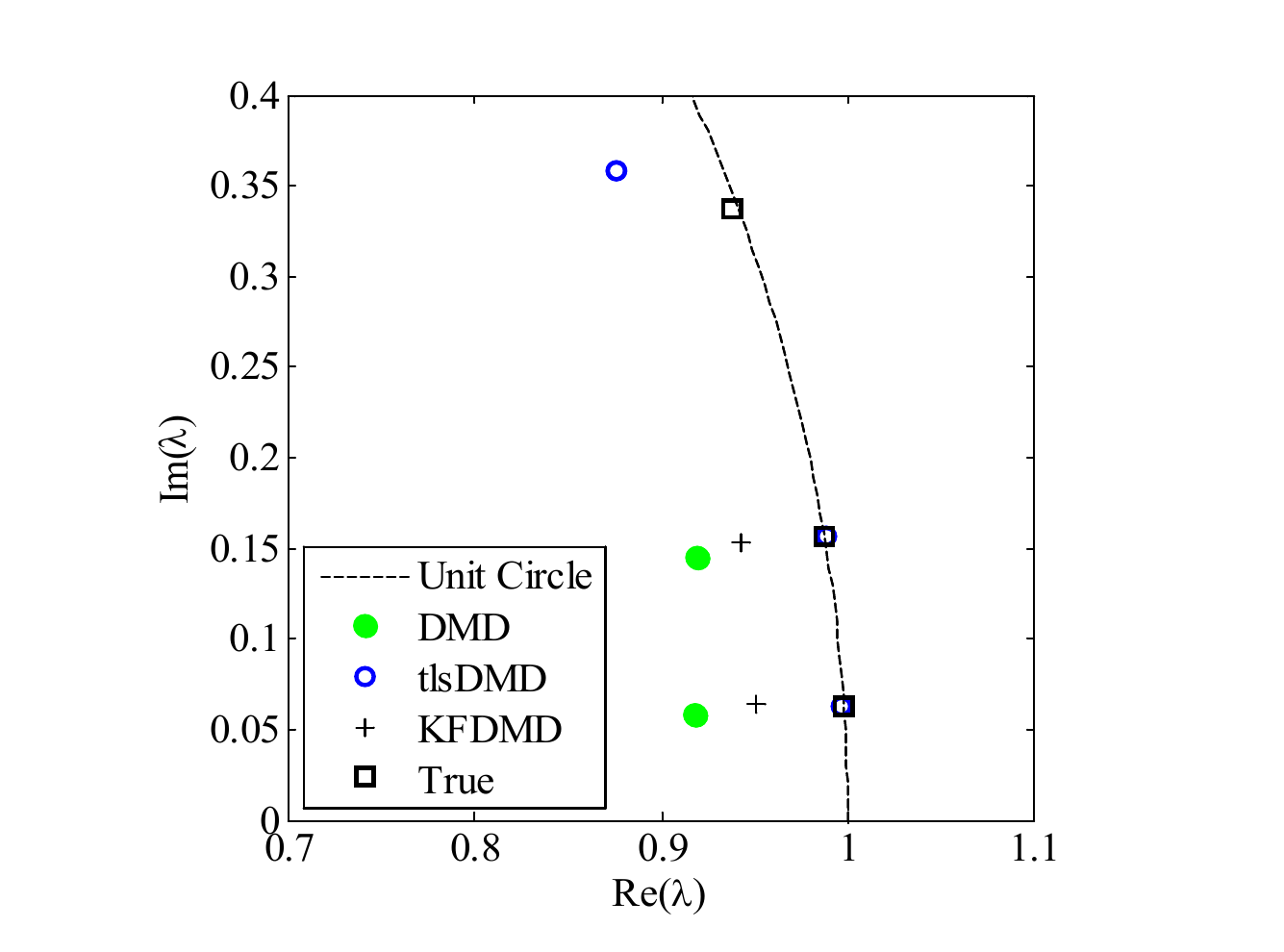}}
	\caption{Representative results of eigenvalues computed in the test problem with quasi-steady noise. }
	\label{fig:steadyeigenexam}
\end{figure}

\begin{figure}
	\subfigure[$\sigma^2=0.0001$]{\includegraphics[width=5cm]{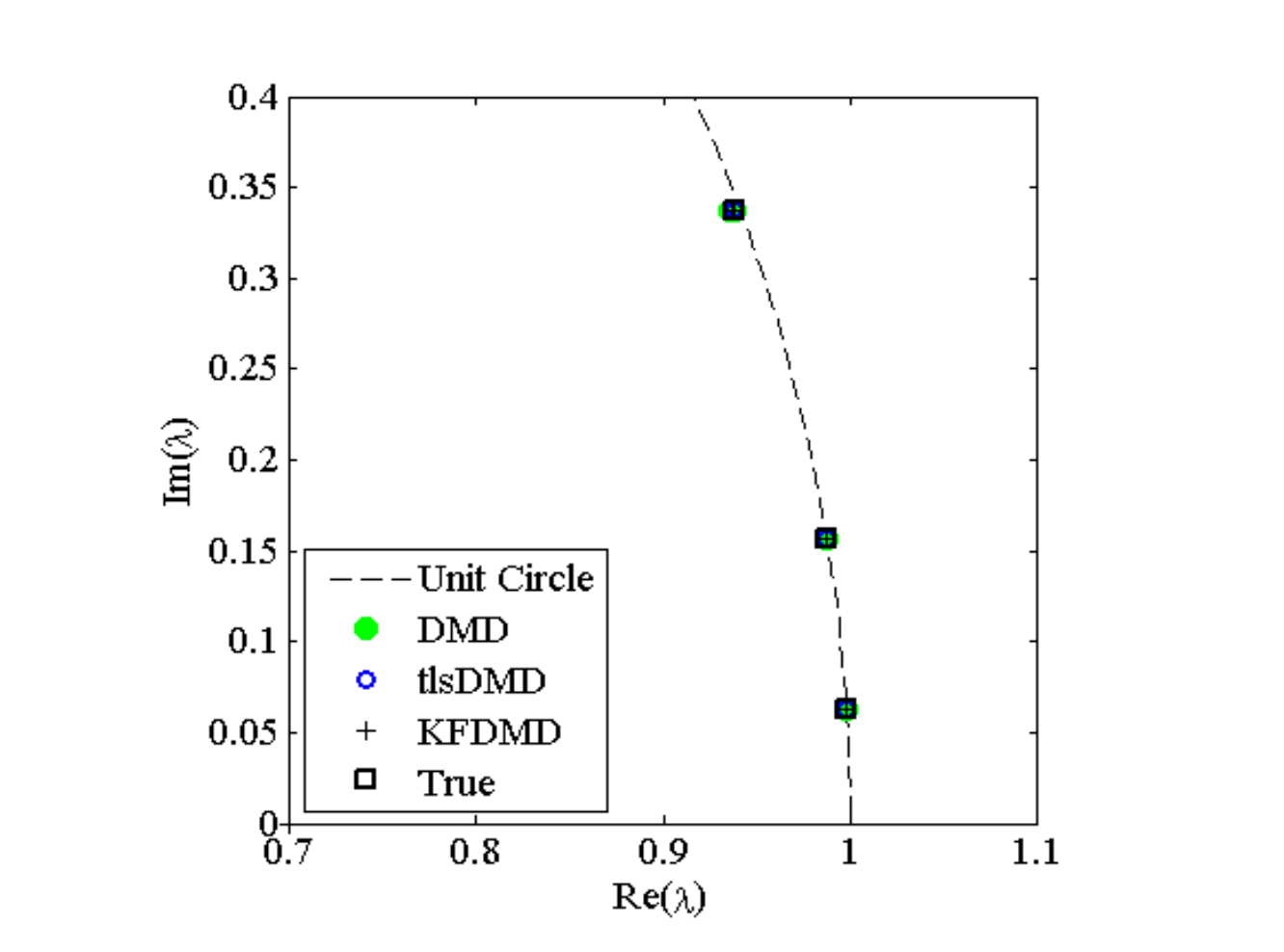}}
	\subfigure[$\sigma^2=0.001$ ]{\includegraphics[width=5cm]{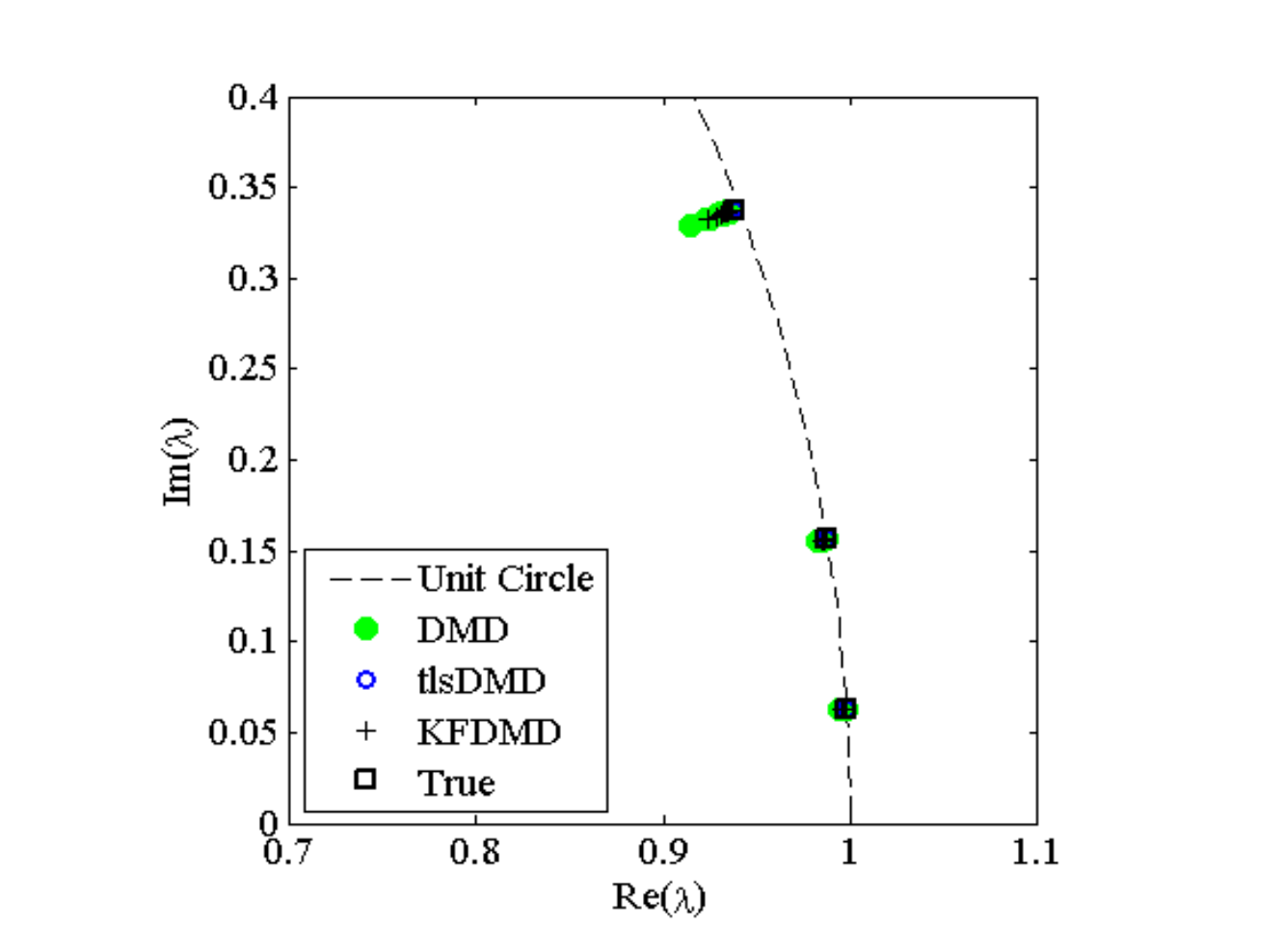}}\\
	\subfigure[$\sigma^2=0.01$  ]{\includegraphics[width=5cm]{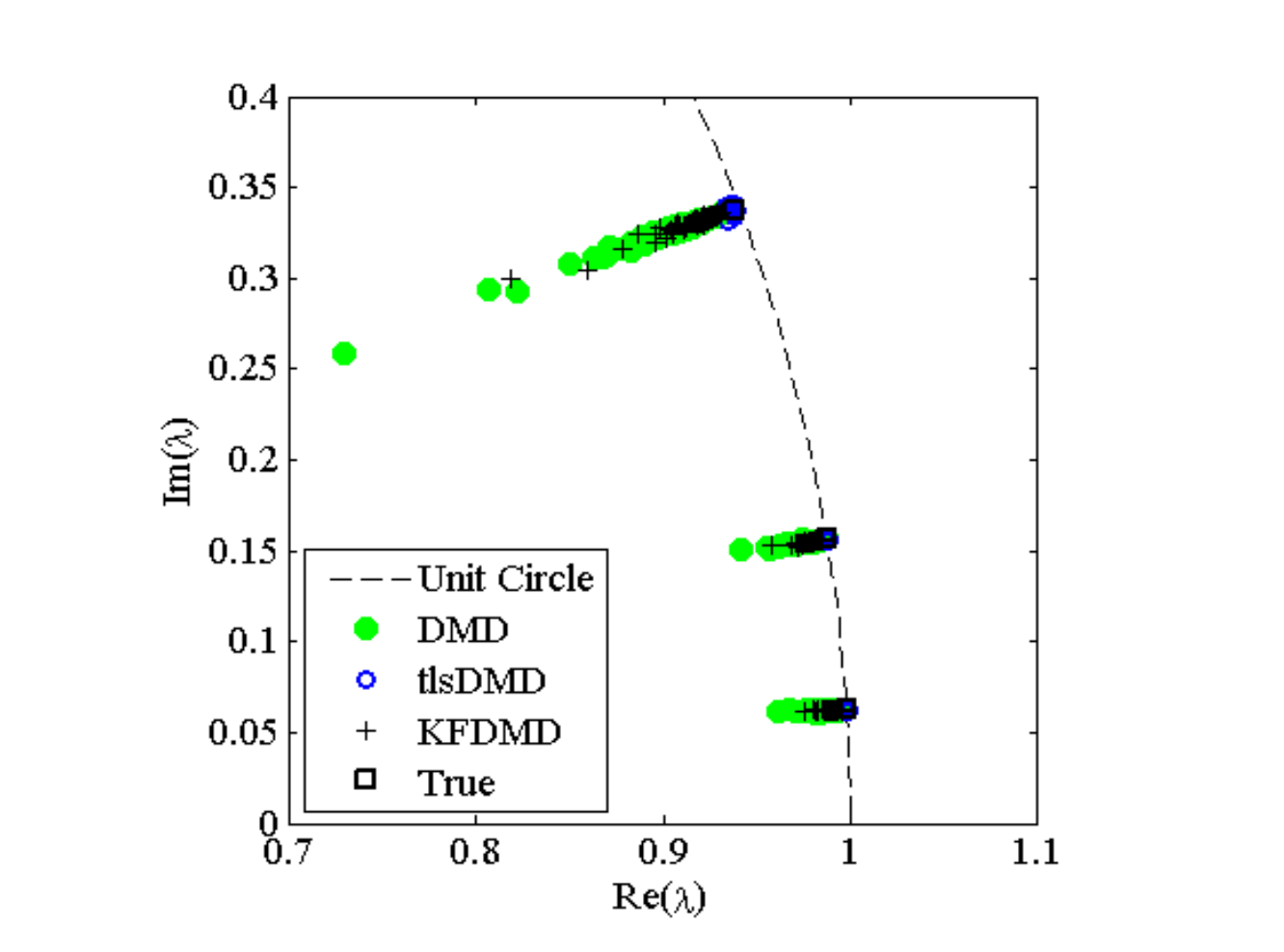}}
	\subfigure[$\sigma^2=0.1$   ]{\includegraphics[width=5cm]{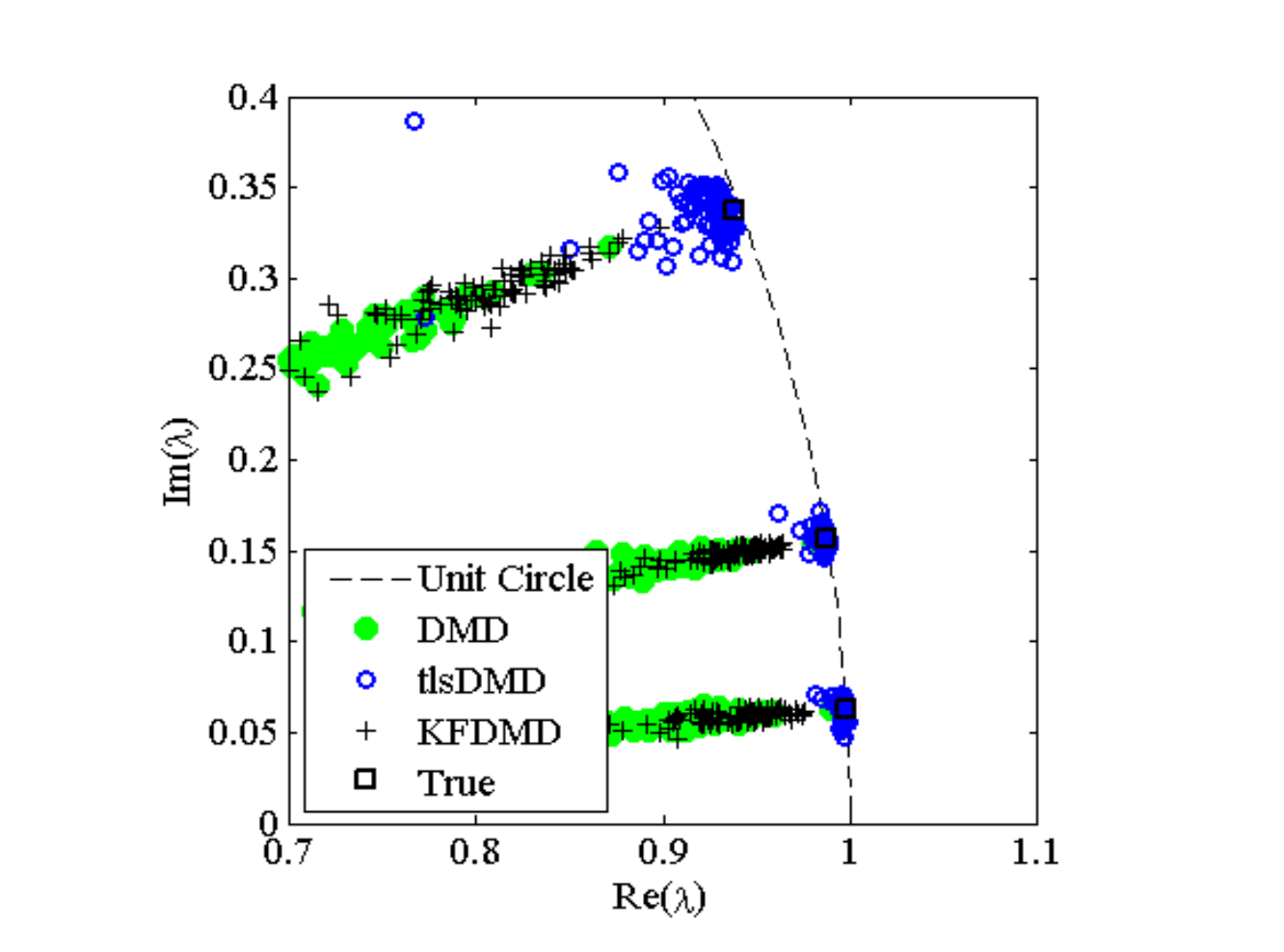}}
    \caption{Results of 100 computations of eigenvalue in the test problem with quasi-steady noise.}	
	\label{fig:steadyeigen}
\end{figure}

\begin{figure}
	\subfigure[$\lambda_1$]{\includegraphics[width=5cm]{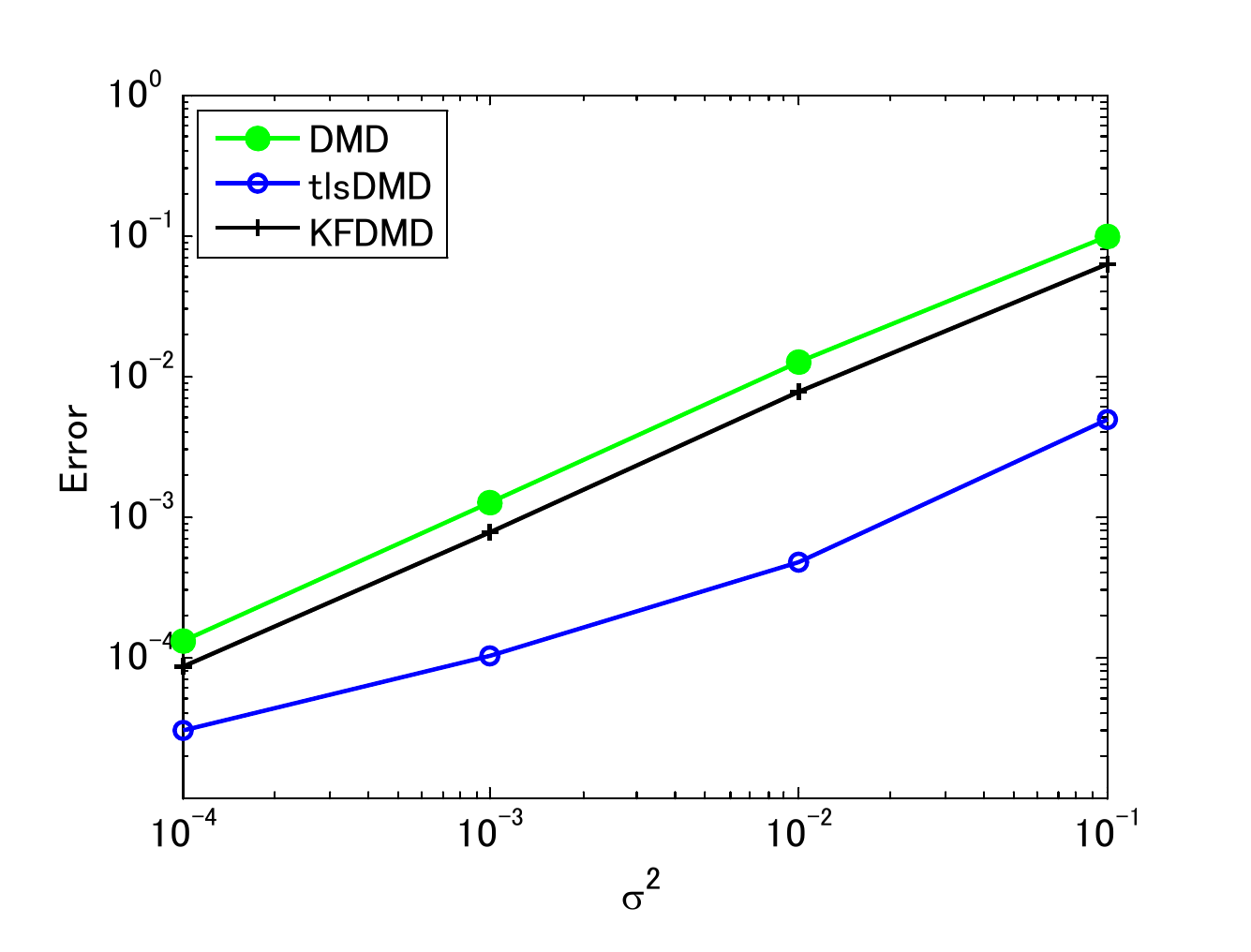}}
	\subfigure[$\lambda_2$ ]{\includegraphics[width=5cm]{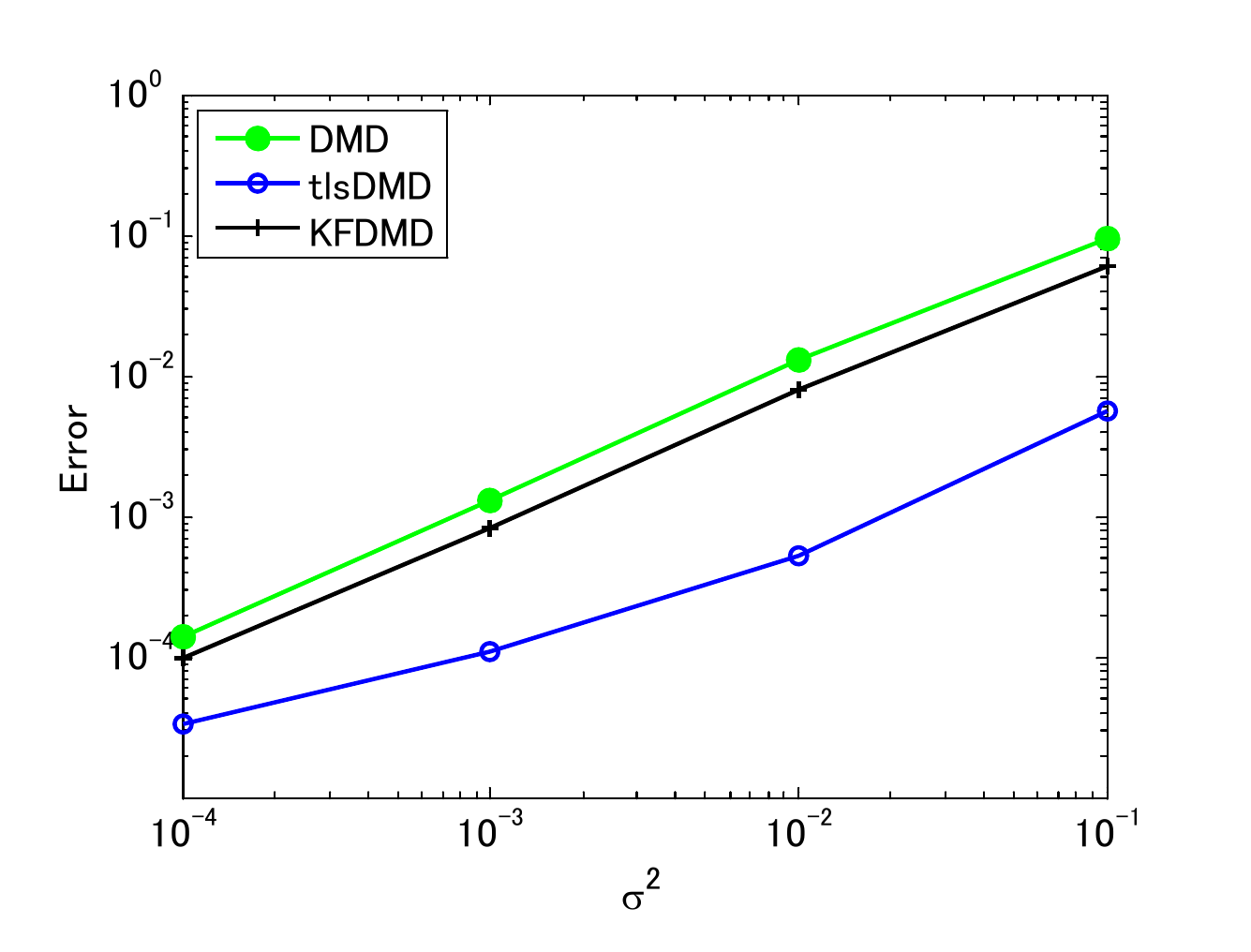}}
	\subfigure[$\lambda_3$  ]{\includegraphics[width=5cm]{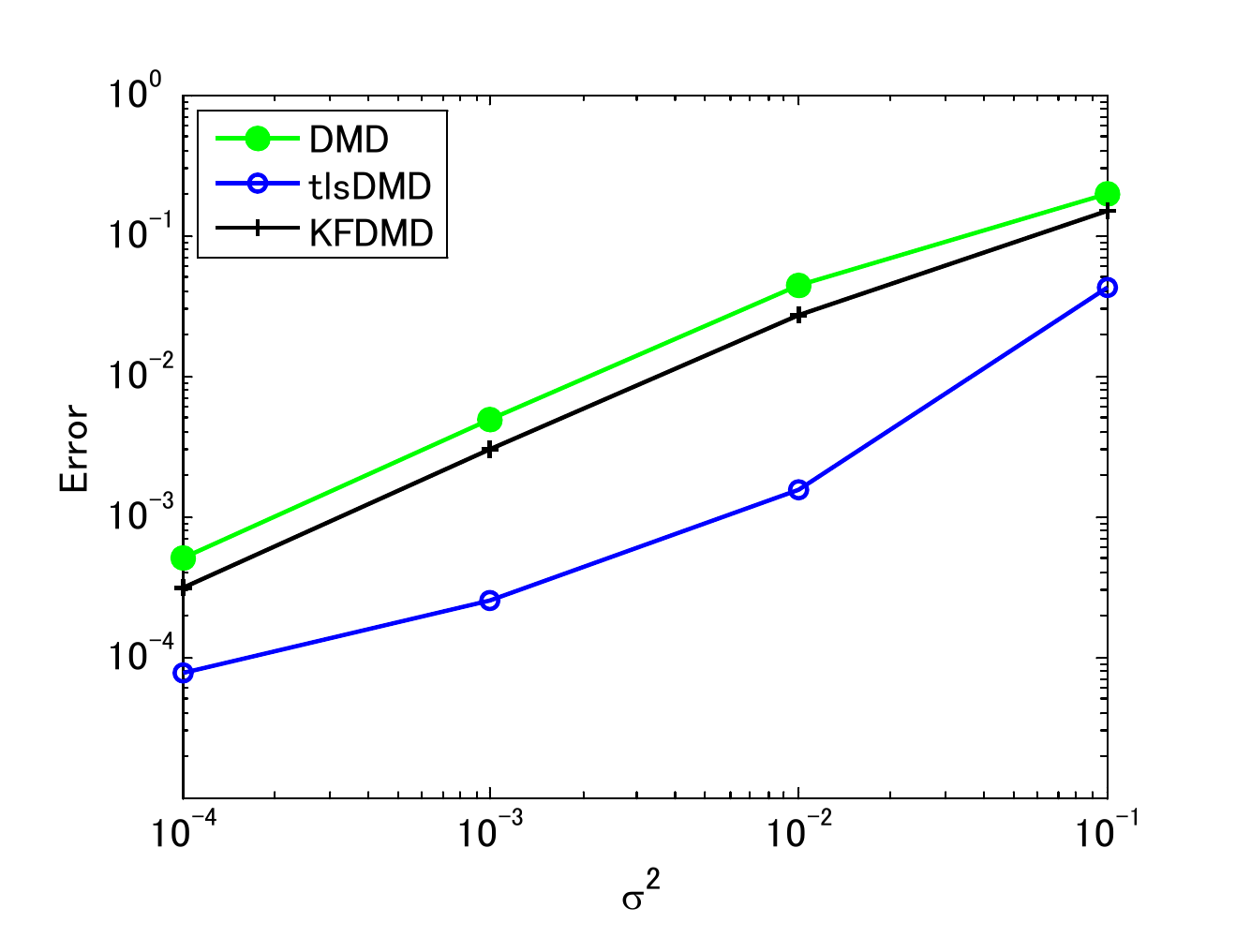}}
   \caption{Errors in the eigenvalues computed in the test problem with quasi-steady noise. Here, $L_2$ error is averaged with 100 test cases. }	
   \label{fig:steadyeigenerror}
\end{figure}

Figures \ref{fig:steadyeigenexam} and \ref{fig:steadyeigen} show the estimated eigenvalues computed using DMD, tlsDMD, and KFDMD, together with the true eigenvalue, for the representative case and for all 100 cases, respectively. The dashed lines in the figures represent the unit circle; the first and second true eigenvalues are on the circle and the third one is located slightly inside the circle. This plot shows that all the methods, including KFDMD, can predict the eigenvalues when the noise level is low. When the noise level is high, KFDMD is more accurate than standard DMD but less accurate than tlsDMD. The errors in the eigenvalues are plotted in Fig. \ref{fig:steadyeigenerror} defined by the norm of the closest computed eigenvalue to the specified true eigenvalue. Here, outliers were not removed and the definition of error given above was straightforwardly applied to all the data.  The error in the eigenvalues decreases with decreasing noise strength for all methods. This plot quantitatively shows that the error for KFDMD is smaller than that for standard DMD but larger than that for tlsDMD. 

In this problem, the noise characteristics do not change and thus KFDMD cannot take advantage of prior knowledge of noise. In addition, KFDMD uses an observation matrix that consists of the $\bm{x}_k$ vector, which includes the sensor noise. This noise is currently not considered in the KFDMD algorithm and biased noise appears, similar to the standard DMD. Therefore, KFDMD does not outperform tlsDMD for this problem.

\subsection{Static System Identification with Noise Known Characteristics}
\label{sec:RESSWN}
\begin{figure}
	\subfigure[$\sigma_0^2=0.0001$]{\includegraphics[width=5cm]{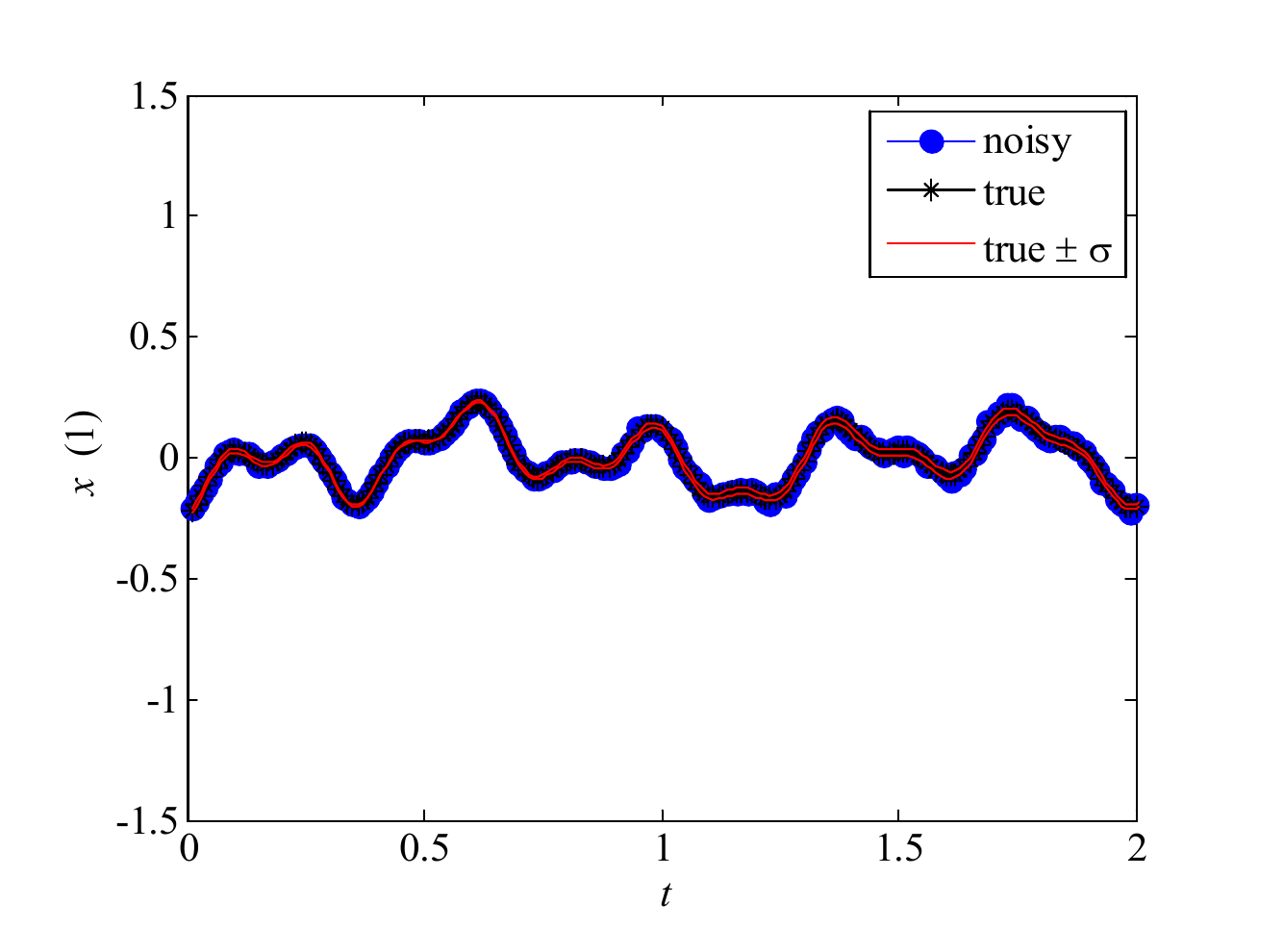}}
	\subfigure[$\sigma_0^2=0.001$ ]{\includegraphics[width=5cm]{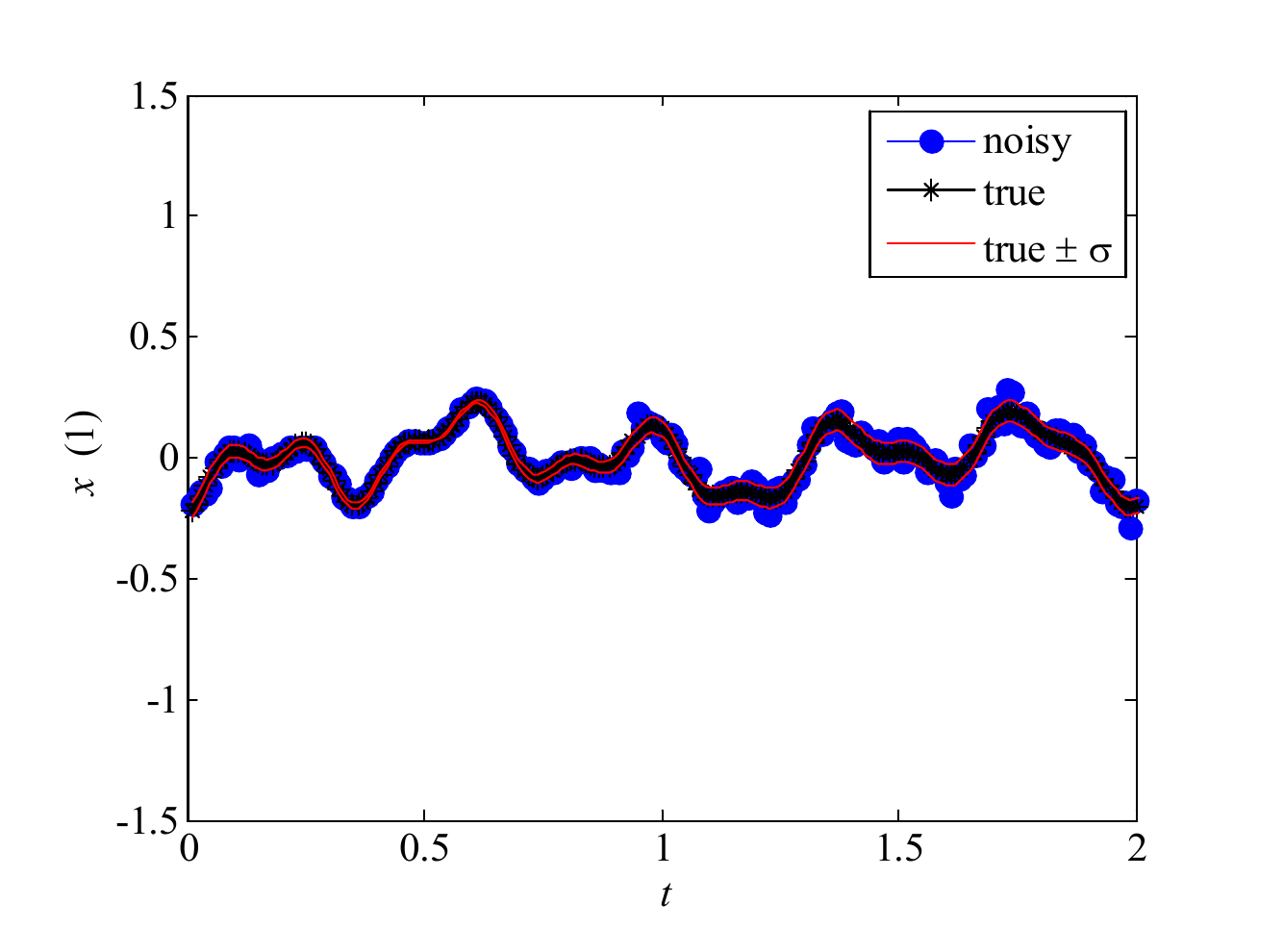}}\\
	\subfigure[$\sigma_0^2=0.01$  ]{\includegraphics[width=5cm]{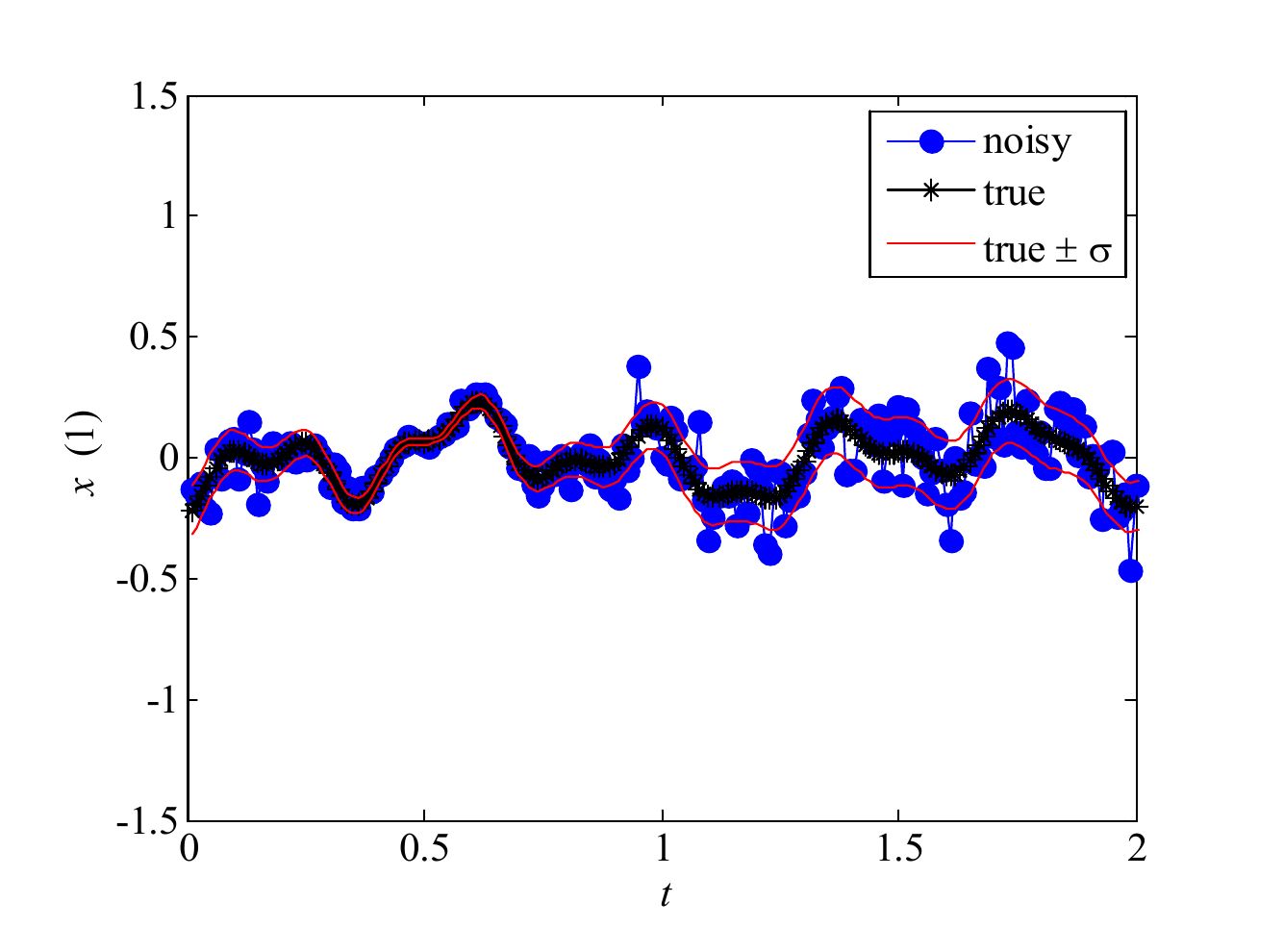}}
	\subfigure[$\sigma_0^2=0.1$   ]{\includegraphics[width=5cm]{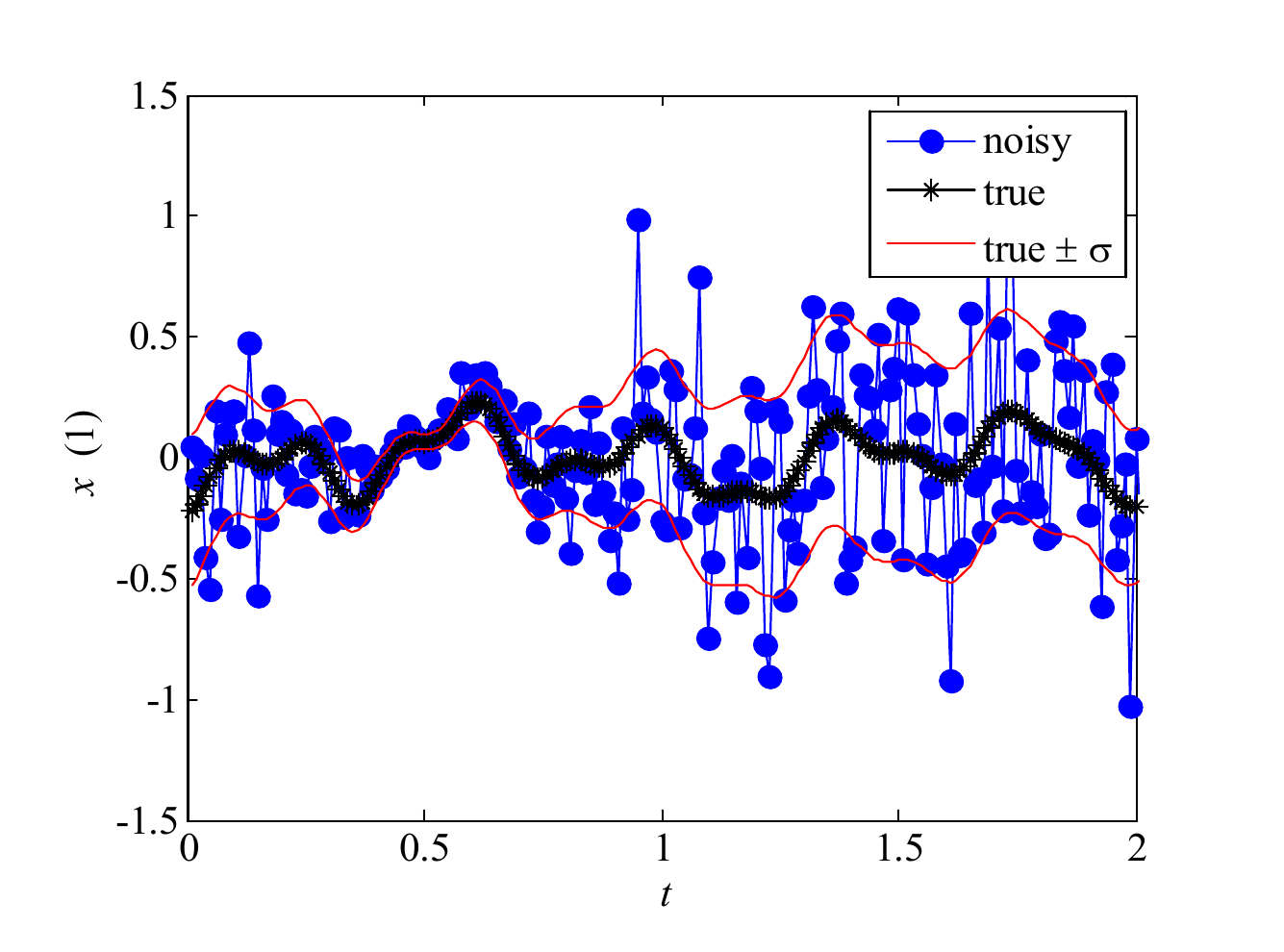}}
	\caption{The noisy and true time history of first node of data matrix for the test problem with time-varying noise. Here, only the first 200 steps of the entire data matrix is illustrated. }
	\label{fig:sinhist}
\end{figure}

Next, the performance of KFDMD for the problem with time-dependent noise with the known characteristics of  $\sigma^2 = \sigma_0^2 \cdot \left( 1.01 - \sin \left( \pi \Delta t k\right)\right)$ is investigated, whereas the other problem settings are the same as in the previous problem. Here, $k$ represents the $k$th time step. The examples of noisy data of first component of the data matrix $X$ and corresponding true data, and standard deviation $\pm \sigma$ are shown in Fig. \ref{fig:sinhist}, similar to the steady noise problem.  Figure \ref{fig:sinhist} shows the strength of the time varying noise and the noise sometimes disappears as defined.   We give the exact information of the noise characteristics to the KFDMD algorithm and set $R=I\sigma^2 = I\sigma_0^2 \cdot \left( 1.01 + \cos\left( \pi \Delta t k\right)\right)$. 

\begin{figure}
	\subfigure[$\sigma_0^2=0.0001$]{\includegraphics[width=5cm]{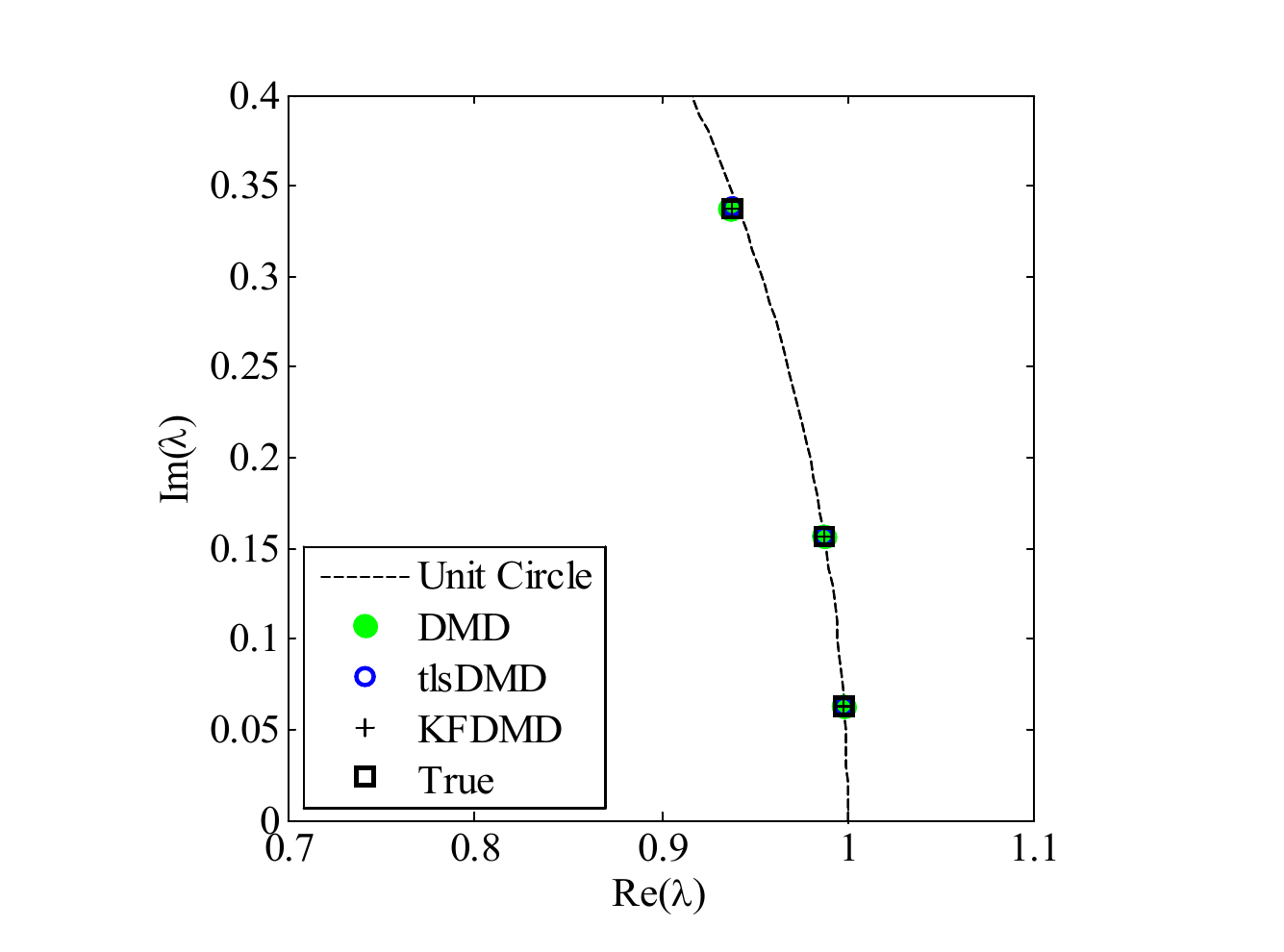}}
	\subfigure[$\sigma_0^2=0.001$ ]{\includegraphics[width=5cm]{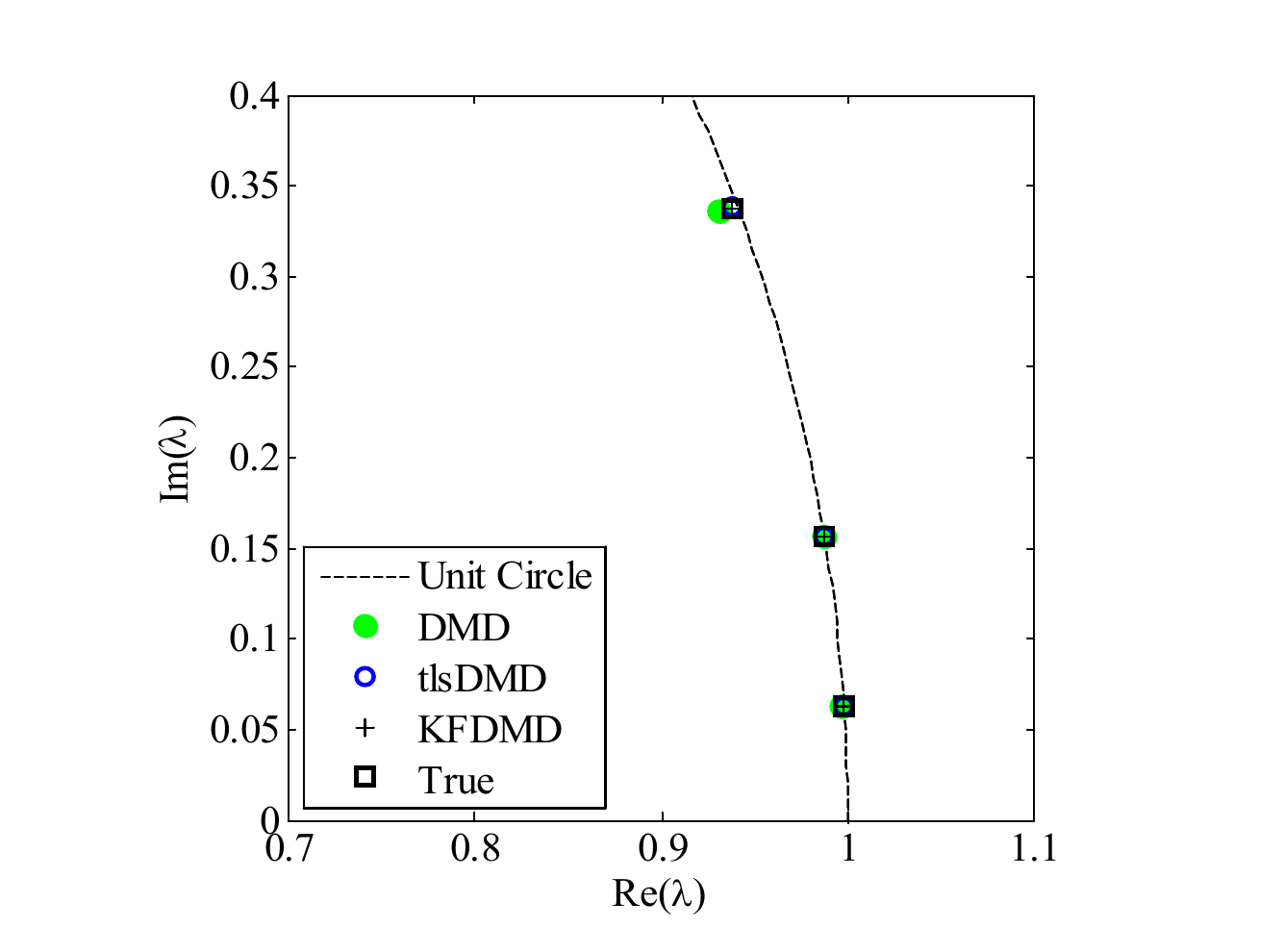}}\\
	\subfigure[$\sigma_0^2=0.01$  ]{\includegraphics[width=5cm]{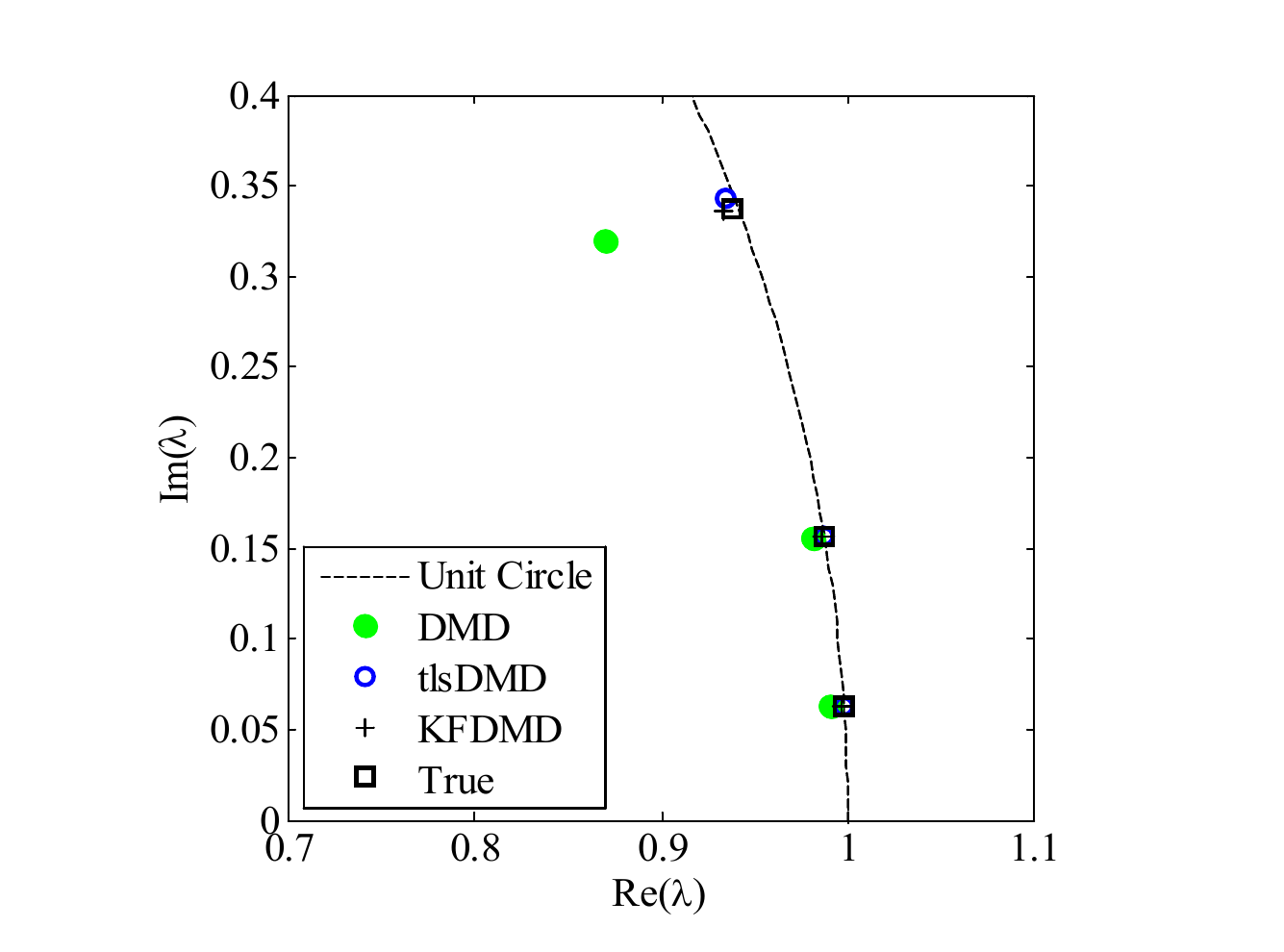}}
	\subfigure[$\sigma_0^2=0.1$   ]{\includegraphics[width=5cm]{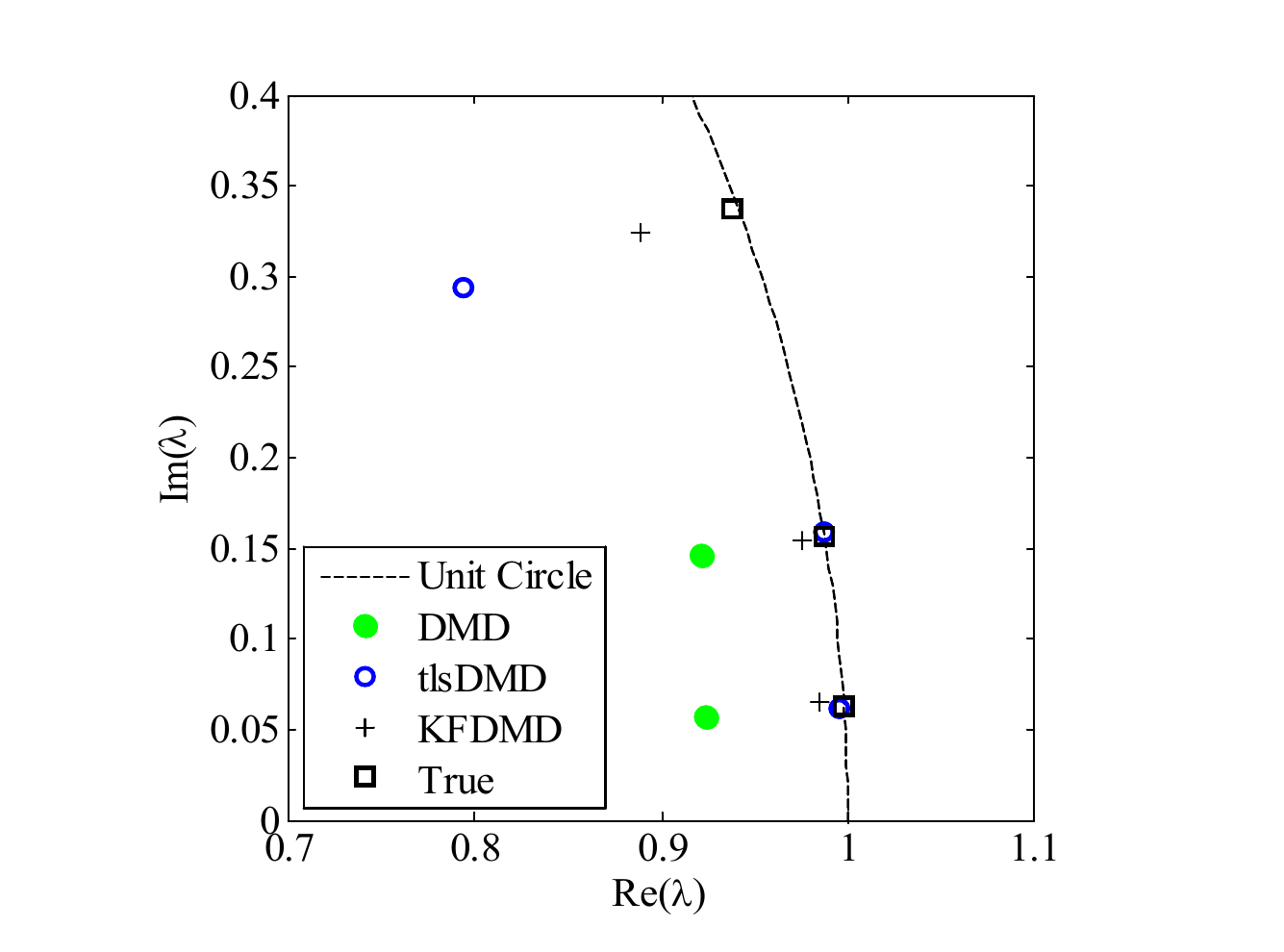}}
	\caption{Representative results of eigenvalues computed in the test problem with time-varying noise. }
	\label{fig:sineigenexam}
\end{figure}

\begin{figure}
	\subfigure[$\sigma_0^2=0.0001$]{\includegraphics[width=5cm]{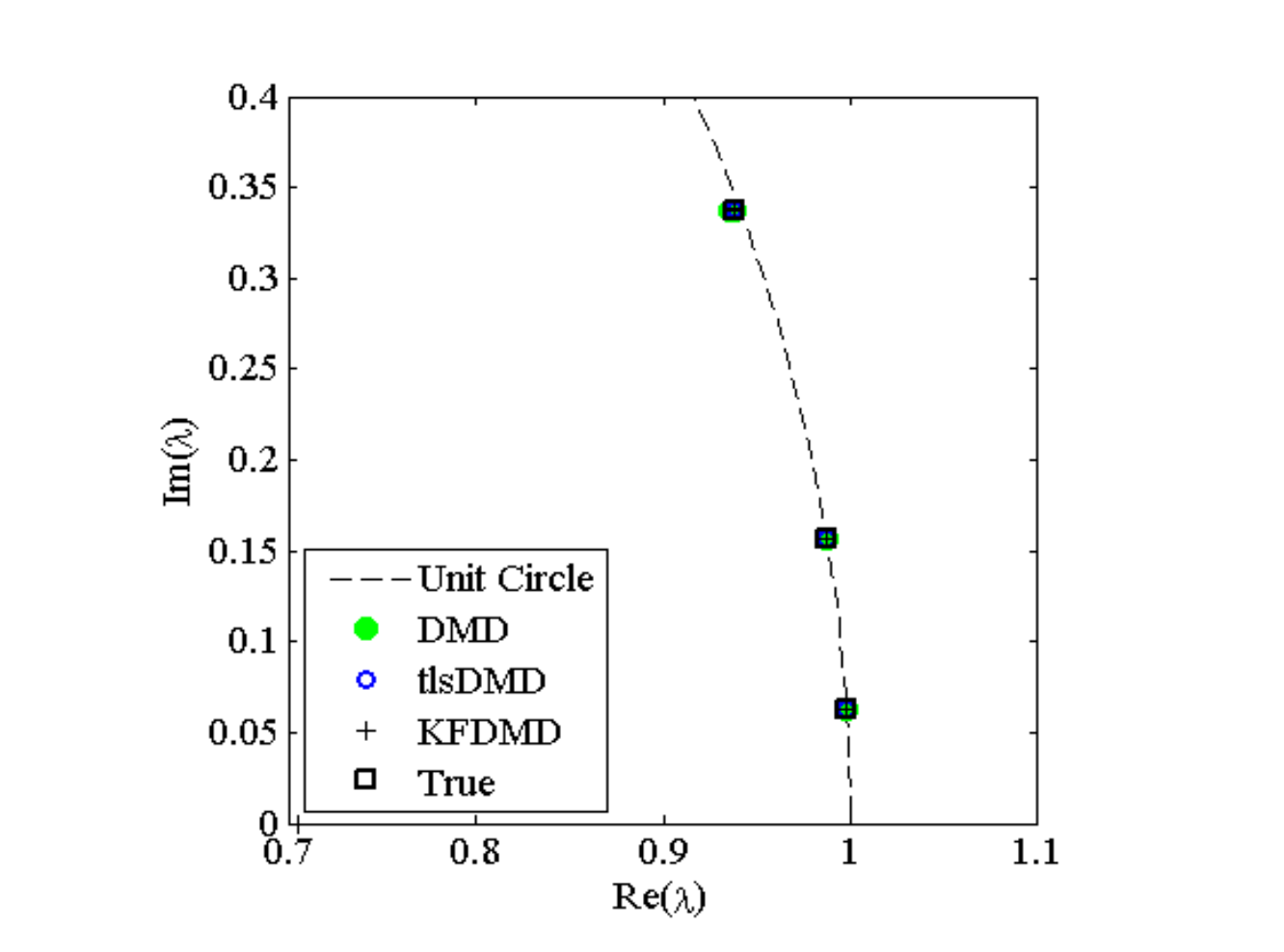}}
	\subfigure[$\sigma_0^2=0.001$ ]{\includegraphics[width=5cm]{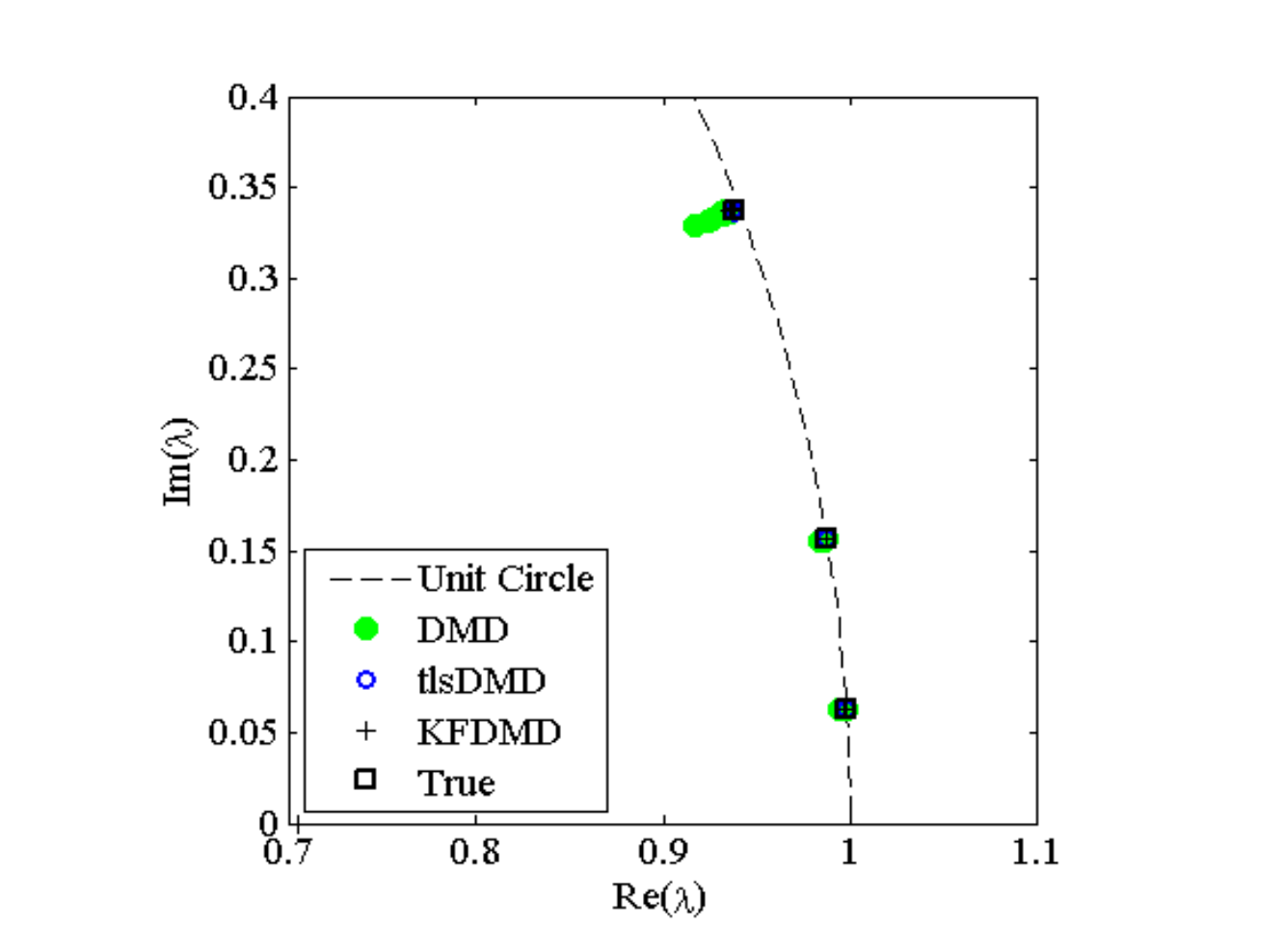}}\\
	\subfigure[$\sigma_0^2=0.01$  ]{\includegraphics[width=5cm]{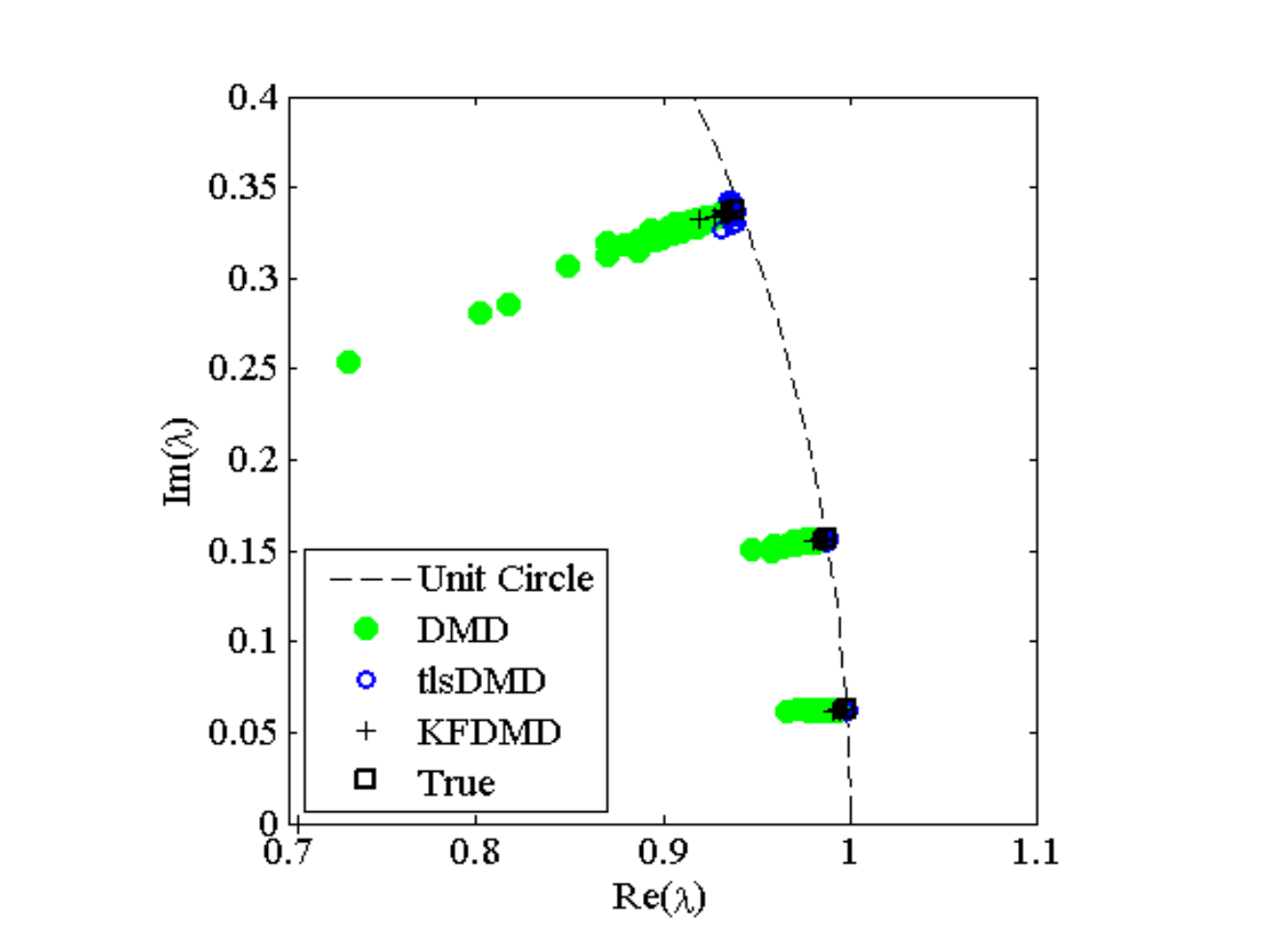}}
	\subfigure[$\sigma_0^2=0.1$   ]{\includegraphics[width=5cm]{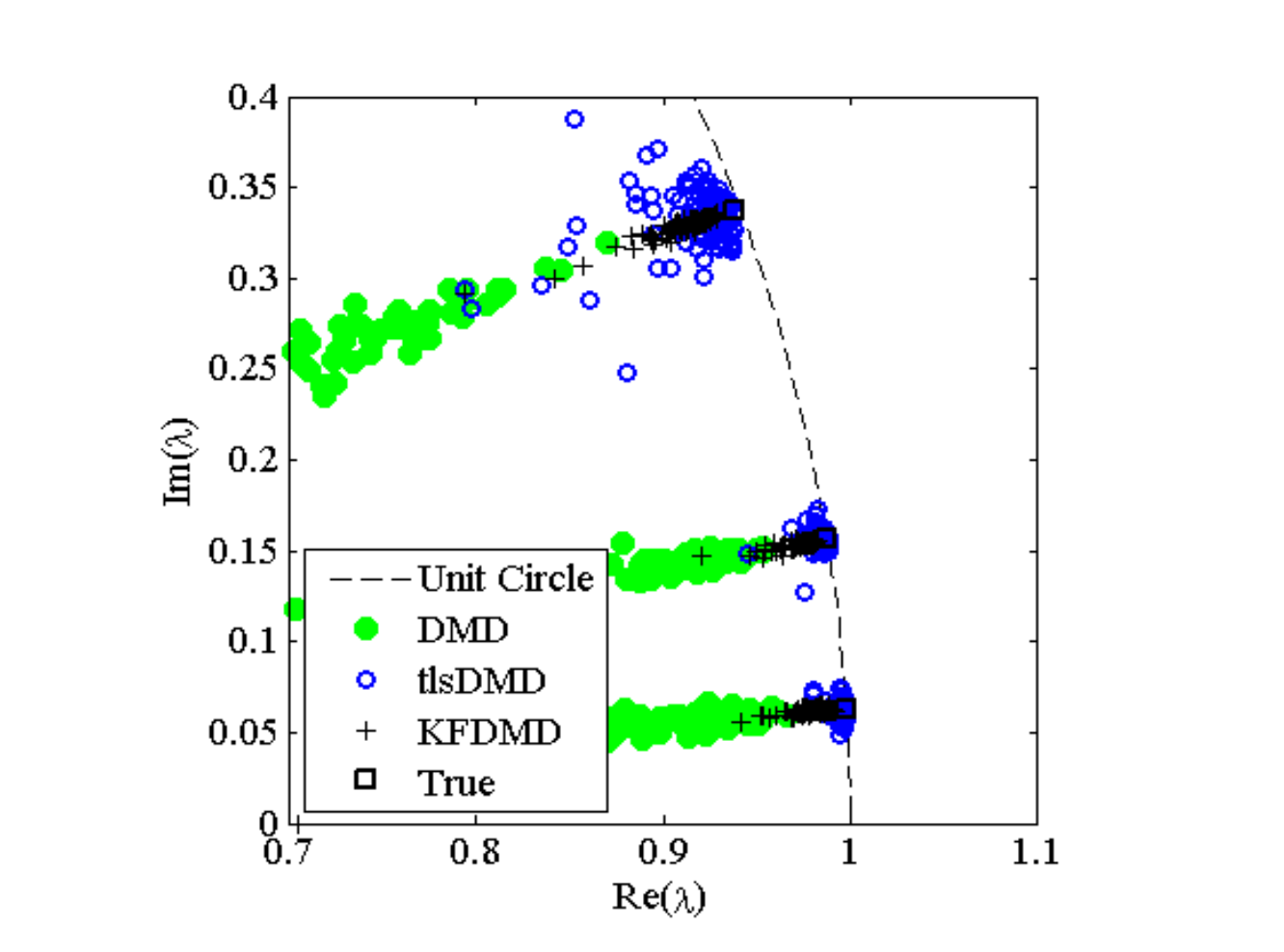}}
    \caption{Results of 100 computations of eigenvalues in the test problem with time-varying noise.}	
	\label{fig:sineigen}
\end{figure}

\begin{figure}
	\subfigure[$\lambda_1$]{\includegraphics[width=5cm]{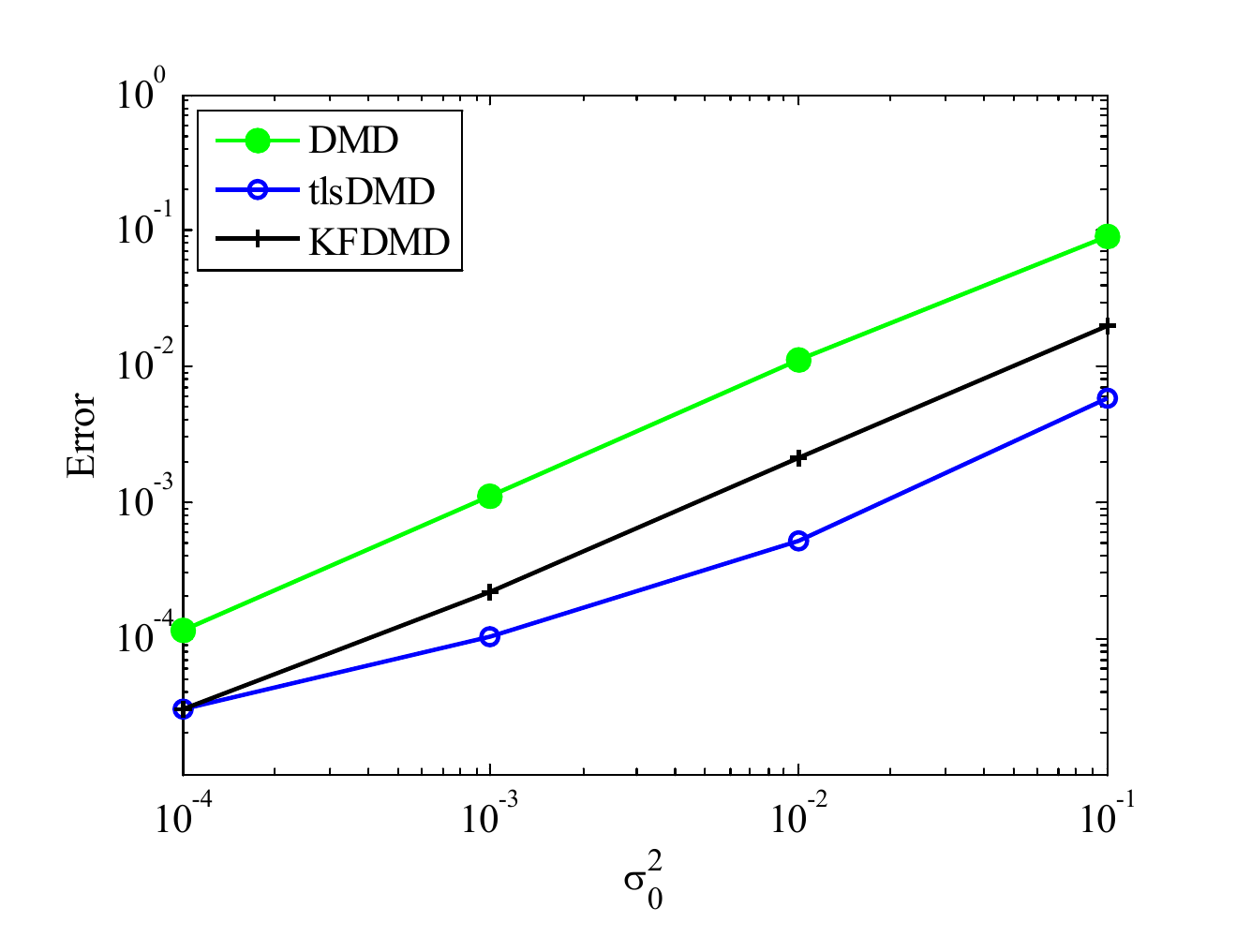}}
	\subfigure[$\lambda_2$ ]{\includegraphics[width=5cm]{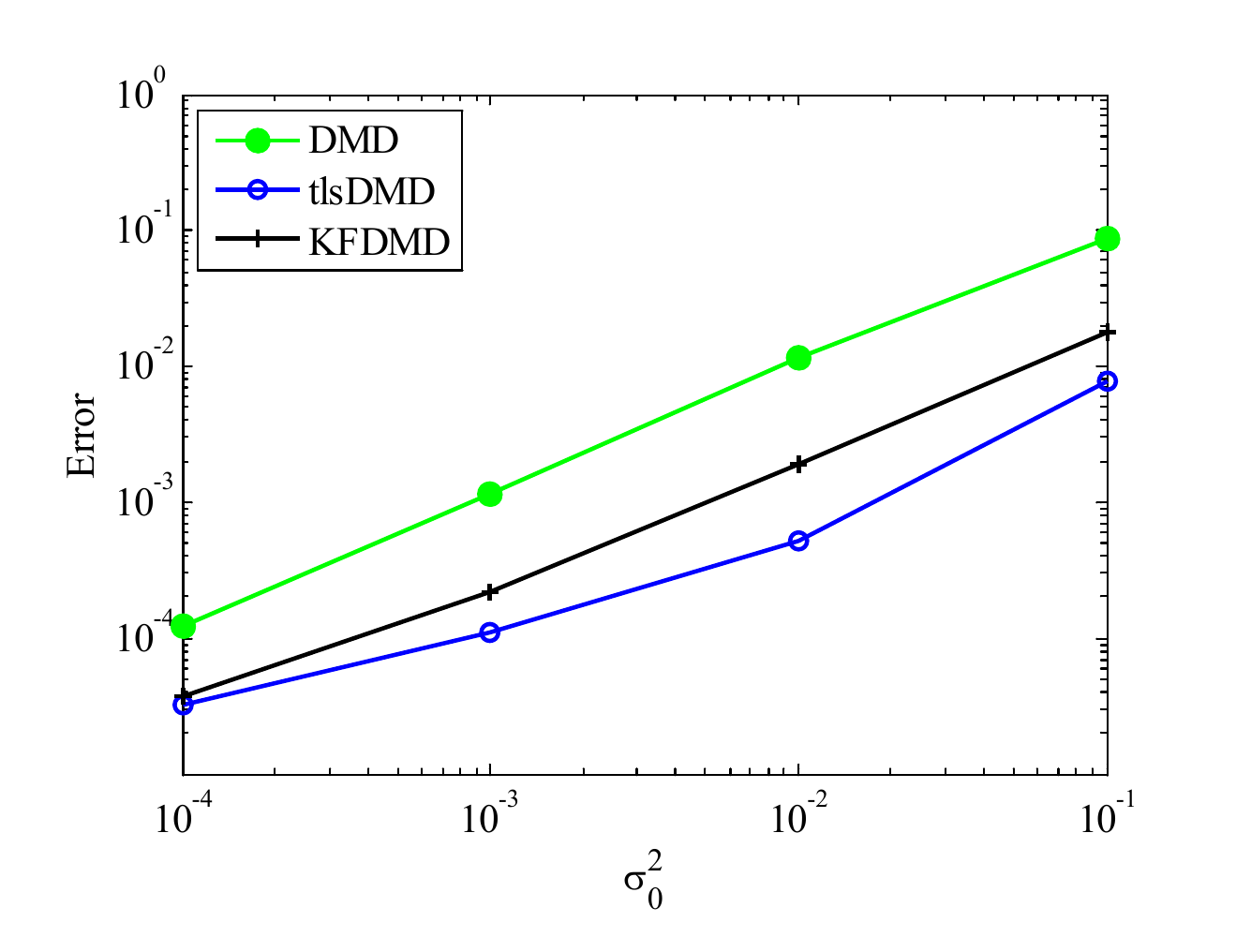}}
	\subfigure[$\lambda_3$  ]{\includegraphics[width=5cm]{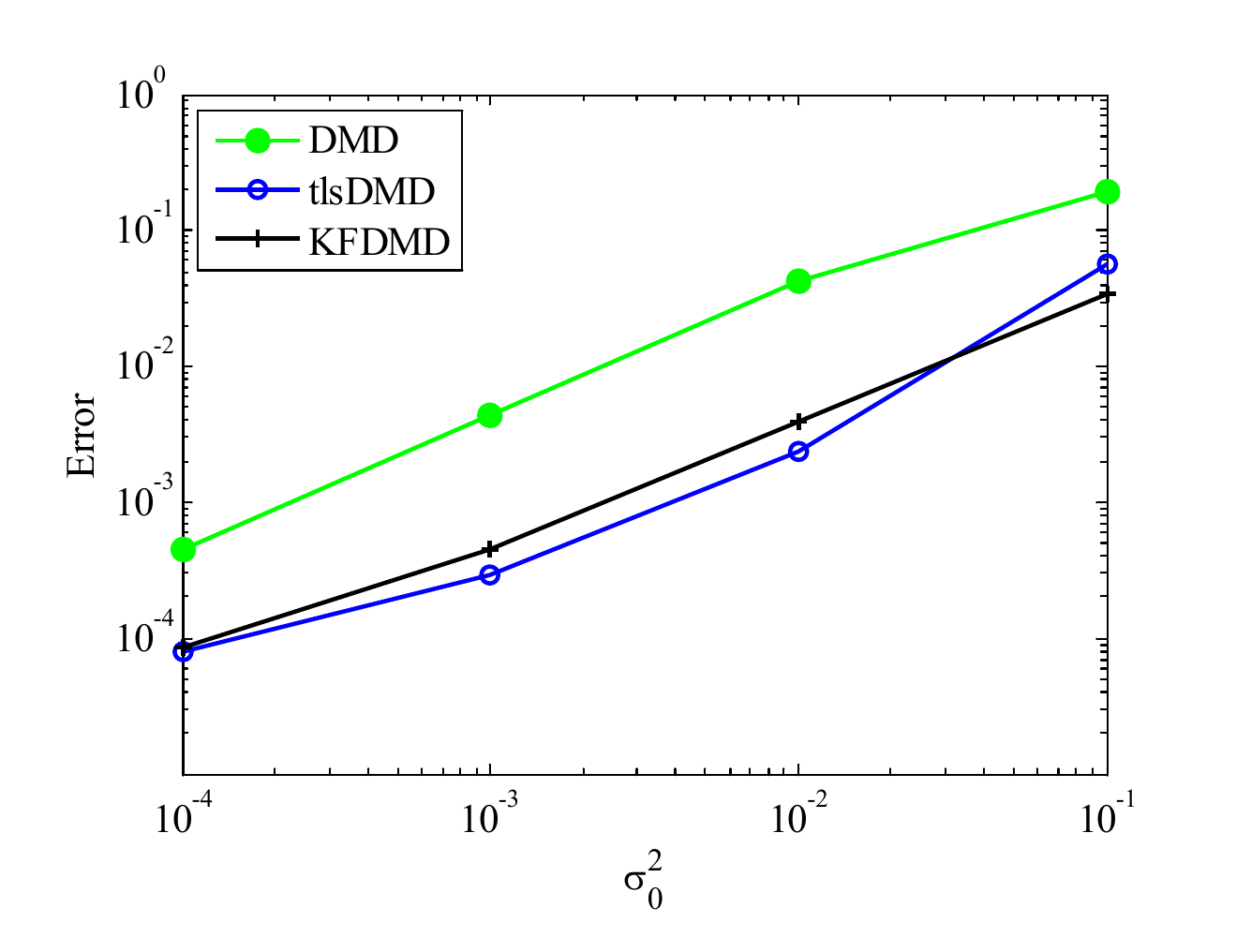}}
   \caption{Errors in the eigenvalues computed in the test problem with time-varying noise. Here, $L_2$ error is averaged with 100 test cases. }	
   \label{fig:sineigenerror}
\end{figure}

Figures \ref{fig:sineigenexam} and \ref{fig:sineigen} show the eigenvalue estimation results with $m=500$ for the representative case and all the 100 cases respectively.
As similar to the previous problem, all the method can predict the eigenvalues when the noise level is low, similar to the previous problem.
For the discussion on the anti-noise characteristics, the condition with high noise level is focused. 
The eigenvalues of standard DMD are widely scattered and there are large discrepancies between true values and values estimated by the standard DMD. This also shows that the standard DMD is not strong for the noise. Although the eigenvalues of tlsDMD shows much better estimations than standard DMD, the third mode is sometimes not captured in the most severe condition ($\sigma^2_0=0.1$), as shown in the Fig. \ref{fig:sineigenexam} and the results of 100 computations in Fig. \ref{fig:sineigen} show that the third eigenvalue estimated by tlsDMD for $\sigma_0^2=0.1$ are substantially scattered. These figures illustrate that the eigenvalues of KFDMD are significantly better than the standard DMD, but slightly worse than tlsDMD for the first and second eigenvalues, and the less severe condition ($\sigma^2_0<0.1$) of the third eigenvalues. On the other hand, the eigenvalues of the third mode estimated by KFDMD  in the most severe condition ($\sigma^2_0=0.1$) are in better agreement with the true values and less scattered than the eigenvalues estimated by tlsDMD. Errors in eigenvalues are plotted in Fig. \ref{fig:sineigenerror} for more quantitative discussion. As similar to the previous errors, errors in eigenvalues decrease with decreasing the noise level. Comparing the methods, errors in KFDMD is intermediate between standard DMD and tlsDMD for the first and second eigenvalues, and the less severe condition ($\sigma^2_0<0.1$) of the third eigenvalues. However, the error of KFDMD becomes smaller for the third eigenvalues of the most severe condition than those of the other methods. 
In KFDMD, the information of the noise characteristics is used, and the eigenvalue is accurately estimated, using the noise information. This implies that the KFDMD likely weights more for the data with low noise level and weights less for the data with high noise level.
This example shows that KFDMD has higher flexibility of treatment for the noise added in the observation as compared to the standard DMD, and KFDMD finds the eigenvalues for quiet high noise level with which tlsDMD even fails to find them.

In the latter of this subsections, the effects of the parameters of KFDMD are investigated. 


\subsubsection{Effects of number of snapshots $m$}
\label{sec:ENS}
\begin{figure}
	\subfigure[$m=200$]{\includegraphics[width=5cm]{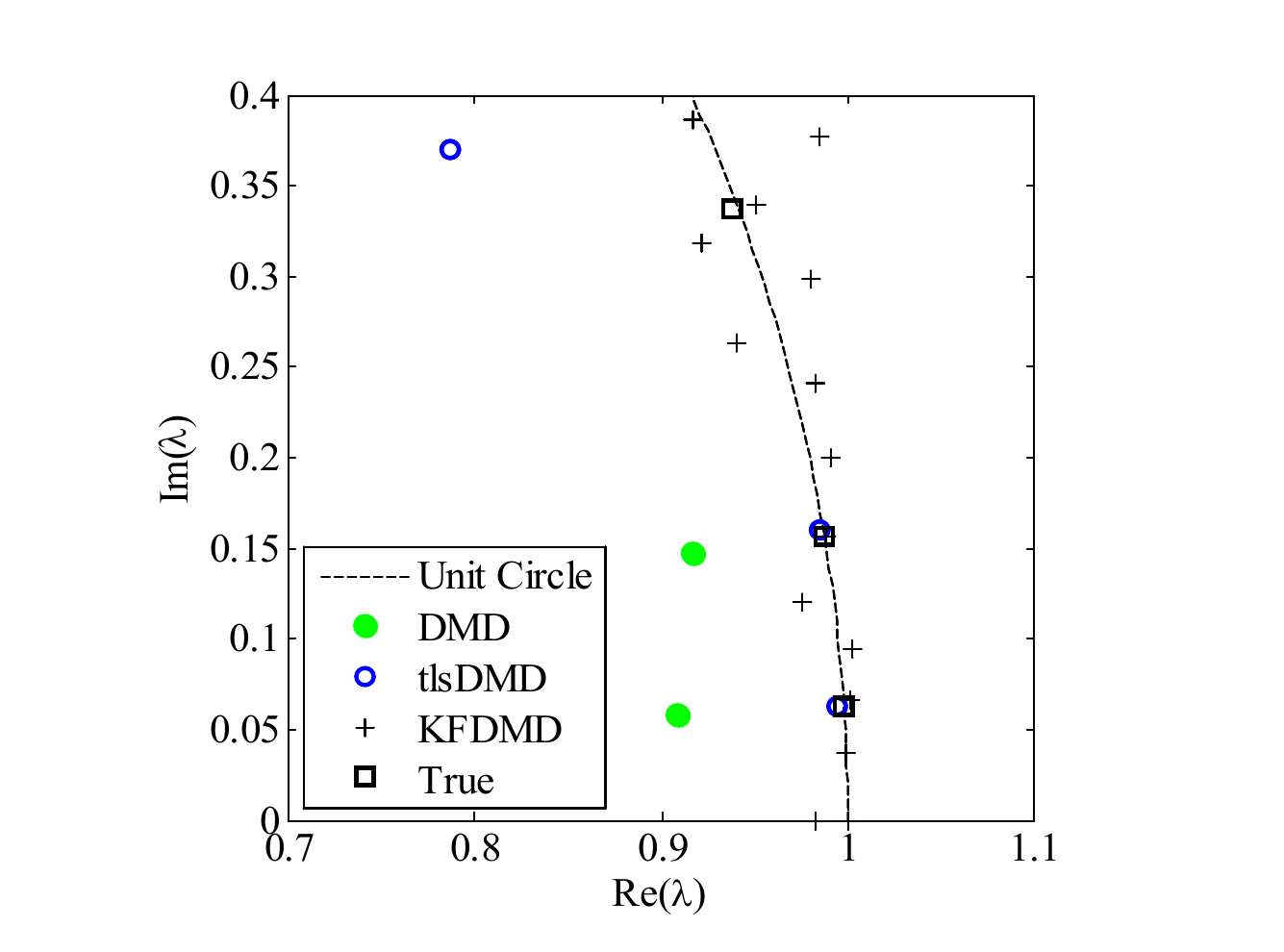}}
    \subfigure[$m=400$ ]{\includegraphics[width=5cm]{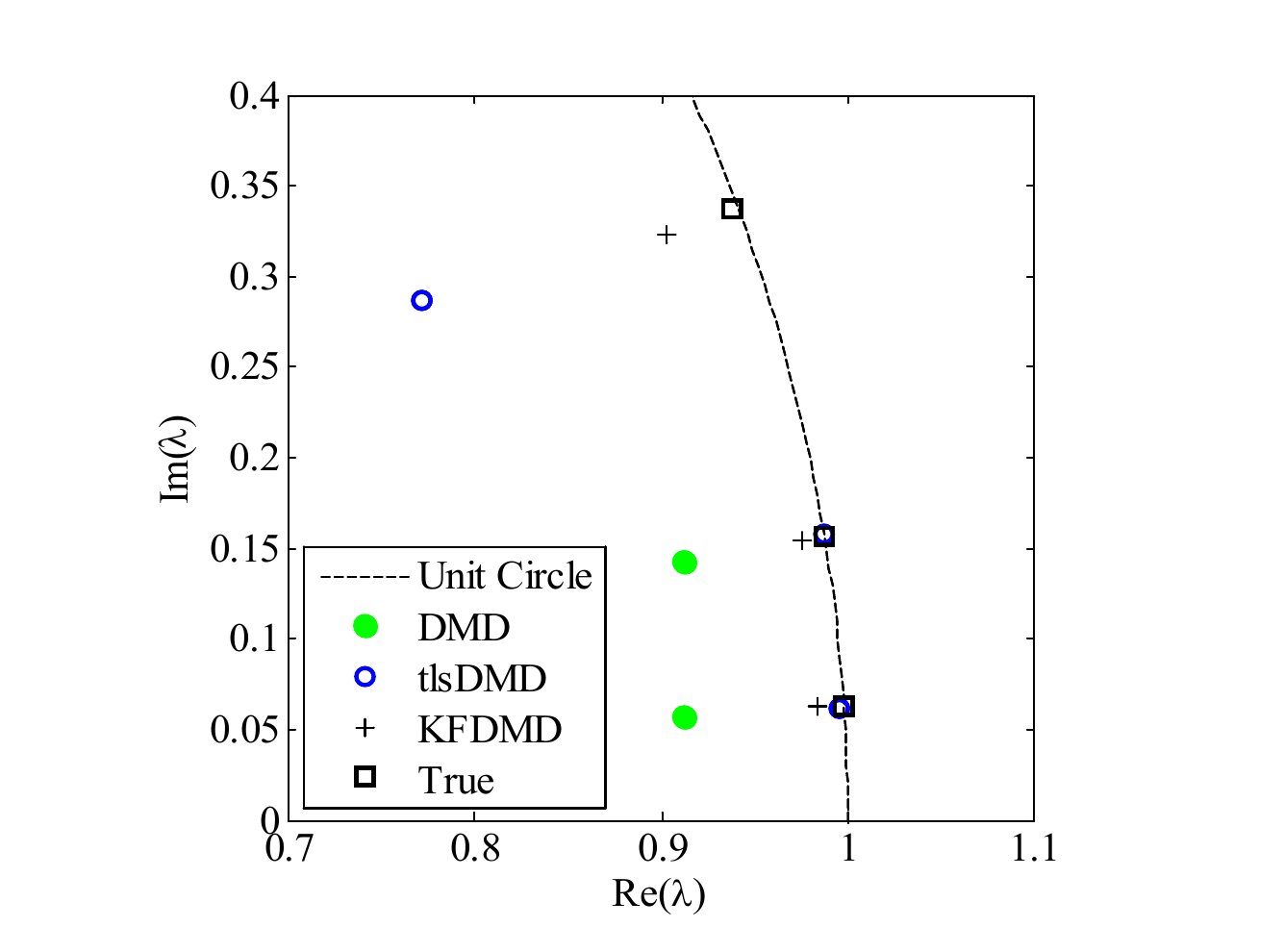}}
    \subfigure[$m=800$ ]{\includegraphics[width=5cm]{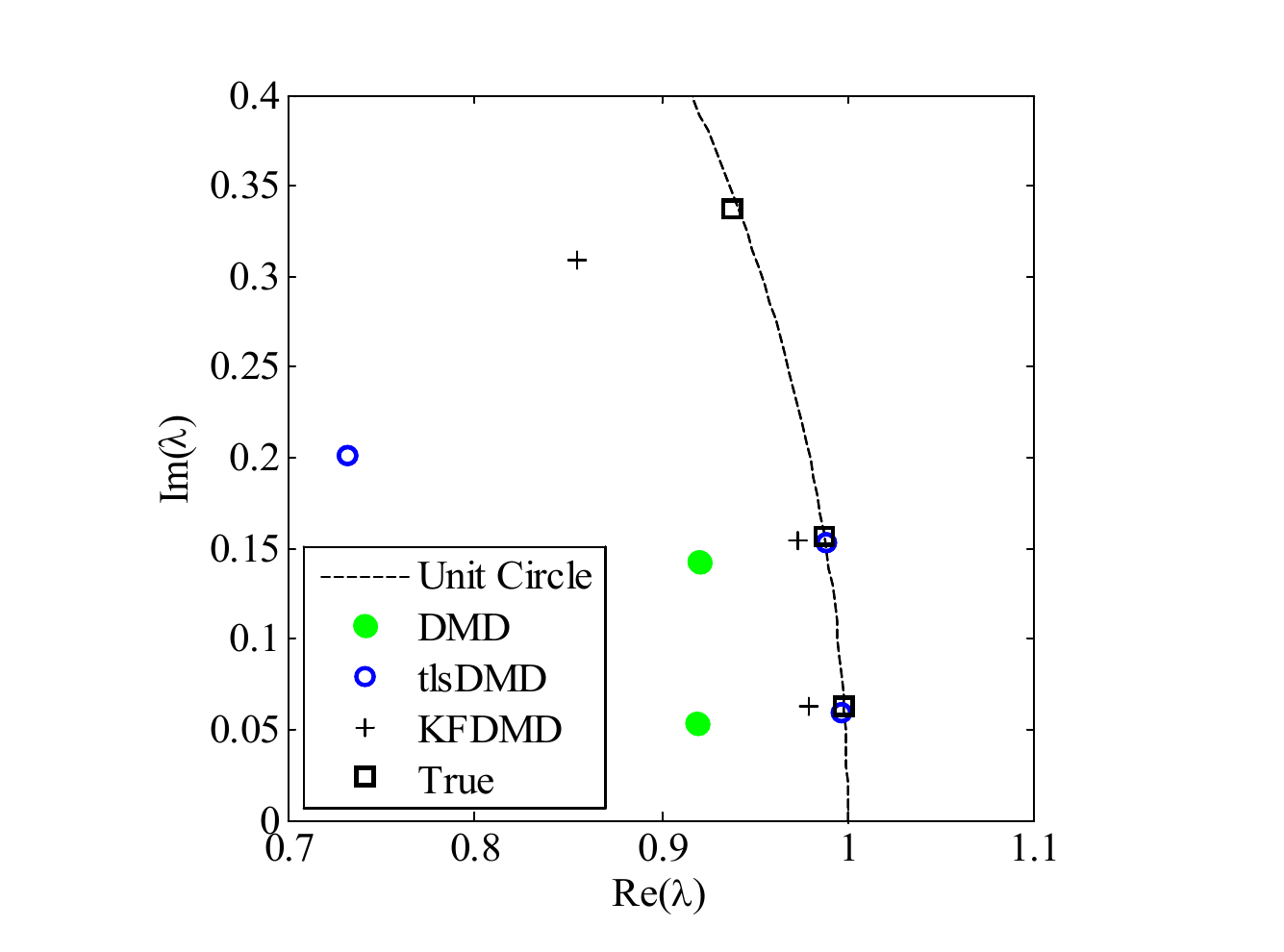}}
	    \caption{Effects of a number of snapshots $m$  on the representative result of eigenvalue computation in the test problem with time-varying noise of $\sigma_0^2=0.1$.}	
	\label{fig:diffm_eigenexam}
\end{figure}

\begin{figure}
	
	\subfigure[$m=200$]{\includegraphics[width=5cm]{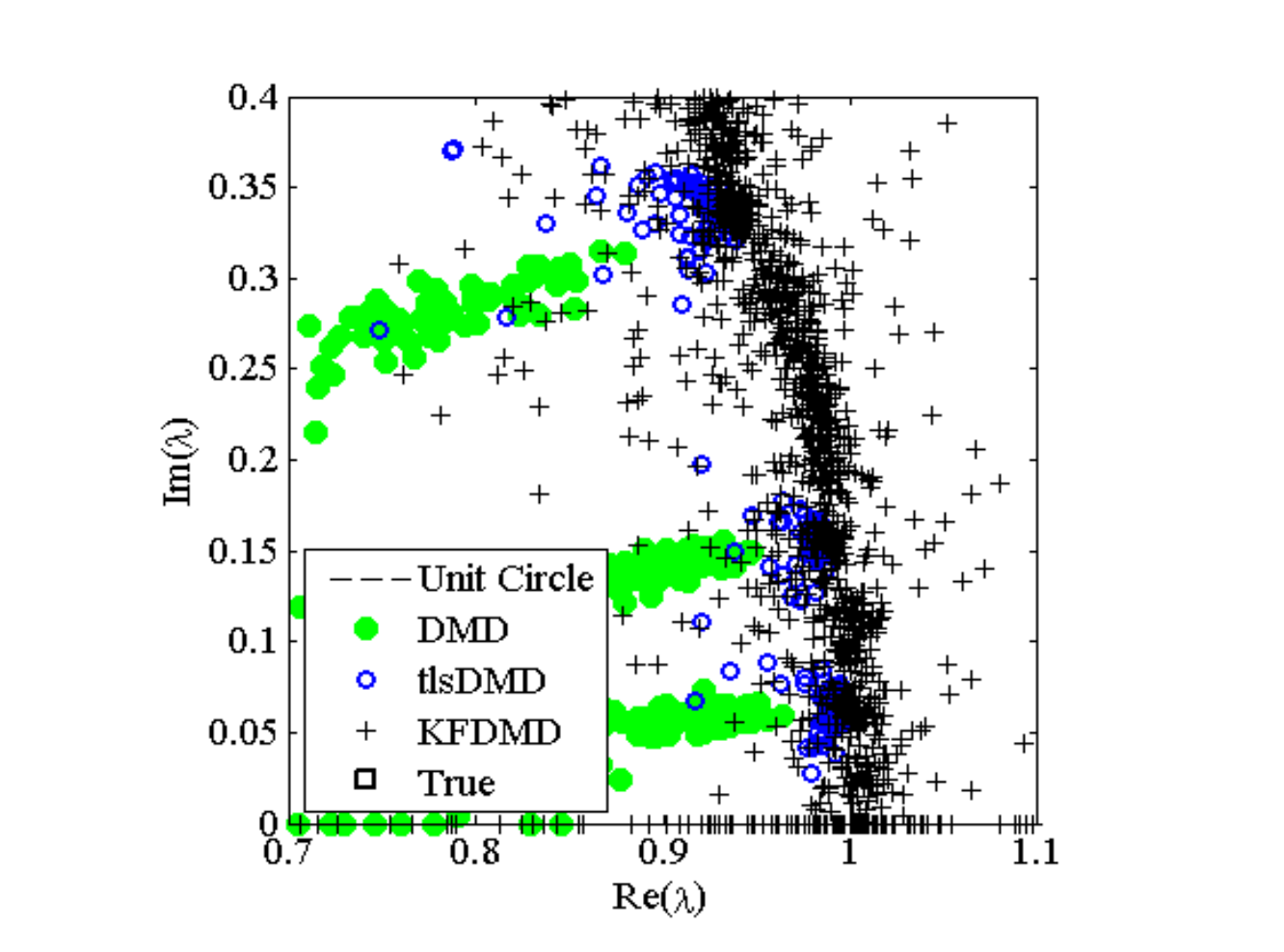}}
    \subfigure[$m=400$ ]{\includegraphics[width=5cm]{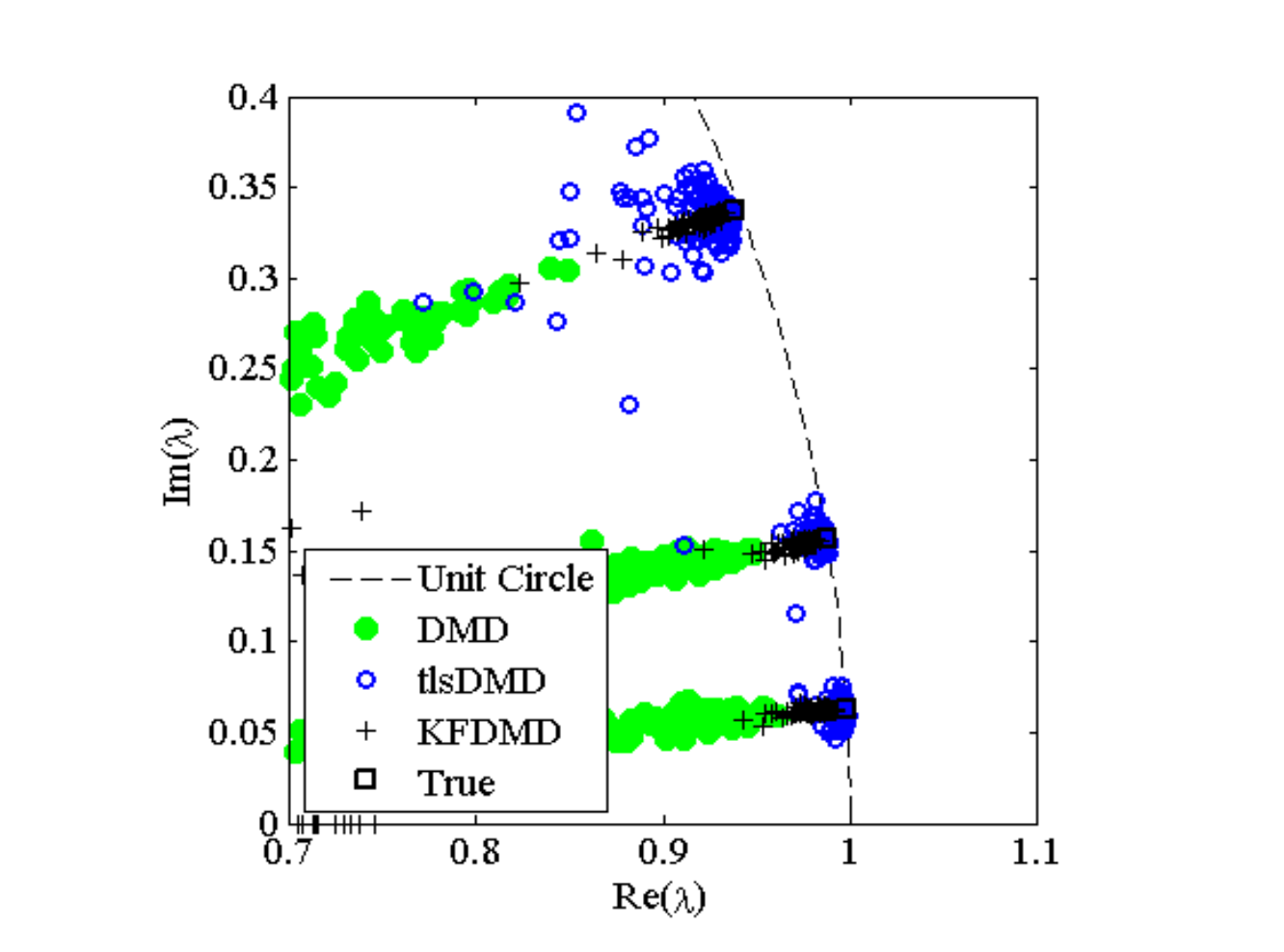}}
    \subfigure[$m=800$ ]{\includegraphics[width=5cm]{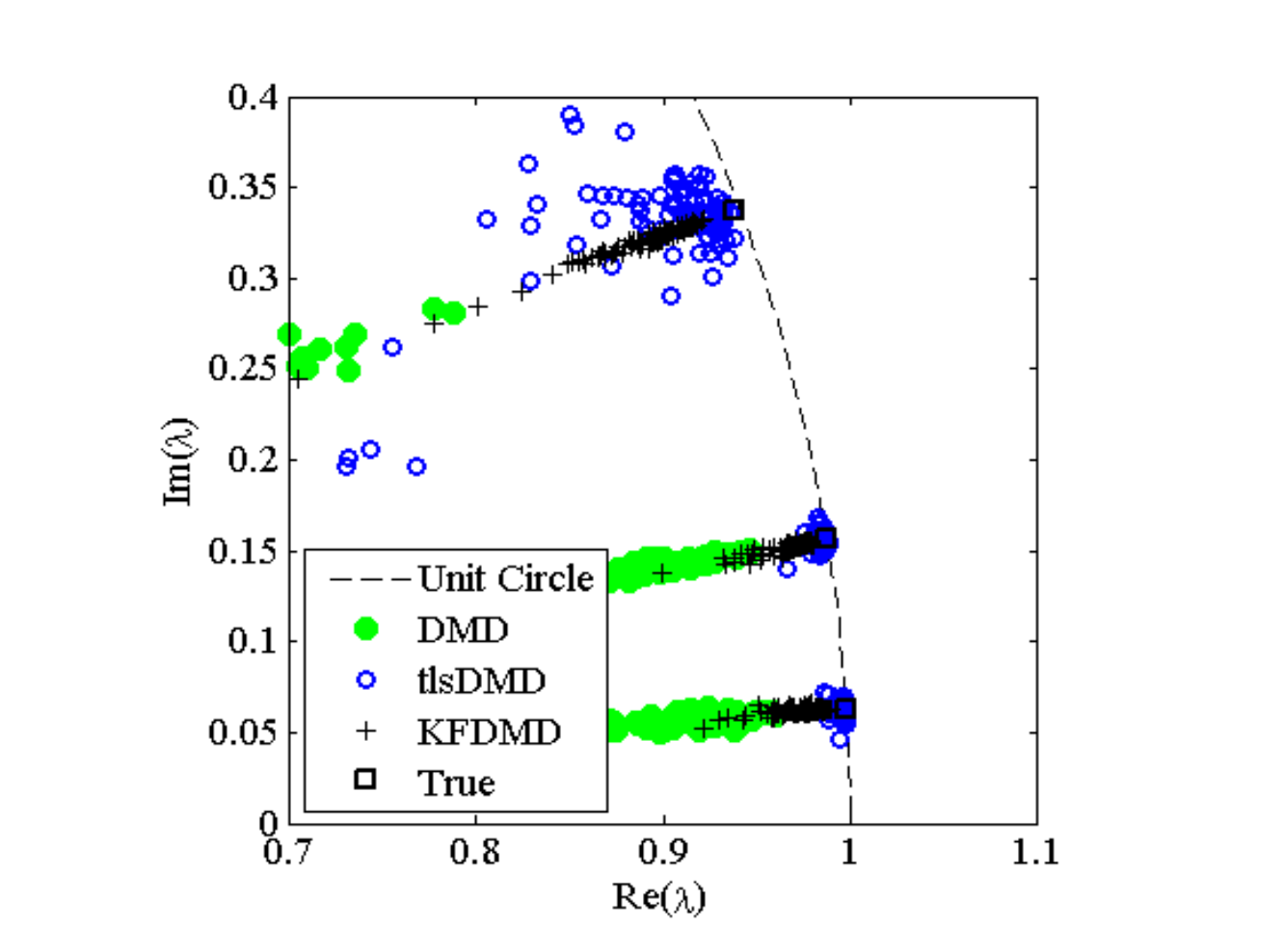}}
	    \caption{Effects of a number of snapshots $m$  on results of 100 computations of eigenvalues in the test problem with time-varying noise of $\sigma_0^2=0.1$.}	
	\label{fig:diffm_eigen}
\end{figure}

\begin{figure}
	\subfigure[$\lambda_1$]{\includegraphics[width=5cm]{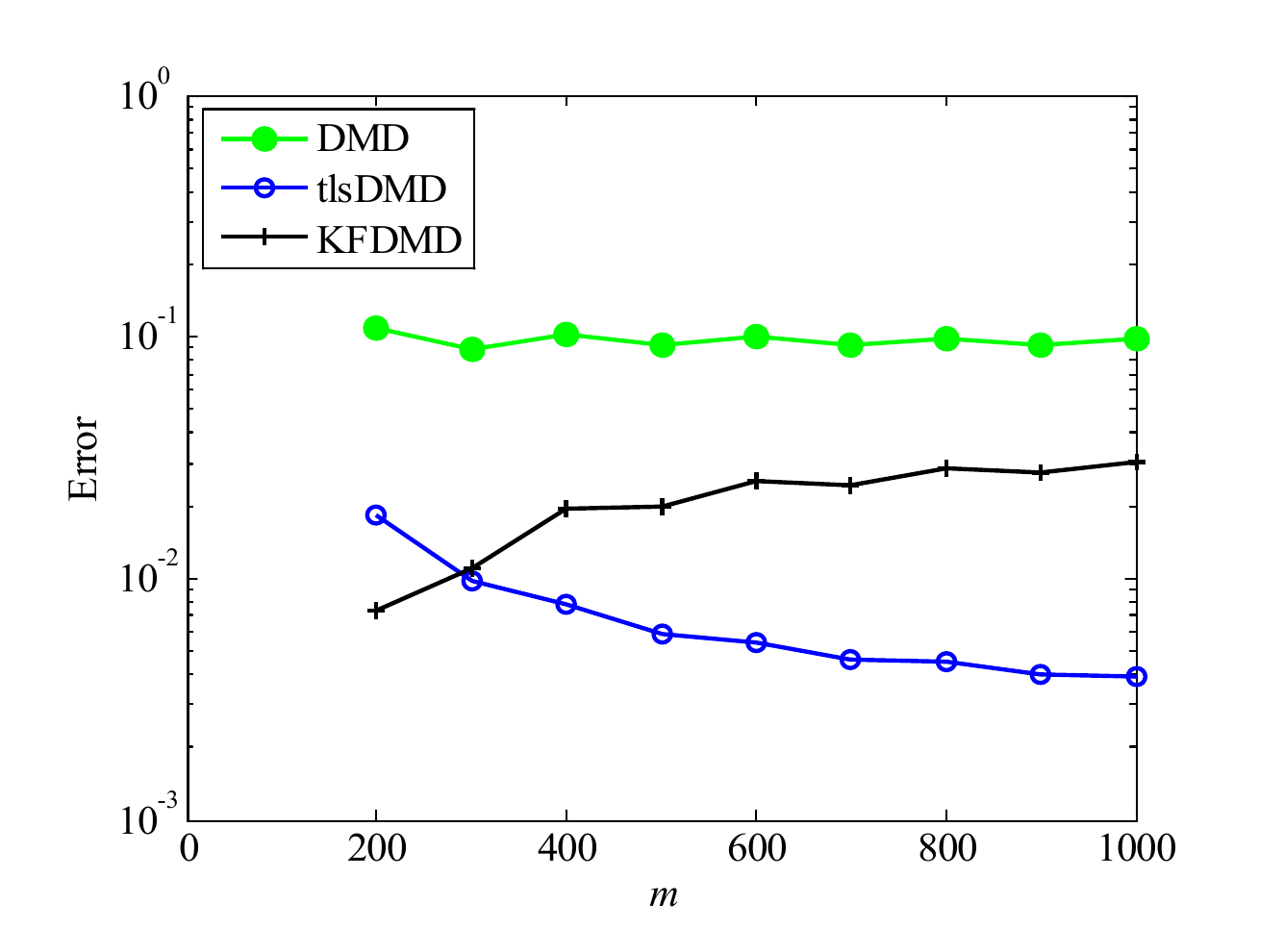}}
	\subfigure[$\lambda_2$ ]{\includegraphics[width=5cm]{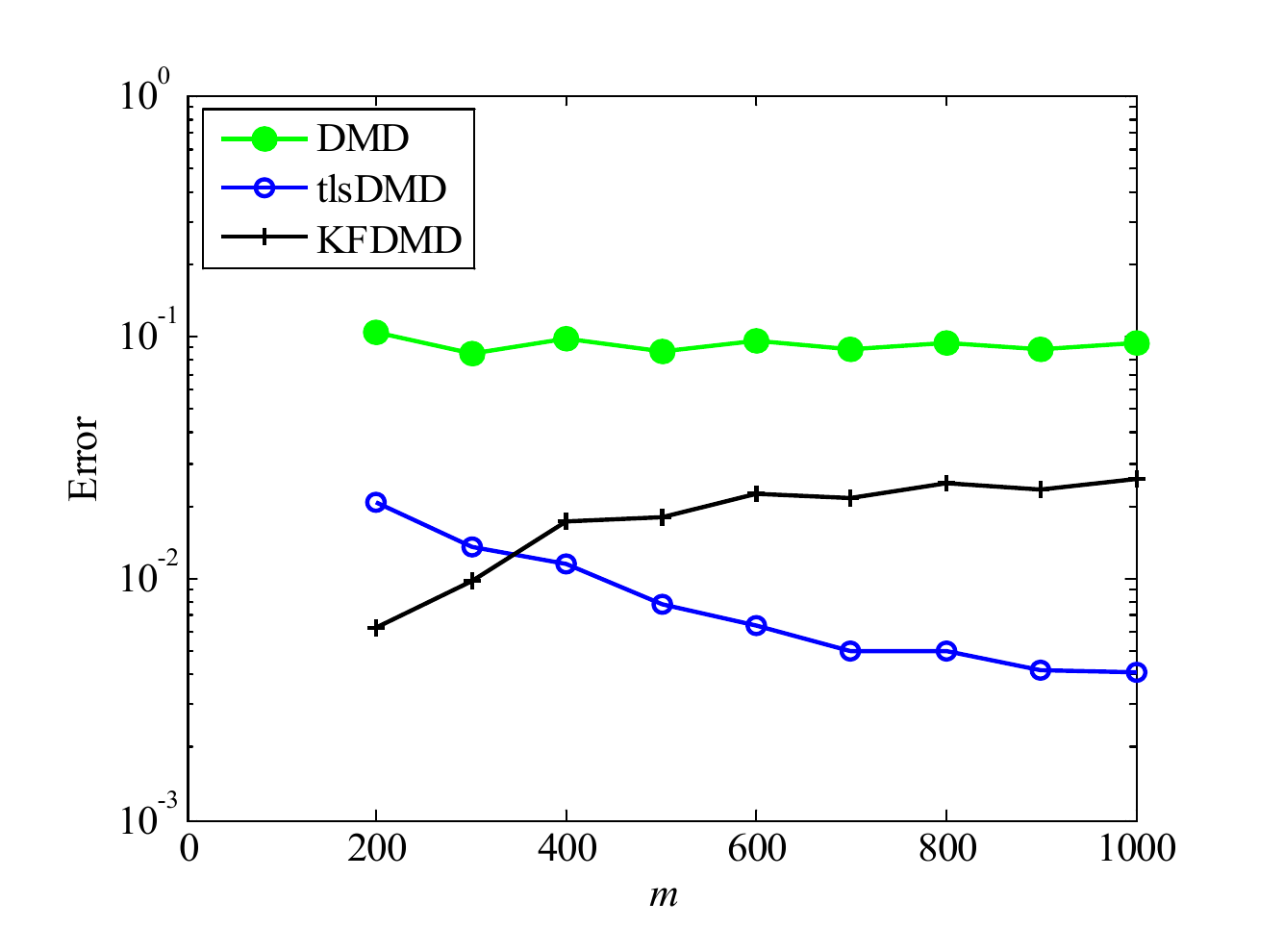}}
	\subfigure[$\lambda_3$  ]{\includegraphics[width=5cm]{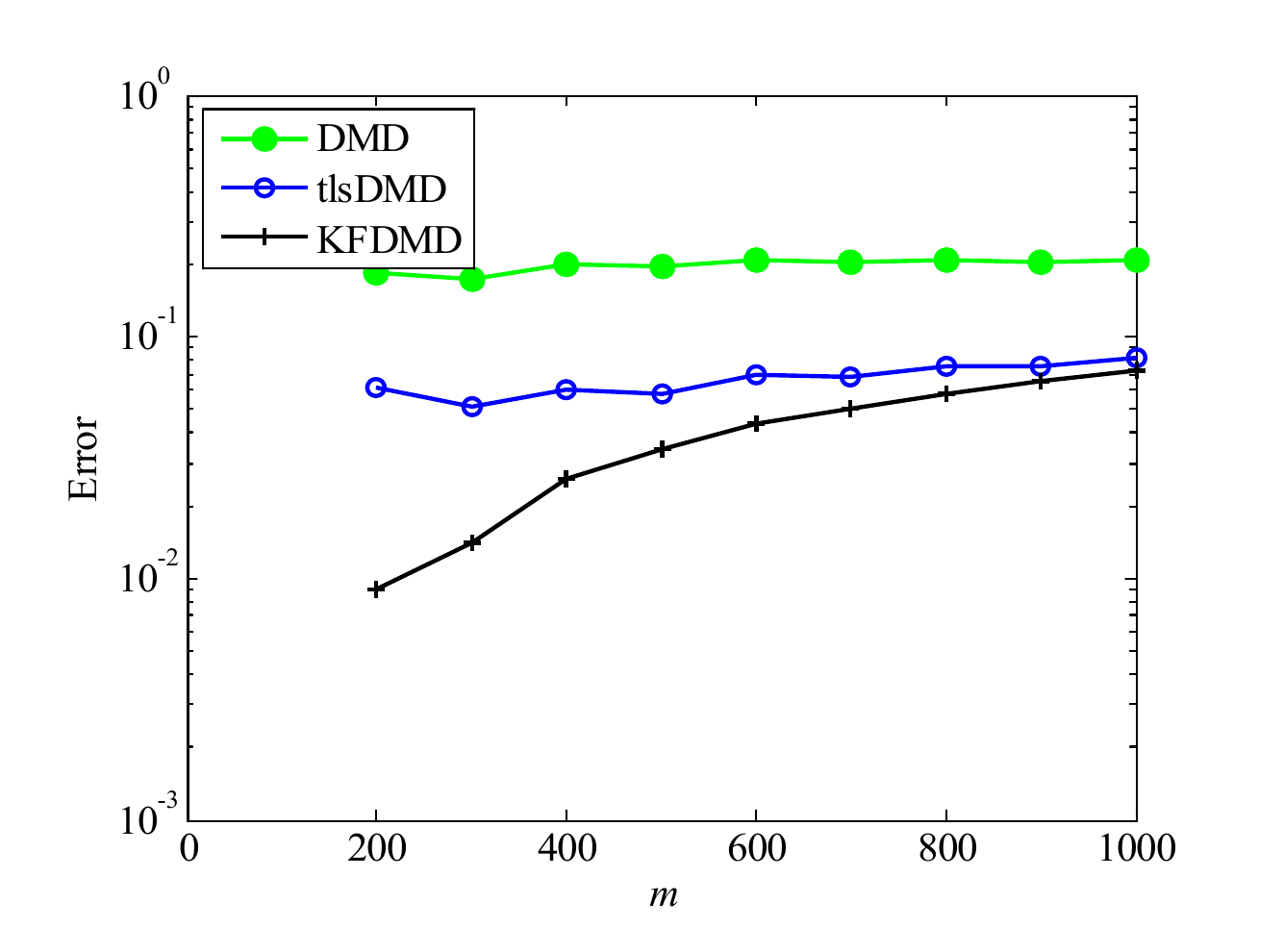}}
   \caption{Effects of a number of snapshots $m$ on the errors in the eigenvalues computed in the test problem with time-varying noise. Here, $L_2$ error is averaged with 100 test cases. }	
	\label{fig:diffm_eigenerror}
\end{figure}

The effects of the number of snapshots are investigated. The number of snapshots $m$ is set to 200, 300, 400, 500 ,600, 700 ,800, 900, and 1000 in the tests. The eigenvalues computed using standard DMD, KFDMD, and tlsDMD are shown in Figs. \ref{fig:diffm_eigenexam} and \ref{fig:diffm_eigen}. The figures show that many spurious eigenvalues of KFDMD appear for $m=200$, but they decrease with increasing $m$. This is supported by the fact that KFDMD for $m=200$ in Fig. \ref{fig:diffm_eigenexam}(a) has more than three eigenvalues in the plot, but Figs.  \ref{fig:diffm_eigenexam}(b) and (c) only have three eigenvalues in each plot.  The effects of $m$ on the errors of the estimated eigenvalues shown in Fig. \ref{fig:diffm_eigenerror} illustrate that the error for the standard DMD does not change with $m$ and that those of $\lambda_1$ and $\lambda_2$ for tlsDMD decrease with $m$. This might be due to the total least squares working better when the number of samples used increases. The error for KFDMD increases with $m$.  For $m=200$ to $400$, there are many spurious eigenvalues and the errors are not correctly estimated. The errors in $\lambda_1$ and $\lambda_2$ do not change much after the spurious eigenvalues disappear for $m \ge 500$. This characteristic at $m \ge 500$ is similar to that for standard DMD. The error in $\lambda_3$ increases with $m$ after the spurious eigenvalues disappear for $m \ge 500$. This might be due to $\lambda_3$ corresponding to the damping mode, which disappears from the data matrix in the latter half. The online algorithm of KFDMD at the latter half (larger $m$) estimates the eigenvalues with more of the latest information, which loses the $\lambda_3$ mode.  This characteristic also leads to the fact that the error in $\lambda_3$ for tlsDMD does not change with $m$, unlike those in $\lambda_1$ and $\lambda_2$. The error in $\lambda_3$ for KFDMD is clearly lower than that for tlsDMD after the spurious eigenvalues disappear for $m \ge 500$ in the range investigated. The proposed algorithm can detect the eigenvalues in highly noisy data by utilizing prior knowledge of noise. 


\subsubsection{Effects of mismatched error level for $R$}
\label{sec:EME}
\begin{figure}
	
	\subfigure[$\sigma_0=0.001$]{\includegraphics[width=5cm]{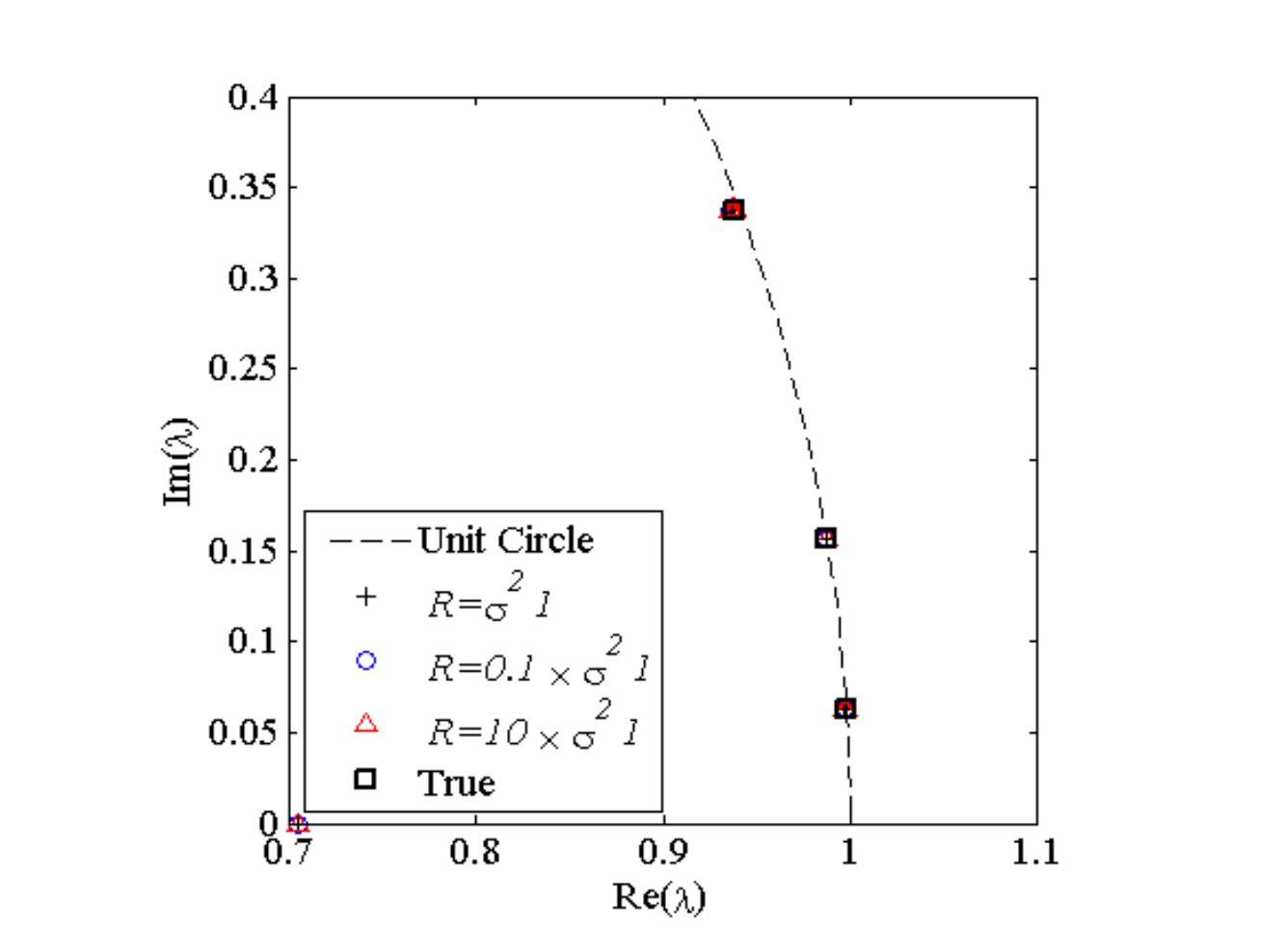}}
	\subfigure[$\sigma_0=0.01$ ]{\includegraphics[width=5cm]{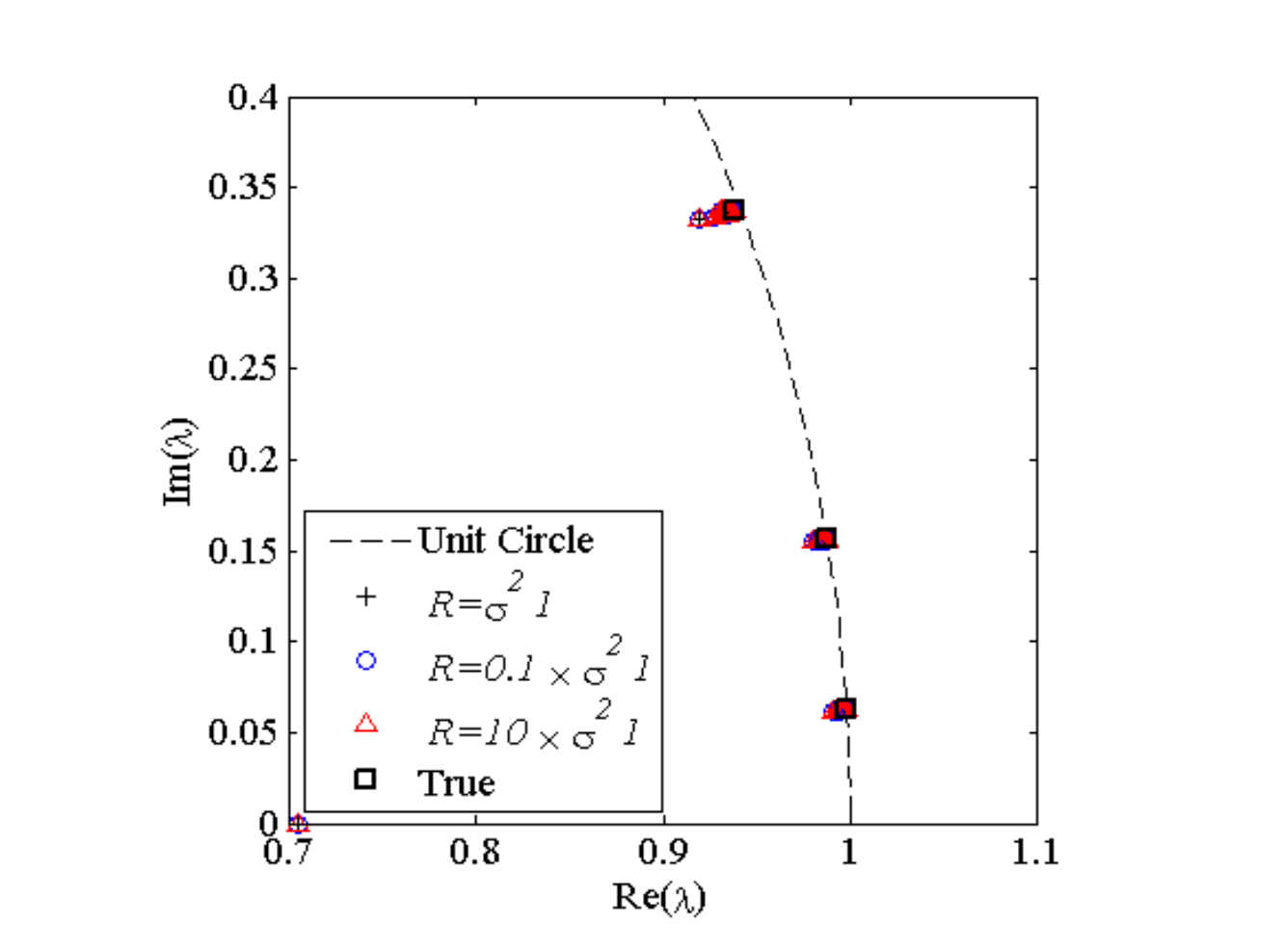}}
	\subfigure[$\sigma_0=0.1$  ]{\includegraphics[width=5cm]{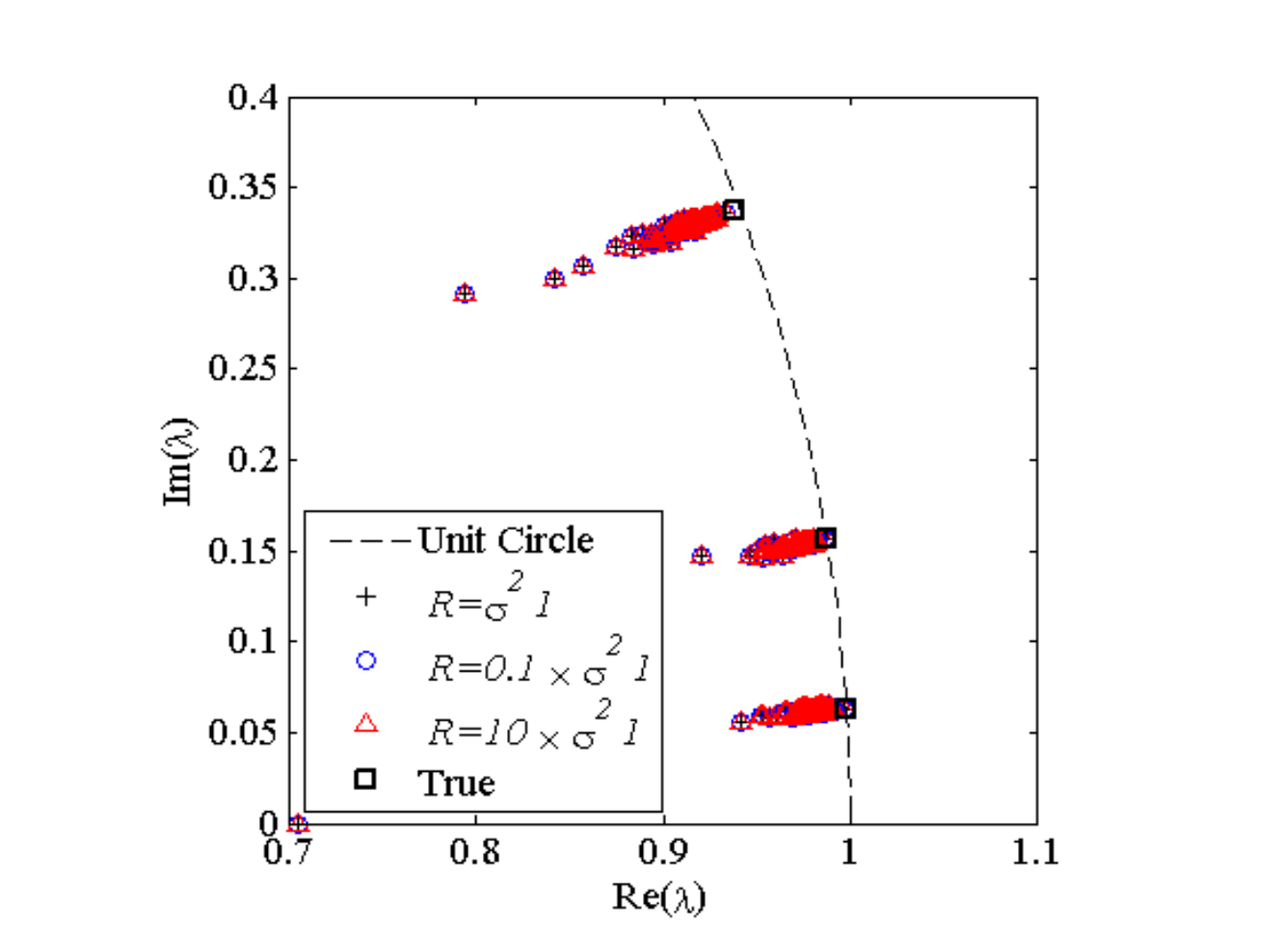}}
    \caption{Effects of mismatched (amplified or attenuated) $R$ on the KFDMD results of 100 computations of eigenvalues in the test problem with time-varying noise.}	
	\label{fig:mis_eigen}
\end{figure}

\begin{figure}
	\subfigure[$\lambda_1$]{\includegraphics[width=5cm]{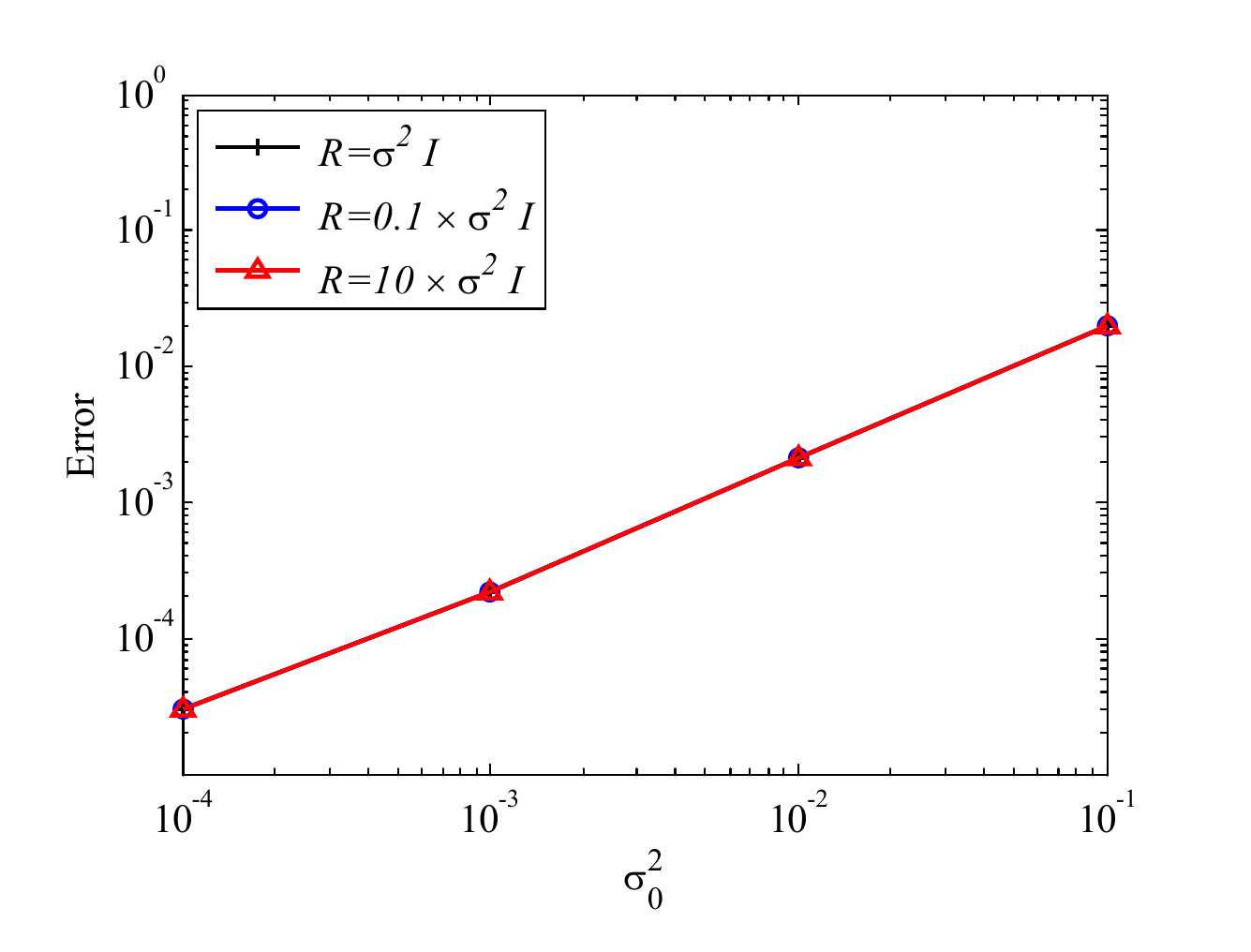}}
	\subfigure[$\lambda_2$]{\includegraphics[width=5cm]{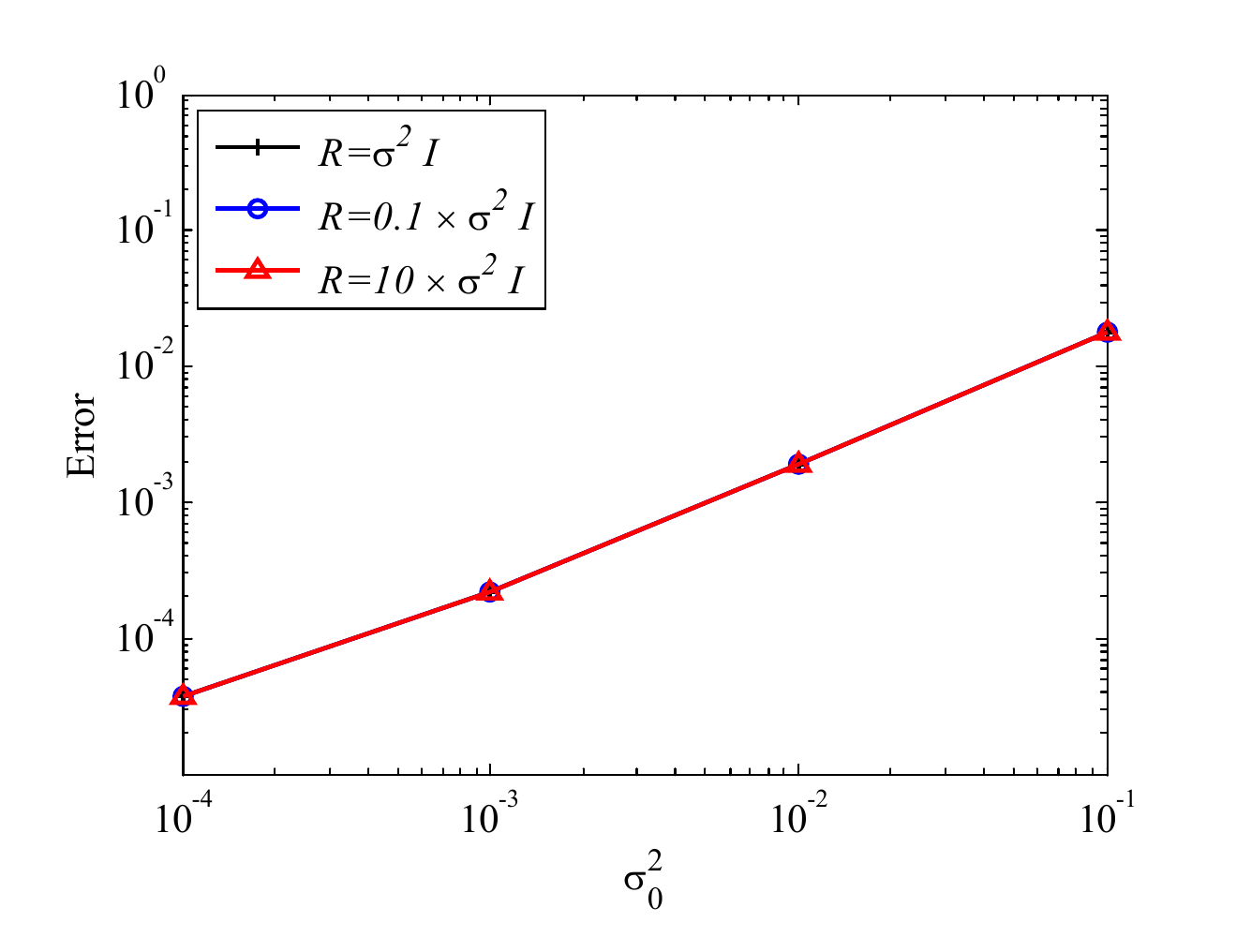}}
	\subfigure[$\lambda_3$]{\includegraphics[width=5cm]{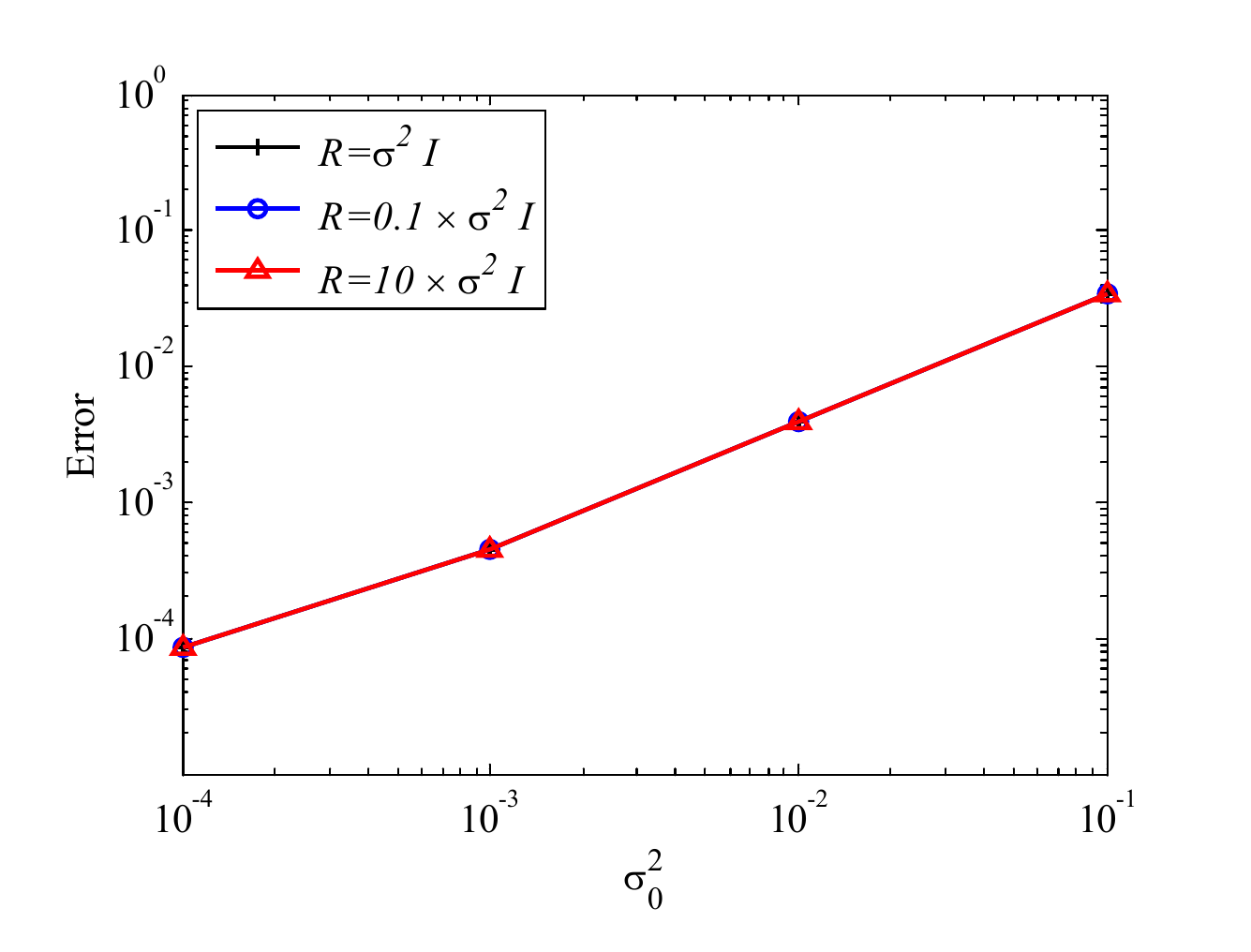}}
	   \caption{Effects of mismatched (amplified or attenuated) $R$ on errors in the eigenvalues computed by KFDMD in the test problem with time-varying noise. Here, $L_2$ error is averaged with 100 test cases. }	
	\label{fig:mis_eigenerror}
\end{figure}

The effects of a hyperparameter $R$ are investigated for the baseline problem in this subsection. First, we amplified the $R$ matrix by 0.1 or 10 times and the results are evaluated, while $Q$ is set to zero.  $R$ is given as  $R=aI\sigma^2 = aI\sigma_0^2 \cdot \left( 1.01 - \sin\left( \pi \Delta t k\right)\right)$, where $a = 0.1$ or $10$ for mismatched cases and $a=1$ for the matched case. Figures \ref{fig:mis_eigen} and \ref{fig:mis_eigenerror} show eigenvalues computed using KFDMD with mismatched and matched $R$ and their errors, respectively. The results are almost the same as the correct settings, as shown in Fig. \ref{fig:mis_eigen}. This implies that the time history of the noise characteristics is required, but not necessarily with the correct amplitude. This might be due to the hyperparameter $R$ and $Q$ being used to balance the system noise and observation noise. Also, in the present problem, the system noise is assumed to be zero and only the observation noise is considered.  For the case of $Q$ being nonzero, the choice of hyperparameter $R$ and $Q$ affects the system identification more. A survey on a wide parameter space is left for a future study.

\begin{figure}
	\subfigure[$\sigma_0^2=0.001$]{\includegraphics[width=5cm]{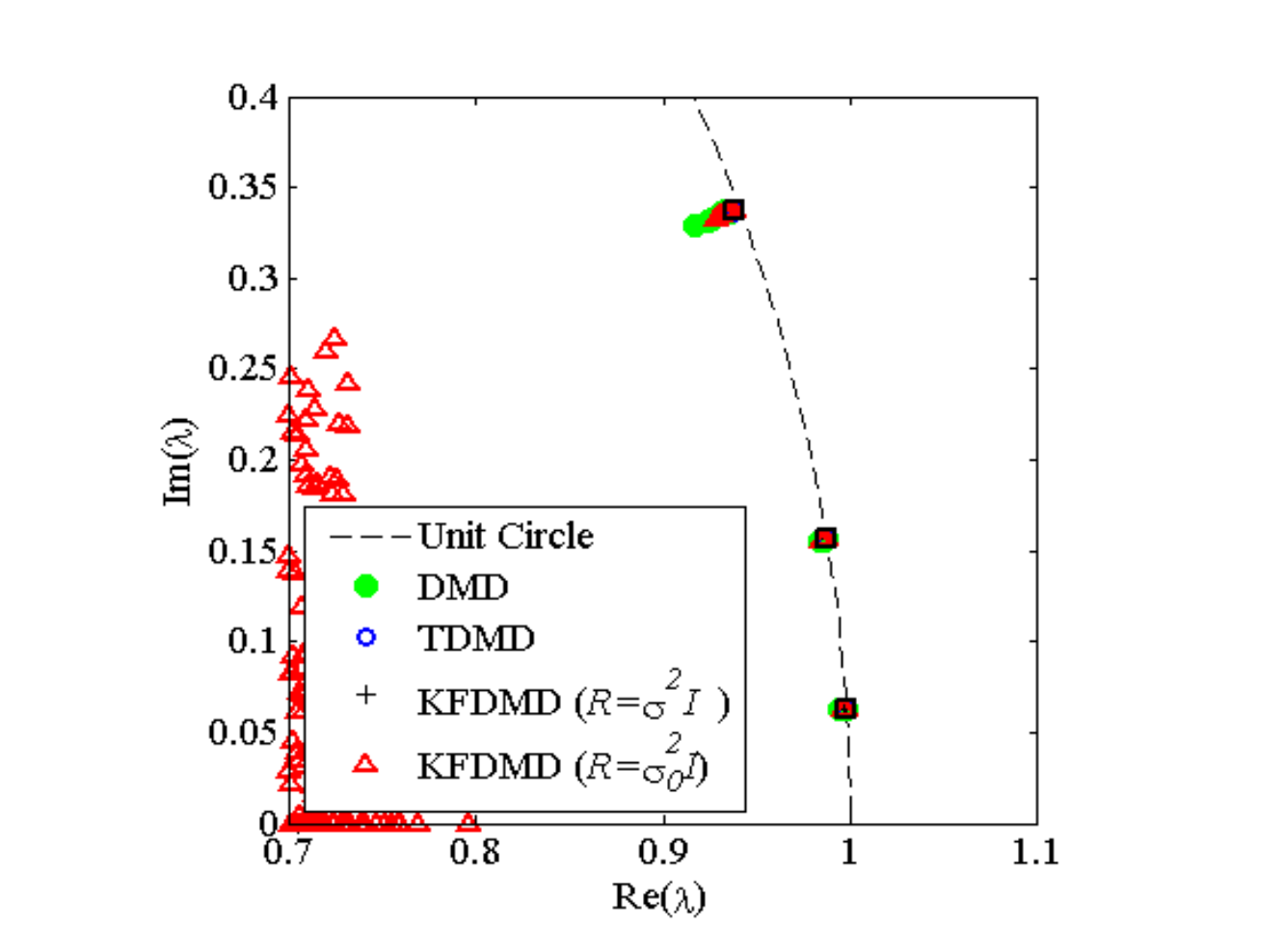}}
	\subfigure[$\sigma_0^2=0.01$ ]{\includegraphics[width=5cm]{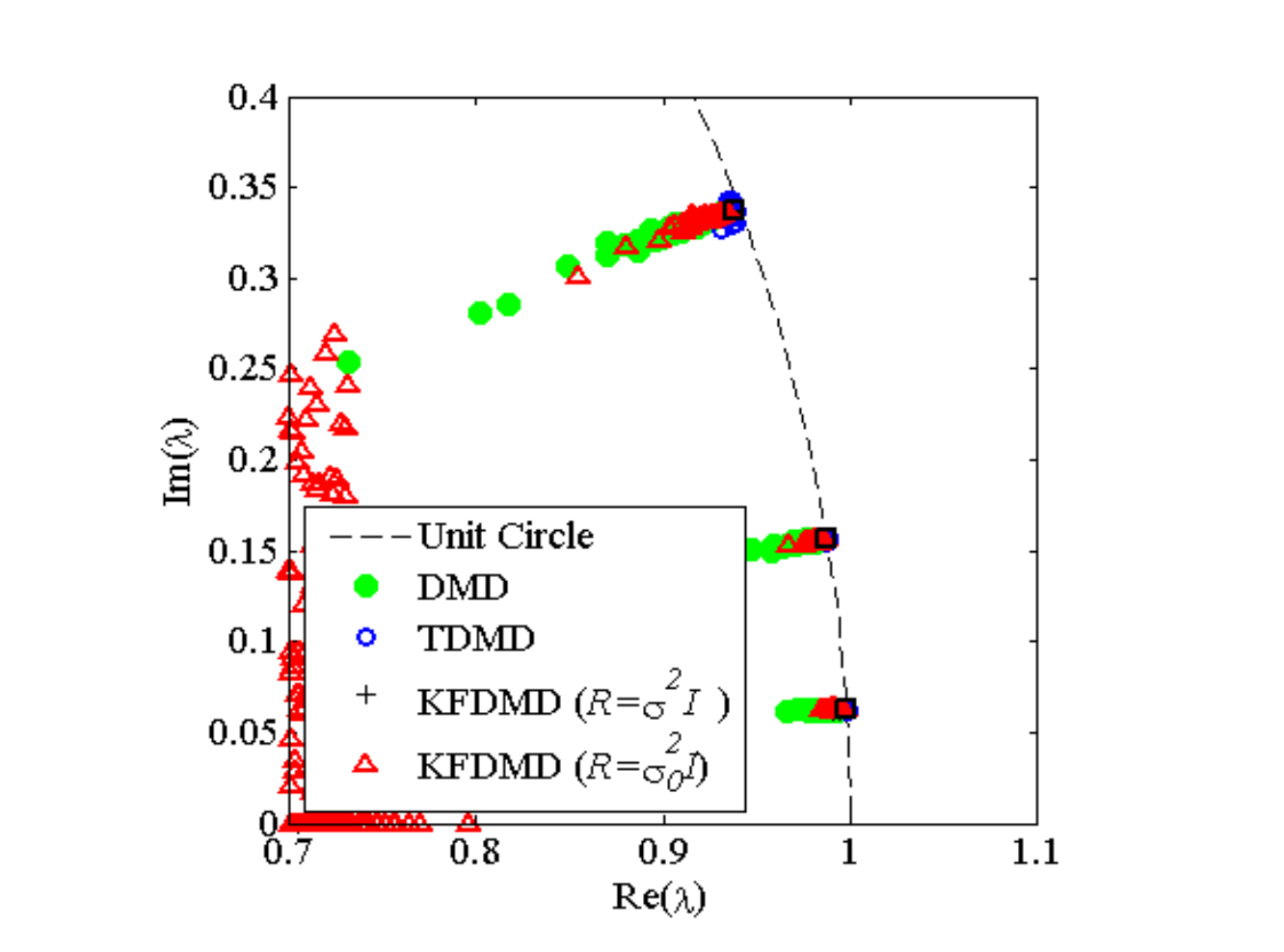}}
	\subfigure[$\sigma_0^2=0.1$  ]{\includegraphics[width=5cm]{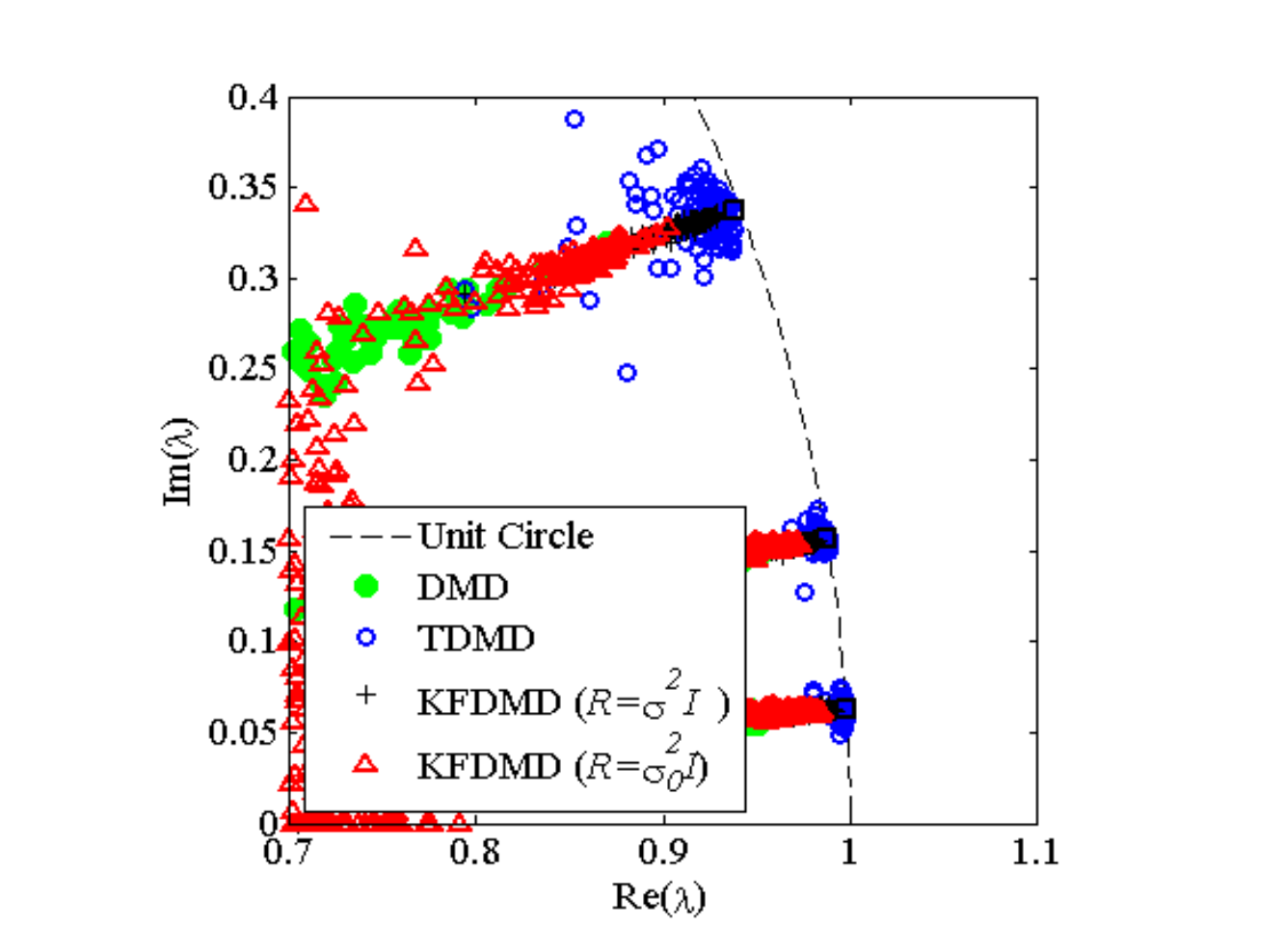}}
    \caption{Effects of mismatched (time-constant) $R$  on the results of 100 computations of eigenvalues in the test problem with time-varying noise.}	
	\label{fig:mis2_eigen}
\end{figure}

\begin{figure}
	\subfigure[$\lambda_1$]{\includegraphics[width=5cm]{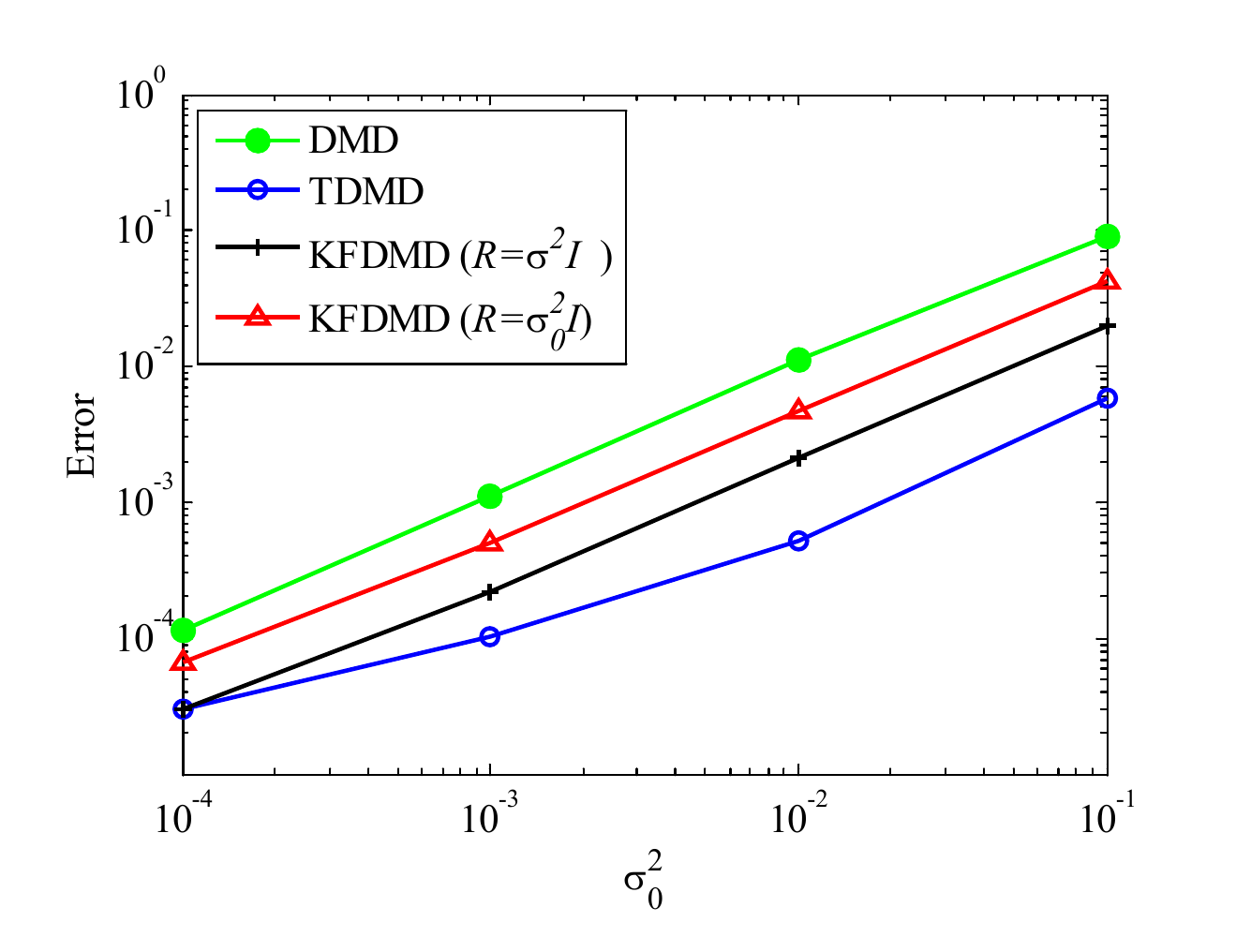}}
	\subfigure[$\lambda_2$]{\includegraphics[width=5cm]{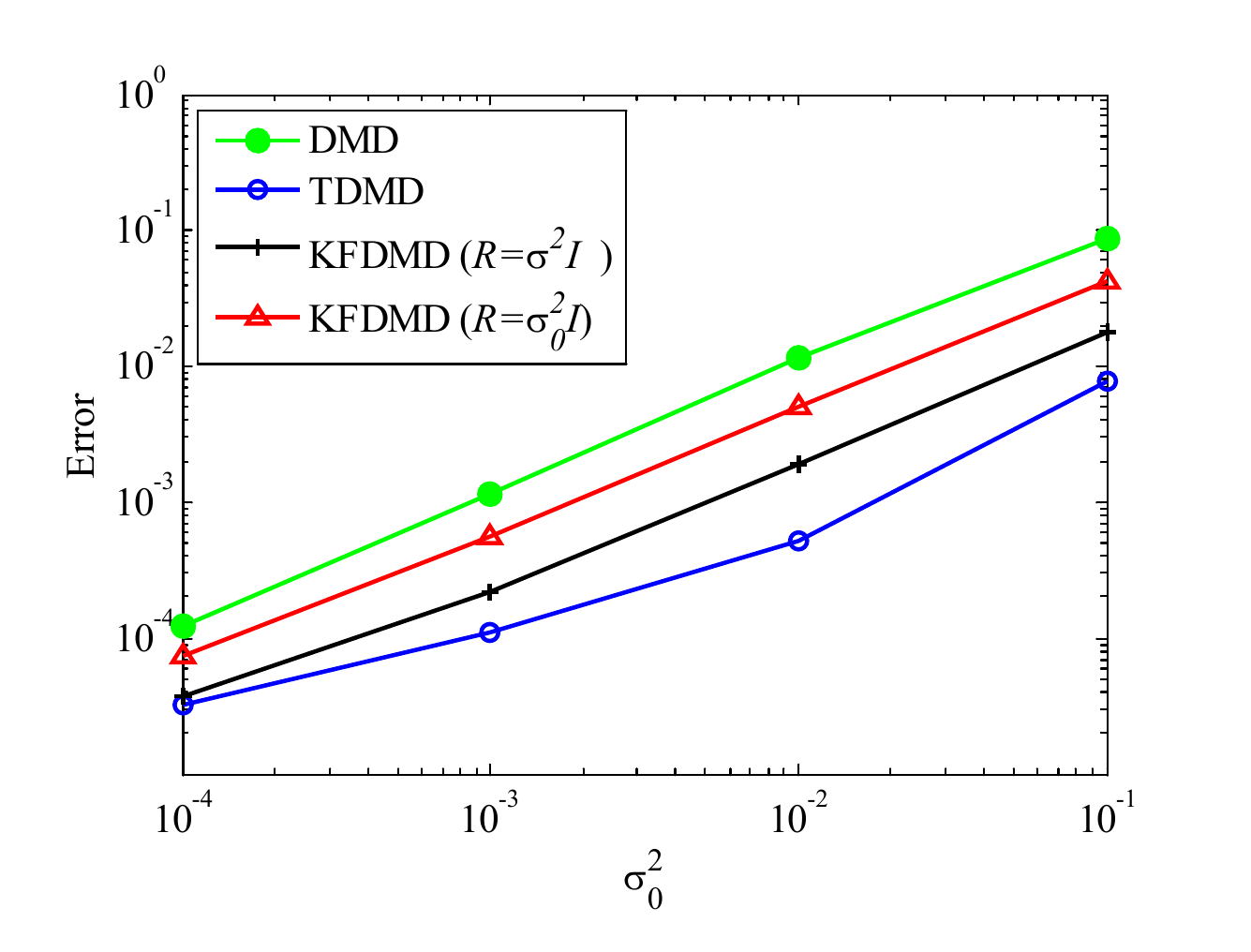}}
	\subfigure[$\lambda_3$]{\includegraphics[width=5cm]{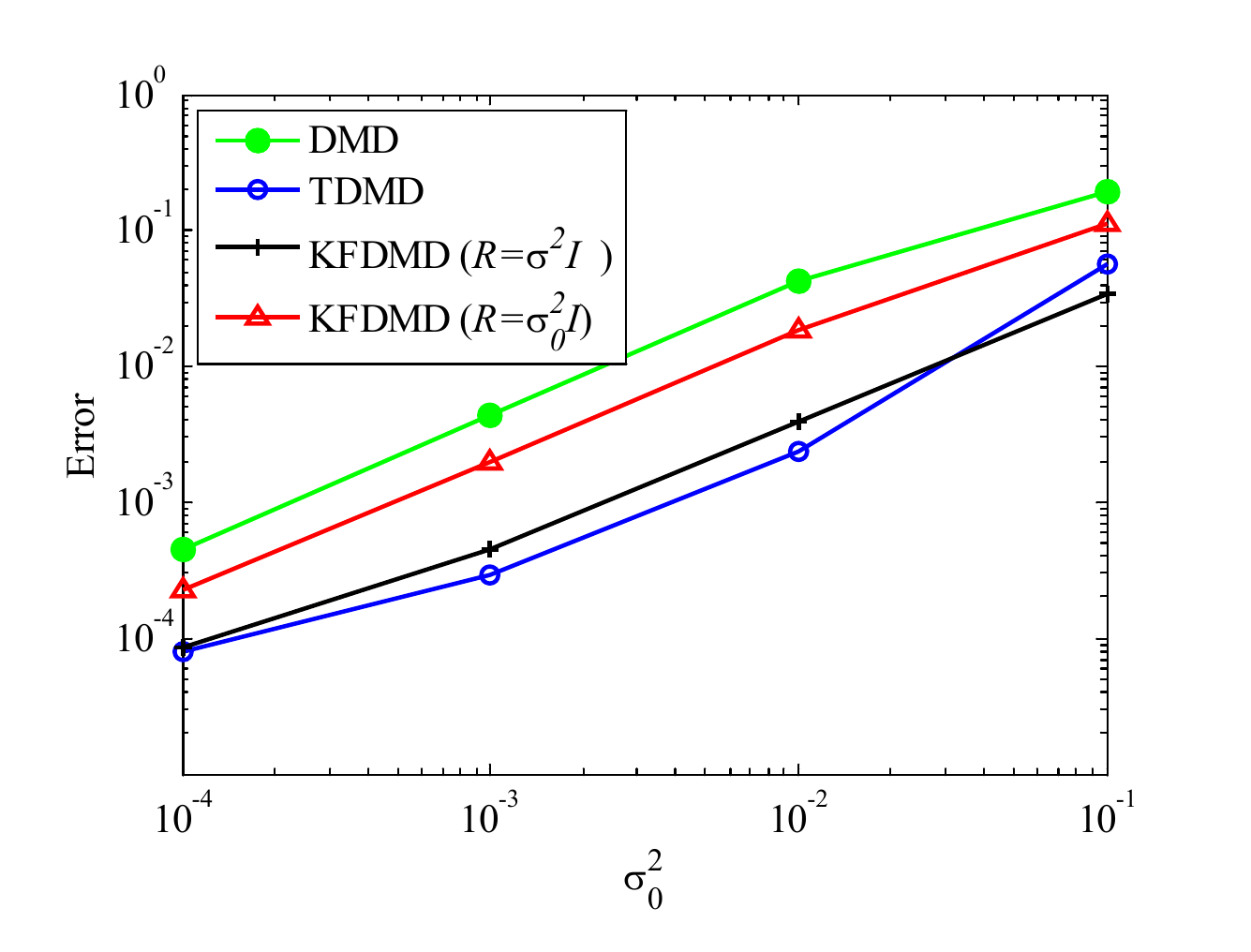}}
	   \caption{Effects of mismatched (time-constant) $R$  on the errors in the eigenvalues computed by KFDMD in the test problem with time-varying noise. Here, $L_2$ error is averaged with 100 test cases. }	
	\label{fig:mis2_eigenerror}
\end{figure}

The constant $R=I\sigma_0^2$ is used for the problem with time-varying noise. The results are shown in Figs. \ref{fig:mis2_eigen} and \ref{fig:mis2_eigenerror}. Here, KFDMD($R=\sigma^2 I$) (the same as in previous results) shows the results with the correct noise information and KFDMD($R=\sigma^2_0 I$) shows the results with the constant noise strength, which is not correct information.  In this case, the KFDMD($R=\sigma_0^2 I$) results are slightly better than, but similar to those for the standard DMD, as expected.  This illustrates that the prior knowledge $R$ of the time variation of noise is effectively utilized by KFDMD. 

\subsubsection{Effects of POD truncation}
\label{sec:EPT}
\begin{figure}
	\subfigure[$r=6$]{\includegraphics[width=5cm]{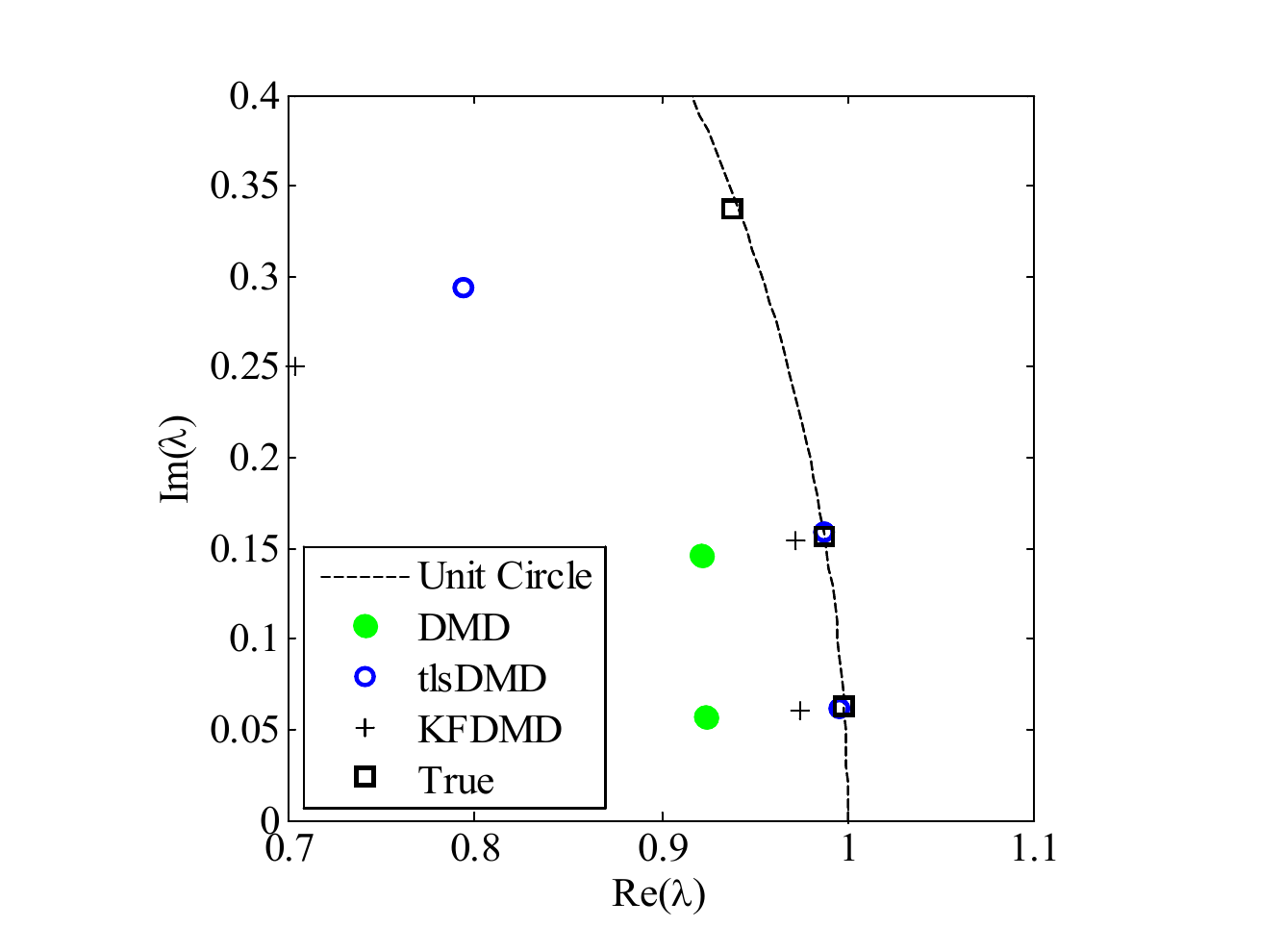}}
    \subfigure[$r=100$ ]{\includegraphics[width=5cm]{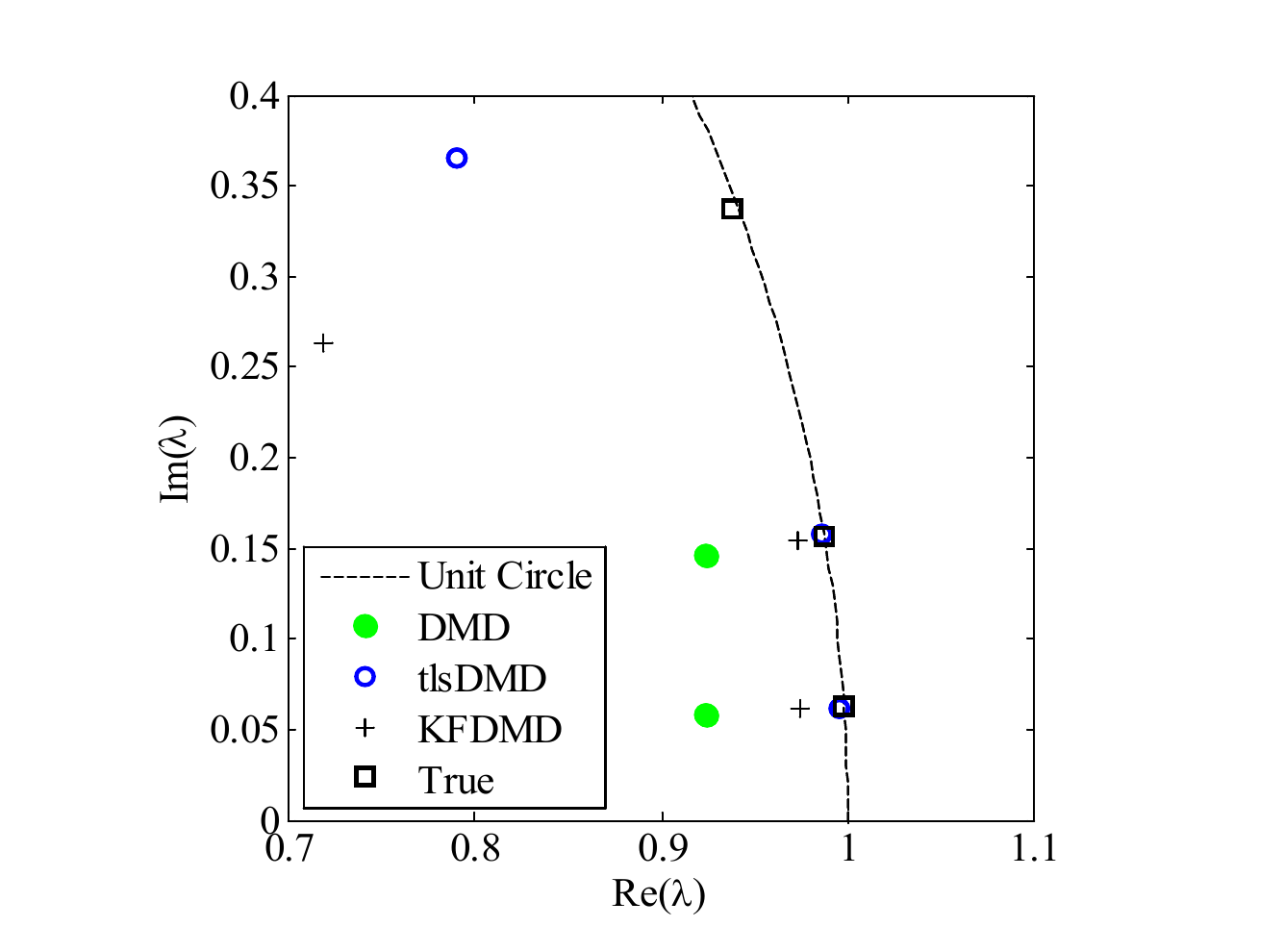}}
    \subfigure[$r=100$ ]{\includegraphics[width=5cm]{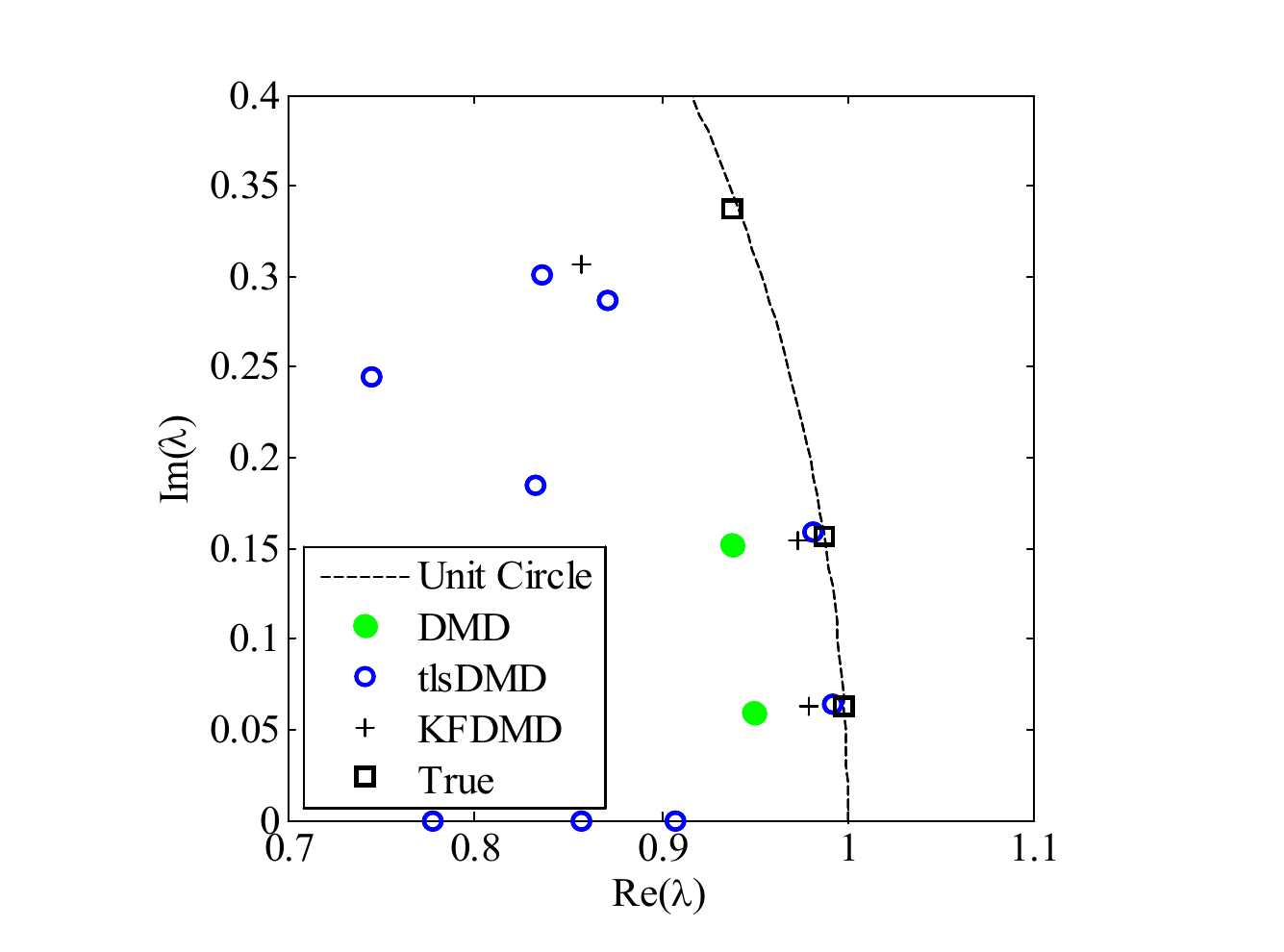}}
	    \caption{Effects of a rank number $r$ of trPOD on the representative results of eigenvalues computed in the test problem with time-varying noise of $\sigma_0^2=0.1$}	
	\label{fig:diffr_eigenexam}
\end{figure}

\begin{figure}
	\subfigure[$r=6$]   {\includegraphics[width=5cm]{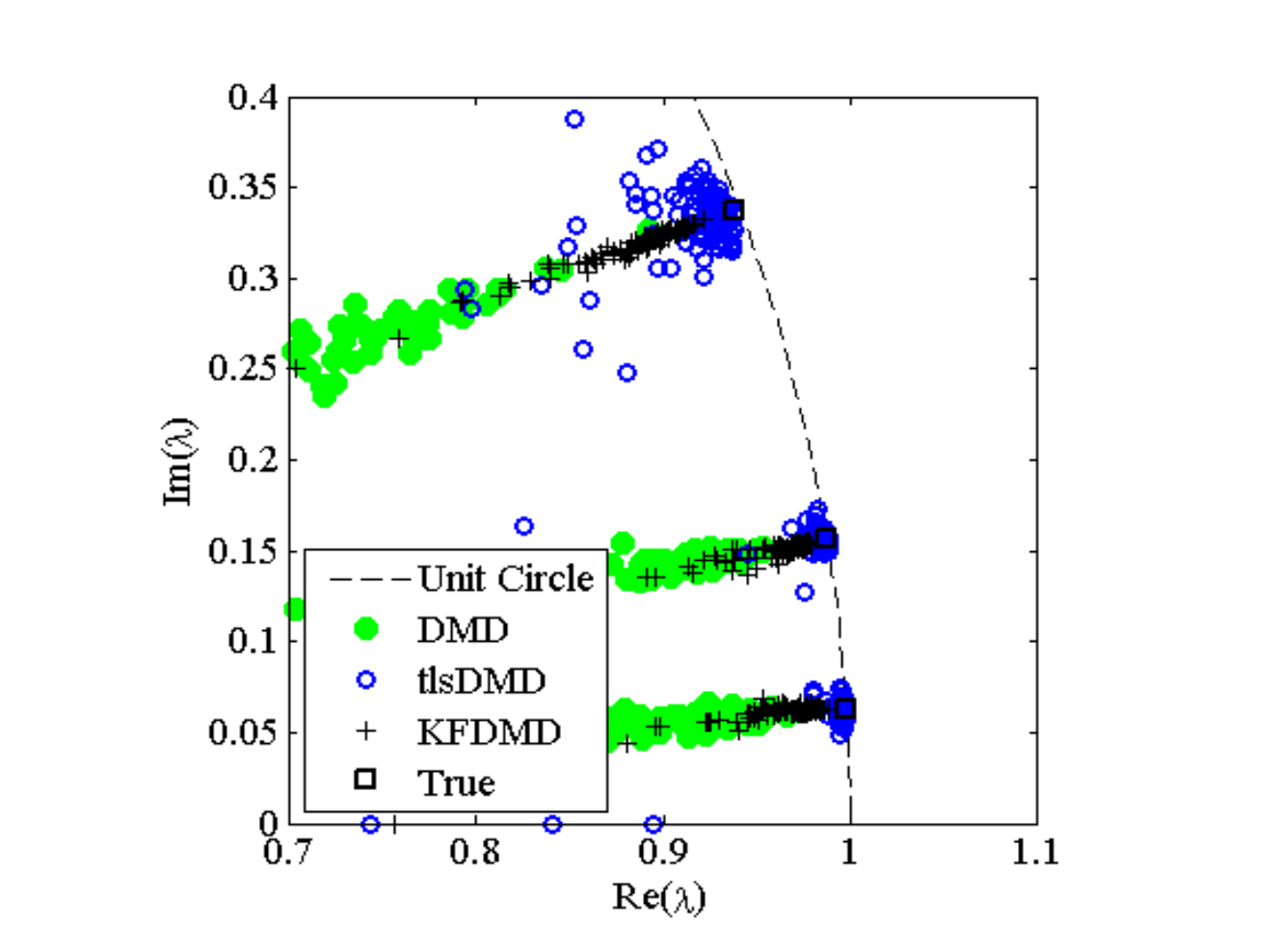}}
    \subfigure[$r=10$] {\includegraphics[width=5cm]{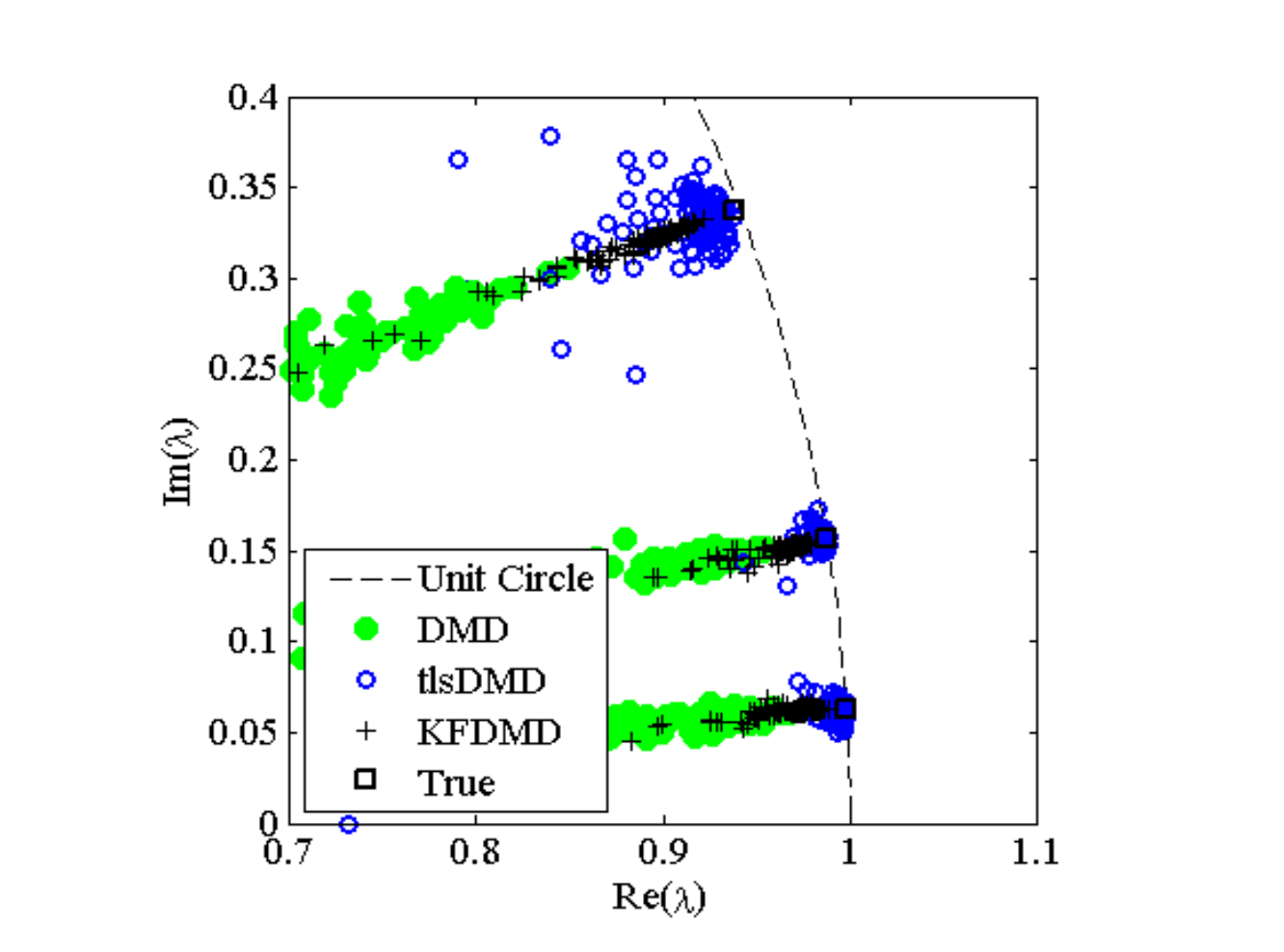}}
    \subfigure[$r=100$]{\includegraphics[width=5cm]{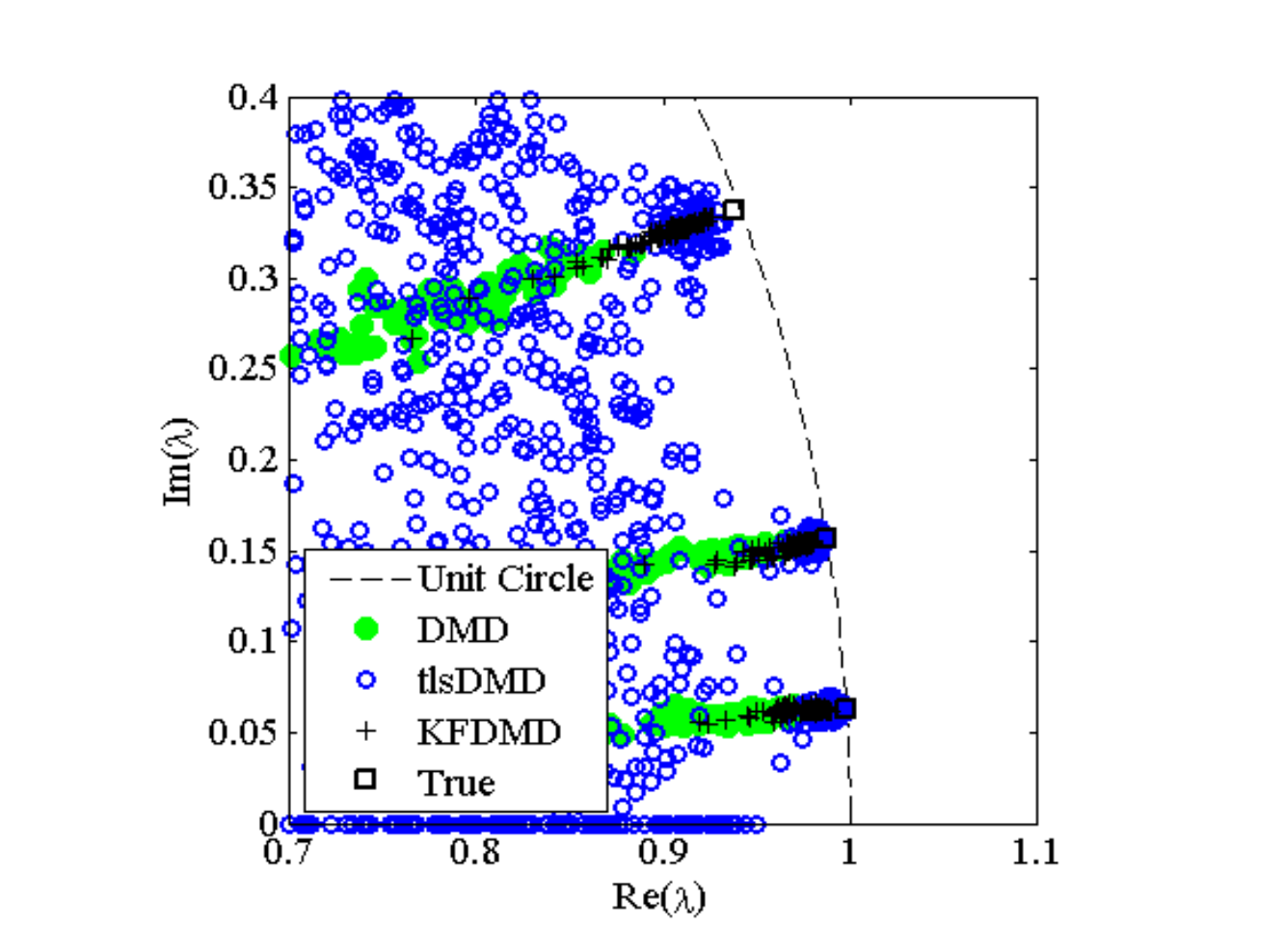}}
	\caption{Effects of a rank number $r$ of trPOD on  results of 100 computations of eigenvalues in the test problem with time-varying noise of $\sigma_0^2=0.1$}	
	\label{fig:diffr_eigen}
\end{figure}

\begin{figure}
	\subfigure[$\lambda_1$]{\includegraphics[width=5cm]{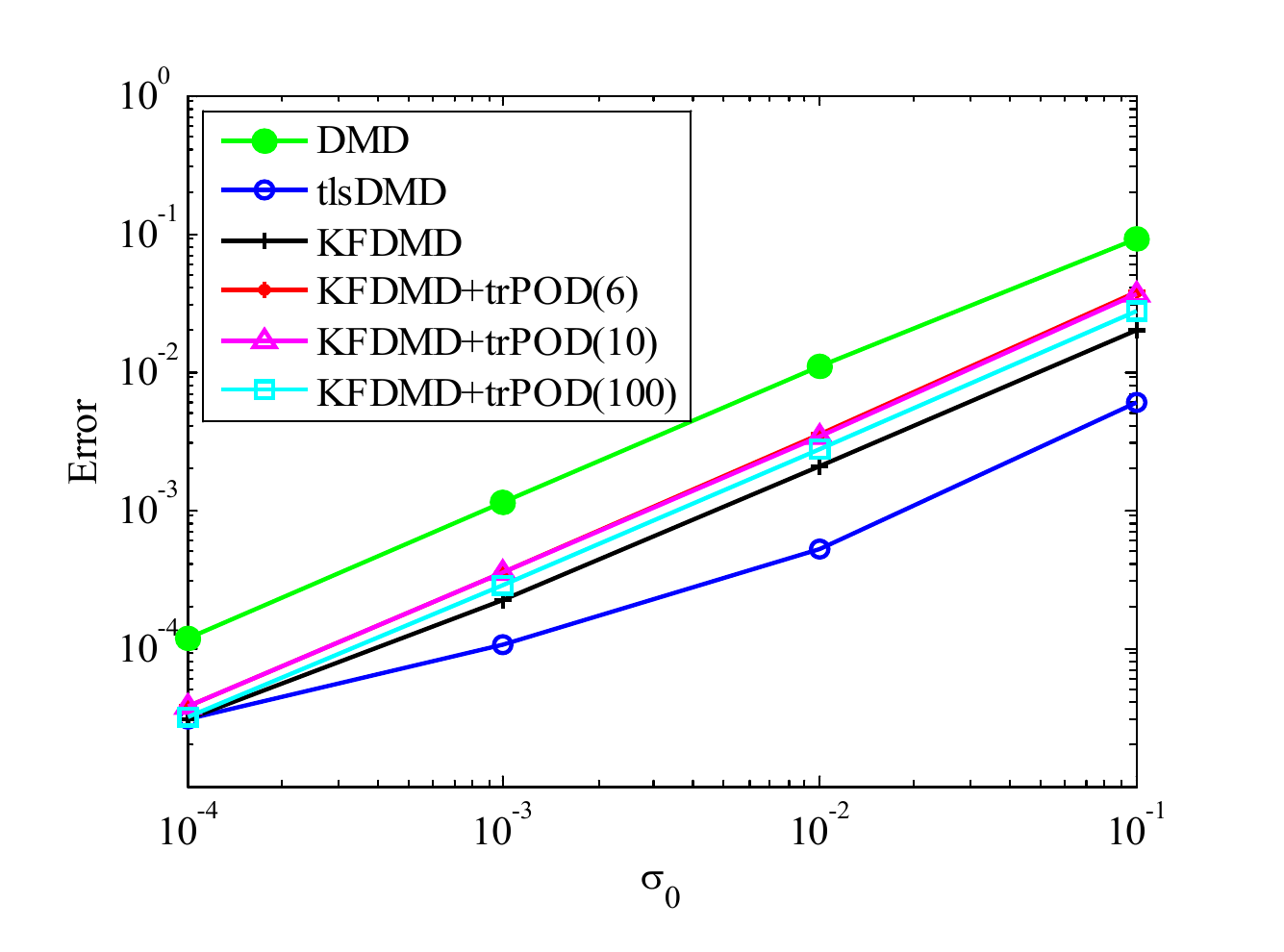}}
	\subfigure[$\lambda_2$ ]{\includegraphics[width=5cm]{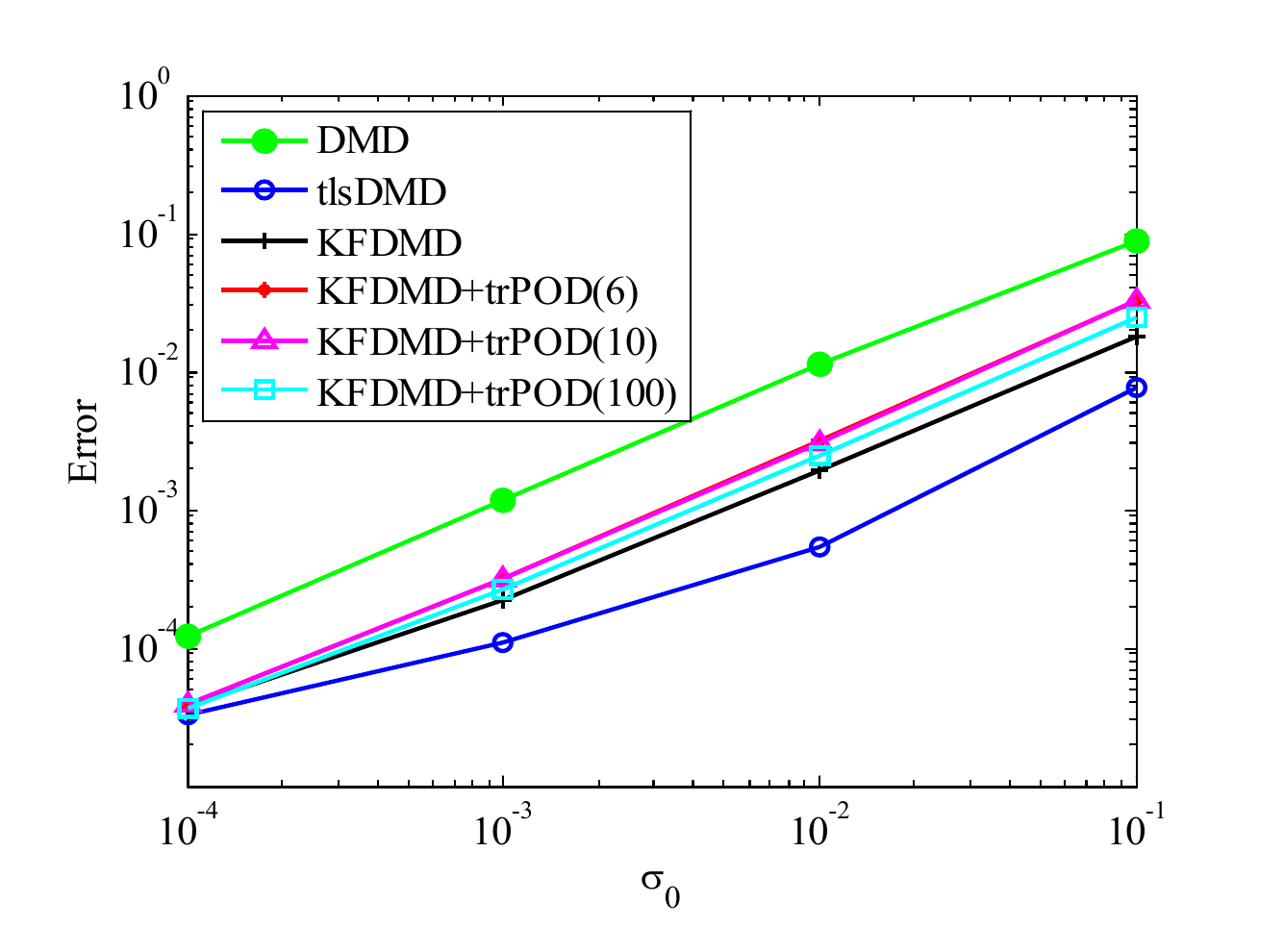}}
	\subfigure[$\lambda_3$  ]{\includegraphics[width=5cm]{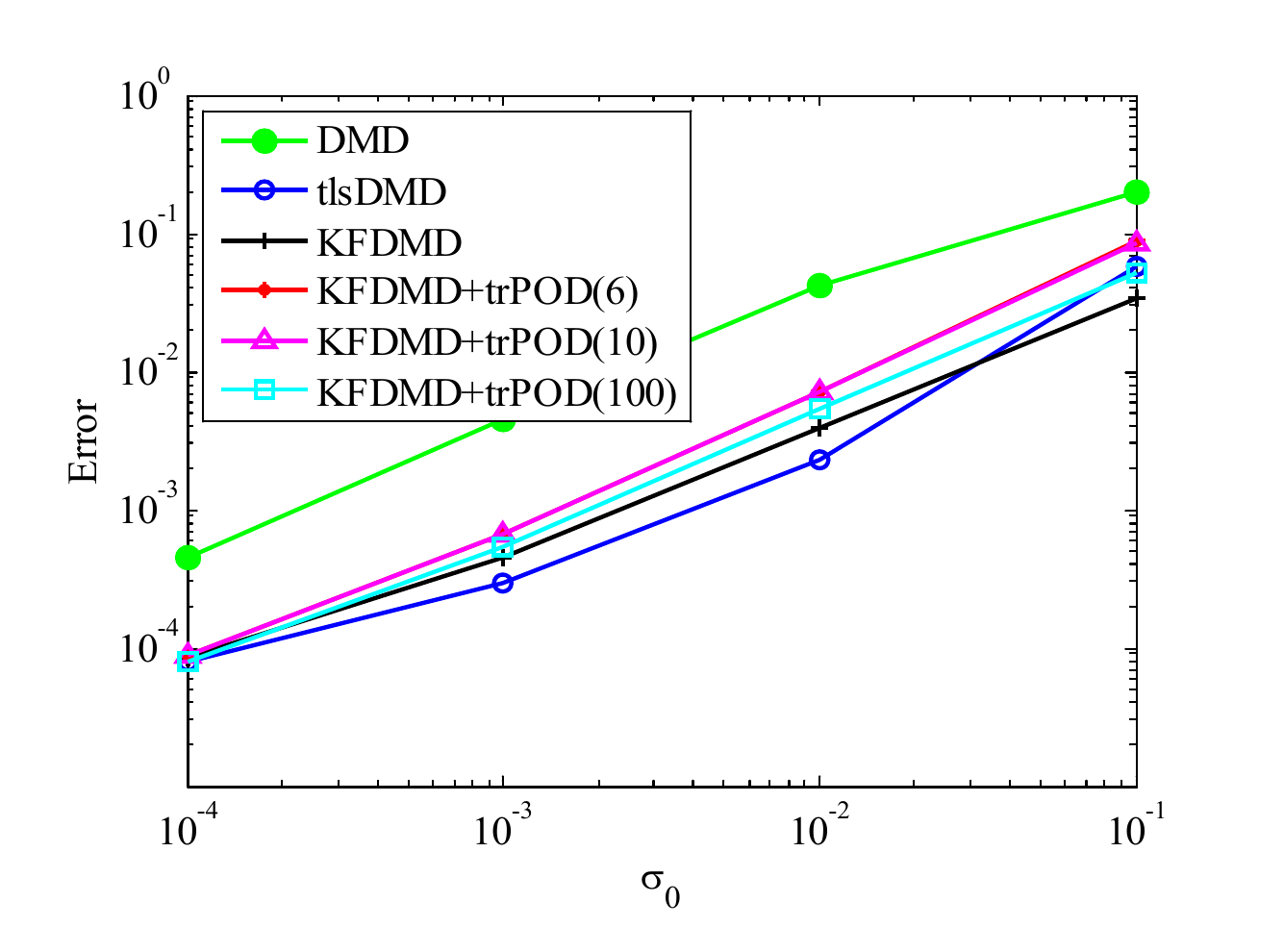}}
   \caption{Effects of a rank number $r$ of trPOD on  the errors in the eigenvalues computed in the test problem with time-varying noise. Here, $L_2$ error is averaged with 100 test cases. Here, KFDMD+trPOD($r$) represents the rank $r$ trPOD is conducted as preconditioning. Lines of KFDMD+trPOD(6) and KFDMD+trPOD(10) are identical in this plot range.}	
   \label{fig:diffr_eigenerror}
\end{figure}

The effects of the POD truncation are investigated by changing the rank number $r$. Here, $r$ is set to 6, 10, and 100. In this part, POD truncation is applied before utilizing the KFDMD algorithm. It should be noted that the rank $r$ used in the POD truncation in standard DMD and tlsDMD algorithms is set to be the same value as that used for preconditioning of KFDMD, except for the error plot shown later. The results of eigenvalues and corresponding errors are shown in Figs. \ref{fig:diffr_eigenexam} and \ref{fig:diffr_eigen}, and Fig. \ref{fig:diffr_eigenerror}, respectively. If $r$ is limited to be 6 or 10, the results are better than those for the standard DMD similar to KFDMD without trPOD, but the third eigenvalues are further from the true location than those for KFDMD without trPOD, as shown in Fig. \ref{fig:sineigenexam}. This implies that the POD filtered out the important signal as well as the noise. However, the results of $100$ computations approach those obtained without POD truncation. These results show that KFDMD performs better with many degrees of freedom and that rank $r$ of POD truncation should be as high as possible to improve accuracy. However increase in rank $r$ also leads to increase in the computational cost. There is thus a trade-off in the practical use of the proposed algorithm.

\clearpage

\subsection{Static Fluid System Identification without Noise}
\label{sec:FluidWON}
\begin{figure}
	\subfigure[$t=41.75 \times D/a_\infty$, 167th snapshot.]{\includegraphics[width=6cm]{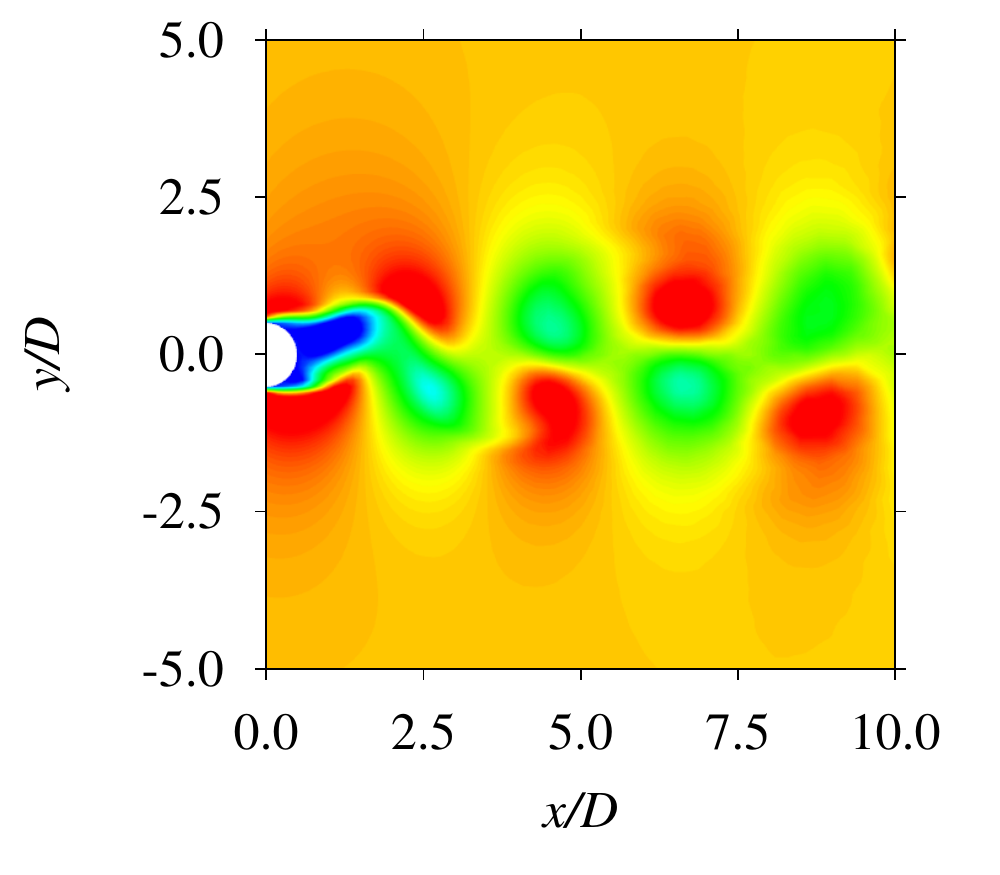}}
	\subfigure[$t=125 \times D/a_\infty$, 500th snapshot.]{\includegraphics[width=6cm]{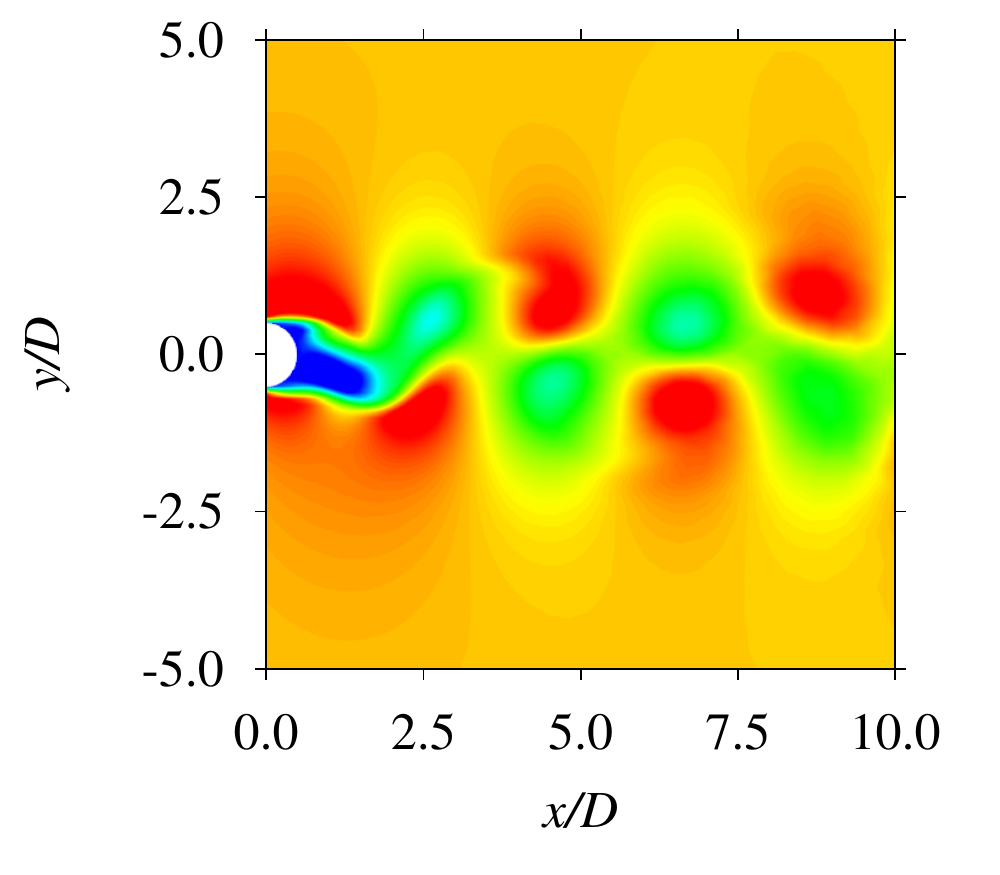}}
	\caption{Snapshots of the flow field without noise. Here, contour color ranges from 0.0 to 1.5$u_\infty$.}
	\label{fig:flowwo}
\end{figure}

Flow simulation is conducted for a two-dimensional flow around a cylinder. The flow Mach number and the Reynolds number based on the cylinder diameter are set to be 0.2 and 300, respectively. LANS3D,\cite{Fujii1990a} an in-house compressible fluid solver, is used for simulation. A computational mesh of 250 $\times$ 111 grid points (radial- and azimuthal-direction grid points, respectively) is used. A sixth-order compact difference scheme\cite{Lele1992} for spatial derivatives and an alternative-directional-implicit symmetric-Gauss-Seidel method\cite{Fujii1999,Nishida2009} for time integration in second-order accuracy are adopted. See Reference\cite{Sato2015b}  for additional details concerning the code of the latest version. The flow variables are nondimentionalized by the density and the sound speed $a_\infty$ of the freestream and the diameter of the cylinder $D$. The cylinder is located at the origin point, and flow fields inside 10$D$ from the origin points are resolved, where $D$ is the diameter of the cylinder. Only the wake region of the velocity fields data at $x=[0,10D], y=[-5D,5D]$ sampled on the $101 \times 101$ uniform mesh is used for the DMD analyses. The data are acquired after the flow enters the quasi-steady condition.
A total of 1,000 samples of ten flow-through data acquired at every $\Delta t=0.25 D/a_{\infty}$ are used for the DMD analyses. Snapshots of the flow fields without noise are shown in Fig. \ref{fig:flowwo}.

\begin{figure}
	\includegraphics[width=6cm]{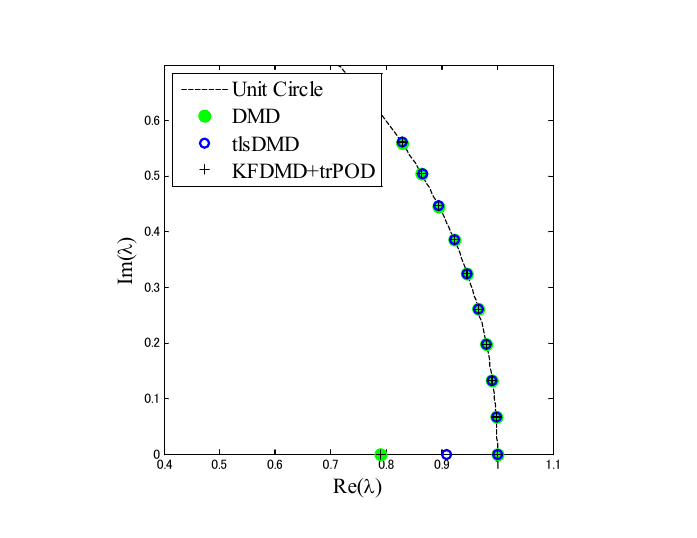}
	\caption{Eigenvalues of a static fluid system without noise.}
	 \label{fig:cyl_eigenwonoise}
\end{figure}

First, the results without noise are processed by standard DMD, tlsDMD and KFDMD, where KFDMD adopts the truncated POD with rank number $r=20$ (Eqs. \ref{eq:xtilde} and \ref{eq:ytilde}) as a preconditioner. The eigenvalue computed by the standard DMD, tlsDMD and KFDMD methods are shown in Fig. \ref{fig:cyl_eigenwonoise}. The eigenvalues computed using the KFDMD method agree well with those computed by the standard DMD method. The lowest frequencies computed by the standard DMD and KFDMD methods correspond to the Strouhal number $St=fD/u_{\infty} \sim 0.2$, which is a well-known characteristic frequency for the K\'arm\'an vortex street of a cylinder wake, where $f$ and $u_\infty$ are the frequency and the freestream velocity, respectively. 

\begin{figure}
	\subfigure[DMD mode 1]{\includegraphics[width=5cm]{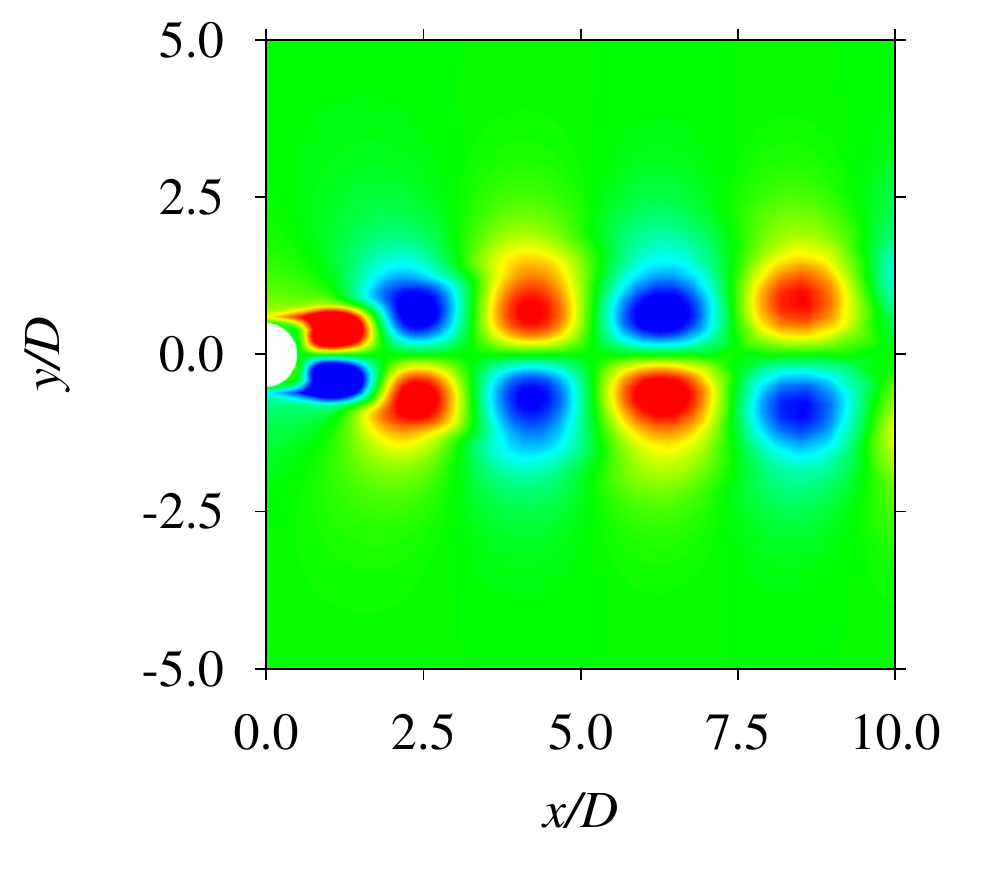}}
	\subfigure[DMD mode 2]{\includegraphics[width=5cm]{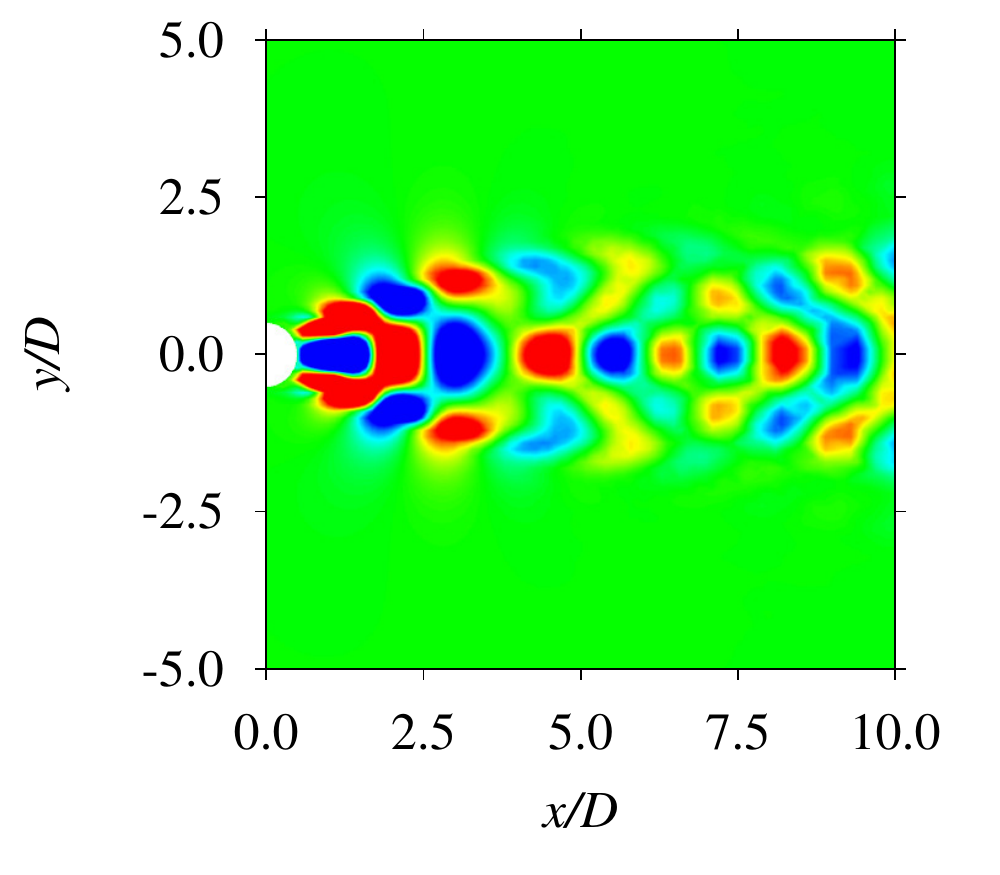}}
	\subfigure[DMD mode 3]{\includegraphics[width=5cm]{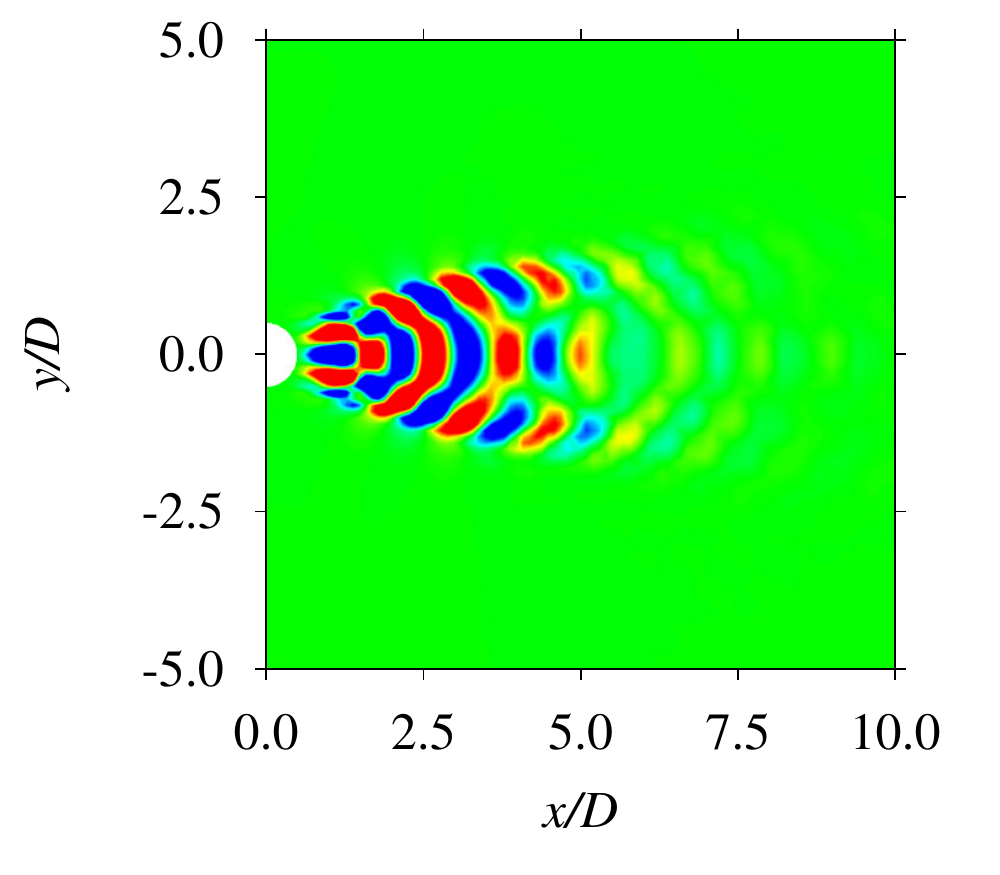}}\\
	\subfigure[tlsDMD mode 1]{\includegraphics[width=5cm]{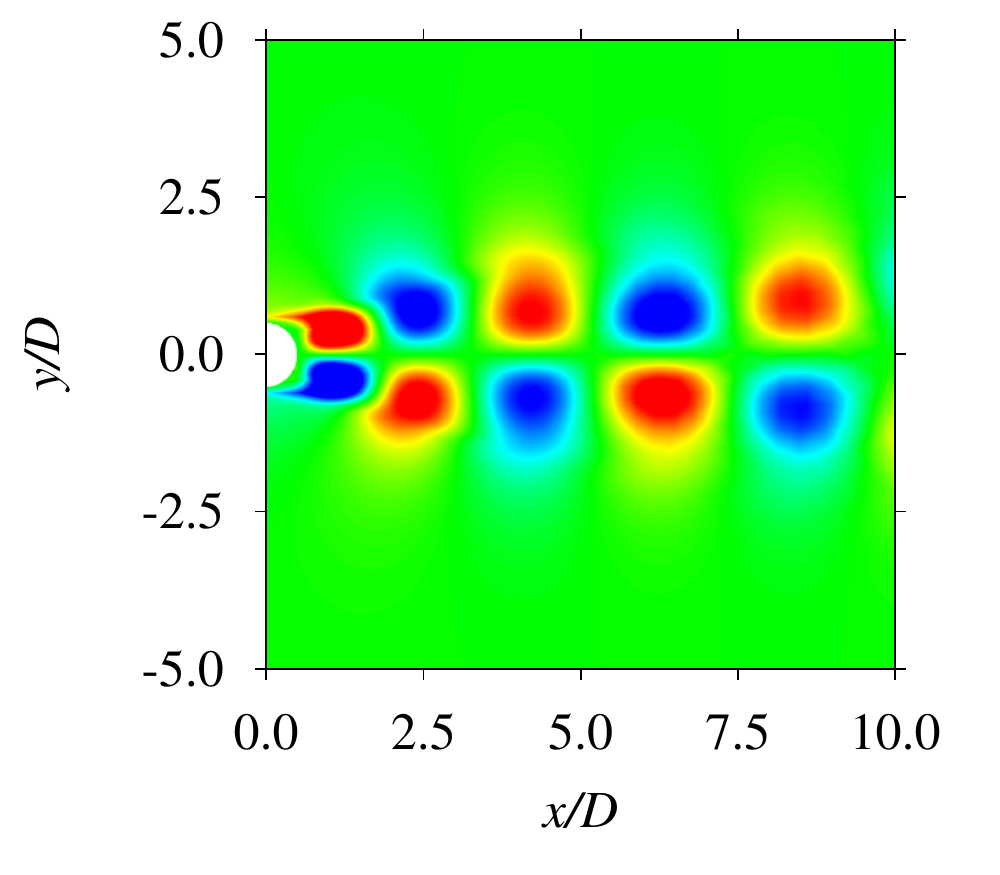}}
	\subfigure[tlsDMD mode 2]{\includegraphics[width=5cm]{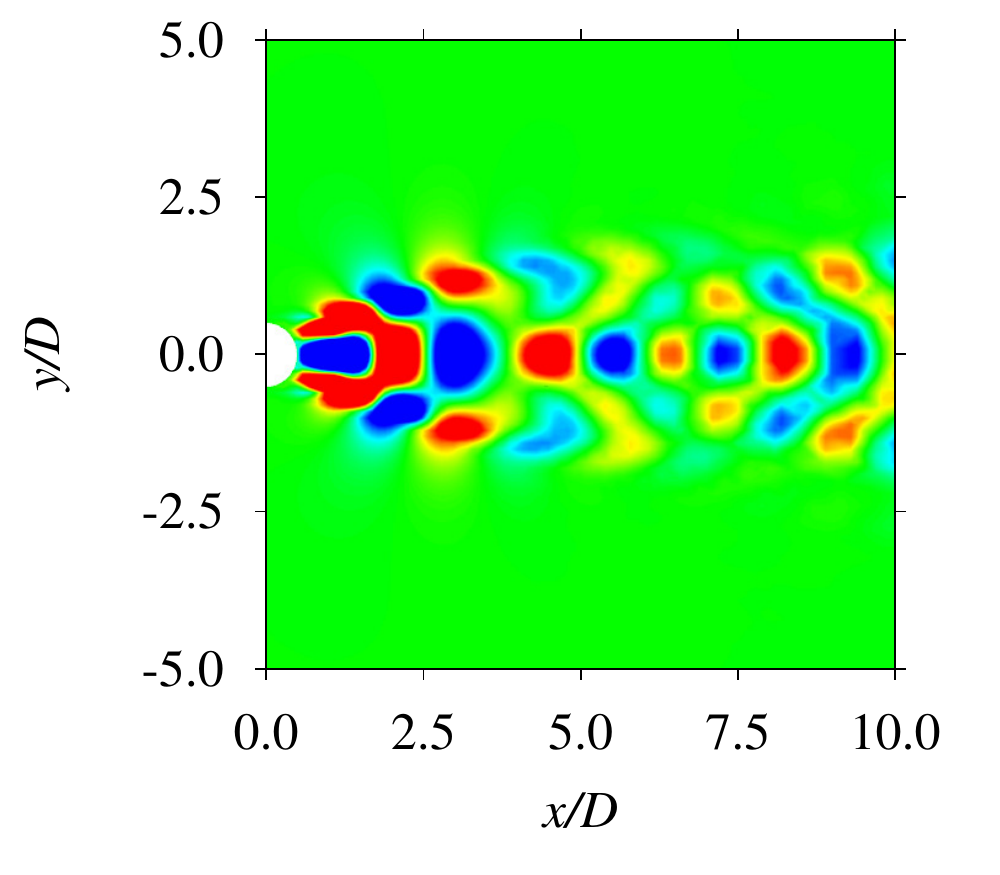}}
	\subfigure[tlsDMD mode 3]{\includegraphics[width=5cm]{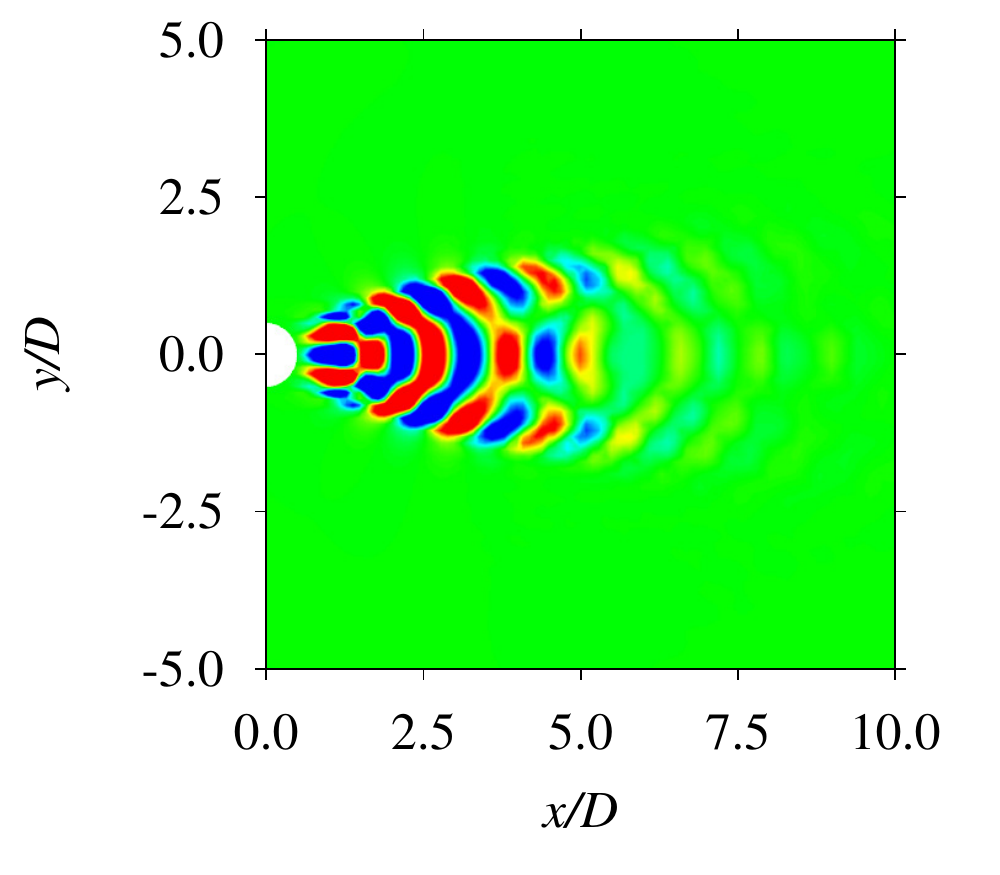}}\\
	\subfigure[KFDMD+trPOS mode 1]{\includegraphics[width=5cm]{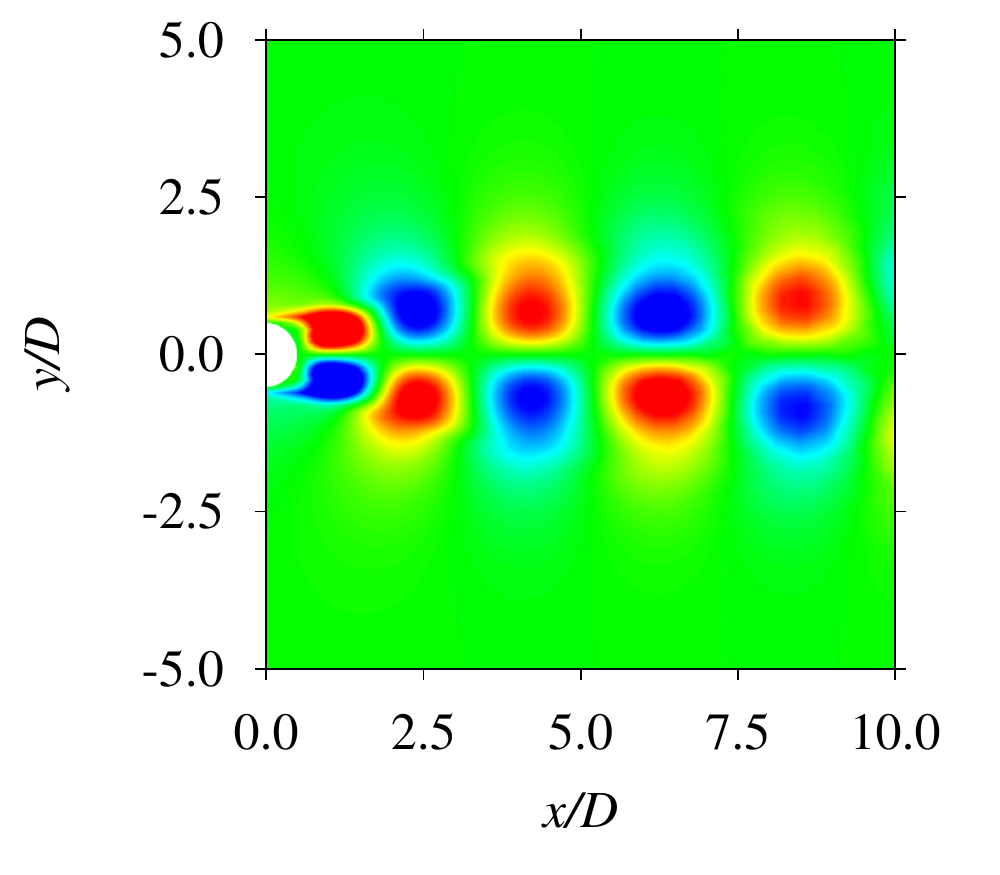}}
	\subfigure[KFDMD+trPOD mode 2]{\includegraphics[width=5cm]{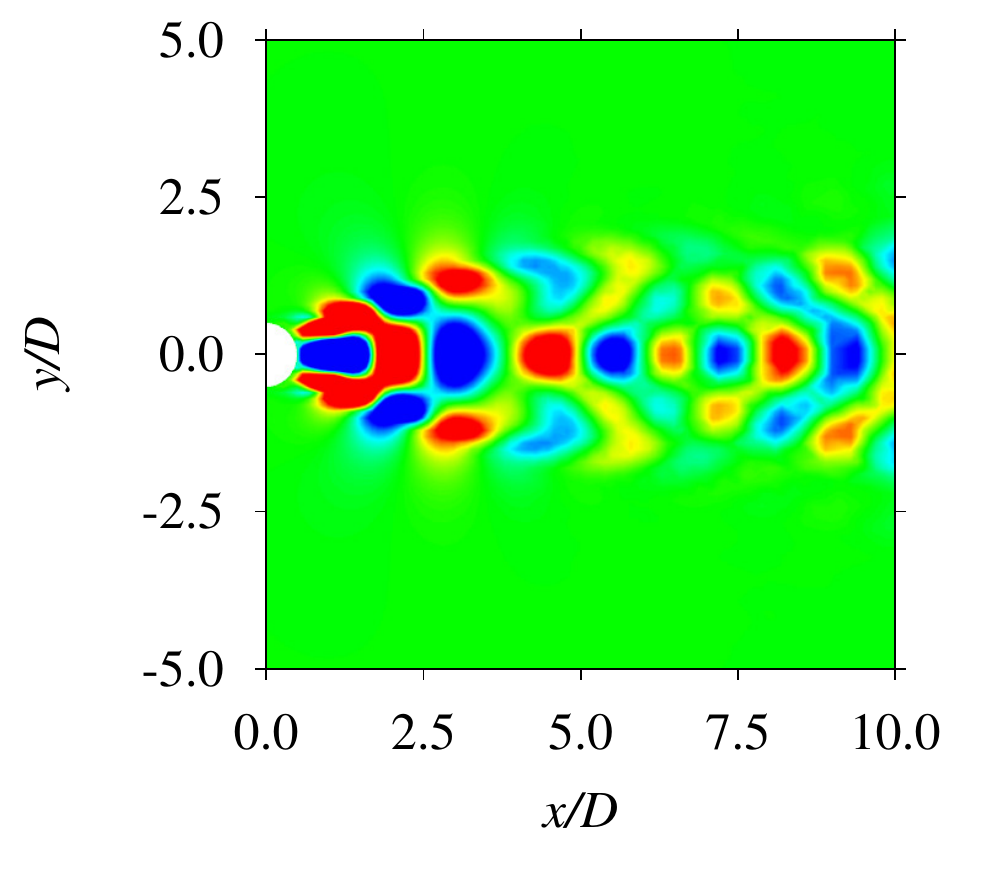}}
	\subfigure[KFDMD+trPOD mode 3]{\includegraphics[width=5cm]{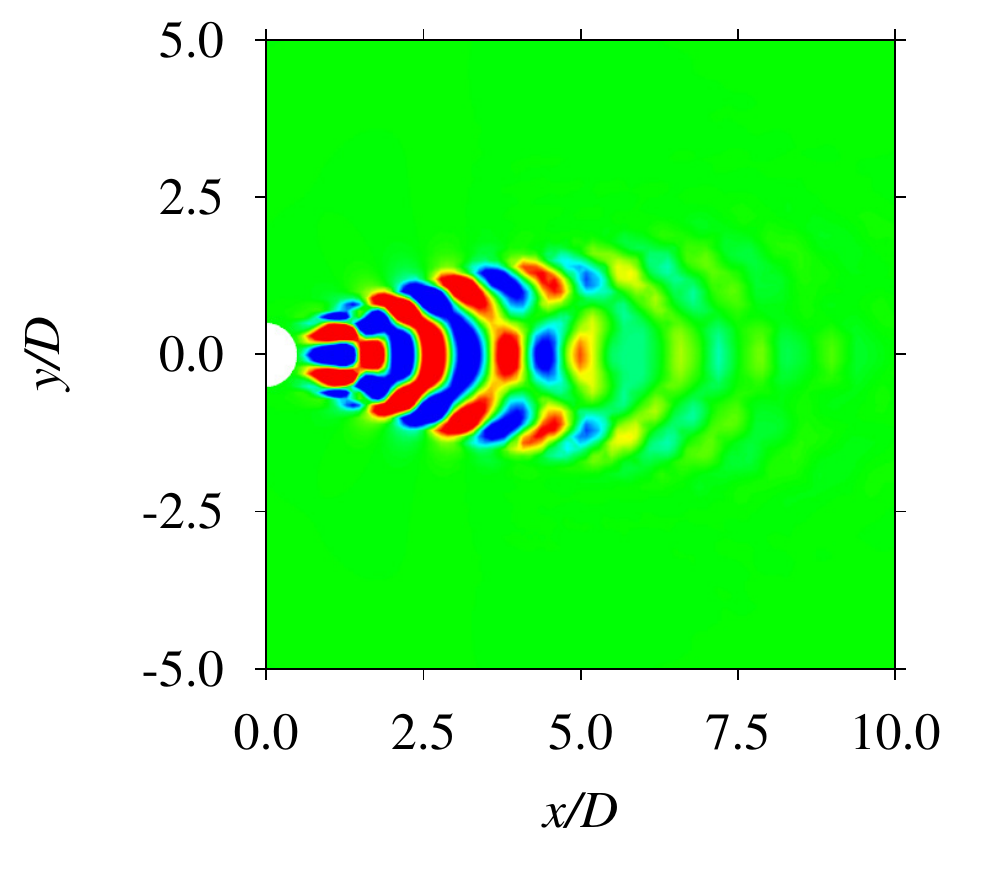}}\\
	\caption{Real part distributions of eigenmodes of a static fluid system without noise. Here, contour color ranges from -0.01 to 0.01.}
	\label{fig:cyl_eigenmodewonoise}
\end{figure}

The real parts of the eigenmode computed by standard DMD, tlsDMD and KFDMD are shown in Fig. \ref{fig:cyl_eigenmodewonoise}. Here, the phase of all the mode shown in this figure are adjusted by multiplying a complex variable so that the inner products of the modes of tlsDMD or KFDMD and the standard DMD mode become a real number. In addition, the all the modes shown here are normalized. All the methods produce the dynamic mode of K\'arm\'an vortex shedding, and these results show that KFDMD can estimate dynamic modes similar to standard DMD and tlsDMD, when noise is absent. 

\subsection{Static Fluid System Identification with Noise Known Characteristics}
\label{sec:FluidWN}
In this subsection, the same data as in the previous problem are used, but with the noise characteristics of which are known. The noise strength of the $k$th time step is determined as $\sigma^2 = \sigma_0^2 \cdot \left( 1.01 - \sin\left( 0.012 \pi \Delta t k\right)\right) $ and the correct information is given to $R$ in the KFDMD algorithm.

\begin{figure}
	\subfigure[$t=41.75 \times D/a_\infty$, 167th snapshot, with minimum covarience of noise.]{\includegraphics[width=6cm]{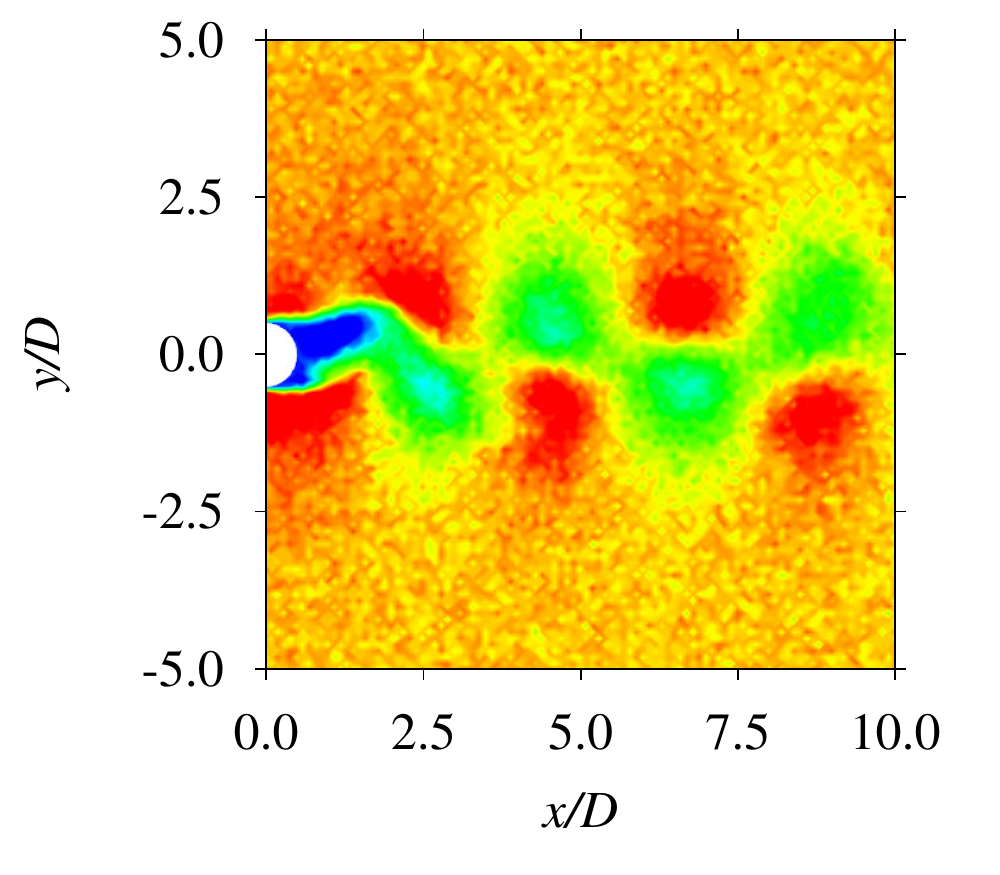}}
    \subfigure[$t=125 \times D/a_\infty$, 500th snapshot, maximum covarience of noise.]{\includegraphics[width=6cm]{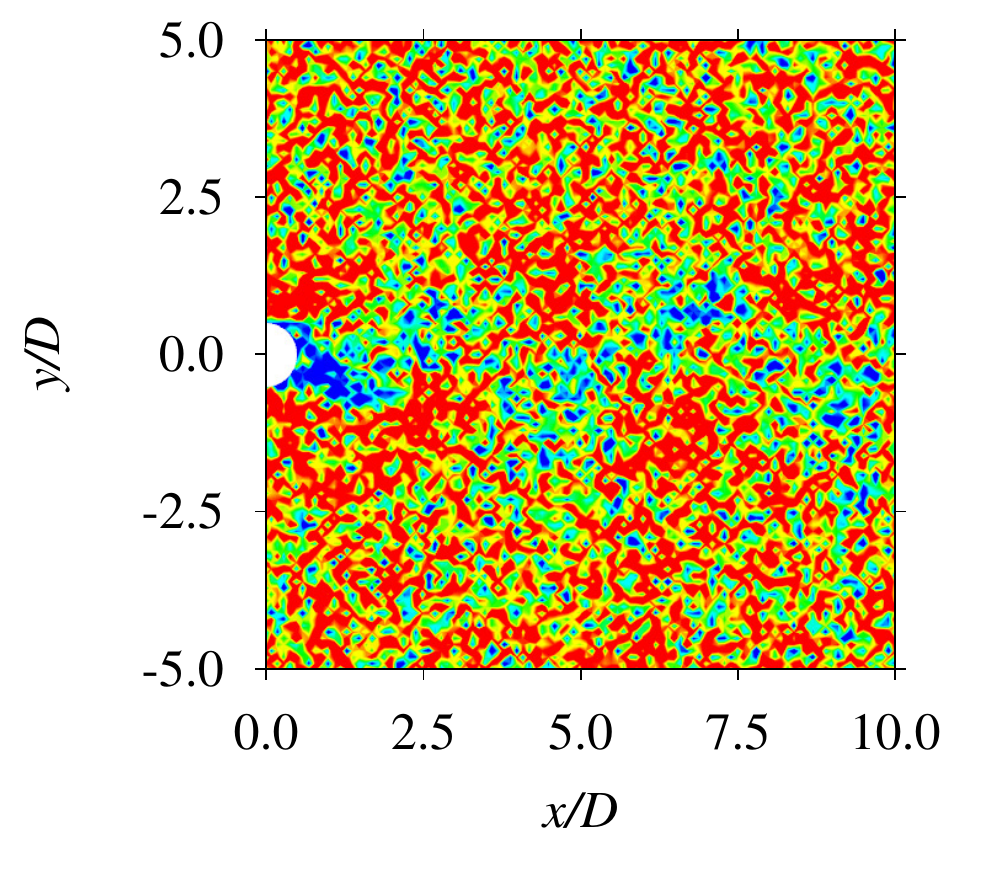}}
	\caption{Snapshots of the flow field with noise at minimum and maximum covarience. See also Fig. \ref{fig:flowwo} for the flow without noise at the exactly the same instance. Here, contour color ranges from 0.0 to 1.5$u_\infty$.}
	\label{fig:flowwnoise} 
\end{figure}

\begin{figure}
	\subfigure[$u$ at $(x,y)=(D,0)$]{\includegraphics[width=5cm]{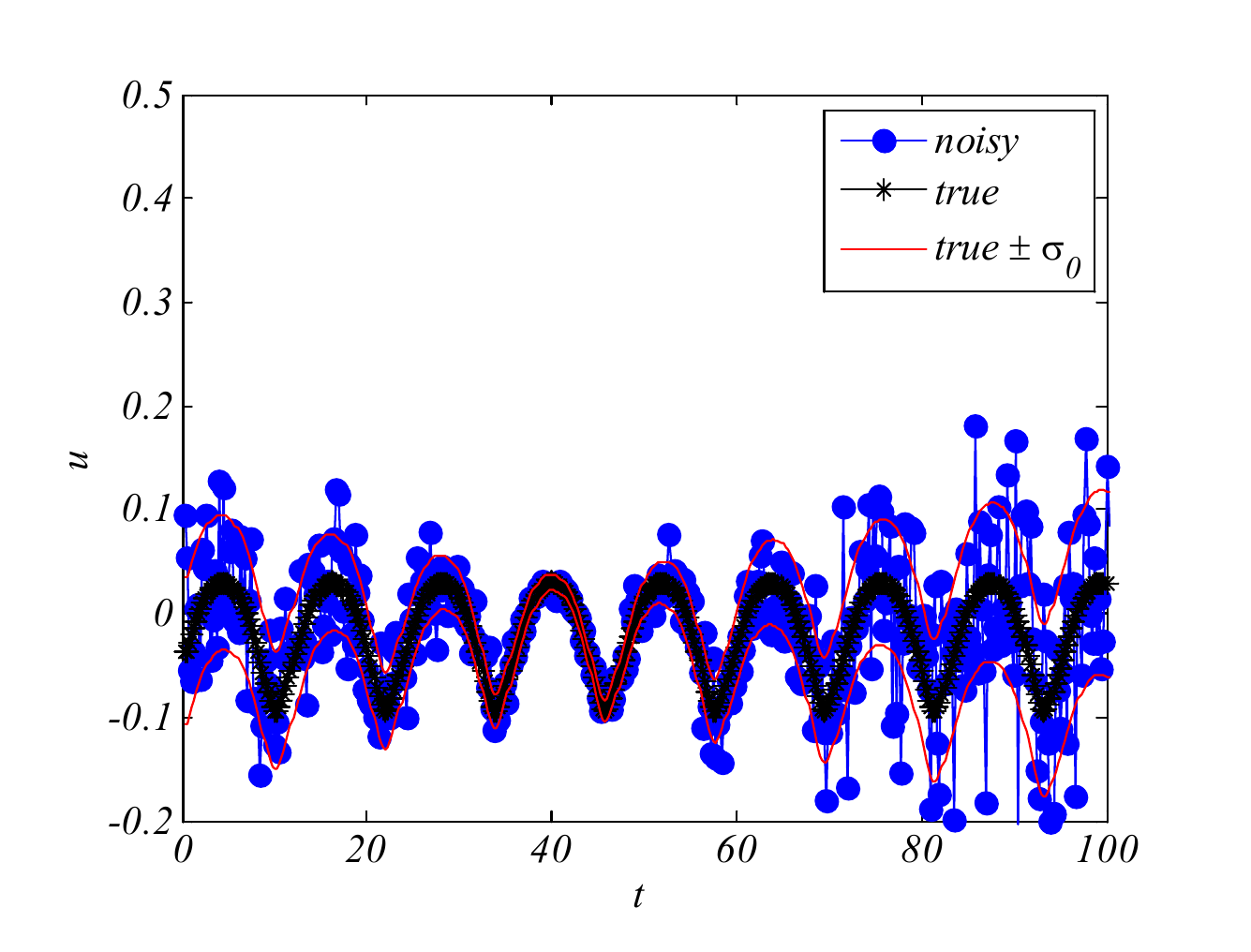}}
	\subfigure[$u$ at $(x,y)=(2D,0)$]{\includegraphics[width=5cm]{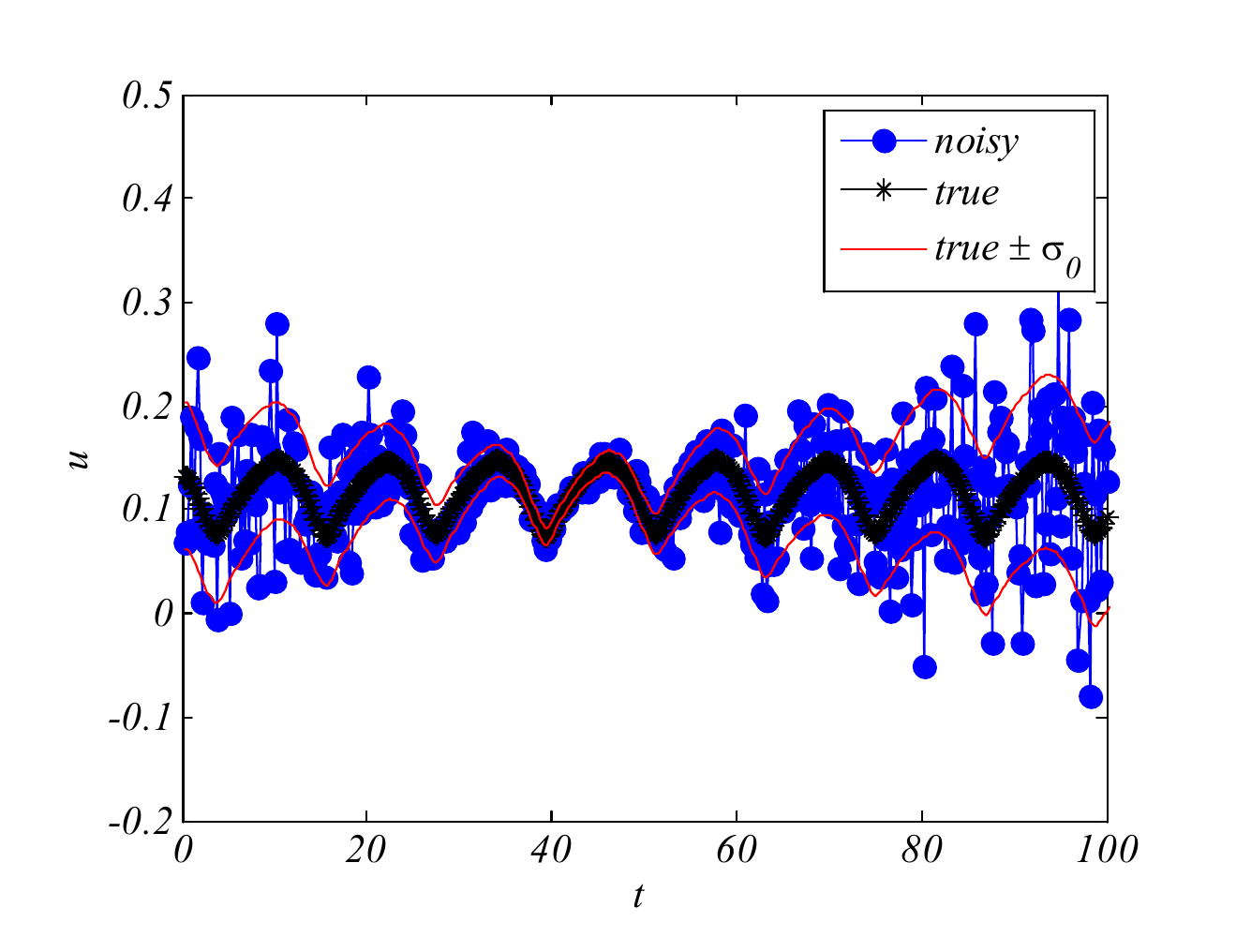}}
	\caption{The noisy and true time history of $x$-velocities at $(x,y)=(D,0)$  and $(x,y)=(2D,0)$ with time-varying noise of $\sigma_0^2=0.05$. Here, only the first 400 steps of the entire data matrix is illustrated. }
	\label{fig:cylhist}
\end{figure}

Here, $\sigma_0^2$ is set to 0.05, 0.1 and 0.2. The flows with the minimum and maximum noise intensities are shown in Figs. \ref{fig:flowwnoise} (a) and (b), respectively. Here, the timing of Fig. \ref{fig:flowwnoise} are set to be the same as in Fig. \ref{fig:flowwo}. The time history of the $x$-direction velocity is also shown in Fig. \ref{fig:cylhist}.

\begin{figure}
	\subfigure[$\sigma_0^2=0.05$]{\includegraphics[width=5cm]{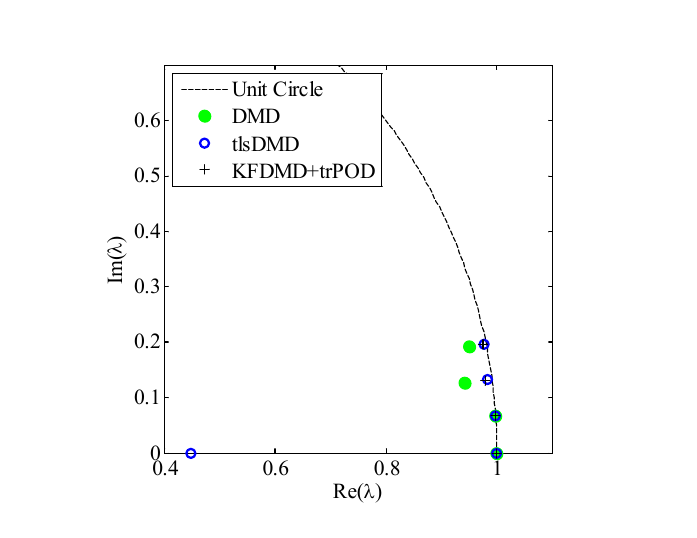}}
	\subfigure[$\sigma_0^2=0.1$]{\includegraphics[width=5cm]{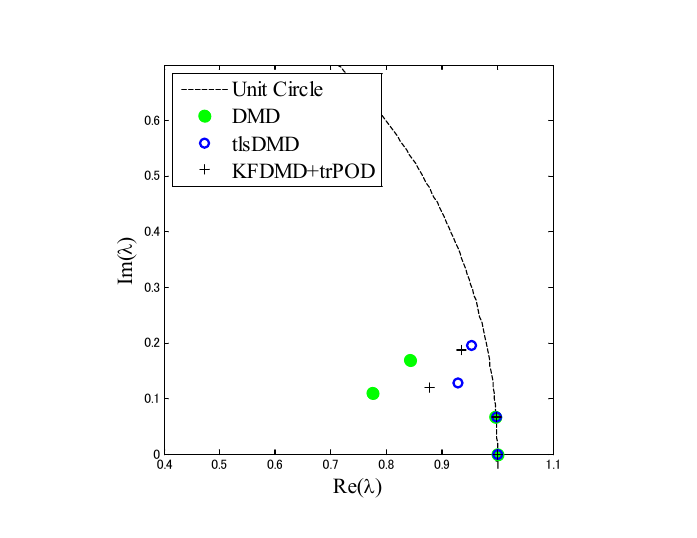}}
	\subfigure[$\sigma_0^2=0.2$]{\includegraphics[width=5cm]{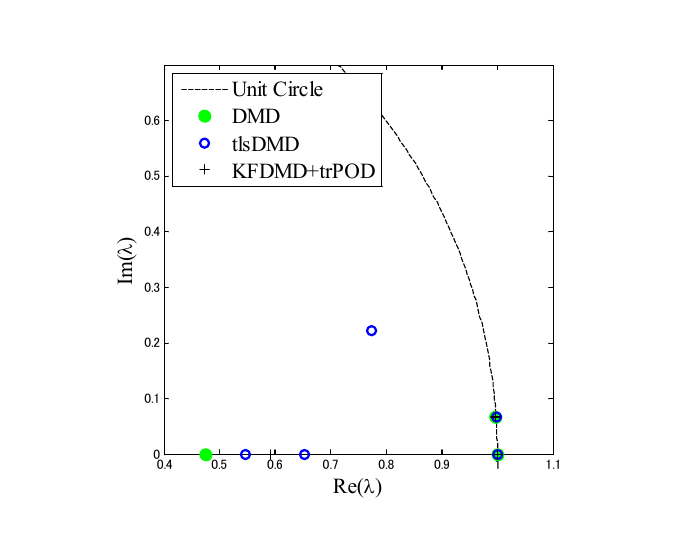}}	
	\caption{Eigenvalues of a static fluid system with noise.}
	\label{fig:cyl_eigenwnoise}
\end{figure}

\begin{figure}
	\subfigure[DMD mode 1]{\includegraphics[width=5cm]{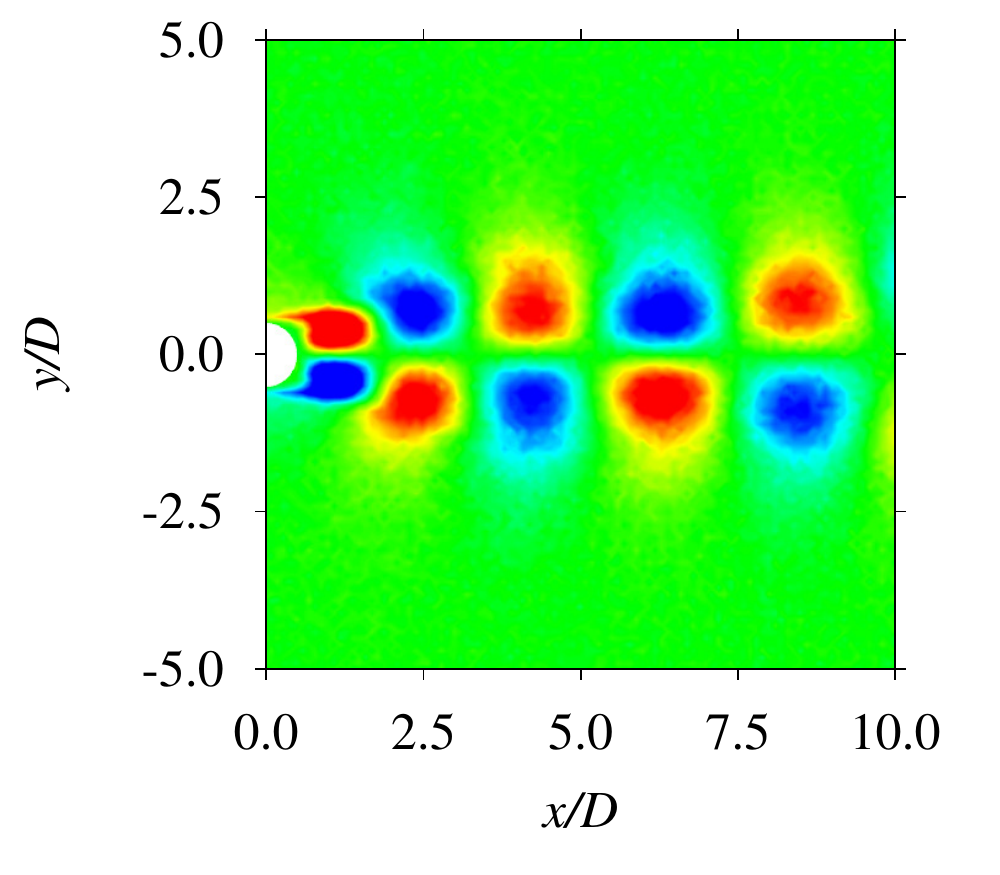}}
	\subfigure[DMD mode 2]{\includegraphics[width=5cm]{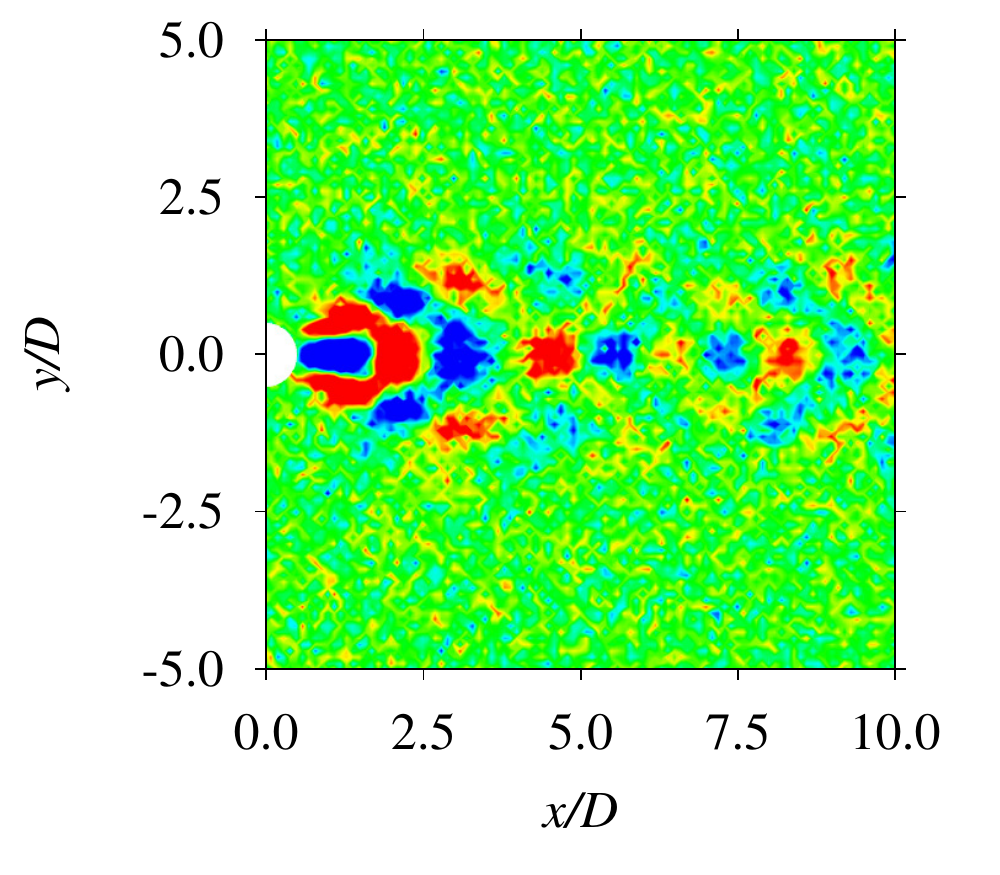}}
	\subfigure[DMD mode 3]{\includegraphics[width=5cm]{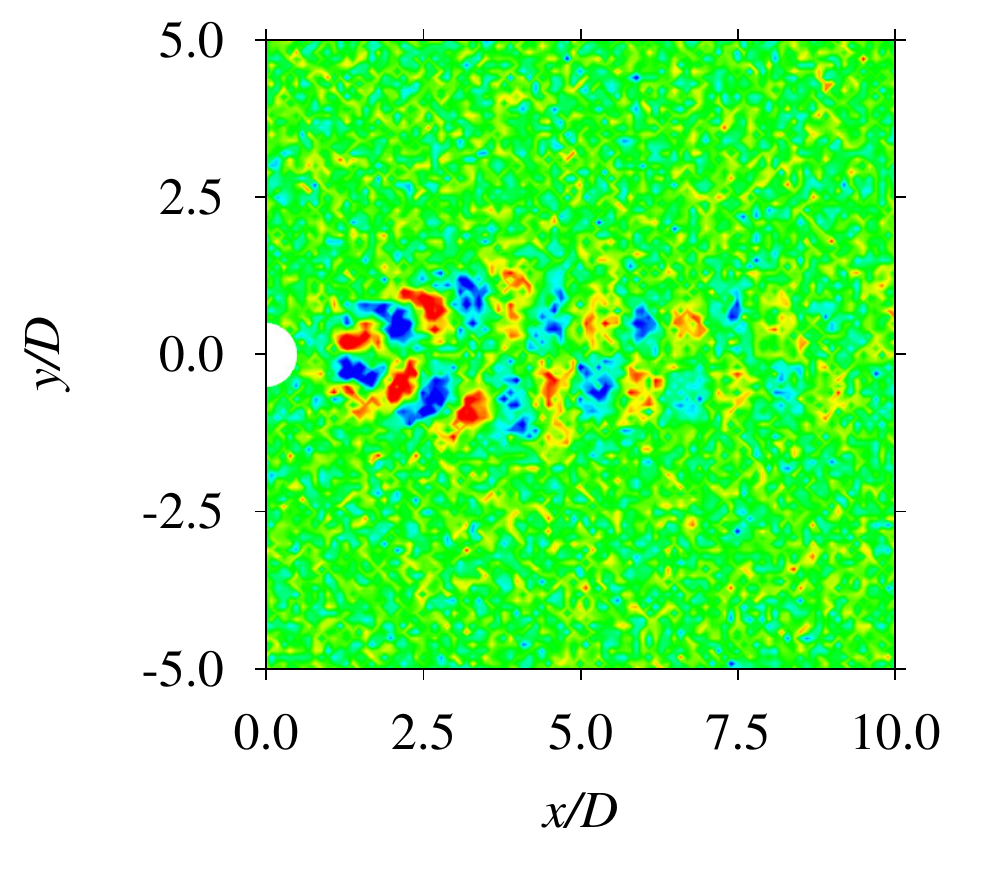}}\\
	\subfigure[tlsDMD mode 1]{\includegraphics[width=5cm]{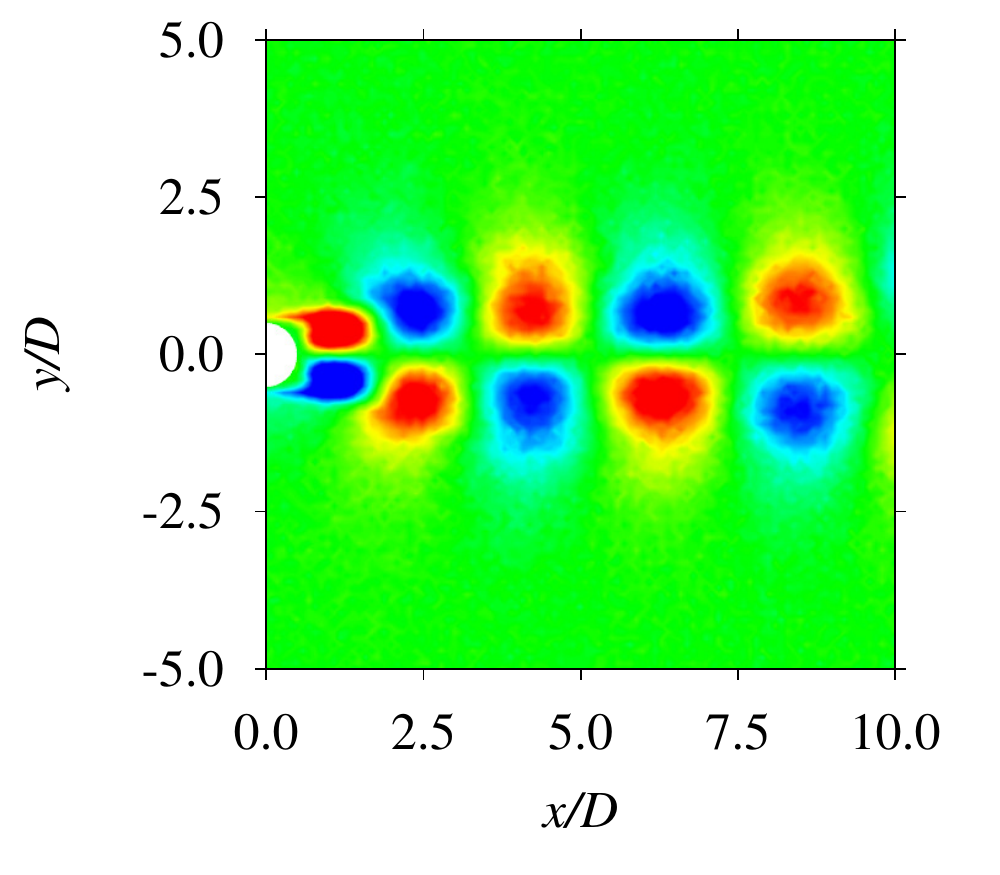}}
	\subfigure[tlsDMD mode 2]{\includegraphics[width=5cm]{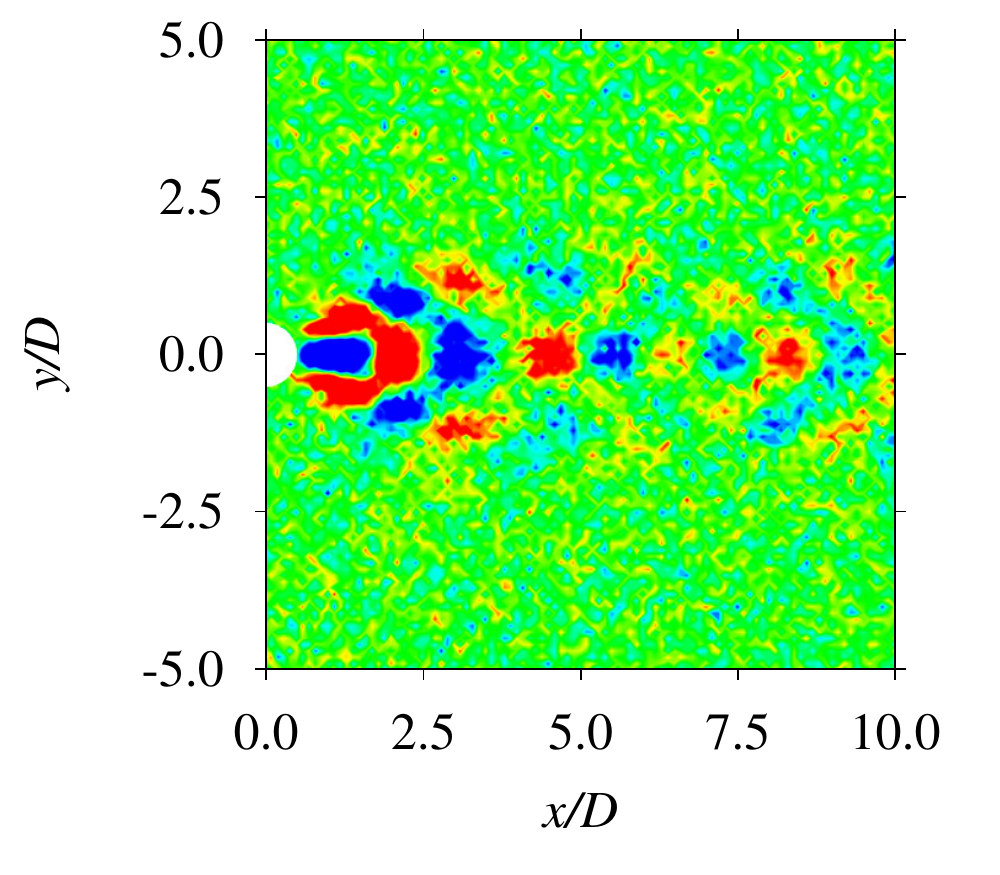}}
	\subfigure[tlsDMD mode 3]{\includegraphics[width=5cm]{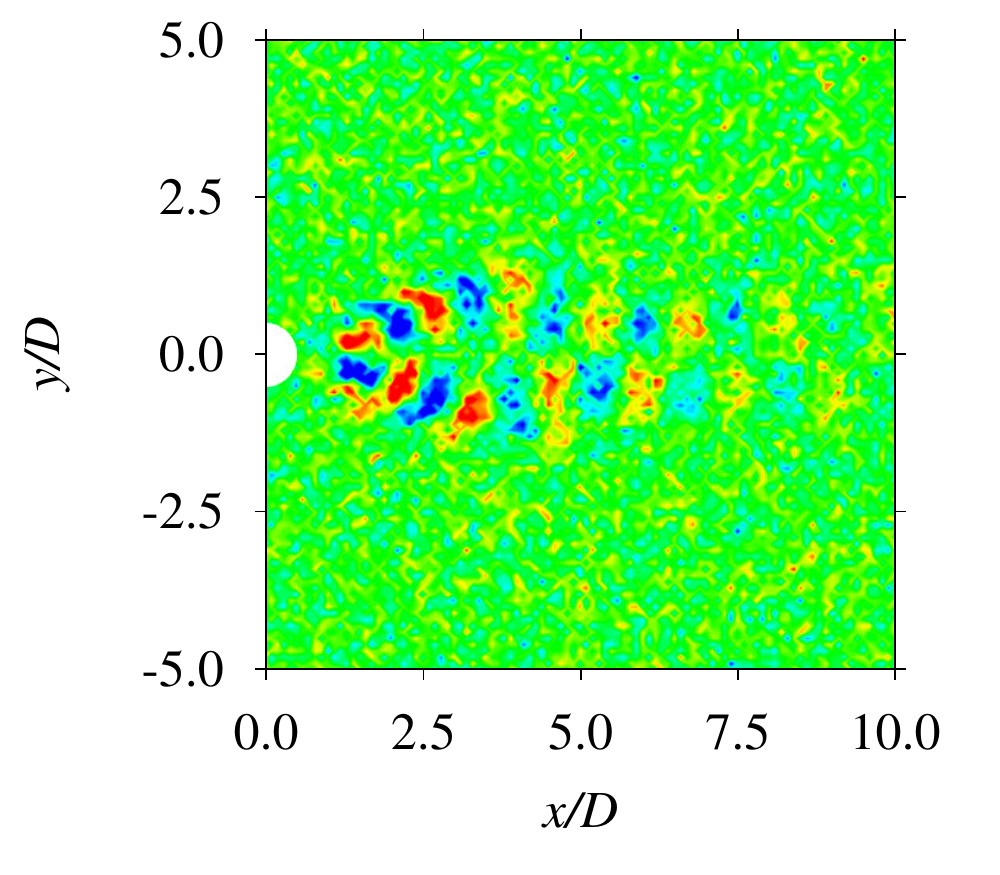}}\\
	\subfigure[KFDMD+trPOS mode 1]{\includegraphics[width=5cm]{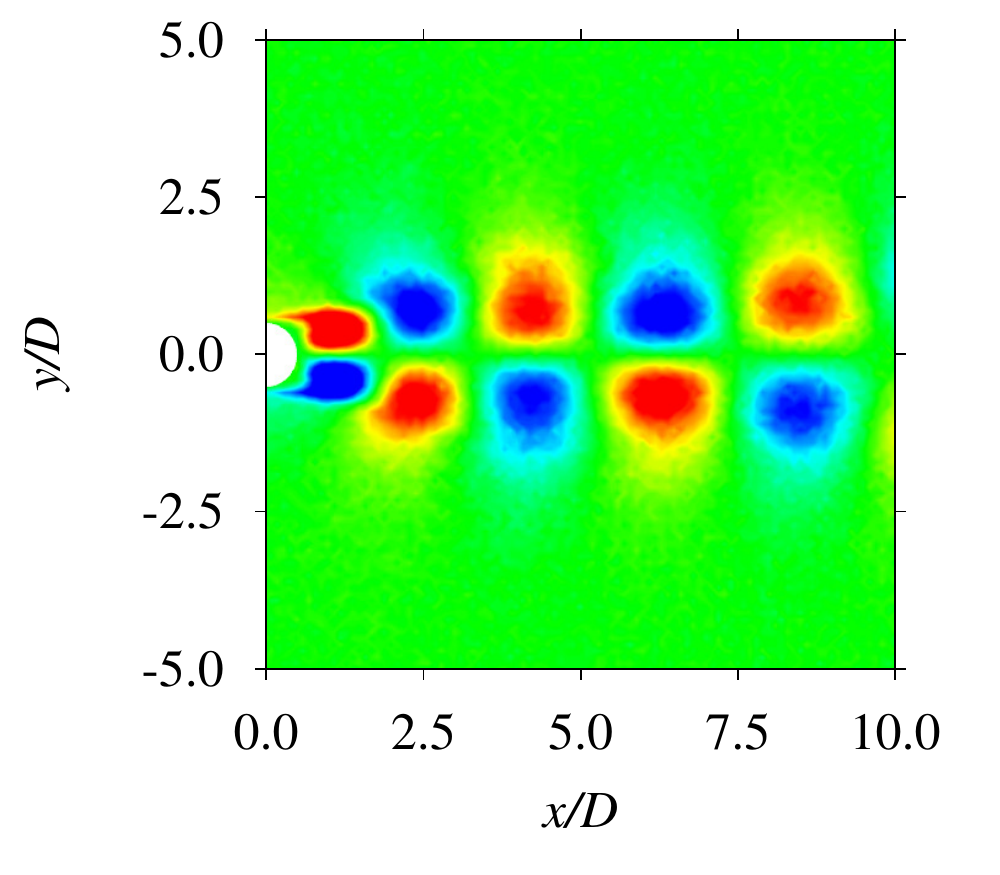}}
	\subfigure[KFDMD+trPOD mode 2]{\includegraphics[width=5cm]{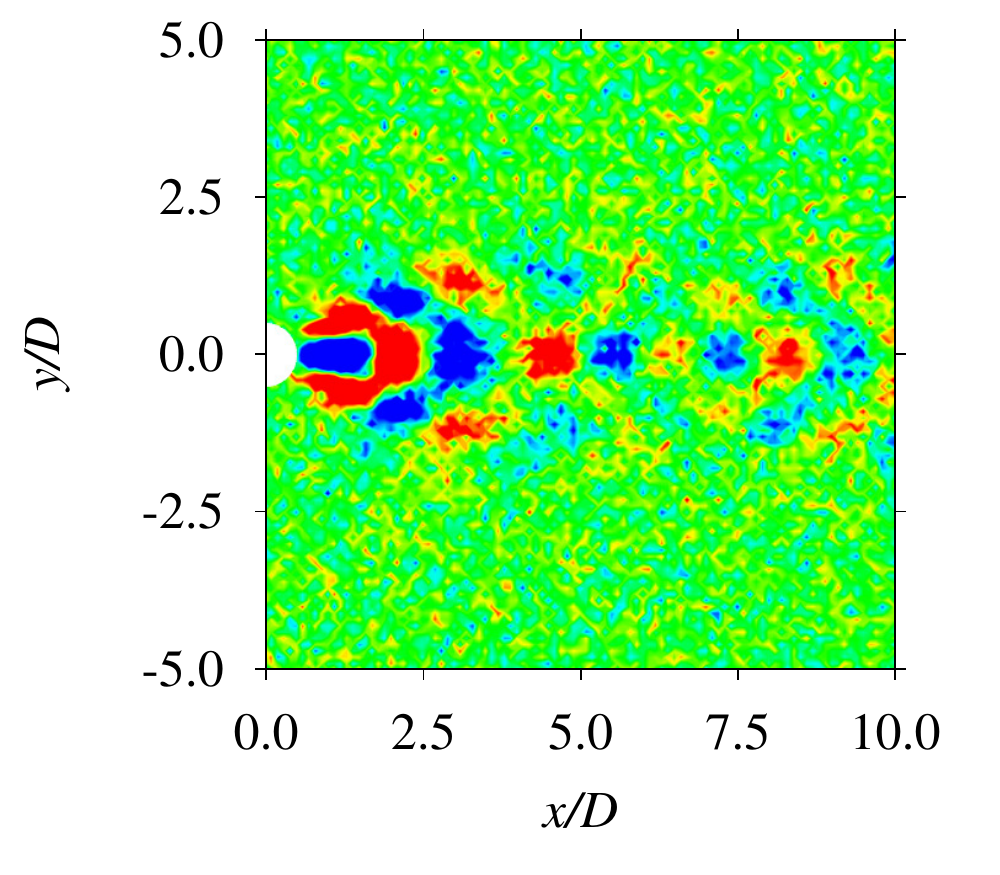}}
	\subfigure[KFDMD+trPOD mode 3]{\includegraphics[width=5cm]{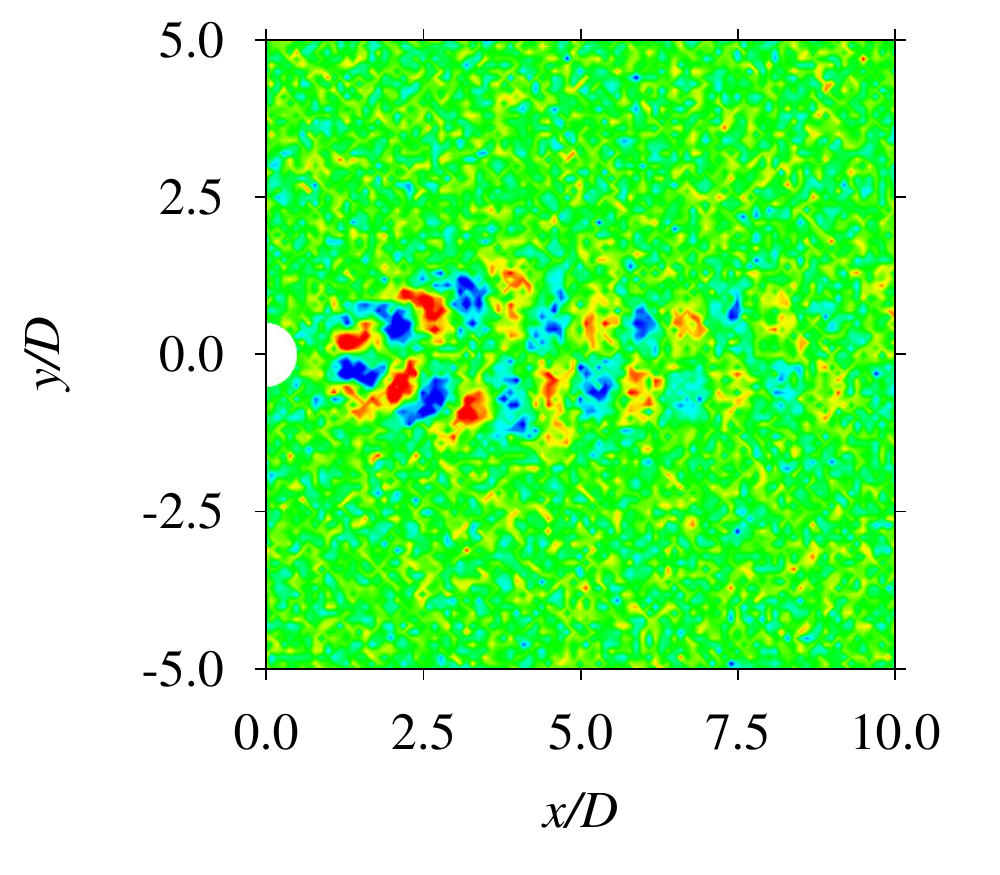}}\\
	\caption{Real part distributions of eigenmodes of a static fluid system with noise of $\sigma_0^2=0.05$. Here, contour color ranges from -0.01 to 0.01.}
	\label{fig:cyl_eigenmodewnoise}
\end{figure}

The results with noise are processed by standard DMD, tlsDMD and KFDMD, where KFDMD adopts the truncated POD (Eqs. \ref{eq:xtilde} and \ref{eq:ytilde}) as a preconditioner similar to the previous subsection. The eigenvalues computed by the standard DMD, tlsDMD and KFDMD methods are also shown in Fig. \ref{fig:cyl_eigenwnoise}. Those method find up to first to third eigenvalues depending on the method or noise strength. The lowest frequencies computed by standard DMD, tlsDMD and KFDMD correspond to the Strouhal number $St=fD/u_\infty \sim 0.2$.  The second and third eigenvalues computed by KFDMD are much closer to the values without noise similar to tlsDMD than  that computed by standard DMD. This result suggests that the eigenvalues computed by KFDMD are more accurate than standard DMD if the noise information is given. However, unfortunately, no superior performance of KFDMD to that of tlsDMD is not shown in Fig. \ref{fig:cyl_eigenmodewnoise}, especially for more severe noise cases ($\sigma_0=0.1$ or $0.2$). This might be because only 20 POD modes are used for KFDMD and its performance is degraded, as discussed in the previous sections. Trade-off in computational costs and accuracy on KFDMD by changing the number of modes used should be addressed in the future study. 

The real parts of the eigenmodes computed by standard DMD, tlsDMD and KFDMD for the data with noise of $\sigma_)=0.05$ are shown in Fig. \ref{fig:cyl_eigenmodewnoise}. Here, the phase of all the mode shown in this figure are adjusted by multiplying a complex variable so that the inner products of the modes shown here and the DMD mode for without noise become a real number. In addition, all the modes shown here are normalized.  All methods produce the first to third dynamic modes of K\'arm\'an vortex shedding, and these results show that estimated modes of all methods are almost similar to each other. Unfortunately, the eigenmodes (DMD mode) of tlsDMD and KFDMD are approximately the same as standard DMD. With regard to KFDMD, this might be because the eigenmodes are greatly affected by the preconditioning POD process, which involves a great deal of noise. This suggests that the present method combined with POD does not improve the eigenmodes, but only the eigenvalues. Including other methods, we need to address the accuracy on the eigenmode of various DMD in the future study. 

\section{Numerical Experiments of Problem for Dynamical System Identification}
\label{sec:NDSI}
Next, dynamical system identification using KFDMD is conducted. The system has only two eigenmodes, the real part of which is zero, and the time-dependent imaginary part is given by $\mathrm{Im}(\lambda) = 2\pi f$,
where the frequency $f$ is given by $f = 1 + k\Delta t$. Actually, the following data is given:
\begin{eqnarray}
x_1&=&\cos{\pi(1+k\Delta t) k\Delta t}\\
x_2&=&\sin{\pi(1+k\Delta t) k\Delta t},
\end{eqnarray}
where $\Delta t$ and $m$ are set to be 0.01 and 500, respectively.
First, this problem with/without quasi-steady noise is solved by KFDMD and the results are compared online DMD (oDMD). 

\begin{figure}
	\includegraphics[width=5cm]{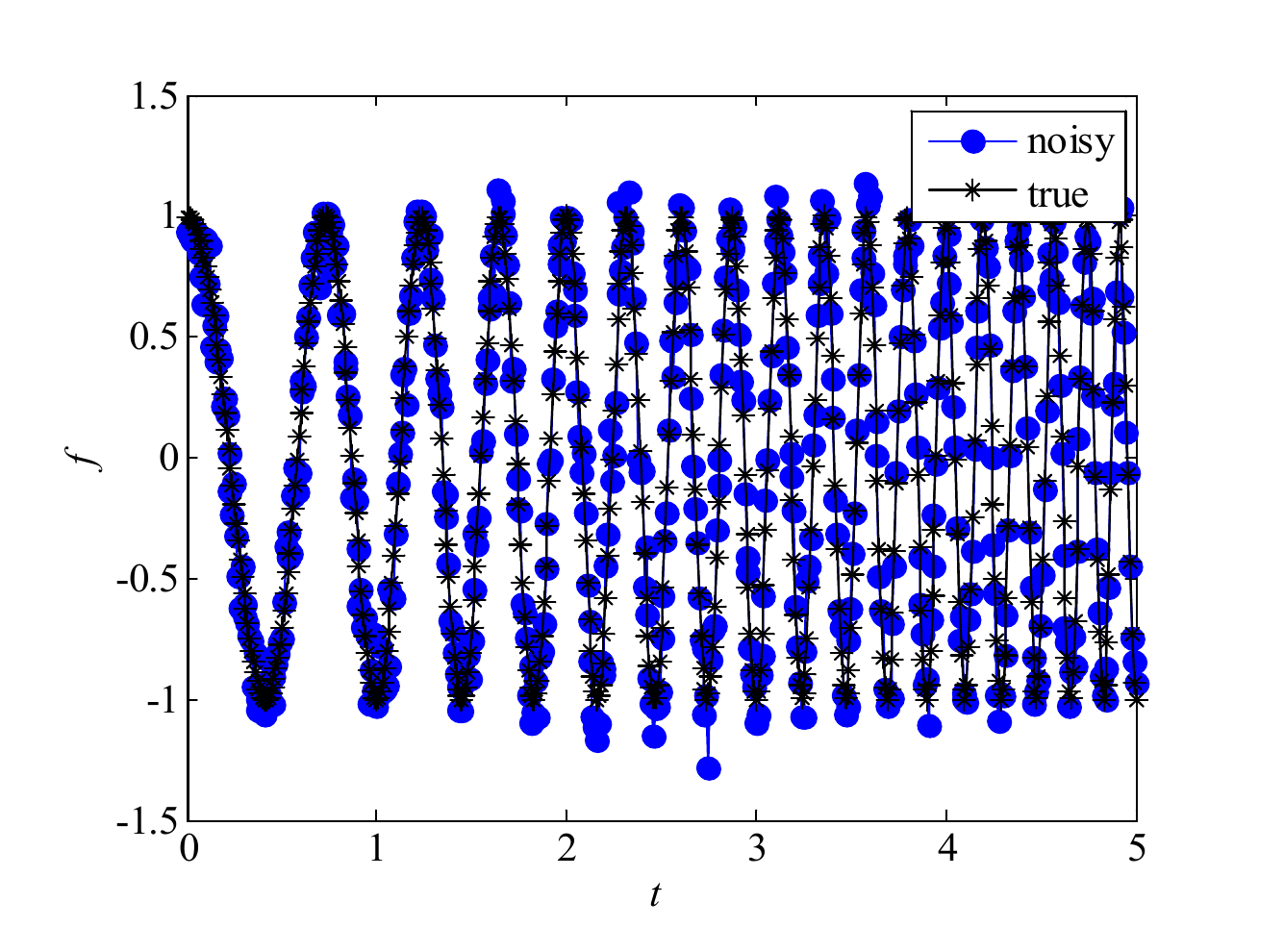}
	\caption{Time history of data with/without noise in dynamical system identification problem.}
	\label{fig:dynamic_history_single_wn}
\end{figure}

\begin{figure}
	\subfigure[oDMD]{\includegraphics[width=5cm]{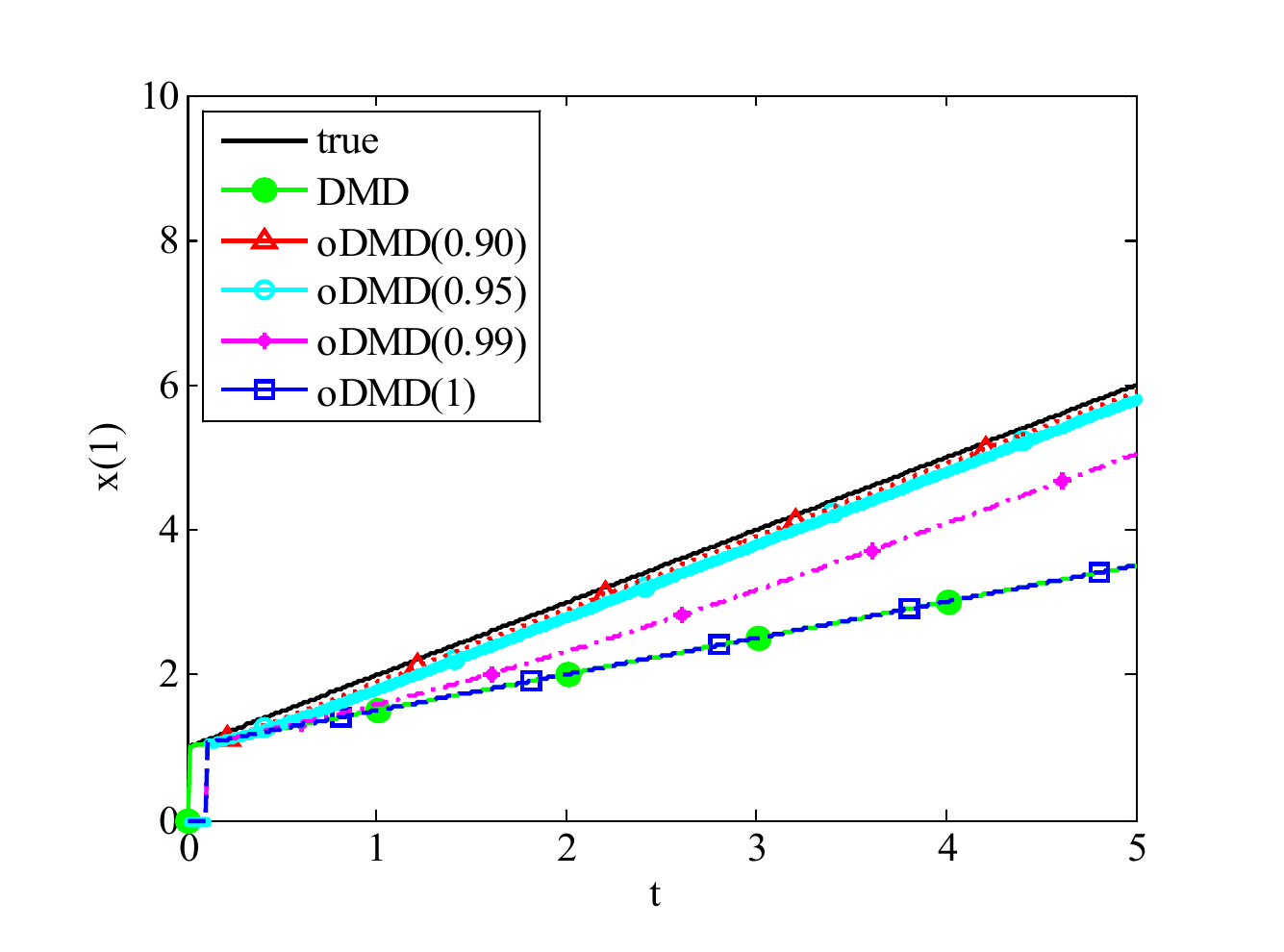}}
	\subfigure[KFDMD]{\includegraphics[width=5cm]{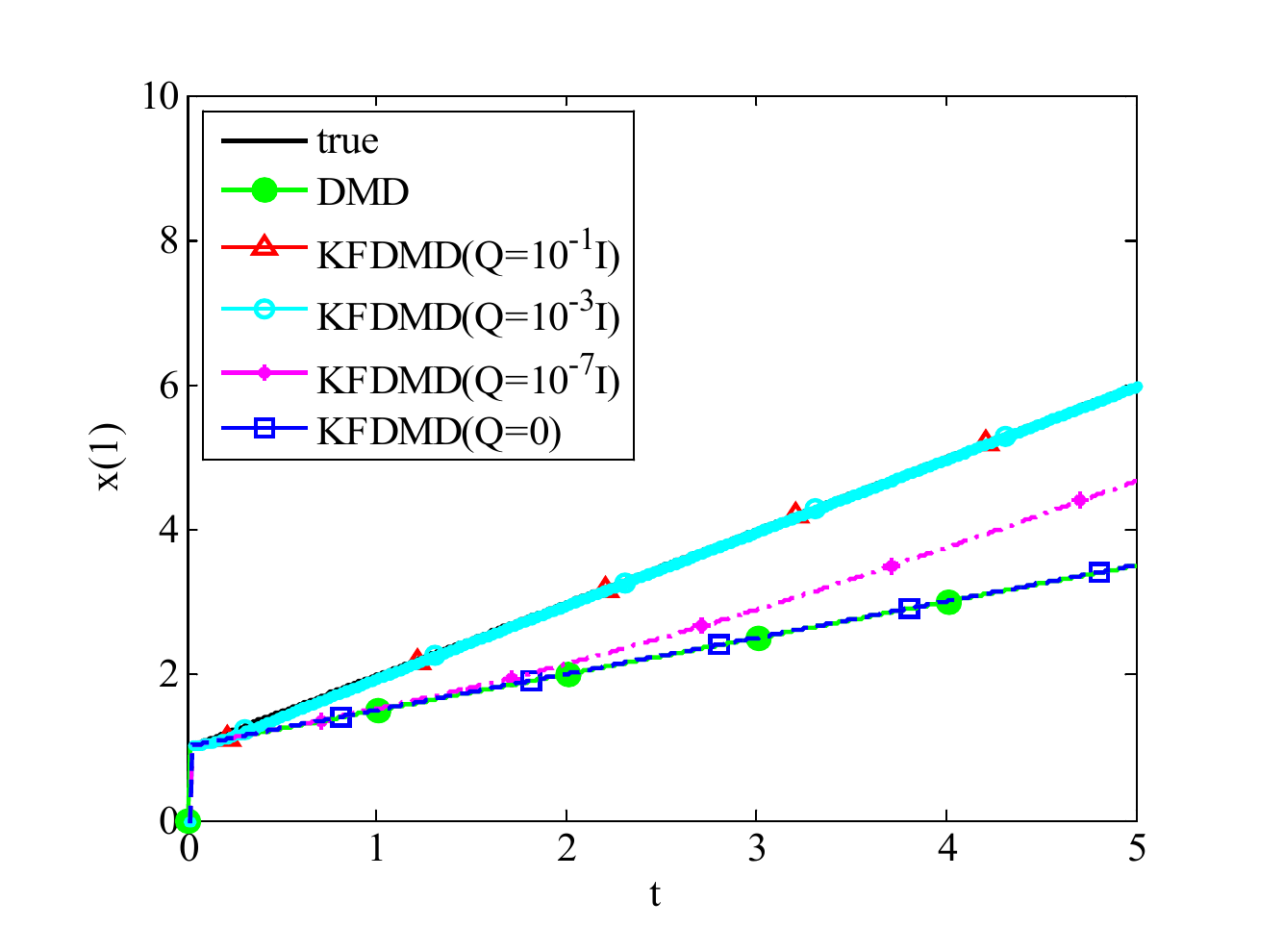}}
	\caption{Dynamical system identification for $n=2$ data without noise.}
	\label{fig:freq_history_wonoise_single}
\end{figure}

The time histories of without noise are shown in a ``true'' plot in Fig. \ref{fig:dynamic_history_single_wn}. For this data, oDMD is applied. Here, oDMD has two parameters, $\rho$ and $q$. Here, $\rho$ is a waiting factor where $\rho=1$ corresponding to equally weighting all the data that the algorithm obtained and less $\rho$ corresponding to weighting more on the recent data. Moreover, $q$ is a number of initial temporal data to guess the initial $A$ matrix.  In this problem, $q$ is set to be 2. The predicted frequencies for the data without noise compared with the true frequency are shown in Fig. \ref{fig:freq_history_wonoise_single}(a). The predicted frequency is based on the eigenvalue obtained by each algorithm, where the closest computed eigenvalue to the exact one is chosen.  Here, $\rho$ in the oDMD($\rho$) in the legend of figure represents the weighting factor. The results of standard DMD and oDMD are corresponding to each other because they use all the data with equal weights.  As explained above, a less weighting factor can predict the time-varying frequency better. Then, the results of KFDMD are discussed. For KFDMD we set the hyperparameter $Q=0, 10^{-6}I,10^{-4}I$ and $10^{-3}I$ and $R=10^{-2}I$, though the observation noises is absent. The results with changing $Q$ and fixed $R$ shown in Fig. \ref{fig:freq_history_wonoise_single}(b) illustrate that the predicted frequencies varies depending on the $Q$ parameter. With $Q=0$, the result collapses with that of standard DMD, because assumption of $Q=0$ is corresponding to that of the constant system and the Kalman filter estimates the system using the all the data obtained so far. On the other hand, increasing the $Q$, the predicted frequency approaches to the true frequency. This is because the Kalman filter assumes the system changing its coefficients more frequently with higher $Q$. In this case, it should be noted that we tried different $R$ conditions and in those case the results are affected by the ratio of $R$ and $Q$ values. These result show that the Kalman filter works as well as oDMD for estimation of time varying system, and the ratio of $R$ and $Q$ is working as well as the weighting parameter $\rho$ in oDMD. 

\begin{figure}
	\subfigure[oDMD]{\includegraphics[width=5cm]{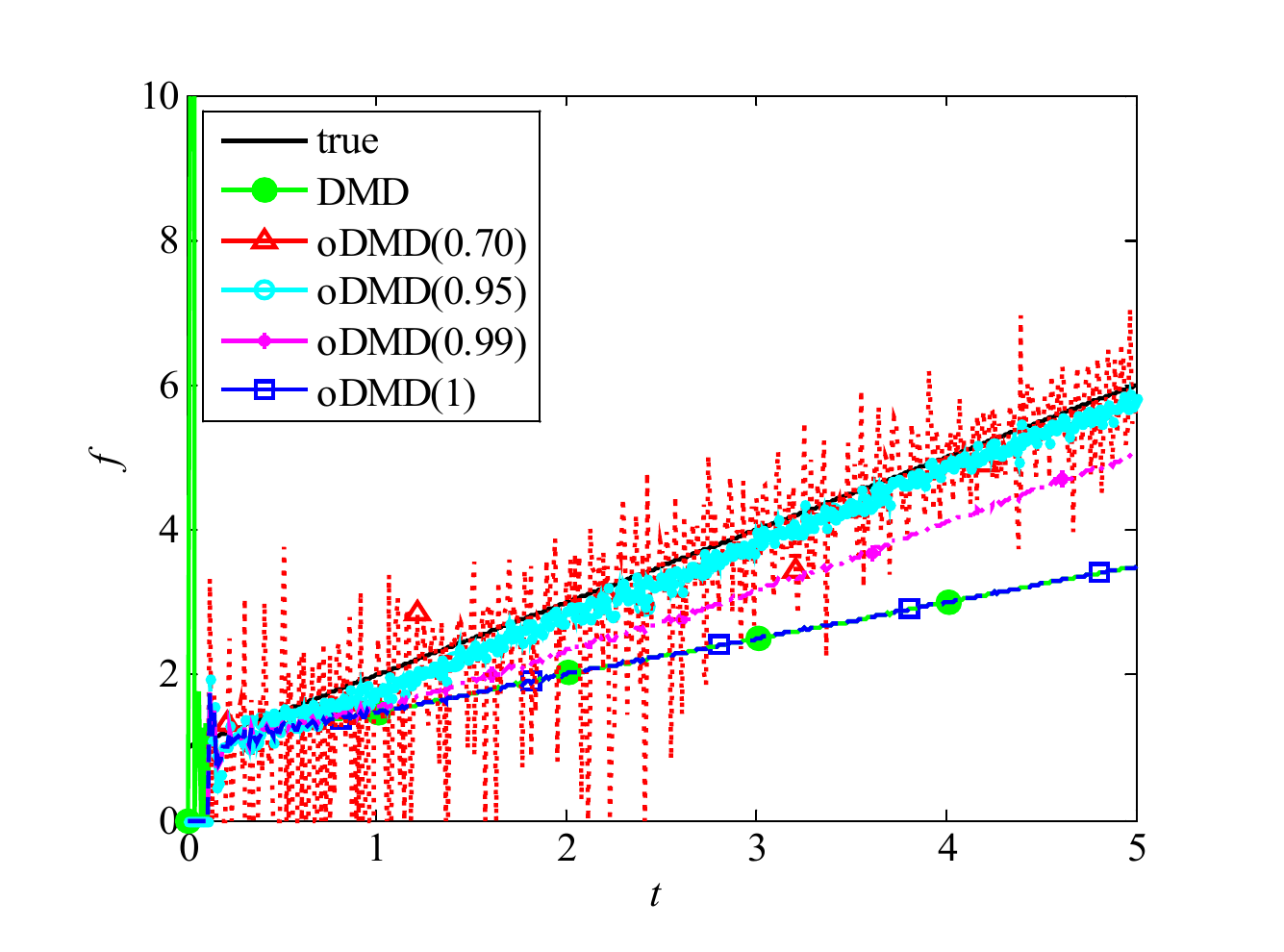}}
	\subfigure[KFDMD]{\includegraphics[width=5cm]{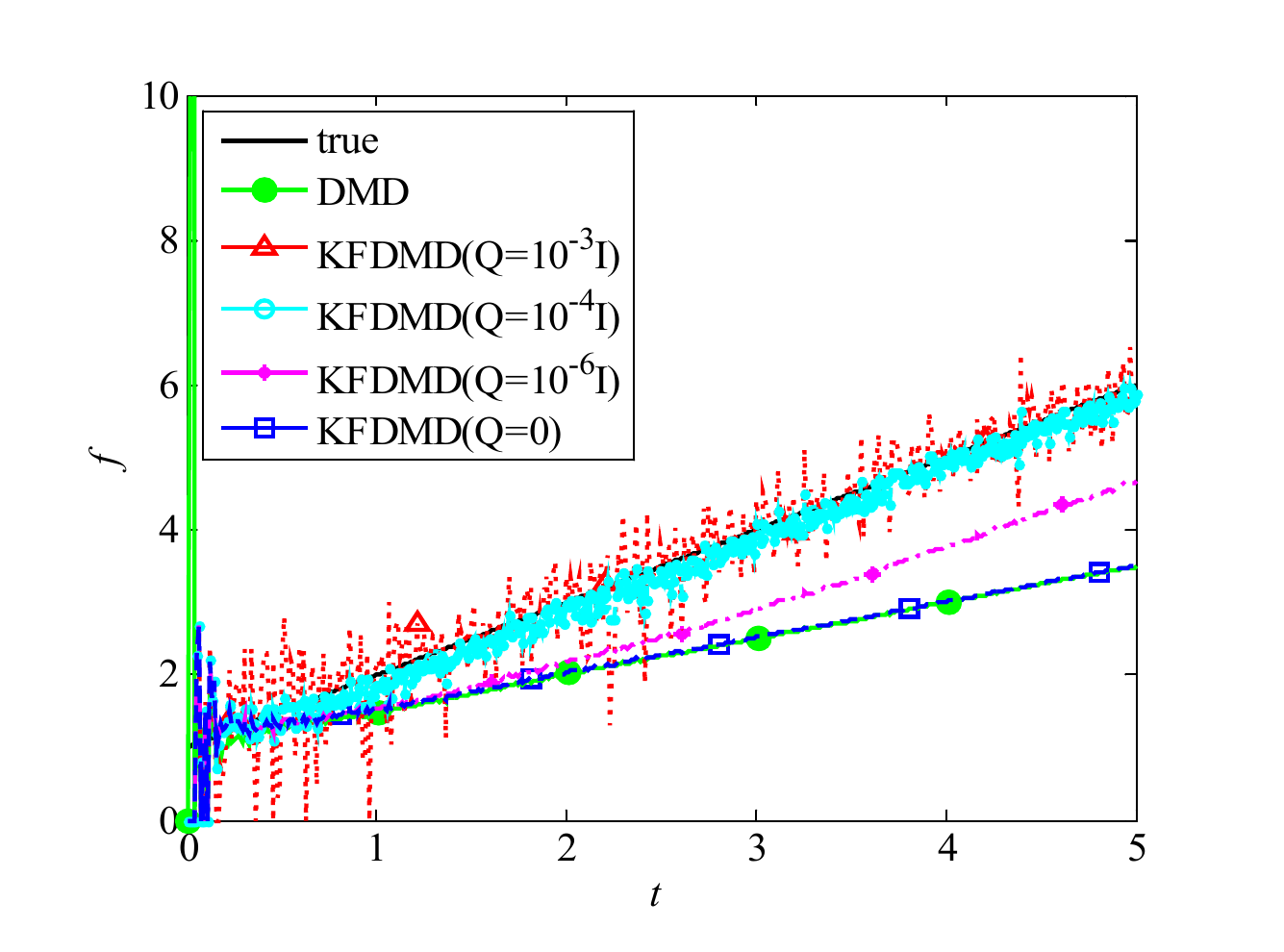}}
	\caption{Dynamical system identification for $n=2$ data with noise of $\sigma^2=0.01$.}
	\label{fig:freq_history_wnoise_single}
\end{figure}

Then, we conducted oDMD and KFDMD processes for the data with noise as shown in a ``noisy'' plot in Fig. \ref{fig:dynamic_history_single_wn}. The noise of $\mathcal{N}(0,\sigma^2)$ is added for both nodes, where $\sigma=10^{-2}$. In this case, $R=10^{-2}I$ is adopted which is correct settings. The predicted frequencies are shown in Fig. \ref{fig:freq_history_wnoise_single}.  In the early stage of $t<0.2$, there is an strong oscillation for all the implementations. This is because there are spurious eigenvalues in the early stage when the noise is added. Figure \ref{fig:freq_history_wnoise_single} shows that both oDMD and KFDMD work well with the parameters $\rho=0.95$ for oDMD and $R=10^{-3}$ for KFDMD.  Results of oDMD with lower $\rho$ or KFDMD with higher $R$ become noisy because those parameters lead to immediate adjustment to the latest dataset and the results are much affected by the noise. On the other hand, results of oDMD with higher $\rho$ or KFDMD with lower $R$ are slowly changed because those parameters lead to the system estimation with longer duration, where this characteristics are the same as the results of the previous test case without noise. The present results illustrate that the parameter should be chosen carefully for dataset with noise. In the parameter tuning process, the criteria of weighting factor in oDMD seem to be depending on empiricism, but the hyperparameters in KFDMD are corresponding to system and observation noises which seems to be natural to give. 

\begin{figure}
	\includegraphics[width=5cm]{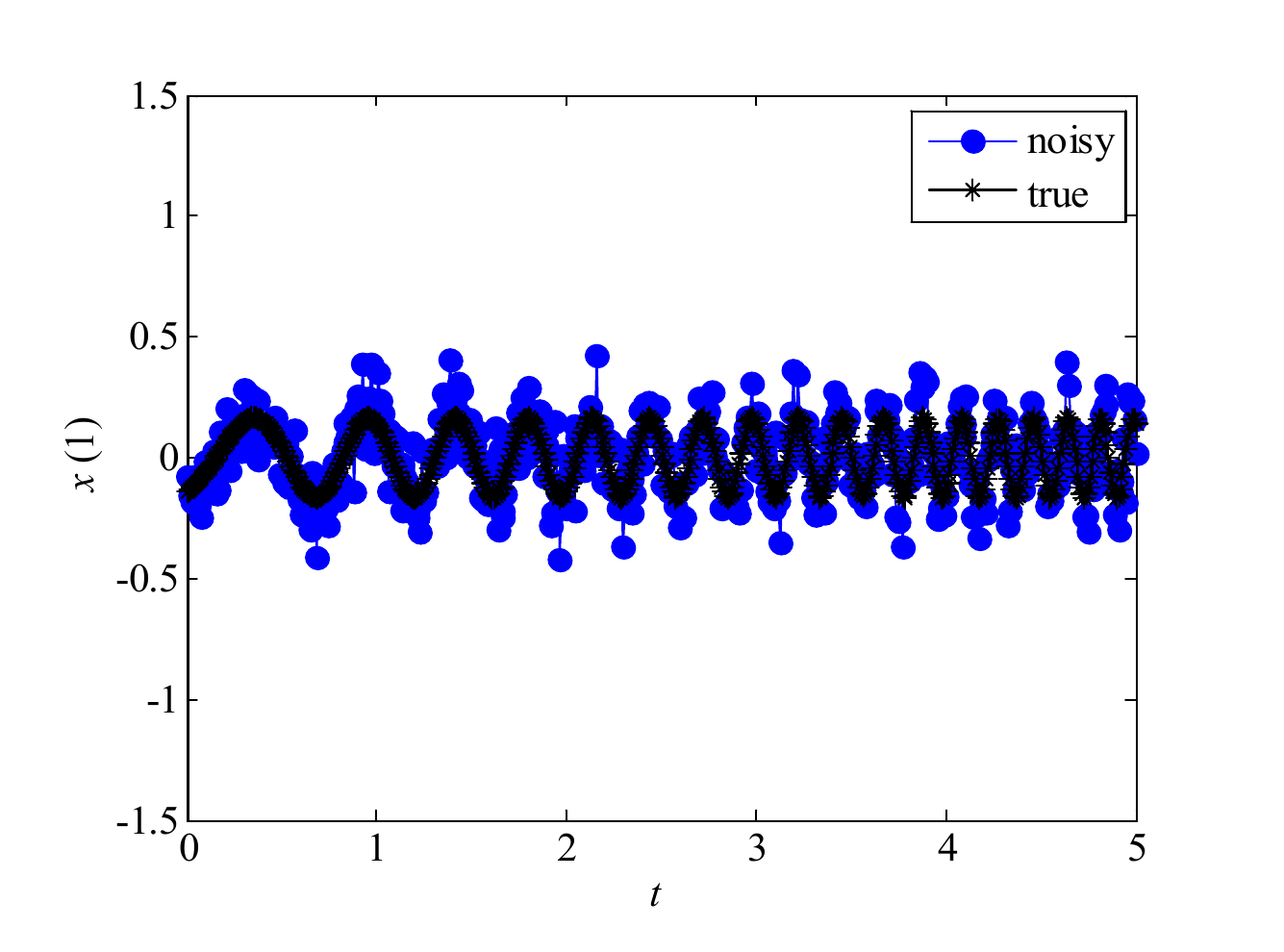}
	\caption{Time history of one of projected data with/without noise in dynamical system identification problem.}
	\label{fig:dynamic_history_wn}
\end{figure}

\begin{figure}
	\subfigure[oDMD]{\includegraphics[width=5cm]{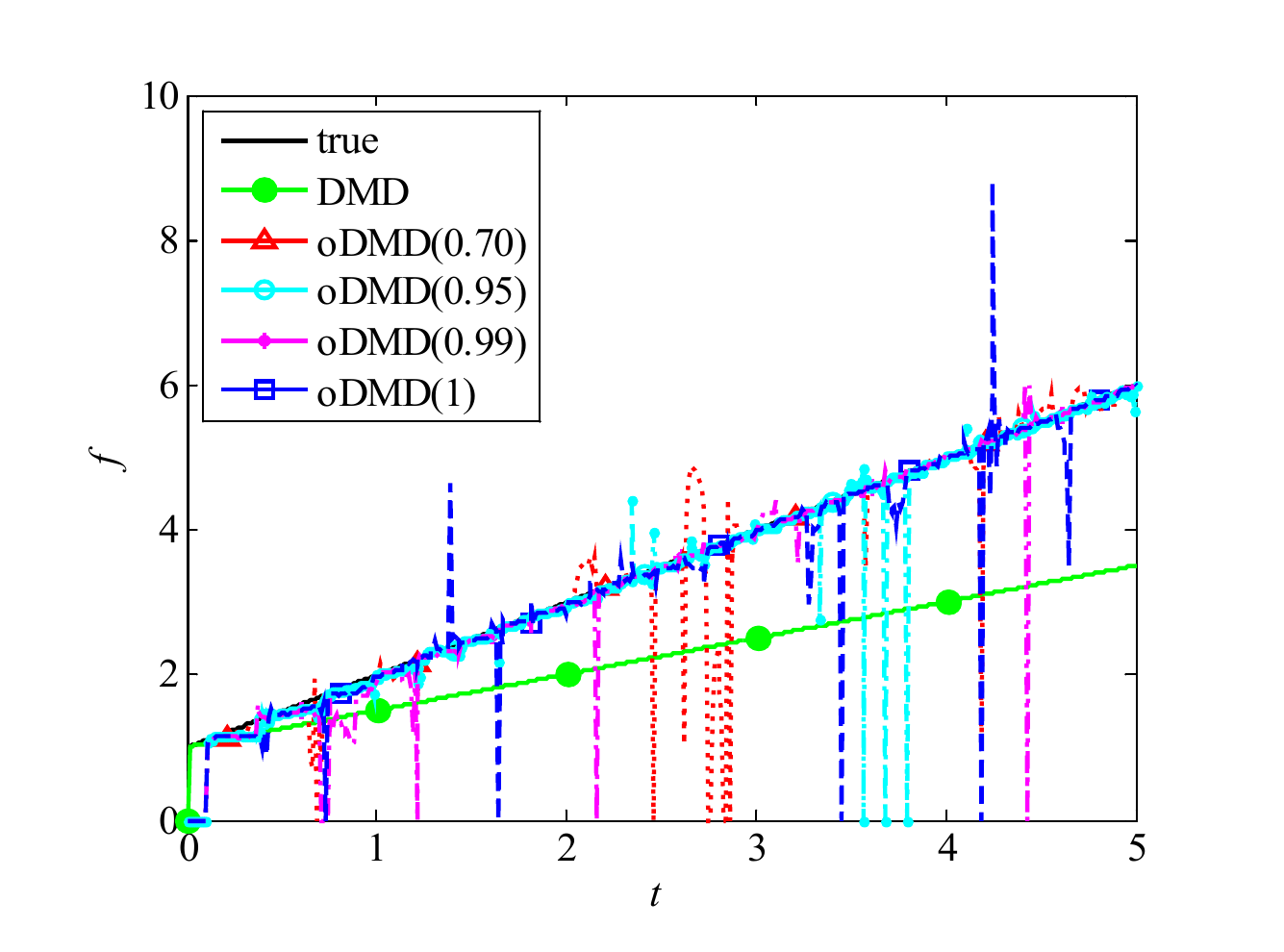}}
	\subfigure[KFDMD]{\includegraphics[width=5cm]{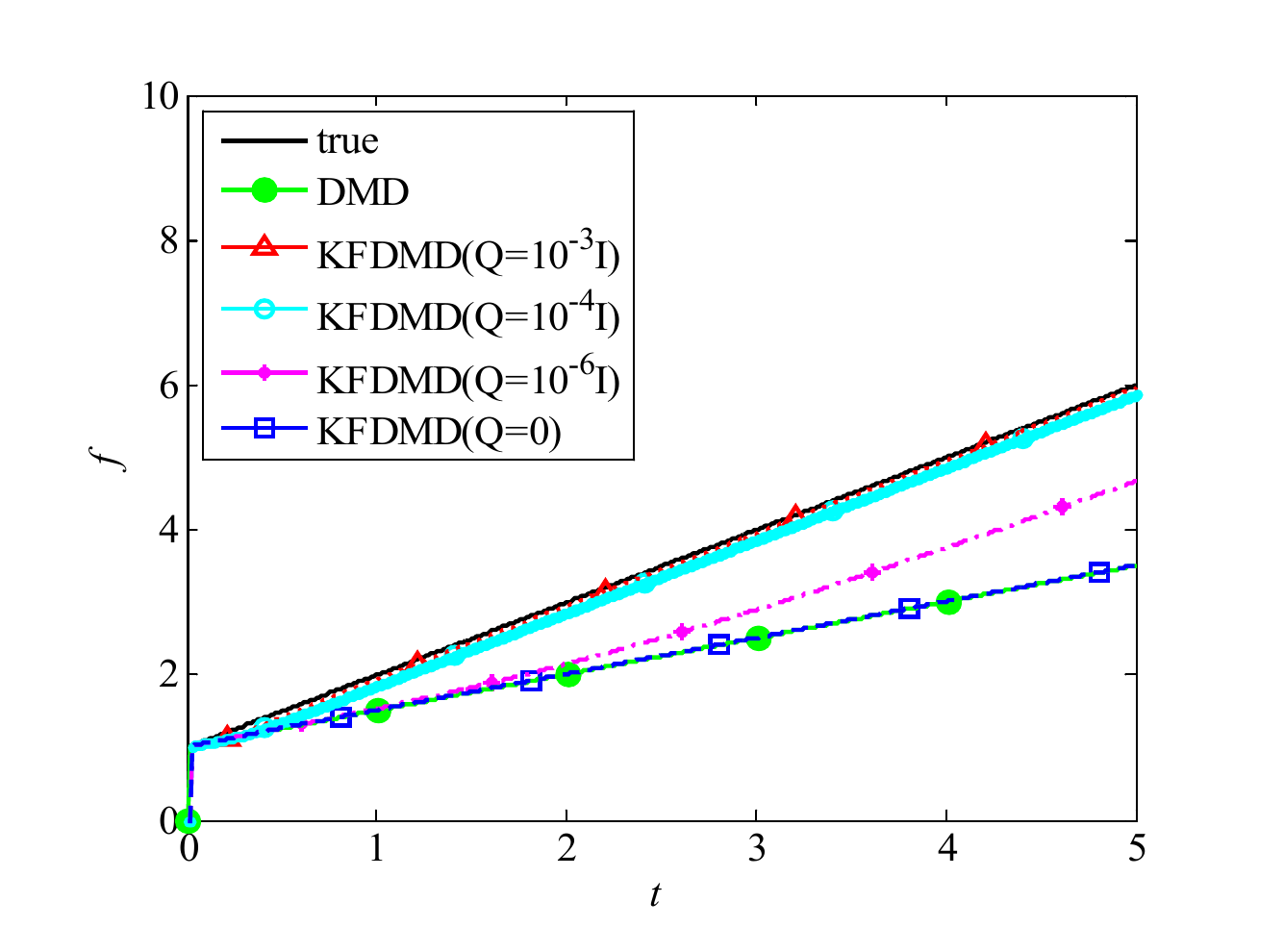}}
	\caption{Dynamical system identification for projected $n=20$ data without noise.}
	\label{fig:freq_history_wonoise}
\end{figure}

Then, the system is projected to a new system of dimension $n=20$ using the same method stated in the Section \ref{sec:NETVN}. Time history of the first node without noise is shown by a ``true'' plot in Fig. \ref{fig:dynamic_history_wn}, where the amplitude of signal is lower than the previous example because of the projection process by the $QR$ decomposition. First, data without noise are processed. In this test case, the same $\rho$ parameters and $q=2$ are used for oDMD and the same $R$ and $Q$ parameters are used for KFDMD. Figure \ref{fig:freq_history_wonoise} shows that oDMD works well for prediction of the frequency in almost all the region, although the frequencies are occasionally lost in short time and recovered after that. On the other hand, Fig. \ref{fig:freq_history_wonoise}(b) shows that KFDMD can predict the frequencies without failing the prediction. Both figures show that the hyperparameters works as well as in the two-degree-of-freedom problem above and trends in the results changed by the hyperparametrs are the same as those in the problem above.

\begin{figure}
	\subfigure[oDMD]{\includegraphics[width=5cm]{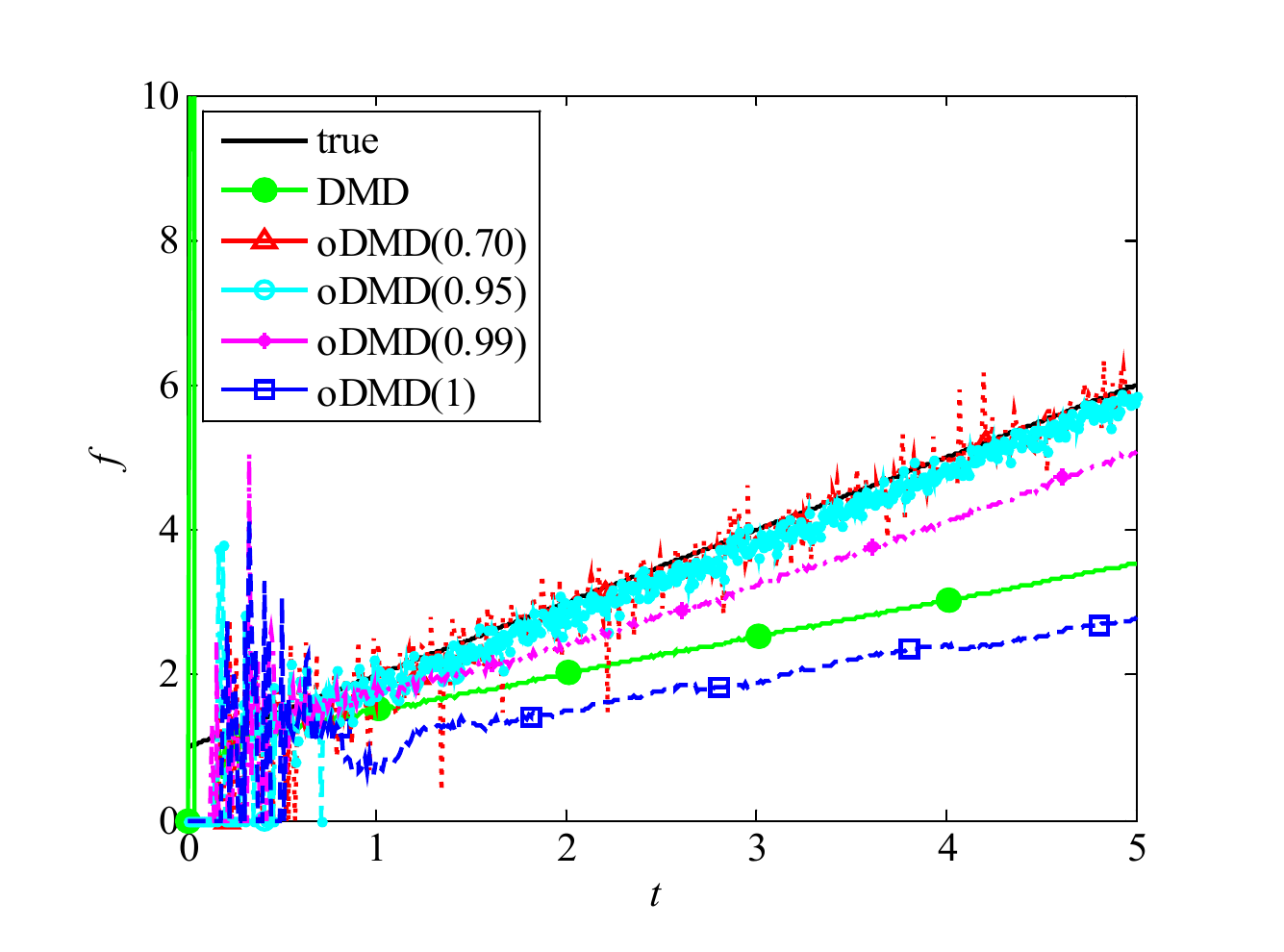}}
	\subfigure[KFDMD]{\includegraphics[width=5cm]{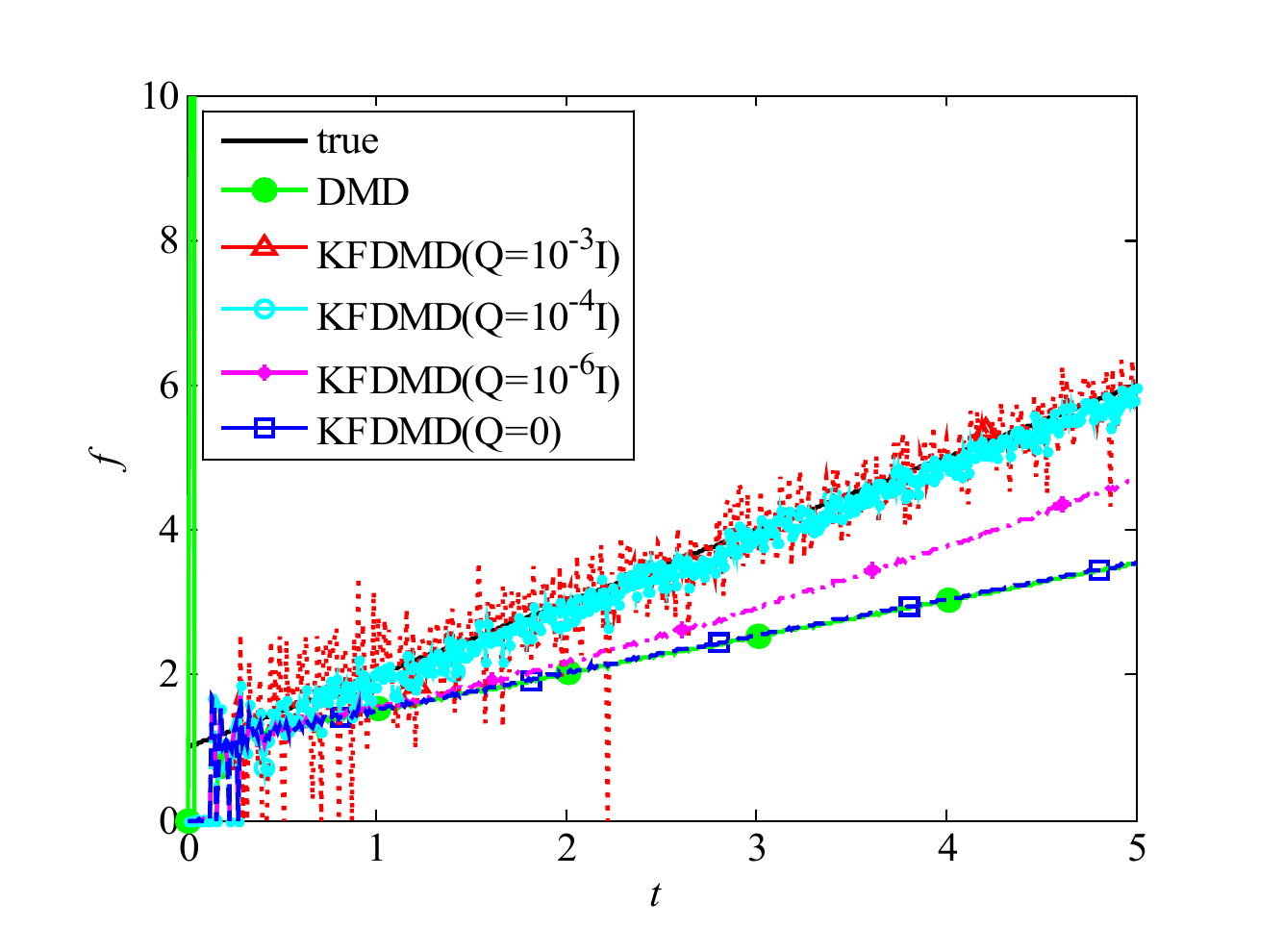}}
	\caption{Dynamical system identification for projected $n=20$ data with noise of $\sigma^2=0.01$.}
	\label{fig:freq_history_wnoise}
\end{figure}

Finally projected $n=20$ data with the noise of $\mathcal{N}(0,\sigma^2)$ are processed, where as $\sigma^2=0.01$. The noisy data are also shown by a ``noisy'' plot in Fig. \ref{fig:dynamic_history_wn}. As noted before, signal level becomes lower but the noise level is the same. Therefore, a resulting signal-to-noise ratio becomes worse for each data on the node, but the more of the data due to the projection can be used for this problem. Figure \ref{fig:freq_history_wnoise} shows the similar trends to those in Figs. \ref{fig:freq_history_wnoise_single}, respectively, and both algorithms work well for the projected data with noise.  

This numerical experiment shows that the proposed method can naturally track the true frequency as well as the recent oDMD algorithm. We also considered to apply KFDMD to transient flow around cylinder which is suggested by an anonymous reviewer, but development of online algorithms for time-varying system is just started and it seems to be out of scope of this paper. Therefore, it is left for the application of KFDMD in the future study. 

\section{Conclusions}
\label{sec:Concl}
A novel dynamic KFDMD was proposed. Our numerical experiments revealed that KFDMD can estimate matrix $A$ more precisely than the standard DMD, but less precisely than tlsDMD. However, the Kalman filter dynamic mode decomposition can find the eigenvalue in the high noise level condition in which the tlsDMD fails to find it.  
This characteristic is fine when the data is contaminated by the strong noise but the characteristics of the noise are known when identification is performed.
Furthermore, the proposed method can identify time-dependent systems in which the matrix is transited with time similar to oDMD, and this expansion is naturally conducted owing to the characteristics of the Kalman filter. Note that all of these properties are preferred in data analysis. The dynamic mode decomposition method based on the Kalman filter is a promising tool for analyzing a noisy dataset. 

In addition, the results in the present paper also illustrates a couple of points to be improved for KFDMD in the future study, as follows. 1)  We proposed a combination with trPOD and KFDMD, but it is sometimes difficult to obtain the spatial POD modes in advance in a purely online situation. In order to solve this problem, combination of KFDMD and online POD should be considered. This is not straightforward as noted in Section \ref{sec:POD} and should be addressed in the future study.
2) In the case of the quasi-steady noise problem, KFDMD always does not work better than the other noise-robust implementations of DMD. Therefore, improvement on KFDMD should be addressed in the future study, especially for the quasi-steady noise problem. 

Finally, it should be noted that the extended Kalman filter for simultaneous system identification and filtering of state variables are constructed based on KFDMD developed in this paper by the present authors,\cite{Nonomura2018} though it is further computational expensive. 

\section*{Acknowledgment}
This work was partially supported by JST Presto ( Grant Number JzMJPR1678). 

\appendix
\section{Combination with KFDMD and tlsDMD Ideas}
As shown in Section \ref{sec:KF} in which Kalman filter DMD is introduced, we consider the noise added in $\bm{y}$, but do not consider the noise added in $\bm{x}$. This leads to biased error as discussed in the previous studies, and the results of the obtained algorithm is expected to have biased error. Therefore, it is required to debias the error. Fortunately, input and output of the KFDMD are the same as those of the other DMD methods and some of the techniques for debiasing could be utilized. 

Here, the extension of trPOD based on tlsDMD is introduced for further improvement of denoising characteristics. We assume the off-line condition. In this case, pair POD which is used in tlsDMD can be used also for KFDMD and the further noise reduction is expected for KFDMD.  
First, we get the pair data matrix as follows:

\begin{eqnarray}
W=\left[\begin{array}{c}
X\\Y
\end{array}\right]
\end{eqnarray}
and POD is applied to pair data matrix:
\begin{eqnarray}
W=\left[\begin{array}{c}U_{X}\\U_{Y}\end{array}\right]D_{W}V^{\text{T}}_{W},
\end{eqnarray}
Then, we get truncated pair POD as follows:
\begin{eqnarray}
\widehat{W}_{1:m}=\left[\begin{array}{c}\widehat{U}_{X}\\\widehat{U}_{Y}\end{array}\right]\widehat{D}_{W}\widehat{V}^{\text{T}}_{W},
\end{eqnarray}
Here, we obtain $\widehat{X}$ and $\widehat{Y}$ as
\begin{eqnarray}
\widehat{X}&=&X\widehat{V}_{W}\widehat{V}_{W}^{\text{T}}\label{eq:xbhat}\\
\widehat{Y}&=&Y\widehat{V}_{W}\widehat{V}_{W}^{\text{T}}\label{eq:ybhat}\\
\widehat{Z}&=&[\widehat{Y} \quad \widehat{Z}]\label{eq:zbhat}
\end{eqnarray}
Hereafter, $\widehat{X}$ and $\widehat{Y}$ and different matrices and should be considered as snapshot pairs, even if the time series data are originally employed. In this case, we can conduct the trPOD procedure of Eq.~\ref{eq:svds} to Eq.~\ref{eq:ytilde} for the  $\widehat{X}$, $\widehat{Y}$ and $\widehat{Z}$.
When adopting the tlsDMD idea into KFDMD, Eqs.~\ref{eq:xbhat} to \ref{eq:zbhat} are used as data matrix instead of original data matrix. 

\begin{figure}
	\subfigure[$\sigma_0^2=0.001$ ]{\includegraphics[width=5cm]{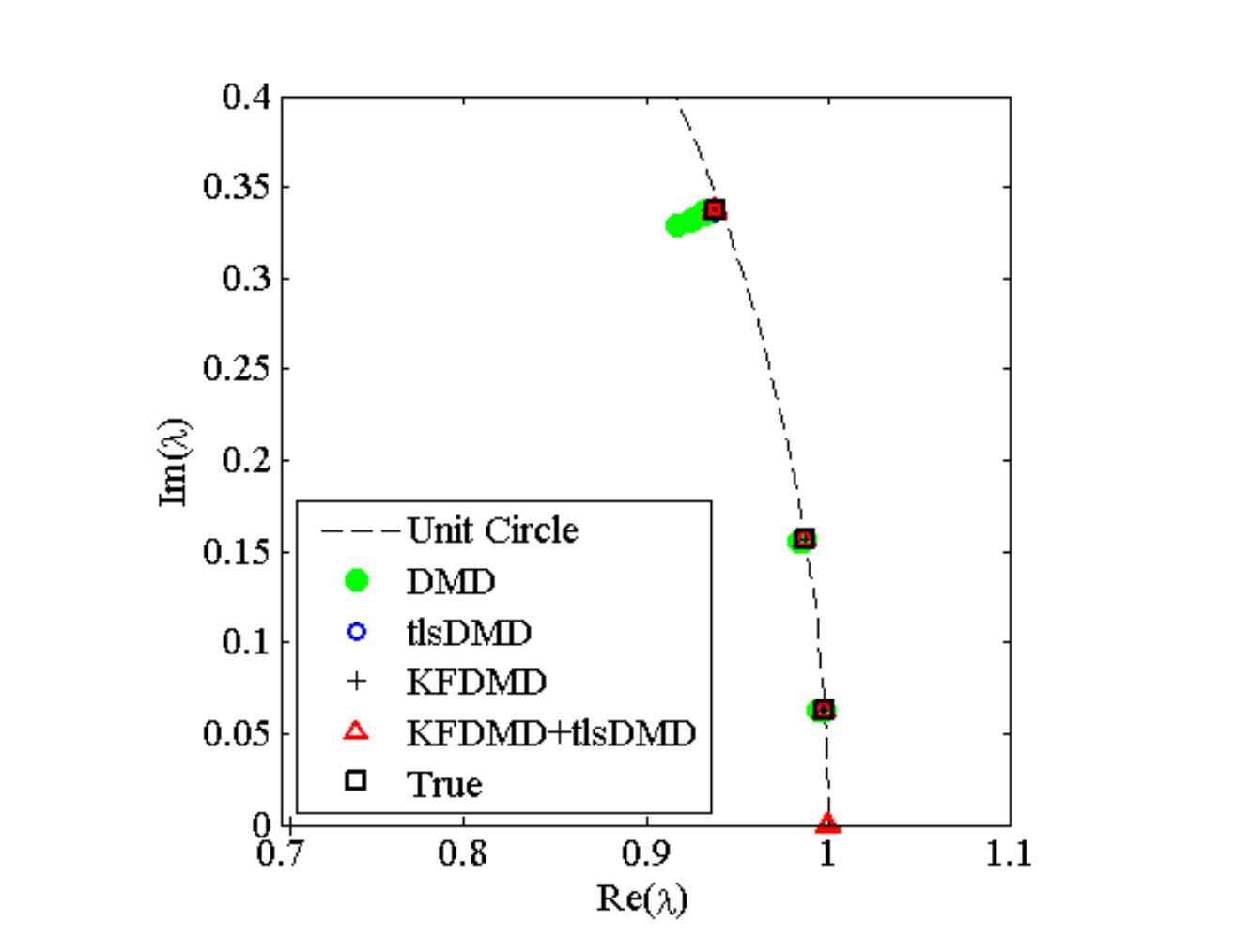}}
	\subfigure[$\sigma_0^2=0.01$  ]{\includegraphics[width=5cm]{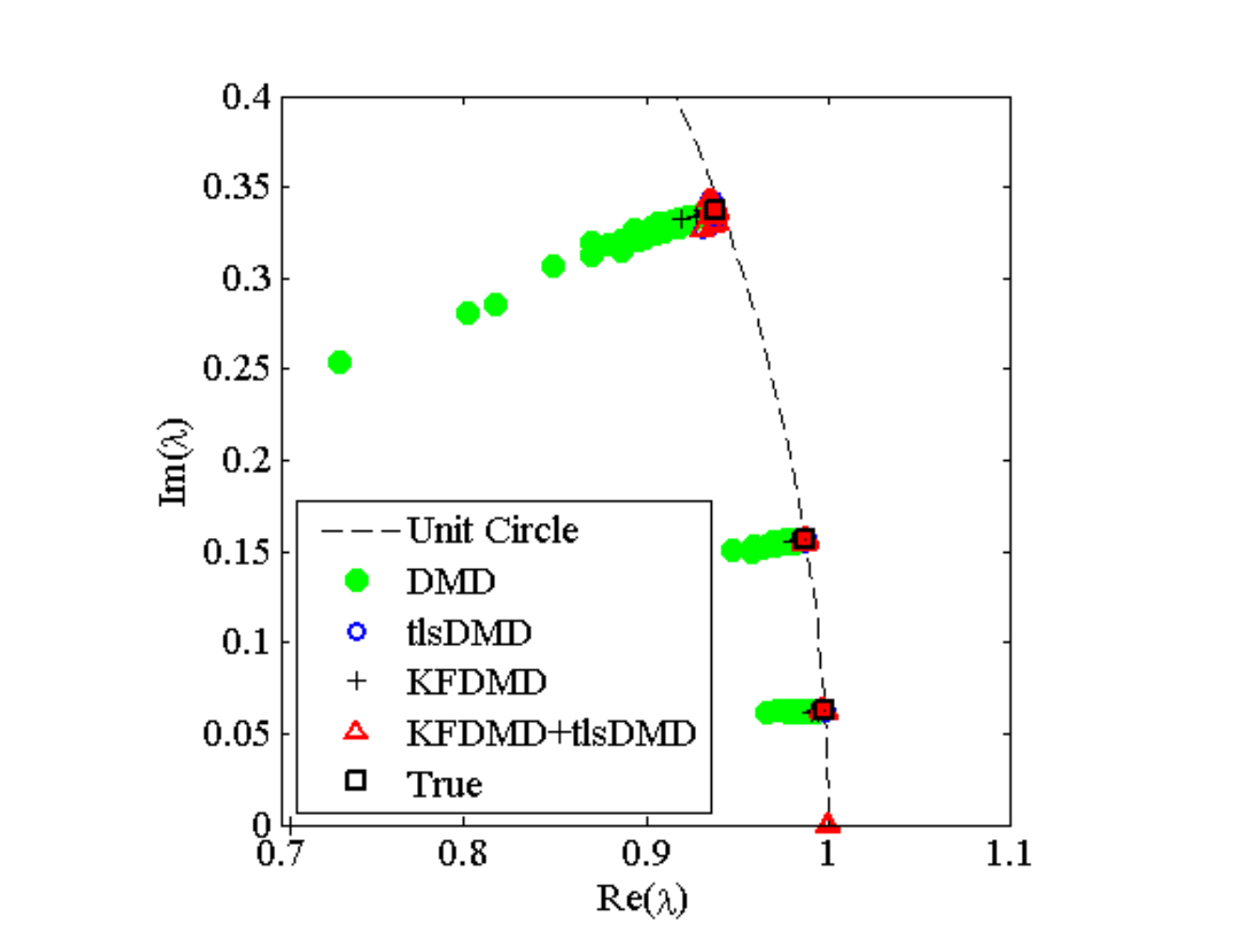}}
	\subfigure[$\sigma_0^2=0.1$   ]{\includegraphics[width=5cm]{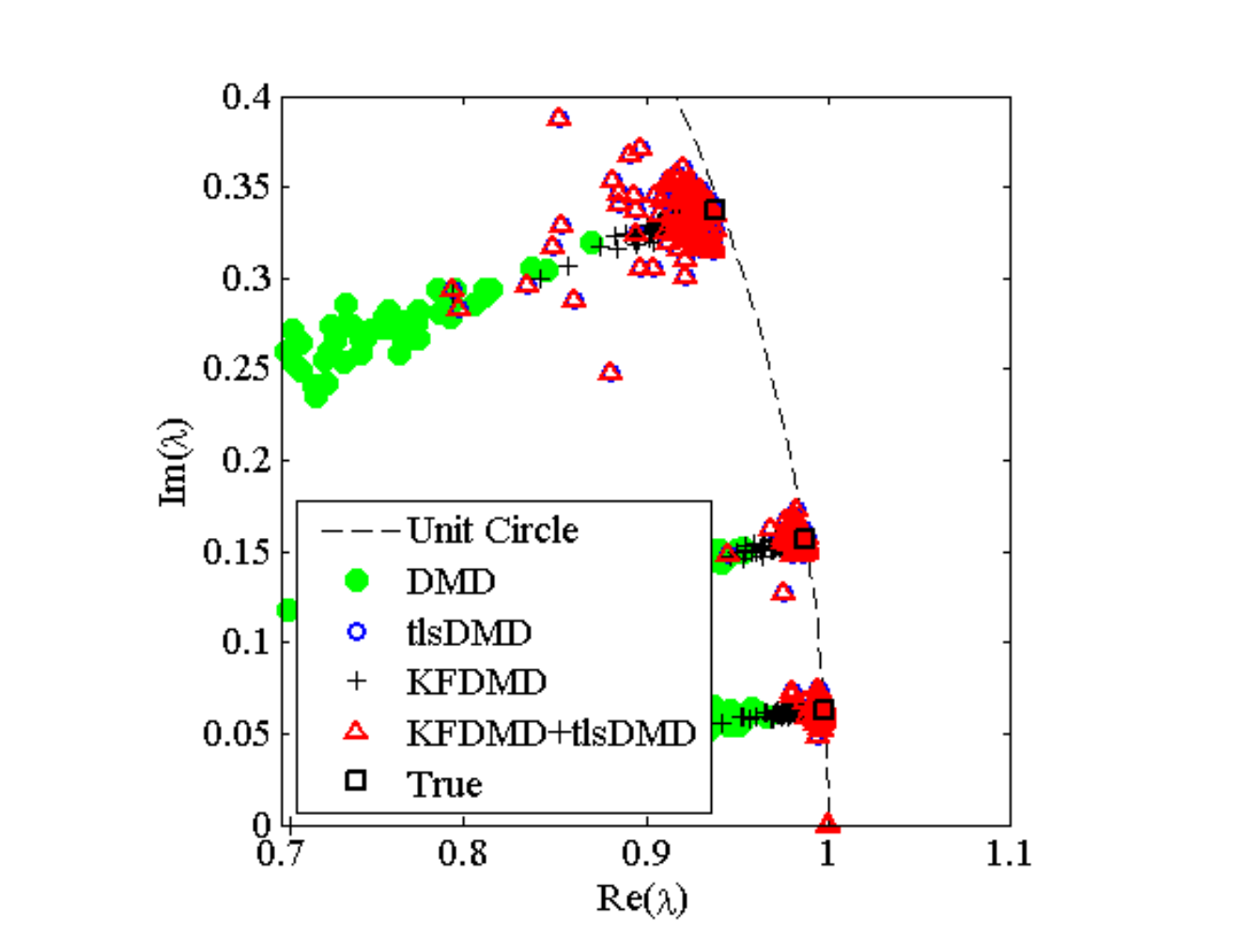}}
    \caption{Results of 100 computations of eigenvalues computed by KFDMD+tlsDMD for the problem same as that in Section \ref{sec:RESSWN}, compared with those by DMD, tlsDMD, and KFDMD.}	
	\label{fig:sinkfteigen}
\end{figure}

\begin{figure}
	\subfigure[$\lambda_1$]{\includegraphics[width=5cm]{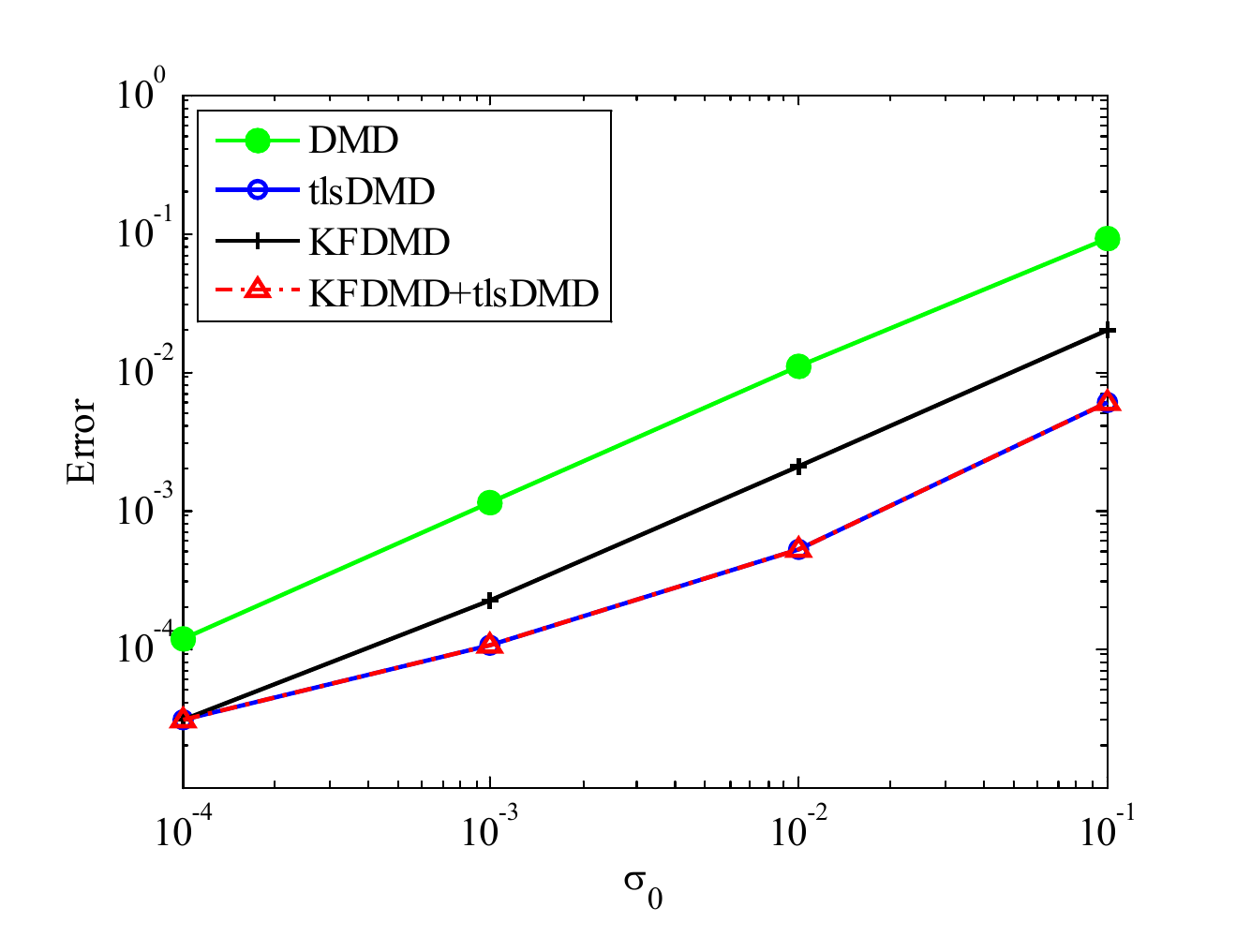}}
	\subfigure[$\lambda_2$ ]{\includegraphics[width=5cm]{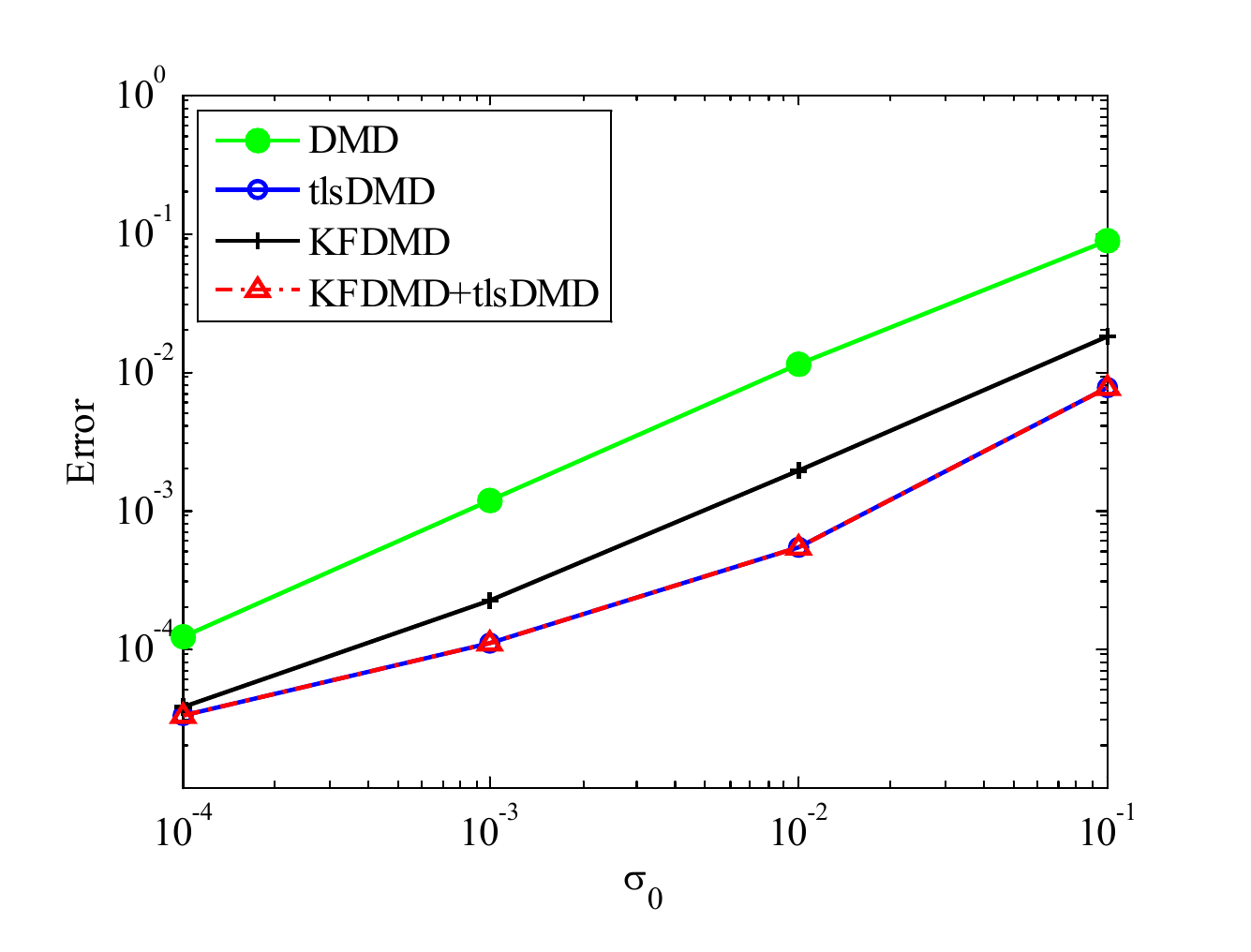}}
	\subfigure[$\lambda_3$  ]{\includegraphics[width=5cm]{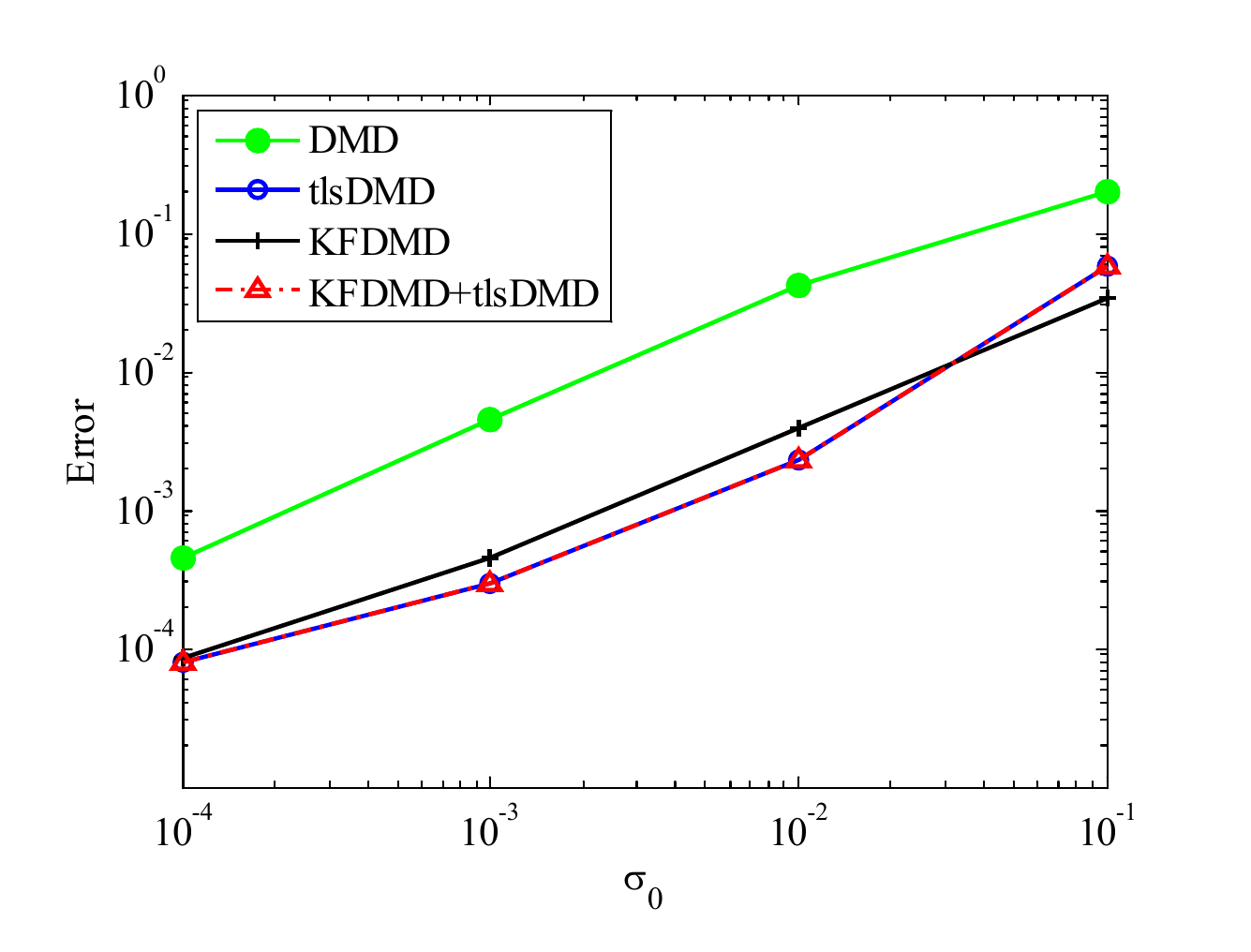}}
   \caption{Errors in the eigenvalues computed by KFDMD+tlsDMD compared with those by standard DMD, tlsDMD and KFDMD. The problem settings are the same as that in Section \ref{sec:RESSWN}. Here, $L_2$ error is averaged with 100 test cases.}	
   \label{fig:sinkfteigenerror}
\end{figure}

The results of KFDMD+tlsDMD are shown in Figs. \ref{fig:sinkfteigen} and \ref{fig:sinkfteigenerror}, when applying the same problem as Section \ref{sec:RESSWN}. Unfortunately, the results are almost the same as tlsDMD, and further improvement by KFDMD is not obtained. This might be because the tlsDMD filtering is very strong and the important information is also lost in the process.  To utilize the both advantages of KFDMD and tlsDMD, further development on algorithms seems to be necessary and this is left for the future study. 


\section{Performance of the other DMD implementations}
\begin{figure}
	\subfigure[$\sigma_0^2=0.001$ ]{\includegraphics[width=5cm]{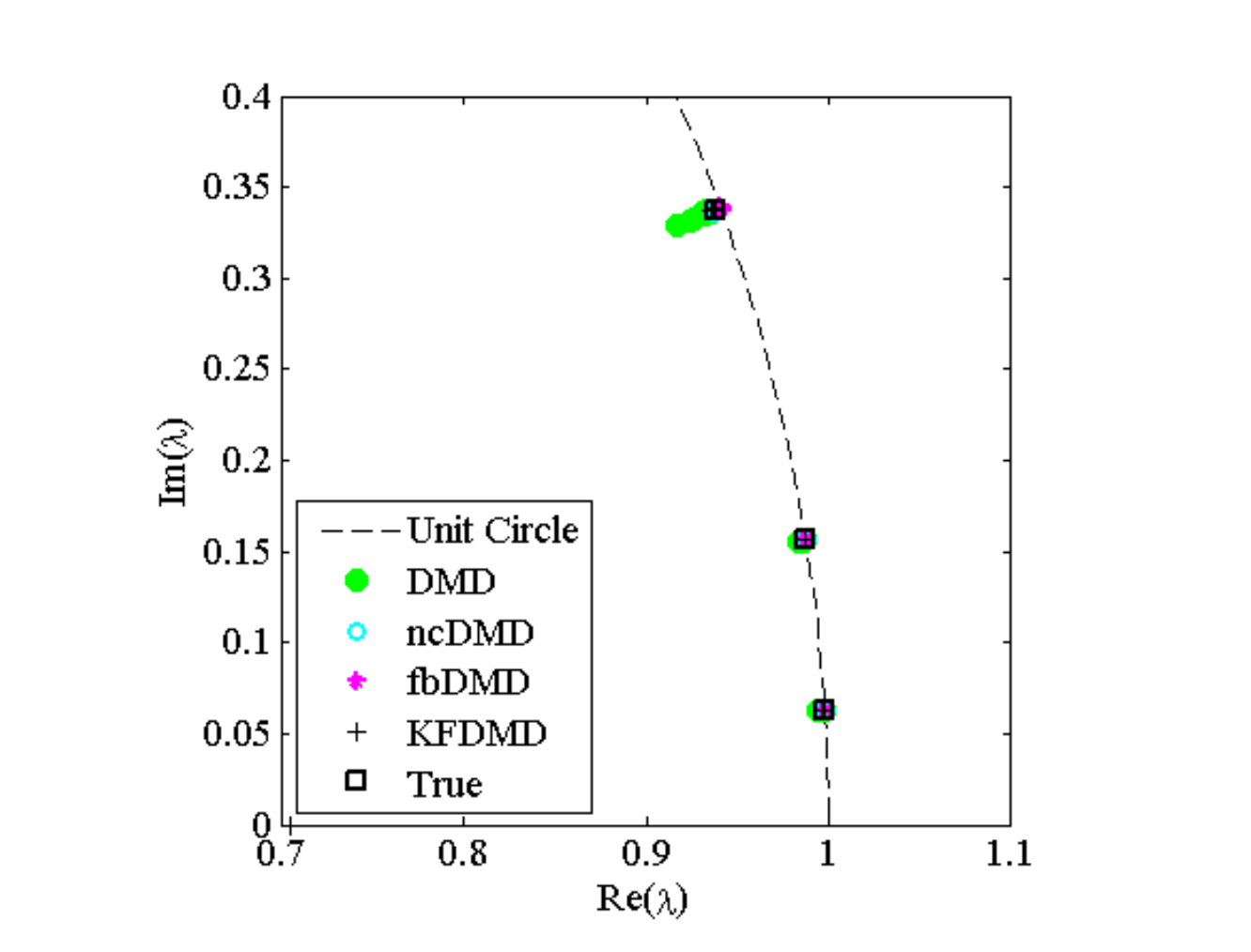}}
	\subfigure[$\sigma_0^2=0.01$  ]{\includegraphics[width=5cm]{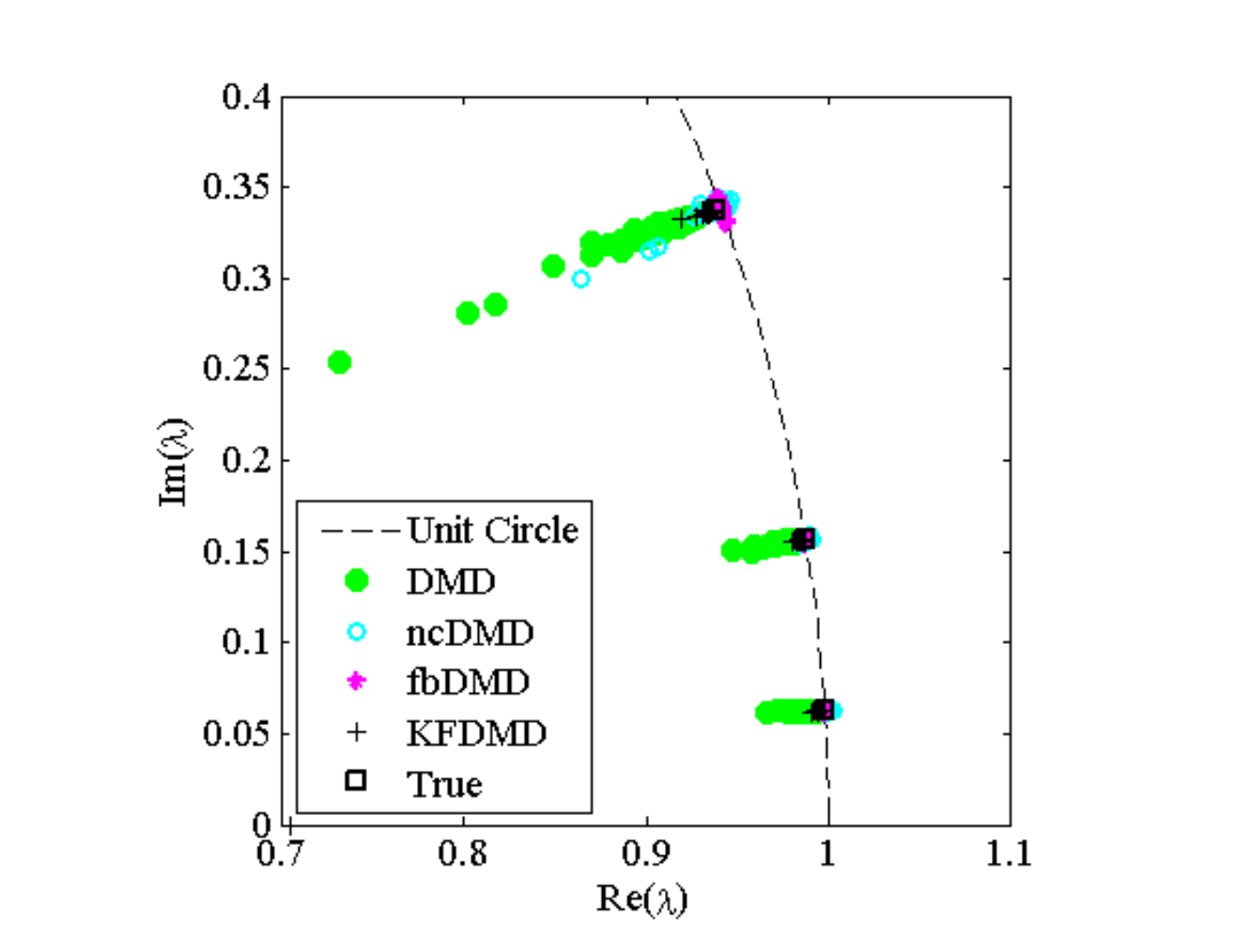}}
	\subfigure[$\sigma_0^2=0.1$   ]{\includegraphics[width=5cm]{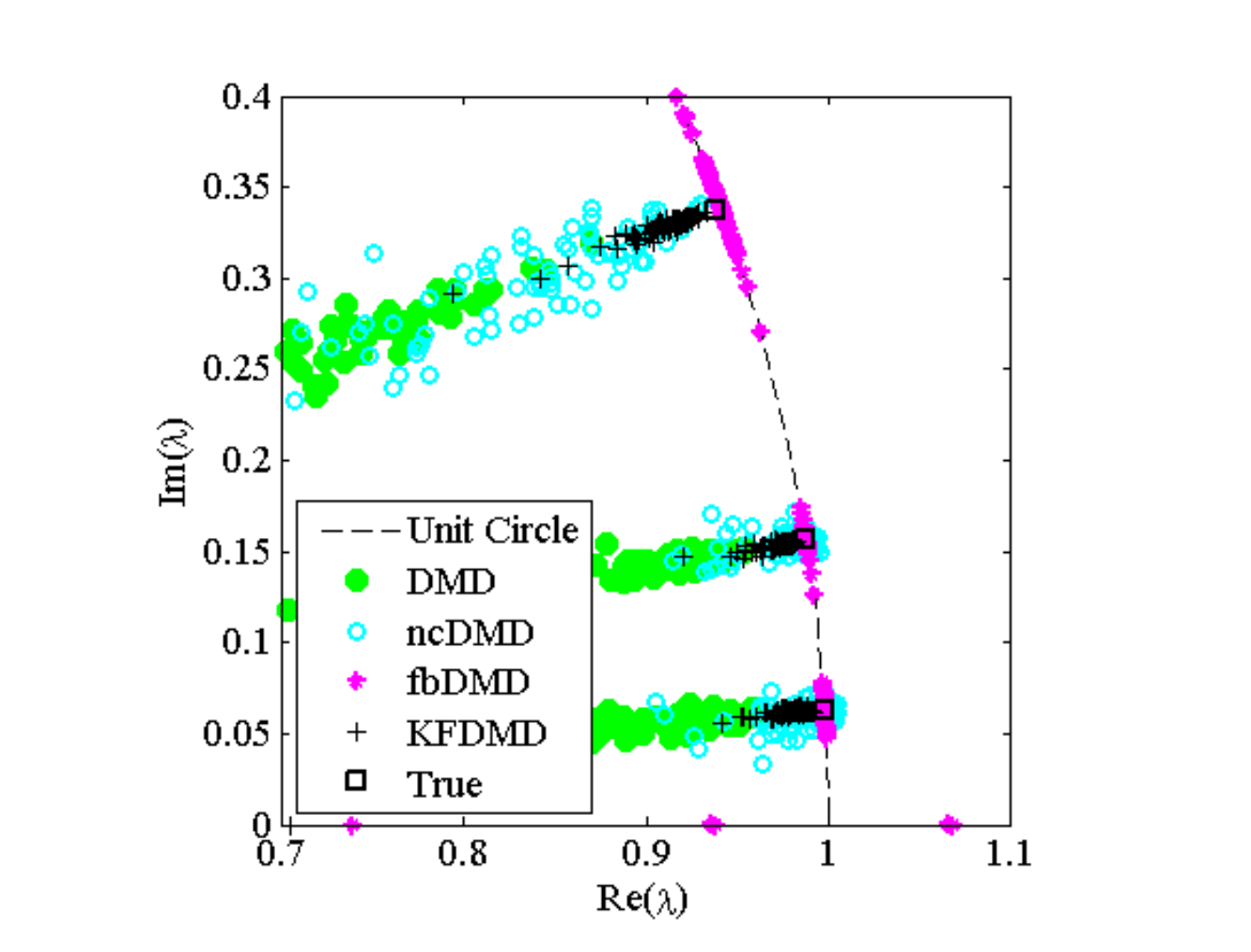}}
    \caption{Results of 100 computations of eigenvalues by DMD, ncDMD, fbDMD and KFDMD in the problem same as that in Section \ref{sec:RESSWN}.}	
	\label{fig:sinothereigen}
\end{figure}

\begin{figure}
	\subfigure[$\lambda_1$]{\includegraphics[width=5cm]{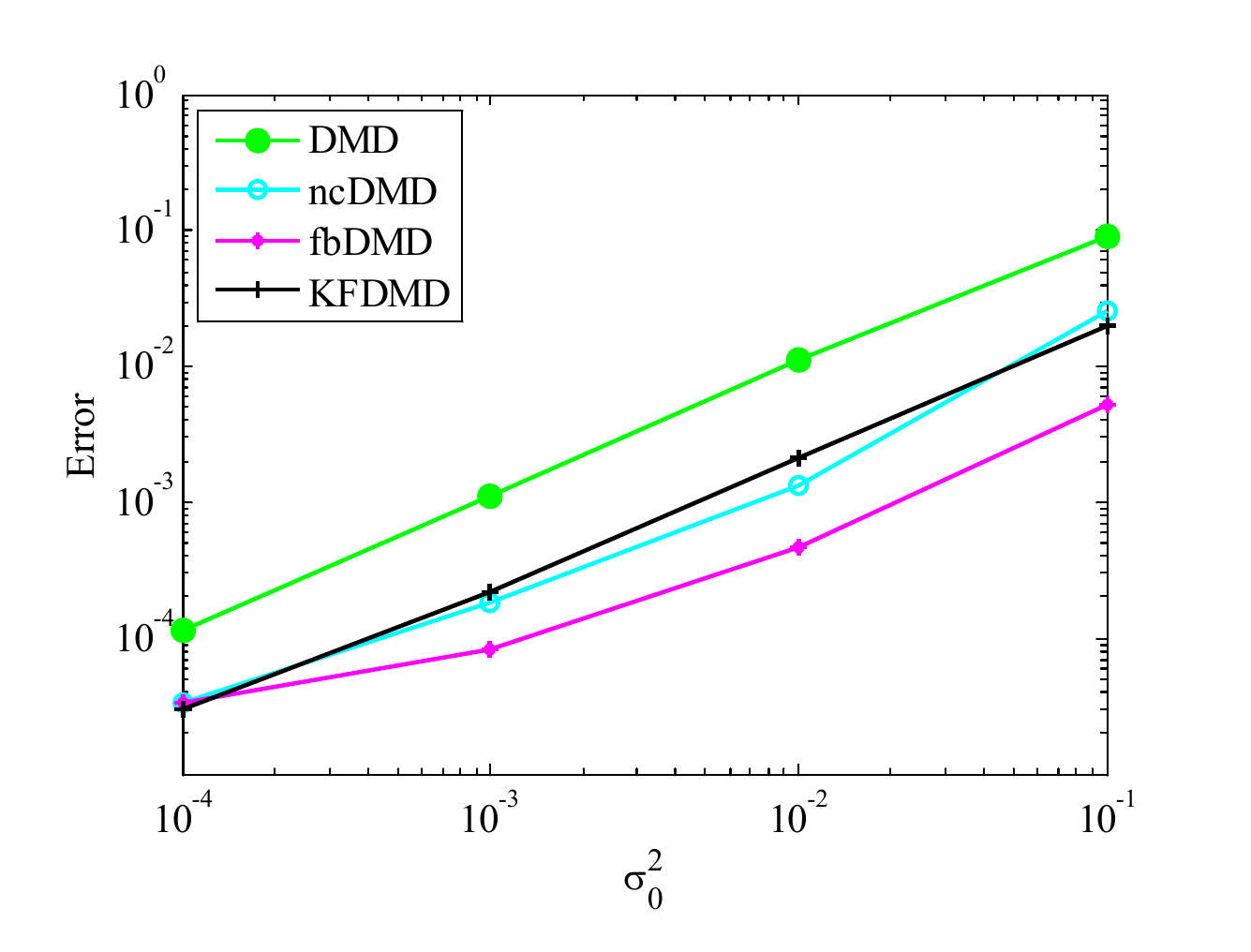}}
	\subfigure[$\lambda_2$]{\includegraphics[width=5cm]{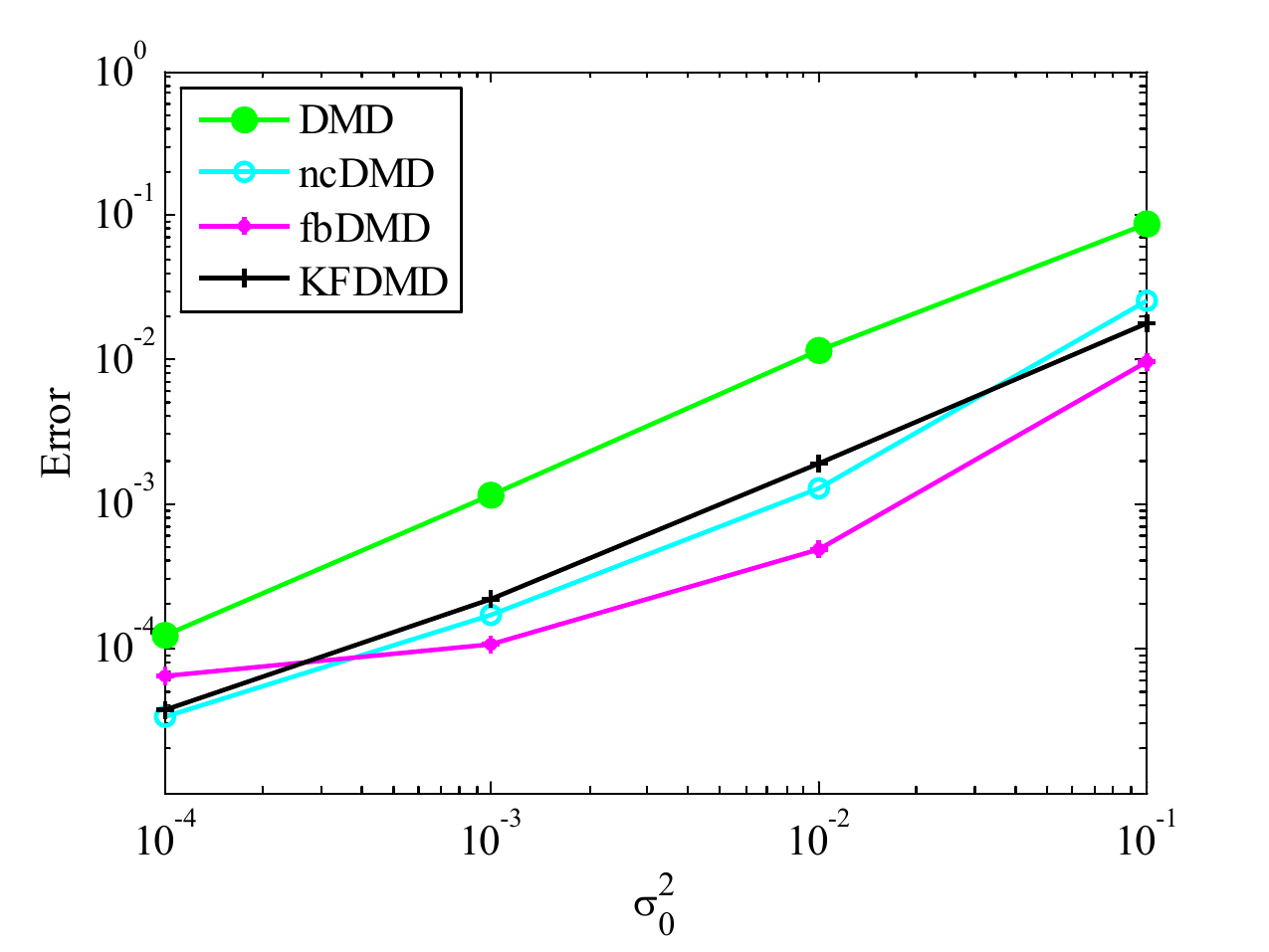}}
	\subfigure[$\lambda_3$]{\includegraphics[width=5cm]{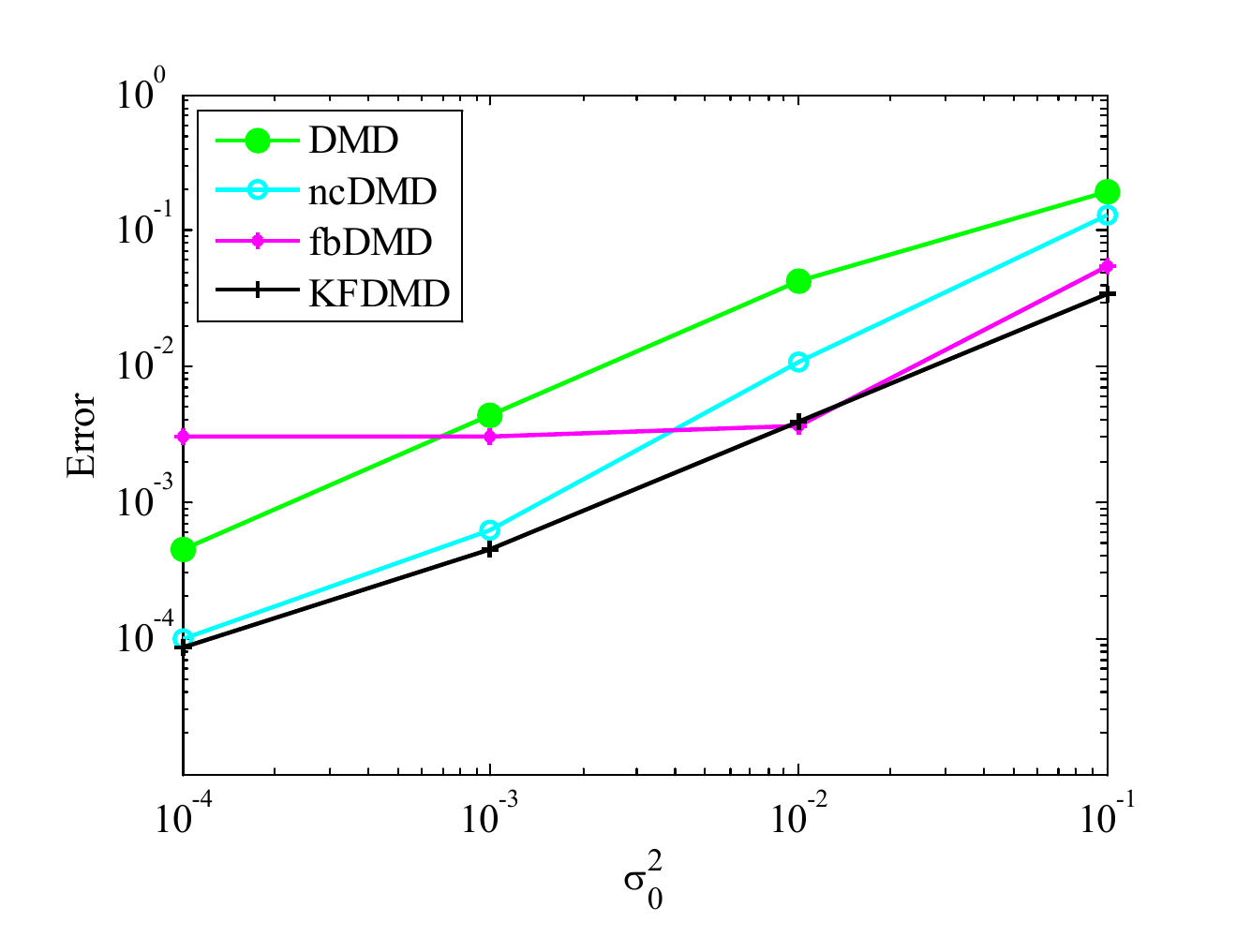}}
   \caption{Errors in the eigenvalues computed by DMD, ncDMD, fbDMD and KFDMD. The problem settings are the same as that in Section \ref{sec:RESSWN}. Here, $L_2$ error is averaged with 100 test cases. }
   \label{fig:sinothereigenerror}
\end{figure}

As we noted, the performances of ncDMD and fbDMD are investigated for the same problem with time-varying noise as in Section \ref{sec:RESSWN}.
The results of ncDMD and fbDMD are shown in Figs. \ref{fig:sinothereigen} and \ref{fig:sinothereigenerror} together with that of KFDMD when applying them to the same problem as in Section \ref{sec:RESSWN}. Interestingly, KFDMD works as well as ncDMD for the first and second eigenvalues and better than ncDMD for the third DMD. On the other hand, KFDMD does not work better than fbDMD for the first and second eigenvalues, but KFDMD works better than fbDMD for the third eigenvalues. The large error in fbDMD is caused by outliers when it fails to find eigenvalues. If we carefully remove outliers from the results, we can get much better results for fbDMD. However, this is not the scope of our study and left for issues in the future studies.


\bibliography{pof}

\begin{thebibliography}{10}%
\makeatletter
\providecommand \@ifxundefined [1]{%
 \ifx #1\undefined \expandafter \@firstoftwo
 \else \expandafter \@secondoftwo
\fi
}%
\providecommand \@ifnum [1]{%
 \ifnum #1\expandafter \@firstoftwo
 \else \expandafter \@secondoftwo
\fi
}%
\providecommand \enquote [1]{``#1''}%
\providecommand \bibnamefont  [1]{#1}%
\providecommand \bibfnamefont [1]{#1}%
\providecommand \citenamefont [1]{#1}%
\providecommand\href[0]{\@sanitize\@href}%
\providecommand\@href[1]{\endgroup\@@startlink{#1}\endgroup\@@href}%
\providecommand\@@href[1]{#1\@@endlink}%
\providecommand \@sanitize [0]{\begingroup\catcode`\&12\catcode`\#12\relax}%
\@ifxundefined \pdfoutput {\@firstoftwo}{%
 \@ifnum{\z@=\pdfoutput}{\@firstoftwo}{\@secondoftwo}%
}{%
 \providecommand\@@startlink[1]{\leavevmode\special{html:<a href="#1">}}%
 \providecommand\@@endlink[0]{\special{html:</a>}}%
}{%
 \providecommand\@@startlink[1]{%
  \leavevmode
  \pdfstartlink
   attr{/Border[0 0 1 ]/H/I/C[0 1 1]}%
   user{/Subtype/Link/A<</Type/Action/S/URI/URI(#1)>>}%
  \relax
 }%
 \providecommand\@@endlink[0]{\pdfendlink}%
}%
\providecommand \url  [0]{\begingroup\@sanitize \@url }%
\providecommand \@url [1]{\endgroup\@href {#1}{\urlprefix}}%
\providecommand \urlprefix [0]{URL }%
\providecommand \Eprint[0]{\href }%
\@ifxundefined \urlstyle {%
  \providecommand \doi [1]{doi:\discretionary{}{}{}#1}%
}{%
  \providecommand \doi [0]{doi:\discretionary{}{}{}\begingroup
  \urlstyle{rm}\Url }%
}%
\providecommand \doibase [0]{http://dx.doi.org/}%
\providecommand \Doi[1]{\href{\doibase#1}}%
\providecommand \selectlanguage [0]{\@gobble}%
\providecommand \bibinfo [0]{\@secondoftwo}%
\providecommand \bibfield [0]{\@secondoftwo}%
\providecommand \translation [1]{[#1]}%
\providecommand \BibitemOpen[0]{}%
\providecommand \bibitemStop [0]{}%
\providecommand \bibitemNoStop [0]{.\EOS\space}%
\providecommand \EOS [0]{\spacefactor3000\relax}%
\providecommand \BibitemShut [1]{\csname bibitem#1\endcsname}%
\bibitem{Taira2017}%
  \BibitemOpen
  \bibfield{author}{%
  \bibinfo {author} {\bibfnamefont{Kunihiko}\ \bibnamefont{Taira}}, \bibinfo
  {author} {\bibfnamefont{Steven~L}\ \bibnamefont{Brunton}}, \bibinfo {author}
  {\bibfnamefont{Scott~TM}\ \bibnamefont{Dawson}}, \bibinfo {author}
  {\bibfnamefont{Clarence~W}\ \bibnamefont{Rowley}}, \bibinfo {author}
  {\bibfnamefont{Tim}\ \bibnamefont{Colonius}}, \bibinfo {author}
  {\bibfnamefont{Beverley~J}\ \bibnamefont{McKeon}}, \bibinfo {author}
  {\bibfnamefont{Oliver~T}\ \bibnamefont{Schmidt}}, \bibinfo {author}
  {\bibfnamefont{Stanislav}\ \bibnamefont{Gordeyev}}, \bibinfo {author}
  {\bibfnamefont{Vassilios}\ \bibnamefont{Theofilis}},\ and\ \bibinfo {author}
  {\bibfnamefont{Lawrence~S}\ \bibnamefont{Ukeiley}},\ }%
  \bibfield{title}{%
  \enquote{\bibinfo {title} {Modal analysis of fluid flows: An overview},}\ }%
  \bibfield{journal}{%
  \bibinfo {journal} {AIAA Journal}\ }%
  \textbf{\bibinfo {volume} {55}},\ \bibinfo {pages} {4013--4041} (\bibinfo
  {year} {2017})\BibitemShut{NoStop}%
\bibitem{Rowley2004}%
  \BibitemOpen
  \bibfield{author}{%
  \bibinfo {author} {\bibfnamefont{Clarence~W.}\ \bibnamefont{Rowley}},
  \bibinfo {author} {\bibfnamefont{Tim}\ \bibnamefont{Colonius}},\ and\
  \bibinfo {author} {\bibfnamefont{Richard~M.}\ \bibnamefont{Murray}},\ }%
  \bibfield{title}{%
  \enquote{\bibinfo {title} {Model reduction for compressible flows using pod
  and galerkin projection},}\ }%
  \bibfield{journal}{%
  \Doi{10.1016/j.physd.2003.03.001}{\bibinfo {journal} {Physica D: Nonlinear
  Phenomena}}\ }%
  \textbf{\bibinfo {volume} {189}},\ \bibinfo {pages} {115--129} (\bibinfo
  {year} {2004}),\ ISSN \bibinfo {issn} {01672789}\BibitemShut{NoStop}%
\bibitem{Berkooz1993}%
  \BibitemOpen
  \bibfield{author}{%
  \bibinfo {author} {\bibfnamefont{Gal}\ \bibnamefont{Berkooz}}, \bibinfo
  {author} {\bibfnamefont{Philip}\ \bibnamefont{Holmes}},\ and\ \bibinfo
  {author} {\bibfnamefont{L.~John}\ \bibnamefont{Lumley}},\ }%
  \bibfield{title}{%
  \enquote{\bibinfo {title} {The proper orthogonal decomposition in the
  analysis of turbulent flows},}\ }%
  \bibfield{journal}{%
  \Doi{10.1146/annurev.fl.25.010193.002543}{\bibinfo {journal} {Annual Review
  of Fluid Mechanics}}\ }%
  \textbf{\bibinfo {volume} {25}},\ \bibinfo {pages} {539--575} (\bibinfo
  {year} {1993}),\ ISSN \bibinfo {issn} {0066-4189},\
  \url{http://citeseerx.ist.psu.edu/viewdoc/download?doi=10.1.1.212.4071{\&}rep=rep1{\&}type=pdf{\%}5Cnhttp://citeseerx.ist.psu.edu/viewdoc/summary?doi=10.1.1.212.4071{\%}5Cnhttp://www.annualreviews.org/doi/pdf/10.1146/annurev.fl.25.010193.002543}\BibitemShut{NoStop}%
\bibitem{Theofilis2011}%
  \BibitemOpen
  \bibfield{author}{%
  \bibinfo {author} {\bibfnamefont{Vassilios}\ \bibnamefont{Theofilis}},\ }%
  \bibfield{title}{%
  \enquote{\bibinfo {title} {Global linear instability},}\ }%
  \bibfield{journal}{%
  \Doi{10.1146/annurev-fluid-122109-160705}{\bibinfo {journal} {Annual Review
  of Fluid Mechanics}}\ }%
  \textbf{\bibinfo {volume} {43}},\ \bibinfo {pages} {319--352} (\bibinfo
  {year} {2011}),\ ISSN \bibinfo {issn} {0066-4189}\BibitemShut{NoStop}%
\bibitem{Shibata2015}%
  \BibitemOpen
  \bibfield{author}{%
  \bibinfo {author} {\bibfnamefont{H.}~\bibnamefont{Shibata}}, \bibinfo
  {author} {\bibfnamefont{Y.}~\bibnamefont{Ohmichi}}, \bibinfo {author}
  {\bibfnamefont{Y.}~\bibnamefont{Watanabe}},\ and\ \bibinfo {author}
  {\bibfnamefont{K.}~\bibnamefont{Suzuki}},\ }%
  \bibfield{title}{%
  \enquote{\bibinfo {title} {Global stability analysis method to numerically
  predict precursor of breakdown voltage},}\ }%
  \bibfield{journal}{%
  \bibinfo {journal} {Plasma Sources Science and Technology}\ }%
  \textbf{\bibinfo {volume} {24}} (\bibinfo {year} {2015}),\ ISSN \bibinfo
  {issn} {13616595 09630252},\ \doi{\bibinfo {doi}
  {10.1088/0963-0252/24/5/055014}}\BibitemShut{NoStop}%
\bibitem{Ohmichi2016}%
  \BibitemOpen
  \bibfield{author}{%
  \bibinfo {author} {\bibfnamefont{Yuya}\ \bibnamefont{Ohmichi}}\ and\ \bibinfo
  {author} {\bibfnamefont{Kojiro}\ \bibnamefont{Suzuki}},\ }%
  \bibfield{title}{%
  \enquote{\bibinfo {title} {Assessment of global linear stability analysis
  using a time-stepping approach for compressible flows},}\ }%
  \bibfield{journal}{%
  \Doi{10.1002/fld.4166}{\bibinfo {journal} {International Journal for
  Numerical Methods in Fluids}}\ }%
  \textbf{\bibinfo {volume} {80}},\ \bibinfo {pages} {614--627} (\bibinfo
  {year} {2016}),\ ISSN \bibinfo {issn} {02712091},\
  \Eprint{http://arxiv.org/abs/fld.1}{arXiv:fld.1 [DOI: 10.1002]},\
  \url{http://doi.wiley.com/10.1002/fld.4166}\BibitemShut{NoStop}%
\bibitem{Schmid2010}%
  \BibitemOpen
  \bibfield{author}{%
  \bibinfo {author} {\bibfnamefont{P.~J.}\ \bibnamefont{Schmid}},\ }%
  \bibfield{title}{%
  \enquote{\bibinfo {title} {Dynamic mode decomposition of numerical and
  experimental data},}\ }%
  \bibfield{journal}{%
  \Doi{10.1017/S0022112010001217}{\bibinfo {journal} {Journal of Fluid
  Mechanics}}\ }%
  \textbf{\bibinfo {volume} {656}},\ \bibinfo {pages} {5--28} (\bibinfo {year}
  {2010}),\ ISSN \bibinfo {issn} {0022-1120},\
  \Eprint{http://arxiv.org/abs/arXiv:1312.0041v1}{arXiv:arXiv:1312.0041v1}\BibitemShut{NoStop}%
\bibitem{Wang2017}%
  \BibitemOpen
  \bibfield{author}{%
  \bibinfo {author} {\bibfnamefont{Lei}\ \bibnamefont{Wang}}\ and\ \bibinfo
  {author} {\bibfnamefont{Li-Hao}\ \bibnamefont{Feng}},\ }%
  \bibfield{title}{%
  \enquote{\bibinfo {title} {Extraction and reconstruction of individual
  vortex-shedding mode from bistable flow},}\ }%
  \bibfield{journal}{%
  \Doi{10.2514/1.J055306}{\bibinfo {journal} {AIAA Journal}},\ \bibinfo {pages}
  {1--13}}%
   (\bibinfo {year} {2017}),\ ISSN \bibinfo {issn} {0001-1452},\
  \url{https://arc.aiaa.org/doi/10.2514/1.J055306}\BibitemShut{NoStop}%
\bibitem{Priebe2016}%
  \BibitemOpen
  \bibfield{author}{%
  \bibinfo {author} {\bibfnamefont{Stephan}\ \bibnamefont{Priebe}}, \bibinfo
  {author} {\bibfnamefont{Jonathan~H.}\ \bibnamefont{Tu}}, \bibinfo {author}
  {\bibfnamefont{Clarence~W.}\ \bibnamefont{Rowley}},\ and\ \bibinfo {author}
  {\bibfnamefont{M.~Pino}\ \bibnamefont{Mart{\'{i}}n}},\ }%
  \bibfield{title}{%
  \enquote{\bibinfo {title} {Low-frequency dynamics in a shock-induced
  separated flow},}\ }%
  \bibfield{journal}{%
  \Doi{10.1017/jfm.2016.557}{\bibinfo {journal} {Journal of Fluid Mechanics}}\
  }%
  \textbf{\bibinfo {volume} {807}},\ \bibinfo {pages} {441--477} (\bibinfo
  {year} {2016}),\ ISSN \bibinfo {issn} {0022-1120},\
  \url{http://www.journals.cambridge.org/abstract{\_}S0022112016005577}\BibitemShut{NoStop}%
\bibitem{Ohmichi2017}%
  \BibitemOpen
  \bibfield{author}{%
  \bibinfo {author} {\bibfnamefont{Yuya}\ \bibnamefont{Ohmichi}}, \bibinfo
  {author} {\bibfnamefont{Takashi}\ \bibnamefont{Ishida}},\ and\ \bibinfo
  {author} {\bibfnamefont{Atsushi}\ \bibnamefont{Hashimoto}},\ }%
  \enquote{\bibinfo {title} {Numerical investigation of transonic buffet on a
  three-dimensional wing using incremental mode decomposition},}\ in\
  \Doi{10.2514/6.2017-1436}{\emph{\bibinfo {booktitle} {AIAA Paper
  2017-1436}}},\ \bibinfo {series and number} {\bibinfo {number} {January}}\
  (\bibinfo {year} {2017})\ pp.\ \bibinfo {pages} {1--8},\ ISBN \bibinfo {isbn}
  {978-1-62410-447-3},\
  \url{http://arc.aiaa.org/doi/10.2514/6.2017-1436}\BibitemShut{NoStop}%
\bibitem{Tu2014}%
  \BibitemOpen
  \bibfield{author}{%
  \bibinfo {author} {\bibfnamefont{Jonathan~H.}\ \bibnamefont{Tu}}, \bibinfo
  {author} {\bibfnamefont{Clarence~W.}\ \bibnamefont{Rowley}}, \bibinfo
  {author} {\bibfnamefont{Dirk~M.}\ \bibnamefont{Luchtenburg}}, \bibinfo
  {author} {\bibfnamefont{Steven~L.}\ \bibnamefont{Brunton}},\ and\ \bibinfo
  {author} {\bibfnamefont{J.~Nathan}\ \bibnamefont{Kutz}},\ }%
  \bibfield{title}{%
  \enquote{\bibinfo {title} {On dynamic mode decomposition: Theory and
  applications},}\ }%
  \bibfield{journal}{%
  \Doi{10.3934/jcd.2014.1.391}{\bibinfo {journal} {Journal of Computational
  Dynamics}}\ }%
  \textbf{\bibinfo {volume} {1}},\ \bibinfo {pages} {391--421} (\bibinfo {year}
  {2014}),\ ISSN \bibinfo {issn} {2158-2491},\
  \Eprint{http://arxiv.org/abs/1312.0041}{arXiv:1312.0041},\
  \url{http://www.aimsciences.org/journals/displayArticlesnew.jsp?paperID=10631}\BibitemShut{NoStop}%
\bibitem{Dawson2016}%
  \BibitemOpen
  \bibfield{author}{%
  \bibinfo {author} {\bibfnamefont{Scott~TM}\ \bibnamefont{Dawson}}, \bibinfo
  {author} {\bibfnamefont{Maziar~S}\ \bibnamefont{Hemati}}, \bibinfo {author}
  {\bibfnamefont{Matthew~O}\ \bibnamefont{Williams}},\ and\ \bibinfo {author}
  {\bibfnamefont{Clarence~W}\ \bibnamefont{Rowley}},\ }%
  \bibfield{title}{%
  \enquote{\bibinfo {title} {Characterizing and correcting for the effect of
  sensor noise in the dynamic mode decomposition},}\ }%
  \bibfield{journal}{%
  \bibinfo {journal} {Experiments in Fluids}\ }%
  \textbf{\bibinfo {volume} {57}},\ \bibinfo {pages} {42} (\bibinfo {year}
  {2016})\BibitemShut{NoStop}%
\bibitem{Hemati2016}%
  \BibitemOpen
  \bibfield{author}{%
  \bibinfo {author} {\bibfnamefont{Maziar}\ \bibnamefont{Hemati}}, \bibinfo
  {author} {\bibfnamefont{Eric}\ \bibnamefont{Deem}}, \bibinfo {author}
  {\bibfnamefont{Matthew}\ \bibnamefont{Williams}}, \bibinfo {author}
  {\bibfnamefont{Clarence~W}\ \bibnamefont{Rowley}},\ and\ \bibinfo {author}
  {\bibfnamefont{Louis~N}\ \bibnamefont{Cattafesta}},\ }%
  \enquote{\bibinfo {title} {Improving separation control with noise-robust
  variants of dynamic mode decomposition},}\ in\ \emph{\bibinfo {booktitle}
  {AIAA-Paper 2016-1103}}\ (\bibinfo {year} {2016})\ p.\ \bibinfo {pages}
  {1103}\BibitemShut{NoStop}%
\bibitem{Hemati2017}%
  \BibitemOpen
  \bibfield{author}{%
  \bibinfo {author} {\bibfnamefont{Maziar~S.}\ \bibnamefont{Hemati}}, \bibinfo
  {author} {\bibfnamefont{Clarence~W.}\ \bibnamefont{Rowley}}, \bibinfo
  {author} {\bibfnamefont{Eric~A.}\ \bibnamefont{Deem}},\ and\ \bibinfo
  {author} {\bibfnamefont{Louis~N.}\ \bibnamefont{Cattafesta}},\ }%
  \bibfield{title}{%
  \enquote{\bibinfo {title} {De-biasing the dynamic mode decomposition for
  applied koopman spectral analysis of noisy datasets},}\ }%
  \bibfield{journal}{%
  \Doi{10.1007/s00162-017-0432-2}{\bibinfo {journal} {Theoretical and
  Computational Fluid Dynamics}},\ \bibinfo {pages} {1--20}}%
   (\bibinfo {year} {2017}),\ ISSN \bibinfo {issn} {14322250},\
  \Eprint{http://arxiv.org/abs/1502.03854}{1502.03854}\BibitemShut{NoStop}%
\bibitem{Jovanovic2014}%
  \BibitemOpen
  \bibfield{author}{%
  \bibinfo {author} {\bibfnamefont{Mihailo~R.}\ \bibnamefont{Jovanovi{\'{c}}}},
  \bibinfo {author} {\bibfnamefont{Peter~J.}\ \bibnamefont{Schmid}},\ and\
  \bibinfo {author} {\bibfnamefont{Joseph~W.}\ \bibnamefont{Nichols}},\ }%
  \bibfield{title}{%
  \enquote{\bibinfo {title} {Sparsity-promoting dynamic mode decomposition},}\
  }%
  \bibfield{journal}{%
  \Doi{10.1063/1.4863670}{\bibinfo {journal} {Physics of Fluids}}\ }%
  \textbf{\bibinfo {volume} {26}},\ \bibinfo {pages} {1--22} (\bibinfo {year}
  {2014}),\ ISSN \bibinfo {issn} {10897666},\
  \Eprint{http://arxiv.org/abs/arXiv:1309.4165v1}{arXiv:arXiv:1309.4165v1}\BibitemShut{NoStop}%
\bibitem{Hemati2014}%
  \BibitemOpen
  \bibfield{author}{%
  \bibinfo {author} {\bibfnamefont{Maziar~S}\ \bibnamefont{Hemati}}, \bibinfo
  {author} {\bibfnamefont{Matthew~O}\ \bibnamefont{Williams}},\ and\ \bibinfo
  {author} {\bibfnamefont{Clarence~W}\ \bibnamefont{Rowley}},\ }%
  \bibfield{title}{%
  \enquote{\bibinfo {title} {Dynamic mode decomposition for large and streaming
  datasets},}\ }%
  \bibfield{journal}{%
  \bibinfo {journal} {Physics of Fluids}\ }%
  \textbf{\bibinfo {volume} {26}},\ \bibinfo {pages} {111701} (\bibinfo {year}
  {2014})\BibitemShut{NoStop}%
\bibitem{Ohmichi2017a}%
  \BibitemOpen
  \bibfield{author}{%
  \bibinfo {author} {\bibfnamefont{Yuya}\ \bibnamefont{Ohmichi}},\ }%
  \bibfield{title}{%
  \enquote{\bibinfo {title} {Preconditioned dynamic mode decomposition and mode
  selection algorithms for large datasets using incremental proper orthogonal
  decomposition},}\ }%
  \bibfield{journal}{%
  \bibinfo {journal} {AIP Advances}\ }%
  \textbf{\bibinfo {volume} {7}},\ \bibinfo {pages} {075318} (\bibinfo {year}
  {2017})\BibitemShut{NoStop}%
\bibitem{Matsumoto2017}%
  \BibitemOpen
  \bibfield{author}{%
  \bibinfo {author} {\bibfnamefont{Daiki}\ \bibnamefont{Matsumoto}}\ and\
  \bibinfo {author} {\bibfnamefont{Thomas}\ \bibnamefont{Indinger}},\ }%
  \bibfield{title}{%
  \enquote{\bibinfo {title} {On-the-fly algorithm for dynamic mode
  decomposition using incremental singular value decomposition and total least
  squares},}\ }%
  \bibfield{journal}{%
  \bibinfo {journal} {arXiv preprint arXiv:1703.11004}}%
   (\bibinfo {year} {2017})\BibitemShut{NoStop}%
\bibitem{Zhang2017}%
  \BibitemOpen
  \bibfield{author}{%
  \bibinfo {author} {\bibfnamefont{Hao}\ \bibnamefont{Zhang}}, \bibinfo
  {author} {\bibfnamefont{Clarence~W}\ \bibnamefont{Rowley}}, \bibinfo {author}
  {\bibfnamefont{Eric~A}\ \bibnamefont{Deem}},\ and\ \bibinfo {author}
  {\bibfnamefont{Louis~N}\ \bibnamefont{Cattafesta}},\ }%
  \bibfield{title}{%
  \enquote{\bibinfo {title} {Online dynamic mode decomposition for time-varying
  systems},}\ }%
  \bibfield{journal}{%
  \bibinfo {journal} {arXiv preprint arXiv:1707.02876}}%
   (\bibinfo {year} {2017})\BibitemShut{NoStop}%
\bibitem{Kalman1960}%
  \BibitemOpen
  \bibfield{author}{%
  \bibinfo {author} {\bibfnamefont{R.~E.}\ \bibnamefont{Kalman}},\ }%
  \bibfield{title}{%
  \enquote{\bibinfo {title} {A new approach to linear filtering and prediction
  problems},}\ }%
  \bibfield{journal}{%
  \Doi{10.1115/1.3662552}{\bibinfo {journal} {Journal of Basic Engineering}}\
  }%
  \textbf{\bibinfo {volume} {82}},\ \bibinfo {pages} {35} (\bibinfo {year}
  {1960}),\ ISSN \bibinfo {issn} {00219223},\
  \url{http://scholar.google.com/scholar?hl=en{\&}btnG=Search{\&}q=intitle:A+New+Approach+to+Linear+Filtering+and+Prediction+Problems{\#}0{\%}5Cnhttp://fluidsengineering.asmedigitalcollection.asme.org/article.aspx?articleid=1430402}\BibitemShut{NoStop}%
\bibitem{Surana2016}%
  \BibitemOpen
  \bibfield{author}{%
  \bibinfo {author} {\bibfnamefont{Amit}\ \bibnamefont{Surana}}\ and\ \bibinfo
  {author} {\bibfnamefont{Andrzej}\ \bibnamefont{Banaszuk}},\ }%
  \bibfield{title}{%
  \enquote{\bibinfo {title} {Linear observer synthesis for nonlinear systems
  using koopman operator framework},}\ }%
  \bibfield{journal}{%
  \bibinfo {journal} {IFAC-PapersOnLine}\ }%
  \textbf{\bibinfo {volume} {49}},\ \bibinfo {pages} {716--723} (\bibinfo
  {year} {2016})\BibitemShut{NoStop}%
\bibitem{Surana2017}%
  \BibitemOpen
  \bibfield{author}{%
  \bibinfo {author} {\bibfnamefont{Amit}\ \bibnamefont{Surana}}, \bibinfo
  {author} {\bibfnamefont{Matthew~O}\ \bibnamefont{Williams}}, \bibinfo
  {author} {\bibfnamefont{Manfred}\ \bibnamefont{Morari}},\ and\ \bibinfo
  {author} {\bibfnamefont{Andrzej}\ \bibnamefont{Banaszuk}},\ }%
  \enquote{\bibinfo {title} {Koopman operator framework for constrained state
  estimation},}\ in\ \emph{\bibinfo {booktitle} {Decision and Control (CDC),
  2017 IEEE 56th Annual Conference on}}\ (\bibinfo {organization} {IEEE},\
  \bibinfo {year} {2017})\ pp.\ \bibinfo {pages} {94--101}\BibitemShut{NoStop}%
\bibitem{Bailey2012}%
  \BibitemOpen
  \bibfield{author}{%
  \bibinfo {author} {\bibfnamefont{Stephen}\ \bibnamefont{Bailey}},\ }%
  \bibfield{title}{%
  \enquote{\bibinfo {title} {Principal component analysis with noisy and/or
  missing data},}\ }%
  \bibfield{journal}{%
  \bibinfo {journal} {Publications of the Astronomical Society of the Pacific}\
  }%
  \textbf{\bibinfo {volume} {124}},\ \bibinfo {pages} {1015} (\bibinfo {year}
  {2012}),\
  \url{http://stacks.iop.org/1538-3873/124/i=919/a=1015}\BibitemShut{NoStop}%
\bibitem{Fujii1990a}%
  \BibitemOpen
  \bibfield{author}{%
  \bibinfo {author} {\bibfnamefont{K.}~\bibnamefont{Fujii}}, \bibinfo {author}
  {\bibfnamefont{H.}~\bibnamefont{Endo}},\ and\ \bibinfo {author}
  {\bibfnamefont{M.}~\bibnamefont{Yasuhara}},\ }%
  \bibfield{title}{%
  \enquote{\bibinfo {title} {Activities of computational fluid dynamics in
  japan: Compressible flow simulations},}\ }%
  \bibfield{journal}{%
  \bibinfo {journal} {High Performance Computing Research and Practice in
  Japan, Wiley Professional Computing, JOHN WILEY\& SONS},\ \bibinfo {pages}
  {139--161}}%
   (\bibinfo {year} {1990})\BibitemShut{NoStop}%
\bibitem{Lele1992}%
  \BibitemOpen
  \bibfield{author}{%
  \bibinfo {author} {\bibfnamefont{Sanjiva~K.}\ \bibnamefont{Lele}},\ }%
  \bibfield{title}{%
  \enquote{\bibinfo {title} {Compact finite difference schemes with
  spectral-like resolution.}.}\ }%
  \bibfield{journal}{%
  \Doi{10.1016/0021-9991(92)90324-R}{\bibinfo {journal} {Journal of
  Computational Physics}}\ }%
  \textbf{\bibinfo {volume} {103}},\ \bibinfo {pages} {16--42} (\bibinfo {year}
  {1992})\BibitemShut{NoStop}%
\bibitem{Fujii1999}%
  \BibitemOpen
  \bibfield{author}{%
  \bibinfo {author} {\bibfnamefont{Kozo}\ \bibnamefont{Fujii}},\ }%
  \bibfield{title}{%
  \enquote{\bibinfo {title} {Efficiency improvement of unified implicit
  relaxation/time integration algorithms},}\ }%
  \bibfield{journal}{%
  \bibinfo {journal} {AIAA Journal}\ }%
  \textbf{\bibinfo {volume} {37}},\ \bibinfo {pages} {125--128} (\bibinfo
  {year} {1999})\BibitemShut{NoStop}%
\bibitem{Nishida2009}%
  \BibitemOpen
  \bibfield{author}{%
  \bibinfo {author} {\bibfnamefont{Hiroyuki}\ \bibnamefont{Nishida}}\ and\
  \bibinfo {author} {\bibfnamefont{Taku}\ \bibnamefont{Nonomura}},\ }%
  \bibfield{title}{%
  \enquote{\bibinfo {title} {{ADI-SGS} scheme on ideal magnetohydrodynamics},}\
  }%
  \bibfield{journal}{%
  \bibinfo {journal} {Journal of Computational Physics}\ }%
  \textbf{\bibinfo {volume} {228}},\ \bibinfo {pages} {3182--3188} (\bibinfo
  {year} {2009})\BibitemShut{NoStop}%
\bibitem{Sato2015b}%
  \BibitemOpen
  \bibfield{author}{%
  \bibinfo {author} {\bibfnamefont{Makoto}\ \bibnamefont{Sato}}, \bibinfo
  {author} {\bibfnamefont{Taku}\ \bibnamefont{Nonomura}}, \bibinfo {author}
  {\bibfnamefont{Koichi}\ \bibnamefont{Okada}}, \bibinfo {author}
  {\bibfnamefont{Kengo}\ \bibnamefont{Asada}}, \bibinfo {author}
  {\bibfnamefont{Hikaru}\ \bibnamefont{Aono}}, \bibinfo {author}
  {\bibfnamefont{Aiko}\ \bibnamefont{Yakeno}}, \bibinfo {author}
  {\bibfnamefont{Yoshiaki}\ \bibnamefont{Abe}},\ and\ \bibinfo {author}
  {\bibfnamefont{Kozo}\ \bibnamefont{Fujii}},\ }%
  \bibfield{title}{%
  \enquote{\bibinfo {title} {Mechanisms for laminar separated-flow control
  using dielectric-barrier-discharge plasma actuator at low reynolds number},}\
  }%
  \bibfield{journal}{%
  \bibinfo {journal} {Physics of Fluids}\ }%
  \textbf{\bibinfo {volume} {27}},\ \bibinfo {pages} {1--29} (\bibinfo {year}
  {2015})\BibitemShut{NoStop}%
\bibitem{Nonomura2018}%
  \BibitemOpen
  \bibfield{author}{%
  \bibinfo {author} {\bibfnamefont{Taku}\ \bibnamefont{Nonomura}}, \bibinfo
  {author} {\bibfnamefont{Hisaichi}\ \bibnamefont{Shibata}},\ and\ \bibinfo
  {author} {\bibfnamefont{Ryoji}\ \bibnamefont{Takaki}},\ }%
  \bibfield{title}{%
  \enquote{\bibinfo {title} {Extended-kalman-filter-based dynamic mode
  decomposition for simultaneous system identification and denoising},}\ }%
  \bibfield{journal}{%
  \bibinfo {journal} {arXiv preprint arXiv:1805.01985}}%
   (\bibinfo {year} {2018})\BibitemShut{NoStop}%
\end{thebibliography}%
\end{document}